\documentclass[a4paper,fleqn,usenatbib]{mnras}

\usepackage{newtxtext,newtxmath}
\usepackage[T1]{fontenc}
\usepackage{ae,aecompl}
\usepackage{longtable}
\usepackage{lipsum}
\usepackage{lscape}
\usepackage{amsmath}
\usepackage{graphicx}	% Including figure files
\usepackage{amsmath}	% Advanced maths commands
\usepackage{amssymb}	% Extra maths symbols
\usepackage{float}
\usepackage{subfig}
\newcommand{\xmm}{{\it XMM-Newton}}

\title[\xmm\, study of X-ray sources in the field of Willman~1 dwarf spheroidal galaxy]{\xmm\, study of X-ray sources in the field of Willman~1 dwarf spheroidal galaxy \thanks{Based on observations obtained with \xmm, an ESA science mission with instruments and contributions directly funded by ESA Member States and NASA.}}

\author[S. Saeedi et al.]{
Sara Saeedi$^{1}$\thanks{E-mail: sara.saeedi@fau.de},
 Manami Sasaki$^{2}$,
\\
$^{1,2}$ Dr. Karl Remeis-Sternwarte, Erlangen Centre for Astroparticle Physics, Friedrich-Alexander-Universit\"at Erlangen-N\"urnberg,\\
 ~~~ Sternwartstrasse 7, D-96049, Bamberg, Germany\\
}

\date{Accepted XXX. Received YYY; in original form ZZZ}

\pubyear{2020}

\begin{document}
\label{firstpage}
\pagerange{\pageref{firstpage}--\pageref{lastpage}}
\maketitle

\begin{abstract}
 We present the results of the analysis of three \xmm\, observations of the Willman~1 dwarf spheroidal galaxy (Wil~1). X-ray sources are classified on the basis of spectral analysis, hardness ratios, X-ray-to-optical flux ratio, X-ray variability, plus cross-correlation with available catalogues in optical and infrared wavelengths. We catalogued 97 sources in the field of Wil~1. Our classification shows the presence of a $\beta$-type symbiotic star in Wil~1.  We classified one M~dwarf foreground star in the field of Wil~1. Moreover, fifty-four sources are  classified as background AGNs and galaxies. Our study shows that the luminosity of the X-ray sources of Wil~1 does not exceed $\sim10^{34}$~erg\,s$^{-1}$ in the energy range of 0.2--12.0 keV, which is similar to observed luminosities of sources in nearby dwarf spheroidal galaxies.   
\end{abstract}


%%%%%%%%%%%%%%%%%%%%%%%%%%%%%%%%%%%%%%%%%%%%%%%%%%

%%%%%%%%%%%%%%%%% BODY OF PAPER %%%%%%%%%%%%%%%%%%

\section{Introduction}
\label{intro}
So far, about thirty dwarf galaxies have been discovered around the Milky~Way, with the majority of them being dwarf spheroidal galaxies\,(dSphs) \citep[e.g,][]{2012AJ....144....4M}. Studies show that almost all nearby dSphs have an old stellar population, mainly late-type stars with low metallicities \citep[e.g,][]{2014ApJ...789..147W, 2009ARA&A..47..371T}. These properties make them ideal targets to study the early stages of galaxy evolution. In X-rays, the population of nearby dSphs has been poorly studied so far in comparison to other types of nearby galaxies \citep{2006ARA&A..44..323F}. The close distance of dSphs together with the low absorption towards most of them give us a unique opportunity to search for the population of soft and low-luminosity ($L_{x}$<$10^{34}$~erg\,s$^{-1}$) X-ray sources, which are not detectable in distant nearby galaxies. Based on the very old stellar population of dSphs, theoretical models suggest that the presence of persistently bright X-ray binaries is rather unlikely in these galaxies  \citep[see e.g,][]{2005MNRAS.364L..61M}. In recent studies no X-ray binary was confirmed in dwarf galaxies \citep[e.g,][]{2014xru..confE.280M,2019MNRAS.485.2259A, 2016A&A...586A..64S}. However, some studies show the presence of low luminosity X-ray sources in these galaxies \citep{2016A&A...586A..64S, 2006A&A...459..777R}. Our deep study using the \xmm\, observations confirmed the presence of four symbiotic stars in the Draco dSph \citep{2019A&A...627A.128S}.  This result shows that in comparison to the X-ray binaries, the presence of accreting white dwarfs is more probable in dSphs, due to the similar time scale of the age of the galaxies and that of the formation of white dwarfs from late-type stars, which are the main population of dSph.  Among the different types of accreting white dwarfs \citep{2017PASP..129f2001M}, the symbiotic stars (i.e, accreting white dwarf or neutron star with a red giant companion) seem to be more likely to be detected in nearby dSphs. In X-rays, different types of symbiotic stars show a wide range of energy (0.1--100~keV)  and luminosity ($10^{31}-10^{38}$ erg\,s$^{-1}$) \citep{2013A&A...559A...6L, 2017PASP..129f2001M}. Usually, X-ray sources with luminosities of $\gtrsim10^{32}$--10$^{33}$~erg\,s$^{-1}$ are detectable in dSphs around the Milky~Way owing to their small distances.  Moreover, the optical brightness of red giants helps us to identify the nature of the companion star. \\
To confirm the results of Draco dSph and to provide a wider view of the population of accreting white dwarfs in dSphs, it is necessary to classify the X-ray sources of more dSphs with different stellar populations. Following this goal, we studied the X-ray sources in Willman~1 dSph\,(Wil~1, hereafter) applying the same multi-wavelength classification methods as for Draco dSph.  \\  
%This result have two aspects. On the one hand, this result can show that in comparison to the X-ray binaries, presence of accreting white dwarfs is more probable in dSphs due to the formation of white dwarfs from late-type stars, which are the main population of dSphs. On the other hand, one can theoretically argue that based on the stellar population synthesis, the main part of low mass X-ray binaries (LMXBs) are expected to be transient sources with a low X-ray luminosity ($\lesssim10^{33}$ erg\,s$^{-1}$) and a long life population \citep[e.g,]{2008ARep...52..299B,  2015A&A...579A..33V}. The donor companion of a symbiotic star (i.e, red giant) is optically much brighter than the companion of a LMXB (i.e, main sequence star). This provides a high chance for symbiotic stars in dSphs to be classified using the multi-wavelength methods in comparison to the transient LMXBs, where the nature of the companion is hard to be identified. The existence of LMXBs in the old population is still under a doubt and only the observational study of   As the methods of \citet{2019A&A...627A.128S} classification of accreting white dwarfs in nearby dSphs needs an intensive multi-wavelength study of the X-ray sources. 
Wil 1 (RA=10h49m21s, DEC=+51$^{\circ}$03.0$\arcmin$0.0$\arcsec$) is an old dSph, discovered in 2005 by \citet{2005AJ....129.2692W}.  This galaxy has a stellar mass of $\sim$1.0$\times10^{3}$ M$_{\sun}$ and is one of the the least massive Milky Way satellites \citep{2007MNRAS.380..281M} located at a distance of 38$\pm$7~kpc \citep{2012AJ....144....4M}. Its half-light radius ($r_{\rm h}$) is 2.3$\arcmin$\,($\sim$20 pc) \citep{2011AJ....142..128W}. The age of the galaxy is estimated to be $\sim$10--14\,Gyr and its metallicity is $-2.0\lesssim$[Fe/H]$\lesssim-1.0$ \citep{2007MNRAS.380..281M, 2011AJ....142..128W}.  Draco and Wil~1 dSphs have similar ages and metallicities \citep{2012AJ....144....4M}. However, with  a stellar mass of $\sim$2.9$\times10^{5}$ M$_{\sun}$ and $r_{\rm h}$ of $\sim$220 pc for Draco dSph \citep{2012AJ....144....4M}, the stellar density of Wil~1 is $\sim40\%$ of that of Draco dSph. \\ 
Wil~1 has been observed three times with \xmm\, in 2010 with the aim to detect an emission line from decaying dark matter \citep{2012ApJ...751...82L}. These observations had never been used for X-ray source classification in the field of Wil~1. The exposure time of \xmm\, observations of Wil~1 was long enough to perform X-ray spectral and timing analyses for the bright X-ray sources in the field of this galaxy.  As \citet{2011AJ....142..128W} have shown, the main stellar population of Wil 1 is located inside 3$r_{\rm h}$  of this dSph. Only 1$\%$ of stars, which belong to  Wil~1 are exponentially distributed beyond 4$r_{\rm h}$. %In Fig.~\ref{rgb-image}, we show the main field of Wil~1 (1$r_{\rm h}$) and also the regions of 3$r_{\rm h}$ and  5$r_{\rm h}$ of Wil~1 over the RGB mosaic image of \xmm. %Therefore, for the classification of X-ray sources in the field of Wil~1, We have focused on the X-ray sources, which are located within 5$r_{\rm h}$ of Wil~1.
In this paper, we report the details of the X-ray analysis together with multi-wavelength studies, which have been performed to classify the X-ray sources in the field of Wil~1. In Sect. \ref{data reduction} we describe the data reduction and analysis of the \xmm\, observations. In Sect. \ref{cross} and Sect.~\ref{x-ray-ana} we present the multi-wavelength studies and X-ray methods, which are used to classify the X-ray sources. In Sect. \ref{diss} we explain the properties and the classification of the detected sources in the field of Wil~1.
 %Symbiotic stars are binary systems consisting of a red giant star and a degenerate object, i.e, a white dwarf\,(WD), or in some cases a neutron star. Some of the symbiotic stars are known to emit X-ray emission caused by the accretion of matter from the red giant companion. According to \citet{1997A&A...319..201M}, there are three main classes of X-ray emitting symbiotic stars, depending on their X-ray spectrum: type $\alpha,$ which shows super-soft emission ($<0.4$~keV) originating from the atmosphere of the accreting compact object, type $\beta$ in which optically thin plasma emits soft X-rays ($<2.4$~keV), and type $\gamma$ with harder emission ($>2.4$~keV), in which the compact object is a neutron star.\citet{2013A&A...559A...6L} 
%suggested two additional types: highly absorbed sources with an accreting WD, and sources with a mixed soft and hard spectrum with a WD accretor.Only six super-soft symbiotic stars are known and have been studied so far in X-rays: StH$\alpha$\,32 \citep{2013A&A...559A...6L}, RX J0550.0-7151 \citep{1996IAUC.6305....2C, 1993ApJ...418L..63C}, RR~Telescopii \citep{1999ApJ...517..925C, 2012MmSAI..83..806G}, AG~Draconis \citep{2008A&A...481..725G}, SMC\,3 \citep{2013ApJ...763....5K}, and Lin~358 \citep{2007ApJ...661.1105O, 2015NewA...36..116S}.
\section{Data reduction}
\label{data reduction}
Wil~1 has been observed in three \xmm\, observations, which are listed in Table~\ref{obs-data}. Full-frame mode and the thin filter were used for cameras EPIC-pn \citep{2001A&A...365L..18S} and EPIC-MOS1,\,2 \citep{2001A&A...365L..27T} in all observations. Data reduction and source detection were performed using the \xmm\, Science Analysis System\,(SAS,\,V.17.0.0).  High background caused by soft proton flares were screened from the event files. Threshold rate of $\leq$ 0.35 count\,s$^{-1}$ for EPIC-MOS and rate $\leq$ 0.4 count\,s$^{-1}$ for EPIC-pn are applied to find good time intervals. Light curves of clean event lists were also checked visually to remove possible background flares. Table\,\ref{obs-data} lists the net exposure time for each observation and EPIC camera. Source detection in the five standard energy-bands of \xmm\, B1\,(0.2--0.5\,keV), B2\,(0.5--1.0\,keV), B3\,(1.0--2.0\,keV), B4\,(2.0--4.5\,keV), B5\,(4.5--12.0\,keV) was performed using the \texttt{SAS} task \texttt{edetect-chain} for each observation. We selected the minimum value of maximum likelihood\,($L$) of 10 for the source detection. Probability of Poisson random fluctuations of the counts\,($p$), which is based on the raw counts of the source and the raw counts of the background maps is used to calculate the detection maximum likelihood $L$=--ln$(p)$. 
\begin{table}%[ht]
    \caption{\xmm\, observations of Wil~1. \label{obs-data}}
    \small
     \begin{tabular}{cccccc}
\hline\hline
OBS-N0 & OBS-ID & OBS-Date &  \multicolumn{3}{c}{EXP.T${^\ast}$ (ks)} \\
       &        &           &  pn&MOS1&MOS2        \\
\hline
       1&  0652810101  &    2010-10-22 &    14.1&     22.5&       22.4  \\
       2&  0652810301  &    2010-10-25 &    20.6&     27.8&       27.5  \\
       3&  0652810401  &    2010-10-31 &    26.5&     32.3&       32.7  \\
      \hline
      \multicolumn{6}{l}{$\ast$: Exposure time of EPICs after screening for high background.}\\
     \end{tabular}
\end{table}
\begin{table}%[ht]
    \caption{Offsets of the \xmm\, observations \label{offset}}
    \small
     \begin{tabular}{clrr}
\hline\hline
OBS-N0 &  EPIC & $\Delta$RA ($\arcsec$) & $\Delta$DEC ($\arcsec$) \\
\hline
1  &PN       &-0.33  $\pm$ 0.69 &  0.48 $\pm$ 0.69\\
   &MOS1     &-0.48  $\pm$ 0.83 & -0.08 $\pm$ 0.83\\
   &MOS2     &-0.29  $\pm$ 0.63 &  0.29 $\pm$ 0.63\\
2  &PN       &-0.20  $\pm$ 0.54 &  0.53 $\pm$ 0.54\\
   &MOS1     &-0.02  $\pm$ 0.57 &  0.68 $\pm$ 0.57\\
   &MOS2     &-0.17  $\pm$ 0.52 &  0.32 $\pm$ 0.52\\
3  &PN       & 0.16  $\pm$ 0.52 & -0.49 $\pm$ 0.52\\
   &MOS1     & 0.18  $\pm$ 0.64 & -0.23 $\pm$ 0.64\\
   &MOS2     & 0.33  $\pm$ 0.50 & -0.01 $\pm$ 0.50\\
      \hline
     \end{tabular}
\end{table}
\section{Source catalogue}
\begin{figure*}
\includegraphics[clip, trim={0.cm 1.6cm 0.0cm 0.cm},width=0.60\textwidth]{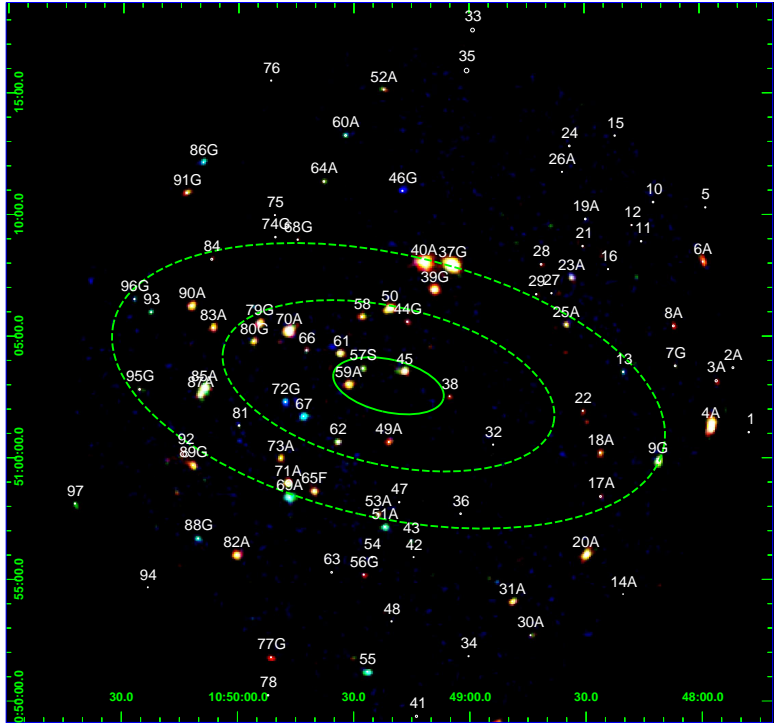}
\caption{Combined X-ray image of \xmm\, observations in the field of Wil~1 in the energy band of 0.2--12.0 keV. Green  hard ellipse and dashed ellipses show 1, 3, and 5 half-light radii of Willman\,1 dwarf galaxy,  respectively.  The  colours  red, green, blue present the  energy  ranges  of   0.2--1.0~keV, 1.0--2.0~keV, and  2.0--12 keV, respectively. The images are smoothed with a 2D Gaussian with a kernel radius of~1.0 pixel. X-ray sources are marked with the  source ID and the class of the sources (AGN:\,A, Galaxy:\,G,  Foreground star:\,F, Symbiotic Star:\,S) in the final catalogue (Table~\ref{catalogue-x-ray}).  \label{rgb-image}}
\end{figure*}

We cross-checked the detected sources of all observations/all EPICS with each other to obtain a final source catalogue. If the positions of the sources in different observations/EPICs were closer than the 3$\sigma$ statistical errors, we considered them as one source. The multiple \xmm\, observations helped to remove the spurious sources caused by bad pixels, hot columns, gaps, edges of the CCD chips \citep{2008A&A...480..611S}. 
 The final catalogue of 97 X-ray sources in the field of Wil~1 is listed in Table~\ref{catalogue-x-ray}. The catalogue shows ID, right ascension\,(RA), declination\,(Dec), position uncertainty, and the flux of different observations for each source. The list of the sources are sorted using their coordinates. In this paper, each source is named by its ID, which is presented in Table~\ref{catalogue-x-ray}.\\
For the astrometrical correction of the position of X-ray sources, we selected 11 X-ray sources, which had bright optical counterpart\,(apparent magnitude <21~mag) and were already classified as AGNs in optical catalgues (see Sect.\ref{AGN-cata}). The weighted mean of the $\Delta$RA and the $\Delta$Dec between the positions of the optical and X-ray sources have been calculated to estimate the error of the position of X-ray sources for each observation (see Table~\ref{offset}). We found no significant instrumental shift in the RA and Dec of X-ray sources in none of the observations. Therefore, the coordinate and the positional error of each source were taken from the observation, in which the source was detected with the highest maximum likelihood.\\
To obtain the flux of the sources in each observation, we assumed an absorbed power law model with the Galactic foreground absorption in the direction of Wil~1 \citep[$N_{\rm H}$= 1.17$\times10^{20}$ ~cm$^{-2}$,][]{2016A&A...594A.116H} and a photon index of $\Gamma$=2. The flux of the sources, for which spectral analysis was performed, are calculated using the best fit model to their spectrum (see Sect.~\ref{spectra-sect}). Table~\ref{catalogue-x-ray} shows the weighted flux of all EPICs measurements for each observation. \\
%We produced a mosaic image using the \texttt{SAS} task \texttt{emosaicproc} of all observations/EPIC images.
The mosaic image were created using the \texttt{SAS} task \texttt{emosaicproc} out of the calibrated event files of all observations/EPICs. The task transforms the event files of different observations/pointings to a common image center, and after filtering the soft proton flares, it creates the exposure maps (including vignetting), and the background maps and performs the source detection. Figure\,\ref{rgb-image} shows the three-colour combined image of all observations. In Fig.~\ref{rgb-image}, we show the main field of Wil~1 (1$r_{\rm h}$) and also the regions of 3$r_{\rm h}$ and  5$r_{\rm h}$ of Wil~1.
\section{Cross-correlation with other catalogues}
We cross-correlated the X-ray source list with catalogues at other wavelengths. A source, which was located within the 3$\sigma$ error circle of an X-ray source position, was considered to be its counterpart. We discuss the catalogues used in this work in the following.
\label{cross}
\subsection{Optical counterparts of X-ray sources}
\begin{figure}
\centering
\includegraphics[clip, trim={3.2cm 0.0cm 0.0cm 0.cm},width=0.43\textwidth]{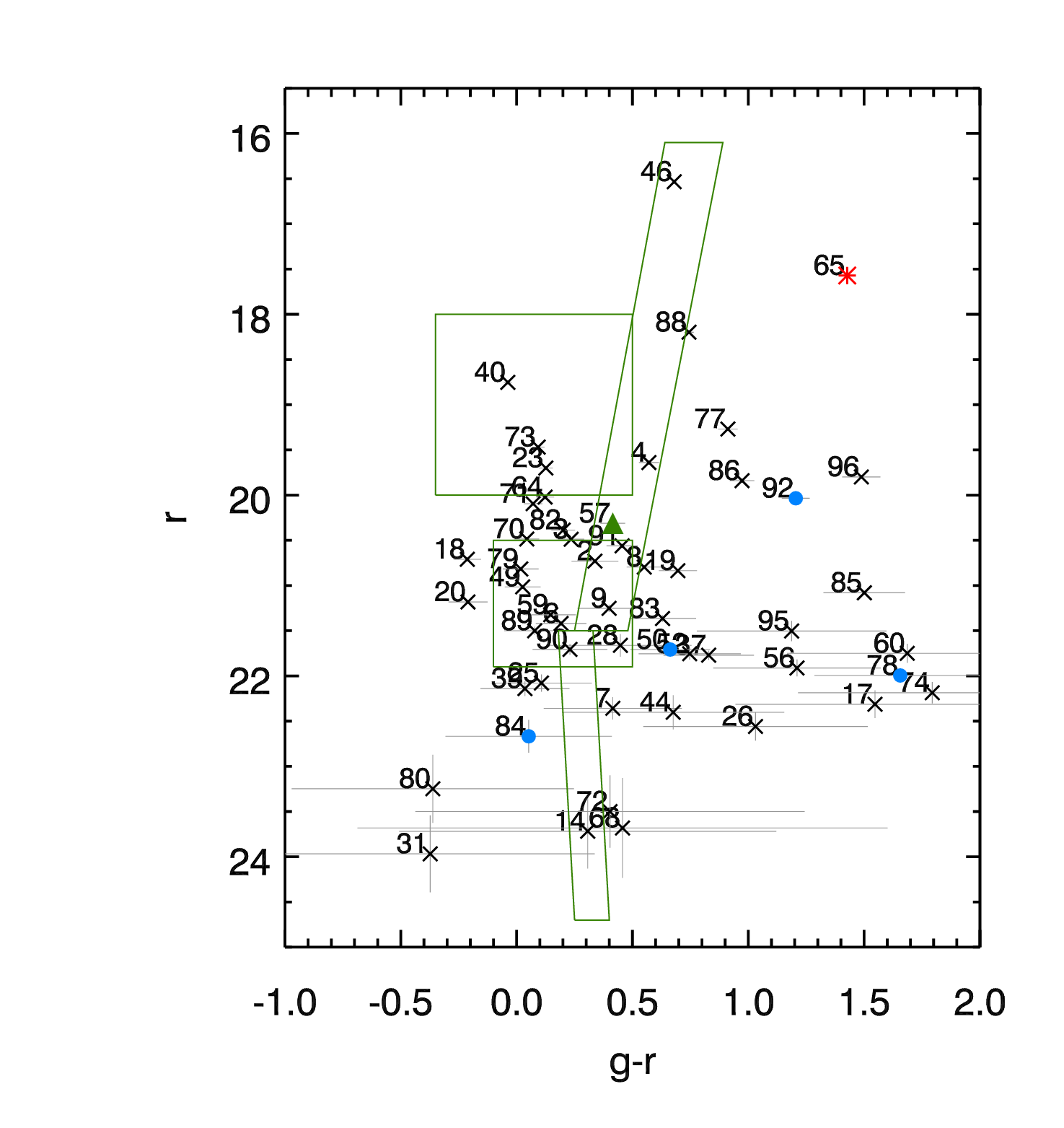}
\caption{Colour-magnitude diagram ($g$ versus $g-r$) for the SDSS12 optical counterparts of the X-ray sources .  The green areas show the regions of the  stellar isochrone of the Willman\,1 dwarf galaxy according to \citet{2011AJ....142..128W}. In all plots classified background sources are marked by black cross, foreground stars by red asterisk, symbiotic stars by green triangle, and unclassified sources by blue circles. \label{opt-counterpart}}
\end{figure}

\begin{figure*}
\includegraphics[clip, trim={2.6cm 0.cm 0.5cm 0.3cm},width=0.43\textwidth]{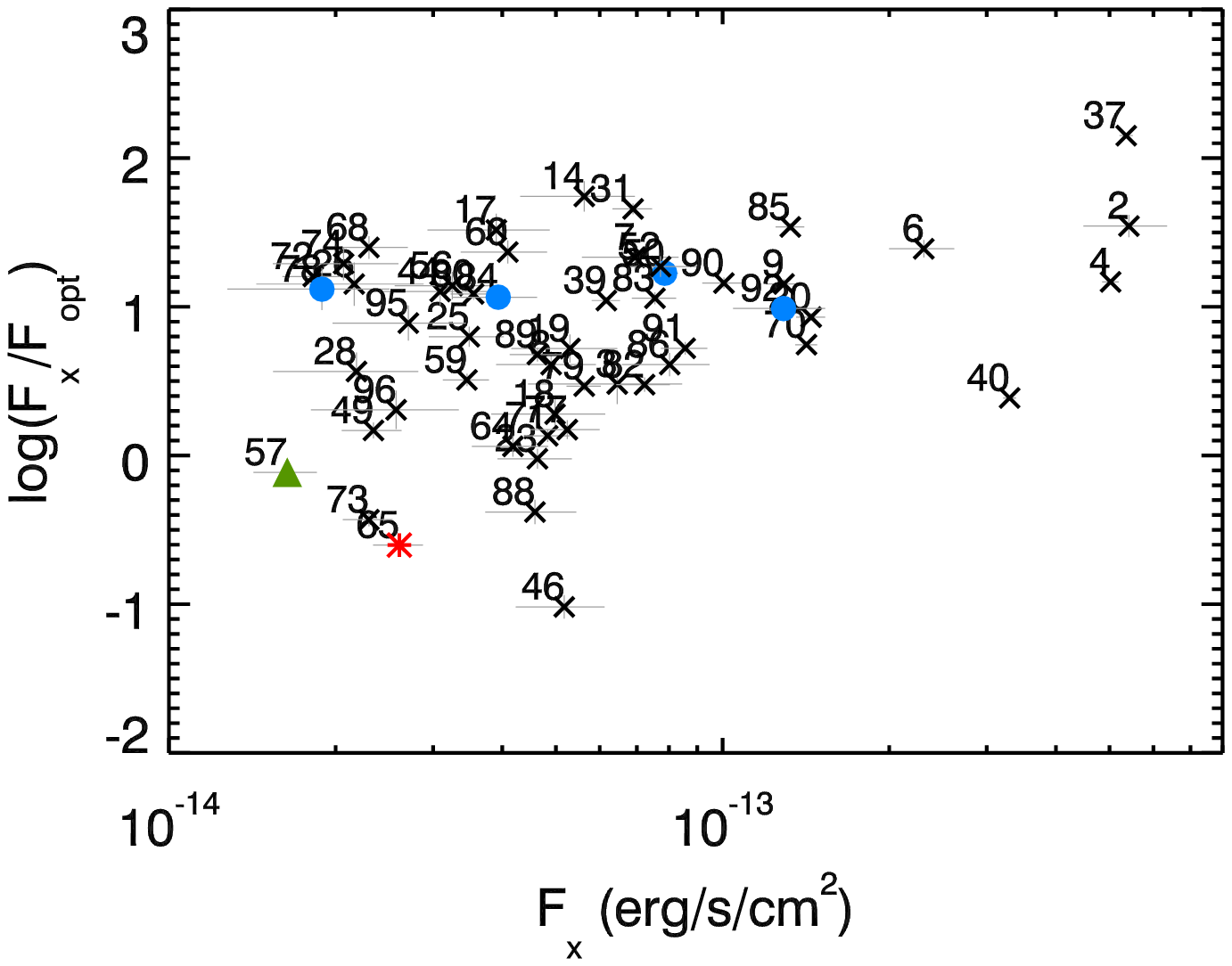}
\includegraphics[clip, trim={2.6cm 0.cm 0.5cm 0.3cm},width=0.43\textwidth]{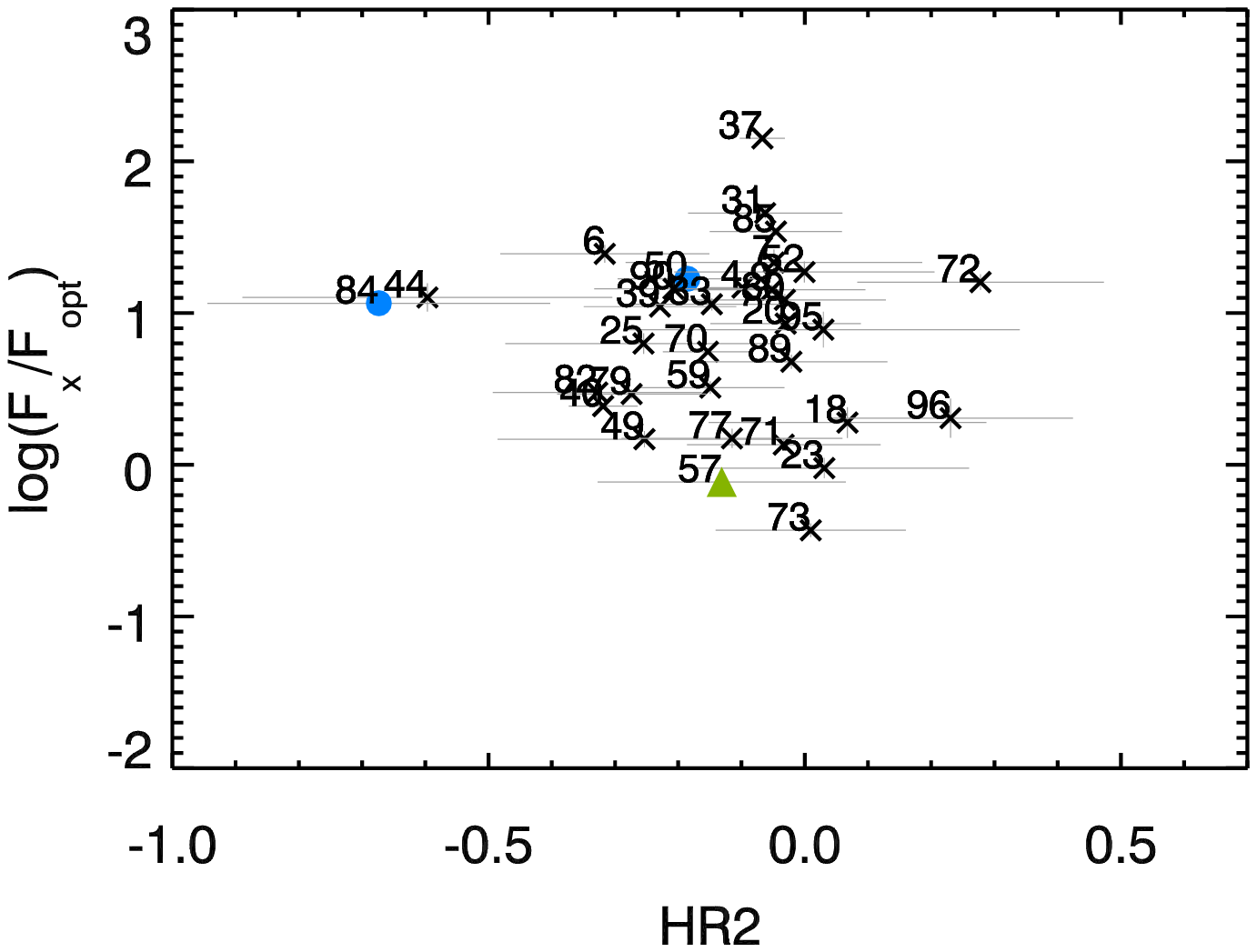}
\caption{Logarithmic X-ray to optical flux ratio  log$(\frac{F_\text{X}}{F_\text{opt}})$ versus the maximum X-ray flux\,({\bf left}) and HR2\,({\bf right}). The symbols are the same as Fig.~\ref{opt-counterpart}. \label{log-x-opt} }
\end{figure*}

To study the optical counterparts of the X-ray sources, we have used the 11th and 12th released data of the Sloan Digital Sky Survey \citep[SDSS12]{2015ApJS..219...12A}. The catalogue includes photometric data in the energy bands from the near ultraviolet\,(UV) to the near infrared ($u$=3551 \AA, $g$=4686 \AA, $r$=6165 \AA, $i$=7481 \AA, $z$=8931 \AA) and allows a spectral study of the optical counterpart. For the $u$, $g$, $r$, $i$, and $z$ bands, the Galactic extinction of 0.04, 0.03, 0.02, 0.01, and 0.01 mag is used in the direction of the Wil~1, respectively \citep{2011ApJ...737..103S}. Table~\ref{opt-count-table} presents the SDSS12 magnitudes of the optical counterparts of the X-ray sources.\\
 Figure~\ref{opt-counterpart} left image, shows the colour-magnitude diagram of the optical counterparts of the X-ray sources in the field of Wil~1. Logarithmic X-ray to optical flux ratio log$(\frac{F_\text{X}}{F_\text{opt}})$, versus the X-ray flux and also hardness ratio (see Sect.~\ref{hrs}) are plotted in Figure~\ref{log-x-opt}. The flux ratio log$(\frac{F_\text{X}}{F_\text{opt}})$ was calculated using the equation of \citet{1988ApJ...326..680M} modified for SDSS bands \citep{2016A&A...586A..64S}:
\begin{equation}
  \label{fx-fopt}
{\rm log}\bigg(\frac{F_ \text{X}}{F_\text{opt}}\bigg)={{\rm log}_{10}(F_\text{X})}+\frac{g+r}{2\times2.5}+5.37,
\end{equation}
where $F_\text{X}$ is the X-ray flux and $g$ and $r$ are the SDSS magnitudes of the optical counterpart associated with the X-ray source. 

\subsection{Infrared counterparts of {\bf X-ray} sources}
\label{inf-count-exp}
\begin{figure}
\centering
\includegraphics[trim={3.2cm 0.0cm 0.0cm 0.cm},width=0.47\textwidth]{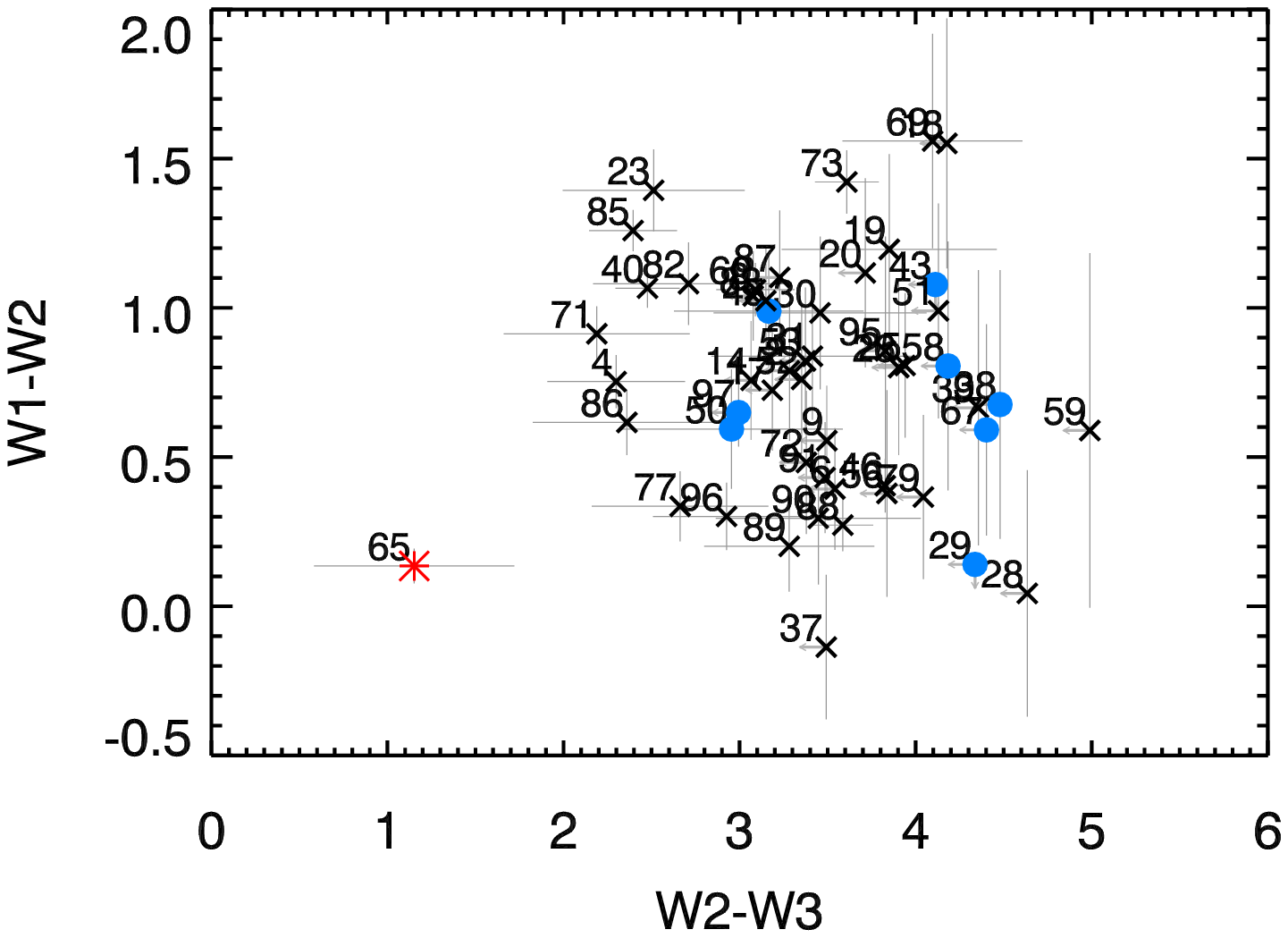}
\caption{Colour-colour diagram of mid-infrared WISE  ($W1(3.4\, \mathrm{\mu m})-W2( 4.6\, \mathrm{\mu m})$ versus ($W2\,(4.6\, \mathrm{\mu m})- W3 \,(12\, \mathrm{\mu m})$). Arrows show instead of error bars for the sources, which have upper limits in the WISE bands (see Table~\ref{inf-count}). The symbols are the same as Fig.~\ref{opt-counterpart}. \label{infra-plot} }
\end{figure}

We searched for mid-infrared counterparts in the WISE All-Sky survey in four energy bands \citep[3.4, 4.6, 12, and 22 $\mathrm \mu$m, named $W1$, $W2$, $W3$, and $W4$, respectively;][]{2014yCat.2328....0C} and near-infrared counterparts in the 2MASS All-Sky Catalogue of Point Sources in the $J$, $H$, $K$ bands \citep{2003yCat.2246....0C}.   The Galactic extinction for the infrared bands in the direction of Wil~1 was negligible \citep{2011ApJ...737..103S}.  Table~\ref{inf-count} list the magnitudes of WISE and 2MASS counterparts of the X-ray sources.  Figure~\ref{infra-plot} shows the colour-colour diagram of the WISE counterparts of the X-ray sources in the field of Wil~1. The colours are selected based on the study of \citet{2010AJ....140.1868W}, which shows that many of background objects are expected to be red ($W2-W3$>1.5) in WISE colour, while stars show ($W2-W3$<1.5). In 2MASS near-infrared counterparts, background sources like e.g, AGNs and quasars are expected to have $J-K>1$ \citep[see e.g,][]{2010PASA...27..302M}. Only three sources had counterparts in the 2MASS near-infrared catalogue (see Table~\ref{inf-count}). These sources are  classified as foreground star and galaxies as explained in Sect.~\ref{backgroundsources} and Sect.~\ref{foregroundstar}.

\subsection{Catalogues of AGNs and galaxies}
\label{AGN-cata}
Most of the background objects are AGNs. Many of these objects were already classified based on spectral properties in the optical and/or infrared surveys. The most updated optical/infrared catalogues of classified AGNs/quasars were considered in this work: The Million Quasars \citep{2019yCat.7283....0F}, WISE AGN candidates catalogs \citep{2018ApJS..234...23A}, QSOs selection from SDSS and WISE \citep{2015ApJS..219...39R}, and  known quasars of the Gaia mission \citep{2019RAA....19...29L}. In addition, we considered the catalogues, which classified galaxies:  \citet{2011AJ....141..189V} used the photometric data of SDSS7 \citep{2009ApJS..182..543A} survey and separated the stars and galaxies based on a functional tree algorithm. We also used the more updated classification of \citet{2015ApJS..219...12A}, which provides a classification of stars and galaxies using redshifts and colour criteria as the details are explained in  \citet[][]{2015ApJS..219...12A} and \citet{2012ApJS..203...21A}. The sources, which have been classified as AGNs or galaxies in above catalogues are labeled in Table~\ref{catalogue-x-ray}. 
  % Sources, which have been classified by comparison with other catalogues are labelled in Table~\ref{catalogue-x-ray}.\\ 
\subsection{ Gaia and GALEX catalogues}
To classify the foreground sources, we used the photometry and parallax measurement of the 2nd release of Gaia \citep{2018A&A...616A...1G} and the GALEX catalogues of UV  sources, which includes Galactic white dwarfs and hot stars \citep{2017ApJS..230...24B}.
%\subsection{ Willman~1 dSph catalogues}
%Deep optical studies of the Willman~1 dSph 
\section{X-ray data analysis}
\label{x-ray-ana}
 We performed X-ray analysis for sources in the field of Wil~1. Following analyses were carried out for these sources.
 \subsection{X-ray timing analysis}
\label{time-sec}
\begin{figure}
\includegraphics[clip, trim={2.5cm 0.cm 0.5cm 0.3cm},width=0.5\textwidth]{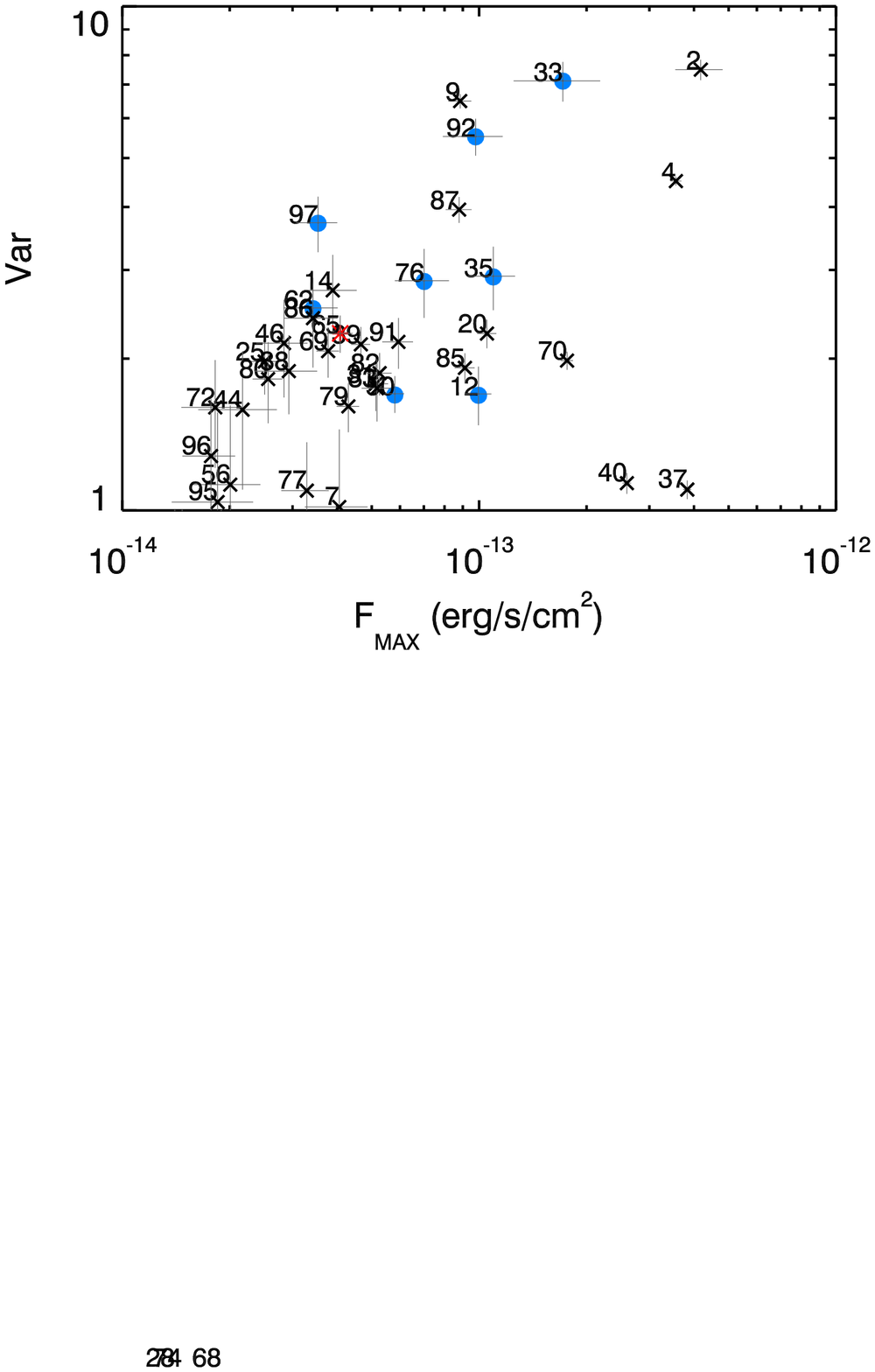}
\caption{Variability factor of sources with significant variability (S>3) in the energy band of 0.2--4.5 keV plotted versus the maximum flux. The symbols are the same as Fig.~\ref{opt-counterpart}.\label{var}}
\end{figure}

 To study the short-term variability of sources we have applied the Lomb-Scargle technique \citep{1982ApJ...263..835S} and the pulsation  Z$^{2}_{n}$  test \citep{1983A&A...128..245B, 1988A&A...201..194B}.  We extracted the light curves of bright sources (i.e, counts\,>\,300 in each observation) in the energy range of 0.2--12~keV, and calculated their Lomb-Scargle periodograms. We could not find a signal of periodicity for any of the sources.  In addition, we applied the Z$^{2}_{n}$ analysis for the barycentrically corrected event files of sources from each observation in the energy range of 0.2--12~keV and extracted the first and second harmonic periodograms.  For none of the sources we found evidence for pulsation. \\
To study the long term variability, we checked the flux variation of sources over three \xmm\, observations. Flux variation and its  significance were calculated using \begin{equation}Var=\frac{F_{\rm max}}{F_{\rm min}}$ and $S=\frac{F_{\rm max}-F_{\rm min}}{\sqrt{EF_{\rm max}^{2}+EF_{\rm min}^{2} }}, \end{equation} respectively {\bf \citep{1993ApJ...410..615P}}.  $F_{\rm max}$ and $F_{\rm min}$ are the maximum and minimum X-ray flux,  and $EF_{\rm min}$  and $EF_{\rm max}$ are their corresponding errors. Due to the high background fluctuations, the energy band 5 (4.5--12.0 kev) was excluded in the calculation of the variability factor. The variability factor was calculated for the sources that were detected in at least two observations (see Table~\ref{catalogue-x-ray}). The variability of a sources considered to be significant if $S$ was higher than 3. Figure~\ref{var} shows the variability factor of the sources with significant variability. As it is shown in the plot, none of sources shows a very high variability during three \xmm\, observations.

\subsection{Hardness ratios}
\label{hrs}
\begin{figure*}
\includegraphics[clip, trim={0.5cm 0.cm 0.5cm 0.3cm},width=0.31\textwidth]{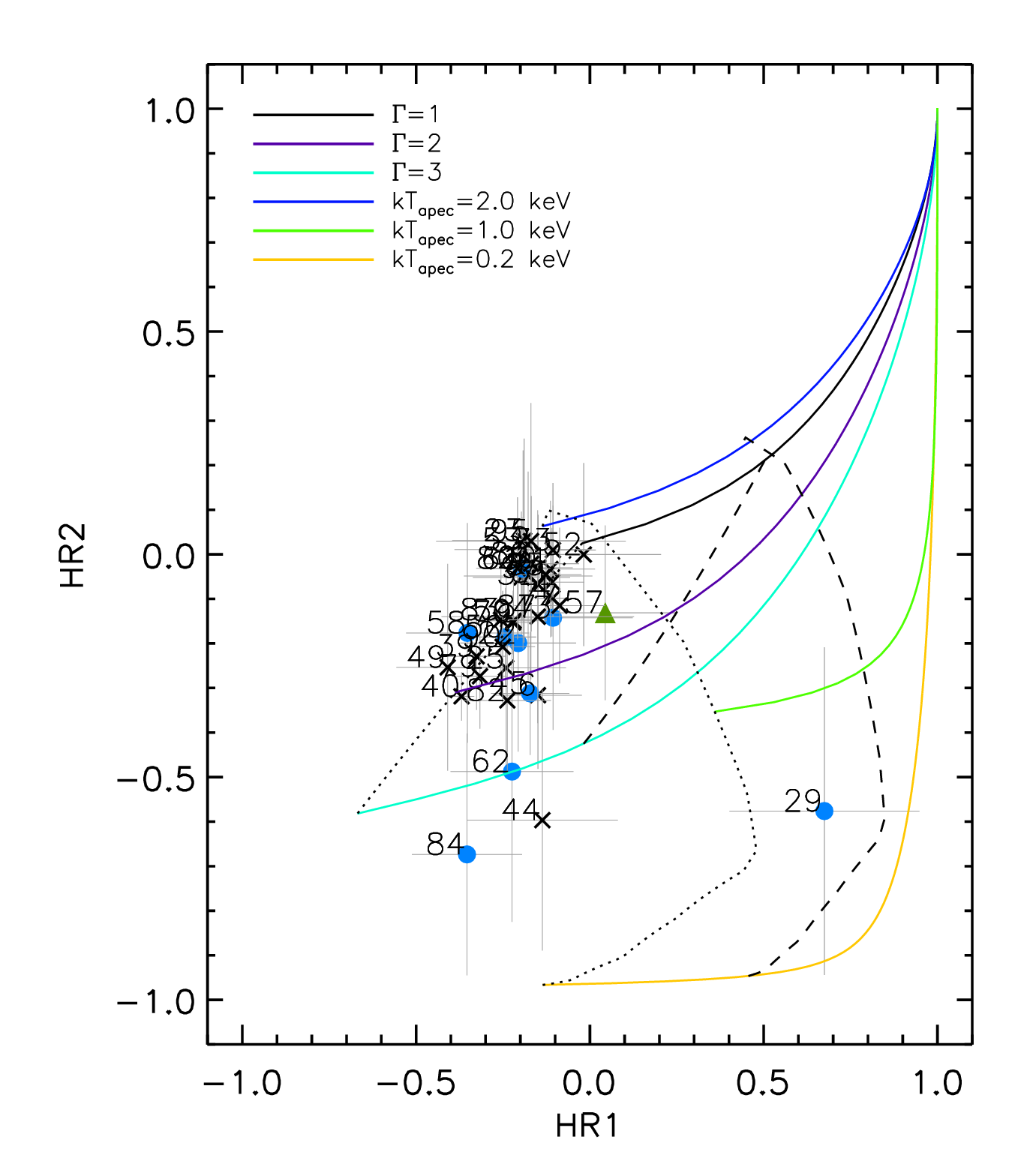}
\includegraphics[clip, trim={0.5cm 0.cm 0.5cm 0.3cm},width=0.31\textwidth]{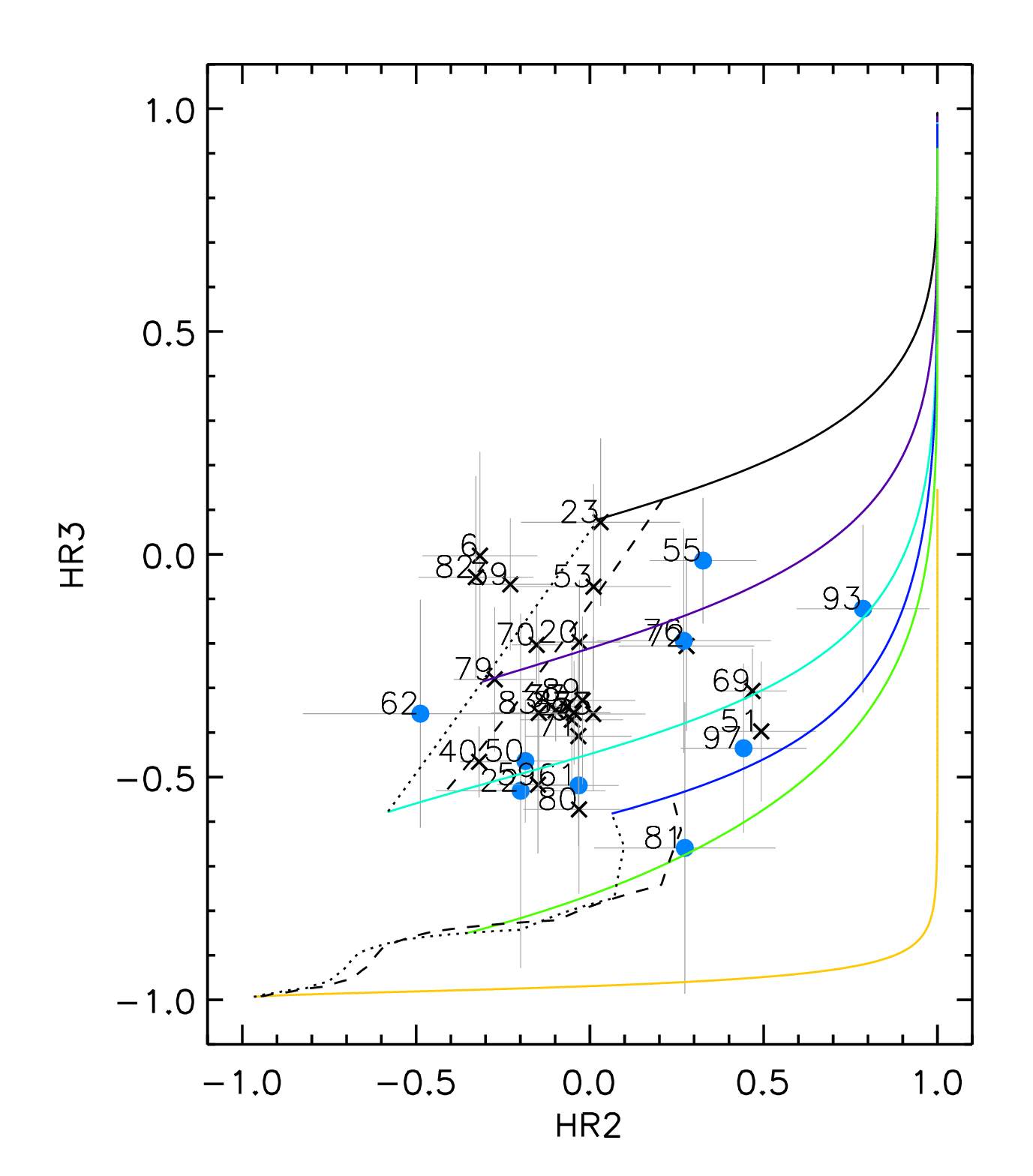}
\includegraphics[clip, trim={0.5cm 0.cm 0.5cm 0.3cm},width=0.31\textwidth]{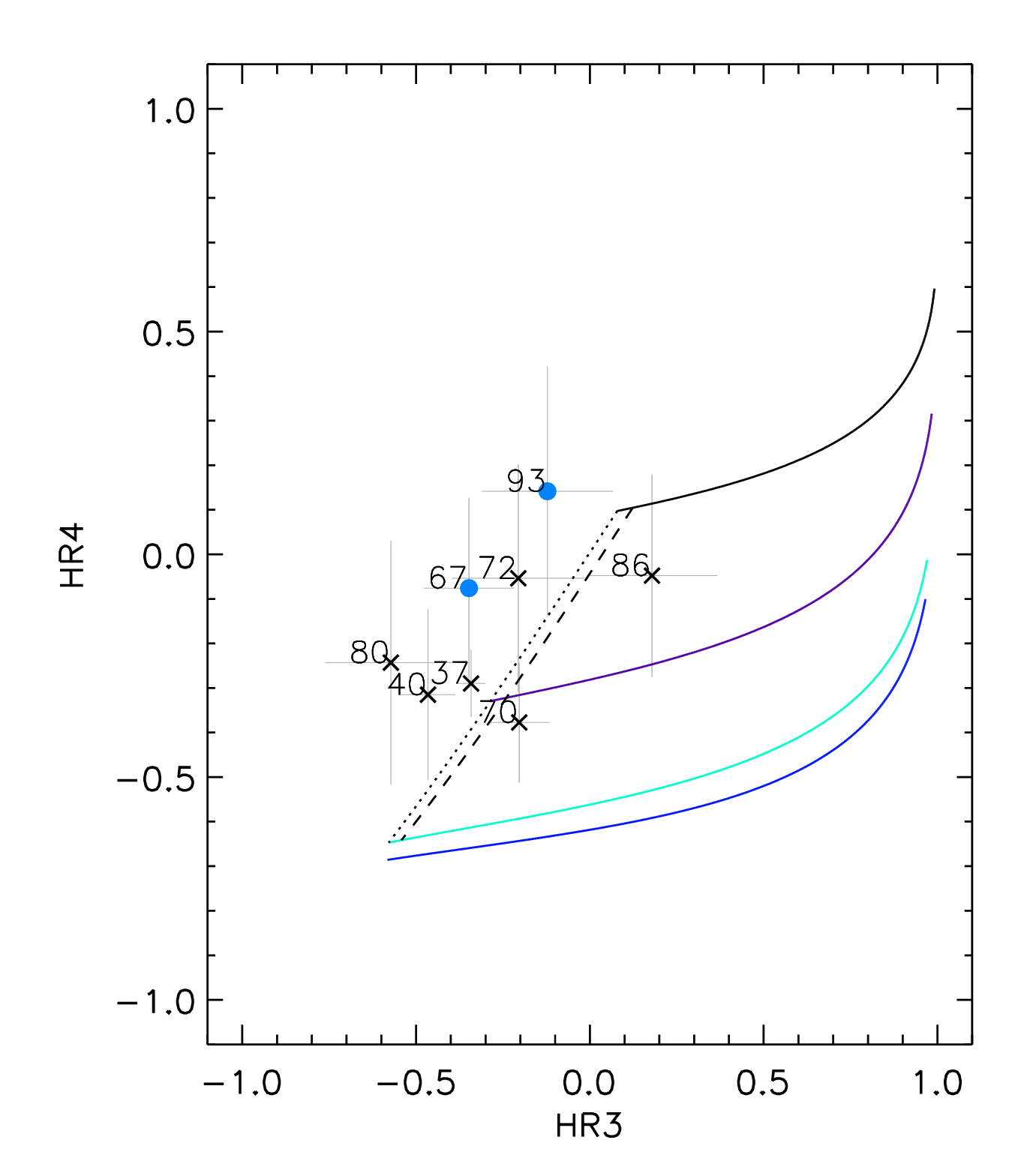}
\caption{Hardness ratio diagrams. The plotted hard lines are the hardness ratios calculated for different spectral models and column densities. The  column densities of $N_{\rm H}$=$10^{20}$\,cm$^{-2}$ (dotted lines) and $N_{\rm H}$=$10^{21}$\,cm$^{-2}$ (dashed lines) are plotted for the power-law models and apec models.  The symbols are the same as Fig.~\ref{opt-counterpart}. \label{hrs-plot}}
\end{figure*}

Hardness ratios\,(HRs) are useful parameters for the study of spectral properties of X-ray sources. The hardness ratio and its error are defined as \begin{equation} HR_\mathrm{i}=\frac{B_\mathrm{i+1}-B_\mathrm{i}}{B_\mathrm{i+1}+B_\mathrm{i}}$ and $EHR_i=2\frac{\sqrt{(B_\mathrm{i+1}EB_\mathrm{i})^2+(B_\mathrm{i}EB_\mathrm{i+1})^2}} {(B_\mathrm{i+1}+B_\mathrm{i})^2}, \end{equation} respectively, where $B_\mathrm{i}$ is the count rate and $EB_\mathrm{i}$ is the corresponding error in the band $i$. For each source we calculated the hardness ratio of the observation, in which the source had the highest detection likelihood.  In the next step, the hardness ratio is calculated only for those energy bands, which had a detection likelihood higher than 6 (>3$\sigma$). Table~\ref{catalogue-x-ray} shows the details of HRs for sources. Figure~\ref{hrs-plot} shows the hardness ratio diagrams. We over-plotted the lines presenting the hardness ratios of different spectral models with various column densities from $N_{\rm H}$=$10^{20}$\,cm$^{-2}$ to $N_{\rm H}$=$10^{23}$\,cm$^{-2}$. We considered three \texttt{power-law} models with photon-index $\Gamma$ of 1, 2, 3 for the hard sources, e.g, X-ray binaries, AGNs, or galaxies. Three \texttt{apec} model with the temperature of $kT$ of  0.2, 1.0, and 2.0 keV correspond to the spectra of soft plasma emissions detected in different sources, e.g, supernova remnant\,(SNR), foreground stars, and symbiotic stars.  The models, which describe the emission of the soft source (i.e, \texttt{apec} model with the temperature of $kT$ of  0.2, 1.0) have a negligible rate in harder energy bands and therefore we do not have these models in $HR3$ and $HR4$. Sources are classified as hard source when it has $HR2-EHR2$>$-0.2$, or only $HR3$ and/or $HR4$ are defined and there is no other classification for the source \citep{2005A&A...434..483P} (see Table~\ref{catalogue-x-ray}).

\subsection{X-ray spectral analysis}
\label{spectra-sect}
\begin{figure*}
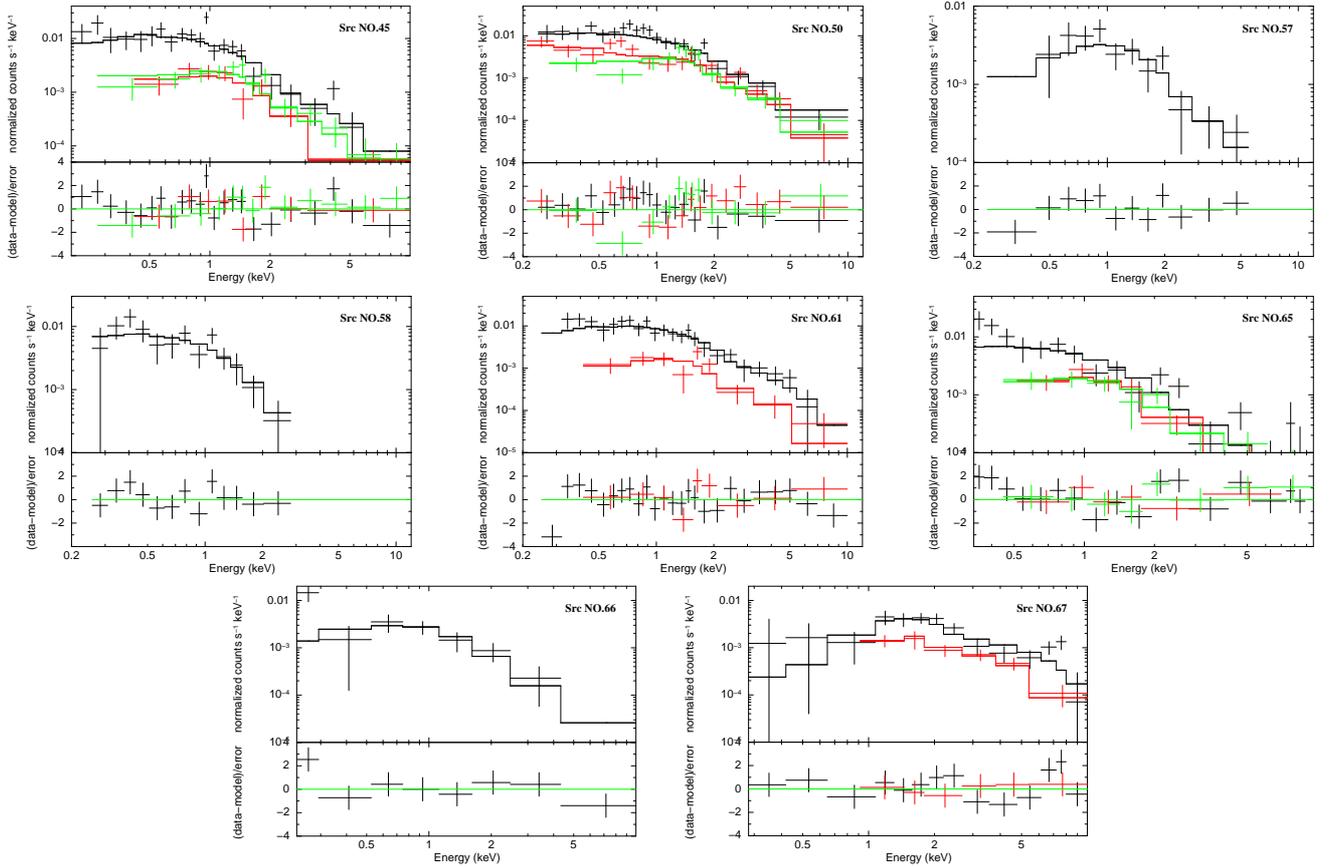

\includegraphics[angle=270, width=0.33\textwidth, trim=1.5cm 0.cm 0.cm 2.0cm]{po-src45.ps}
\includegraphics[angle=270, width=0.33\textwidth, trim=1.5cm 0.cm 0.cm 2.0cm]{spec-src-50.ps}
\includegraphics[angle=270, width=0.33\textwidth, trim=1.5cm 0.cm 0.cm 2.0cm]{spec-src-57.ps}\\
\includegraphics[angle=270, width=0.33\textwidth, trim=1.5cm 0.cm 0.cm 2.0cm]{spec-src-58.ps}
\includegraphics[angle=270, width=0.33\textwidth, trim=1.5cm 0.cm 0.cm 2.0cm]{spec-src-61.ps}
\includegraphics[angle=270, width=0.33\textwidth, trim=1.5cm 0.cm 0.cm 2.0cm]{spec-src-65.ps}\\
\includegraphics[angle=270, width=0.33\textwidth, trim=1.5cm 0.cm 0.cm 2.0cm]{po-src66.ps}
\includegraphics[angle=270, width=0.33\textwidth, trim=1.5cm 0.cm 0.cm 2.0cm]{po-src67.ps}
\caption{Combined spectrum of all \xmm\, observations of sources in the FOV  of Willman~1 dSph, which have enough statistics for spectral analysis: EPIC-pn\,(black), EPIC-MOS1 (red), and EPIC-MOS2 (green), together with the residuals in units of the standard deviation. \label{spec.fig}} 
\end{figure*}
 
%\begin{figure*}
%\includegraphics[angle=270, width=0.4\textwidth, trim=1.5cm 0.cm 0.cm 2.0cm]{spec-src-65.ps}
%\includegraphics[angle=270, width=0.4\textwidth, trim=1.5cm 0.cm 0.cm 2.0cm]{spec-src-66.ps}
%\includegraphics[angle=270, width=0.4\textwidth, trim=1.5cm 0.cm 0.cm 2.0cm]{spec-src-67.ps}
%\caption{Continued Fig.~\ref{spec.fig}. \label{spec.fig1}} 
%\end{figure*}
We extracted and analysed the X-ray spectra of bright sources in the field of Wil~1, which have not been classified as background objects (see Sect.~\ref{diss}).  We extracted the spectra of the sources that had a net source counts of >500 in total. To improve the statistics of the spectrum of a source, we merged  the spectra of all observations, in which the source was detected using  the SAS task \texttt{epicspeccombine}. Before merging the spectra of different observation, we checked the variability of the source to exclude the spectrum of observation, in which the source showed a significantly different flux (see table~\ref{catalogue-x-ray}).  Figure~\ref{spec.fig} shows the spectrum of the X-ray sources and Table~\ref{spectral-Table} the details of the models fitted to the spectrum of sources.  We fitted an absorbed \texttt{power-law} model or an absorbed collisionally-ionised thermal gas model \citep[\texttt{APEC},][]{2000HEAD....5.2701B} to the spectrum of sources. The model for each source is selected based on the best fit. The details of the spectral study of each source are discussed in Sect.~\ref{diss}. For those sources, which were too faint for the spectral analysis, the hardness ratios can be  used to characterise the spectrum of the sources  (see Sect.~\ref{hrs}). 
\begin{table*}
\centering
\caption{Best-fit parameters of the X-ray spectra. Errors are at the 90$\%$ confidence level. \label{spectral-Table}}
\small
\centering
\addtolength{\tabcolsep}{-0.1cm}   
\begin{tabular}{llccccccc}
\hline\hline
Src-No & Model & $N_{\rm H}$&Photon index& $kT$ & Abundance&$\chi^2$ (d.o.f) &  Absorbed $F_{\rm X}$&$L_{\rm X}^{(1)}$\\
&&$10^{22}$ cm$^{-2}$&&keV&&&$10^{-15}$erg\,s$^{-1}$\,cm$^{-2}$&erg\,s $^{-1}$\\
\hline
%9&\texttt{tbabs$\times$(apec)}&<0.08&5.06$^{7.31}_{-2.91}$&frozen to 1.0 &0.97 (12)&64.42$^{+8.55}_{8.58}$&1.23$\times10^{34}$\\
%\vspace{1mm}
%37&\texttt{tbabs$\times$(apec+apec)}&<0.01&0.26$^{0.07}_{-0.06}$&frozen to 1.0 &1.2 (158)&232.86$^{+9.35}_{7.90}$&4.5$\times10^{34}$\\
%\vspace{1mm}
%  &                                 &     &4.83$^{0.88}_{-0.89}$ & \\
%  \vspace{1mm}
%39&\texttt{tbabs$\times$(apec+apec)}&<0.07&0.19$^{+0.18}_{-0.10}$&frozen to 1.0 & 1.17 (53)&29.44$^{+2.79}_{-2.78}$&5.6$\times10^{33}$\\
%\vspace{1mm}
%  &                                 &     &3.17$^{+3.86}_{-1.42}$ &              &          &                   &       \\ 
%  \vspace{1mm}
%45&\texttt{tbabs$\times$(apec)}&<0.03&3.65$^{+1.70}_{-1.43}$&<0.4&1.064 (38)&13.10$^{1.70}_{-1.72}$&2.5$\times10^{33}$\\
%\vspace{1mm}
45&\texttt{tbabs$\times$(po)}&<0.13&2.00$^{+0.46}_{-0.32}$&&&1.07 (39)&21.48$^{+2.83}_{-2.82}$&4.1$\times10^{33}$\\
\vspace{2mm}
50$^{(2)}$&\texttt{tbabs$\times$(apec+apec)}&<0.06&&$kT1$=0.17$^{+0.10}_{-0.12}$&<0.18&1.35 (44) &41.49$^{+3.96}_{-3.94}$&7.9$\times10^{33}$\\
\vspace{1mm}
  &                                 &     &&$kT2$=5.27$^{+3.53}_{-1.62}$ &             &          &                   &       \\ 
  \vspace{1mm}
57$^{(3)}$&\texttt{tbabs$\times$tbabs(apec)}&$N_{\rm H}1$=0.012 frozen&&3.60$^{+9.59}_{-2.12}$&<1.0&1.10 (10)&10.43$^{+2.31}_{-8.07}$&2.0$\times10^{33}$\\
\vspace{1mm}
  &                            & $N_{\rm H}2$=0.14$^{+0.17}_{-0.09}$     &&&             &          &                   &       \\ 
  \vspace{1mm}
58&\texttt{tbabs$\times$(apec)}&<0.06&&1.71$^{+1.21}_{-0.73}$& <0.90 &0.91 (9)&6.80$^{+1.30}_{-1.38}$&1.6$\times10^{33}$\\
\vspace{1mm}
61&\texttt{tbabs$\times$(apec)}&0.04$^{+0.05}_{-0.03}$&&4.36$^{+4.96}_{-1.84}$&<1.01&1.07 (30)&20.7$^{+2.71}_{-2.76}$&4.0$\times10^{33}$\\
\vspace{1mm}
65&\texttt{tbabs$\times$(apec)}&<0.02&&2.97$^{+2.33}_{-1.20}$&<0.22&1.13 (26)&10.33$^{+1.93}_{-3.85}$&3.3$\times10^{29}$$^{(4)}$ \\ 
\vspace{1mm}
66&\texttt{tbabs$\times$(po)}&<0.58&2.49$^{+2.55}_{-1.05}$&&&1.86 (5)&4.45$^{+2.72}_{-1.92}$&8.5$\times10^{32}$\\
%66&\texttt{tbabs$\times$(apec)}&<0.26&3.33$^{+12.24}_{-2.57}$&frozen to 0.1&1.81 (5)&5.10$^{+4.80}_{-2.91}$&9.8$\times10^{32}$ \\
\vspace{1mm}
67&\texttt{tbabs$\times$(po)}& 0.54$^{+0.45}_{-0.31}$&1.34$^{+0.54}_{-0.47}$&&&1.02 (16) &45.71$^{+11.50}_{-8.72}$&8.7  $\times10^{33}$\\
%67&\texttt{tbabs$\times$(apec)}&0.88$^{+0.39}_{-0.31}$&5.74$^{9.47}_{-2.19}$&frozen to 1.0& 0.98 (17)&47.59$^{+6.00}_{-5.99}$&1.00$\times10^{34}$\\
\hline
\hline
      \multicolumn{9}{l}{(1): We assumed a distance of Wil~1 ($\sim$40 kpc) to estimate the luminosity of the sources (see Sect.~\ref{intro}).}\\
   \multicolumn{9}{l}{(2):  For the source No.\,50 there are two temperatures of two  \texttt{apec} models: $kT1$ and $kT2$.}\\
   \multicolumn{9}{l}{(3):  Source No.\,57 has two absorption models, which show the Galactic absorption($N_{\rm H}1$) and the intrinsic absorption ($N_{\rm H}2$).}\\
\multicolumn{9}{l}{(4): For the source No.\,65, which is classified as a foreground M~dwarf, the distance of 441 parsec is taken to estimate its}\\
      \multicolumn{9}{l}{~~~ luminosity (see Sect.~\ref{diss}).} \\
\end{tabular}
%\vspace{-2mm}
%\tablefoot{}
 \end{table*}

\section{classification of X-ray sources in the field of Wil~1}
\label{diss}
In following we explain the details of source classification in the field of Wil~1.  Table~\ref{critria} summarises the criteria of the source classification.
\subsection{Classified AGN and galaxies}
\label{backgroundsources}
\begin{figure*}
\includegraphics[clip, trim={0.0cm 0.cm 0.0cm 0.0cm},width=0.33\textwidth]{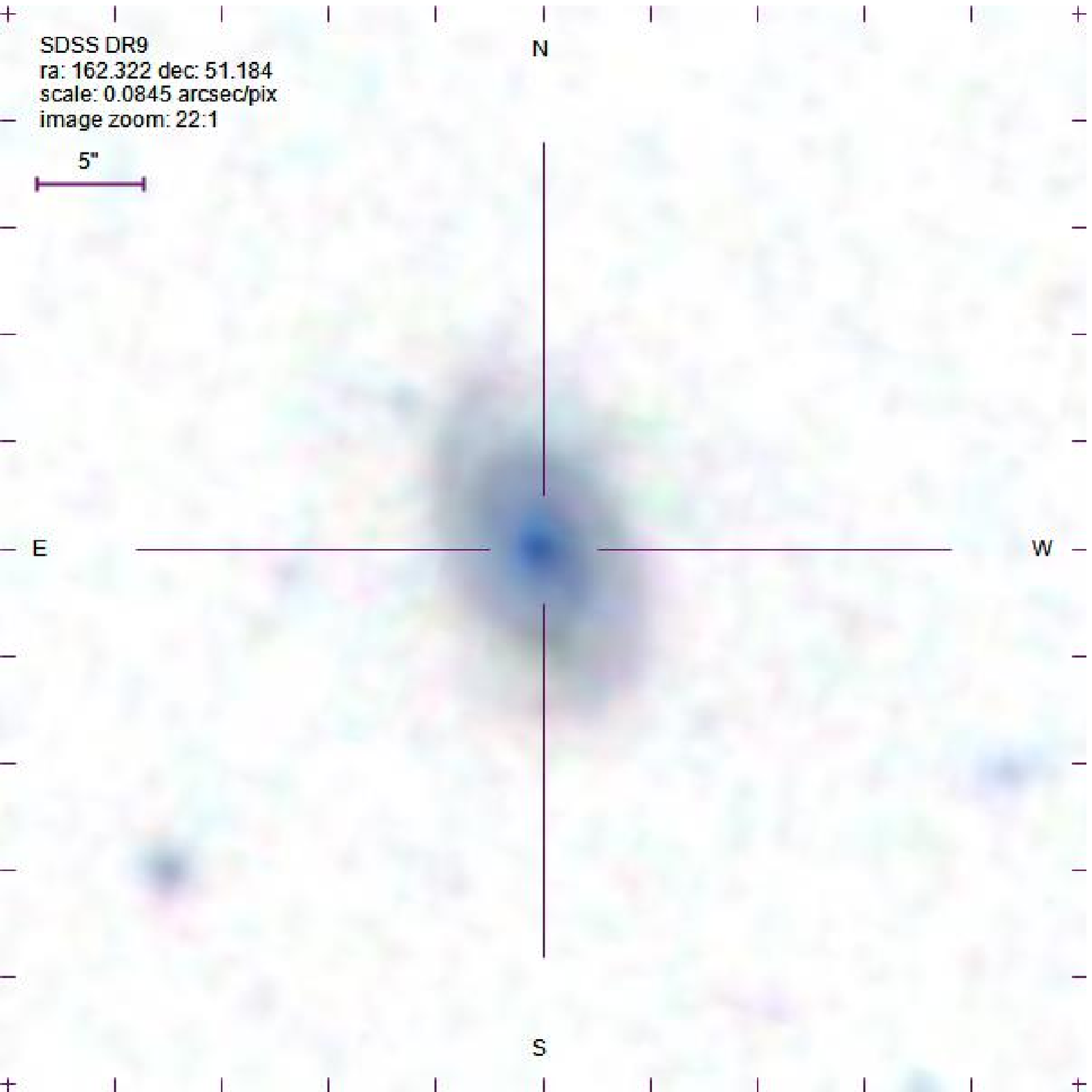}
\includegraphics[clip, trim={0.0cm 0.cm 0.0cm 0.0cm},width=0.33\textwidth]{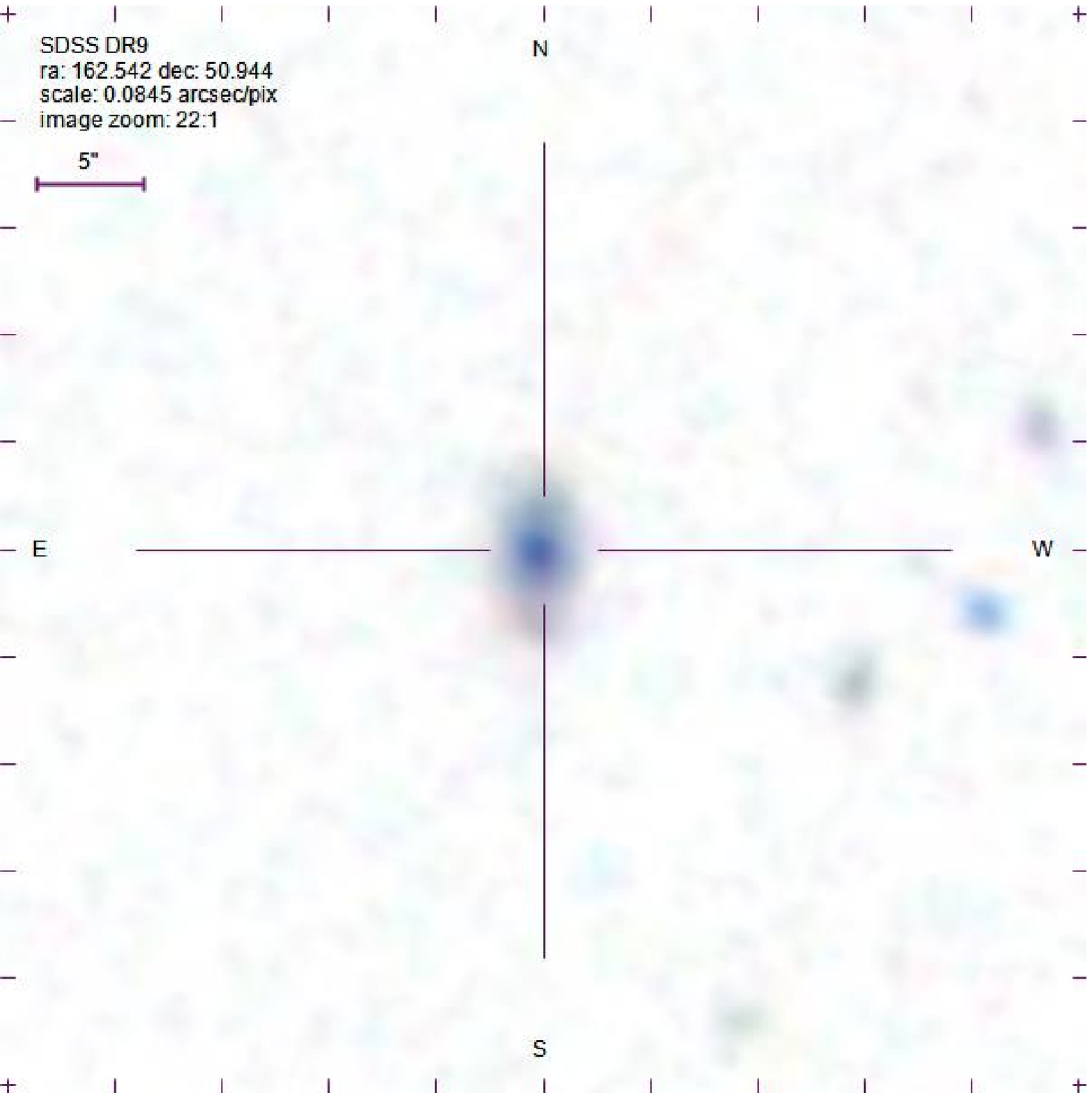}
\caption{Optical SDSS9 image of counterparts of sources No.\,46\,(Left) and No.\,88\,(right), which are classifed as background galaxies. \label{gg}}
\end{figure*}
As Figure~\ref{infra-plot} shows, sources, which have been classified as background galaxies or AGNs (see Table.~\ref{catalogue-x-ray}) show consistency with the properties of background objects in deep infrared studies as explained in Sect.\,\ref{inf-count-exp}. %As figure~\ref{infra-plot} shows with $W2-W3$>2.0, they are consistent with the colours of the background objects as it was suggested by \citet{2010AJ....140.1868W} and later applied for the WISE infrared counterparts of X-ray sources by \citet{2018MNRAS.473.4937S}. 
Only sources No.\,46 and No.\,88 have a 2MASS near-infrared counterpart, which has a similar colour of background sources in 2MASS survey (see Sect.~\ref{inf-count-exp}). These two sources are the brightest background galaxies in the field of Wil~1 and their extent is visually observable in their optical counterpart (see Fig.~\ref{gg}). The classified background sources with optical counterpart have X-ray to optical flux ratio  ${\rm log}\bigg(\frac{F_ \text{X}}{F_\text{opt}}\bigg)>-1.0$ (see Fig.~\ref{log-x-opt}). It is also similar to the ratio of background sources in other studies \citep[see e.g,][]{2013A&A...553A...7D,2016A&A...586A..64S}. As Fig.~\ref{log-x-opt} shows, for the majority of background sources the X-ray emission is more dominant than the optical emission, which is different from foreground stars or symbiotic stars. 
 The HR diagrams of Fig.~\ref{hrs-plot} show that most of the classified background sources have a hard spectrum (located in the area of power-law models with $\Gamma$=1, 2). %Also, in the HR diagrams they are mainly located where the models have a low column density. This is expected from the sources in the field of Wil~1 due the very low Galactic absorption in the direction of Wil~1 (see Fig.~\ref{hrs-plot}).

%Many sources in the field of Will~1, which have been observed by \xmm, are already classified as AGNs or background galaxies in other catalogues (see Sect.~\ref{AGN-cata}). All these sources have optical and/or infrared counterparts. Our study confirms the classification of these sources.  They all satisfy two criteria for optical/infrared counterparts of the background sources, which we have also applied to our previous studies \citep{2016A&A...586A..64S, 2019A&A...627A.128S}. All these sources have ${\rm log}\bigg(\frac{F_ \text{X}}{F_\text{opt}}\bigg)>-1.0$ and $W1-W3$<3 (see Fig.~\ref{log-x-opt} and Fig.\ref{infra-plot}). 
% In following, we discuss the classification of X-ray sources, which are located within 5$r_{\rm h}$  of Wil~1.
%most of the X-ray source in the field of Wil~1 are background objects. All the objects, which have been classifed as AGNs or galaxies in this work, have been prevously classified in  
\begin{table*}
\centering
\caption{\bf Summary of critria of source classification in field of Wil~1.\label{critria}}
\small
\centering
\addtolength{\tabcolsep}{-0.1cm}   
\begin{tabular}{ll}
\hline\hline
source type& classification critra\\
\hline
foreground stars& log$(\frac{F_\text{X}}{F_\text{opt}})$<0, infrared counterpart with $W2-W3<1.5$ and/or $J-K<1$, optical counterpart brighter\\
&or redder\,(M dwarfs) than the steller isochrone of Wil~1, main X-ray emission <2.0 keV\\
AGNs& -1<log$(\frac{F_\text{X}}{F_\text{opt}})$<2,  infrared counterpart with $W2-W3>1.5$ and/or $J-K>1$, HR2>-0.5\\
galaxies&-1<log$(\frac{F_\text{X}}{F_\text{opt}})$<2, infrared counterpart with $W2-W3>1.5$ and/or $J-K>1$, optically classifed as a\\
& galaxy.\\ 
symbiotic stars& log$(\frac{F_\text{X}}{F_\text{opt}})$<0,  infrared counterpart  of an stellar object ($W2-W3<1.5$) and optical counterpart  on the isochrone\\
& of  Wil 1 dSph or confirmed as member of Wil 1 dSph, X-ray emmsion similar to one of the types of \\
&symbiotic stars {\citep{2013A&A...559A...6L}}\\
hard sources& $HR2-EHR2>-0.2$, or only $HR3$ and/or $HR4$ are defined\\
\hline
\end{tabular}
%\vspace{-2mm}
%\tablefoot{}
 \end{table*}

\subsection{Sources No.\,65, a foreground M~dwarf}
\label{foregroundstar}
Source No.\,65 has optical and infrared counterparts with $W2-W3<2$, $J-K<1$, and log$(\frac{F_\text{X}}{F_\text{opt}})$<0. This makes its classification as a foreground star likely \citep{2016A&A...586A..64S,2013A&A...553A...7D,2018MNRAS.473.4937S}. The optical and infrared colours of the source  with  $g-r$=1.40$\pm$0.01 $r-i$=0.98$\pm$0.01, $r-z$=0.52$\pm$0.01, and $z-J$=1.29$\pm$0.05, $J-H$=0.57$\pm$0.09, and $H-K_{s}$=0.28$\pm$0.09 agree very well with the M~dwarfs  of the spectral type M2 according to \citet{2011AJ....141...97W}. It is located at the distance of  $437^{+34}_{-30}$ pc based on the parallax measurement of \citet{2018AJ....156...58B}\footnote{ Gaia source ID: 835973362206830080}. The spectrum of the source is fitted with an absorbed \texttt{apec} model (see Table~\ref{spectral-Table} and Fig.~\ref{spec.fig})  with a temperature of $kT$=2.97$^{+2.33}_{-1.20}$ keV, which is slightly higher than the typical temperature of classified M~dwarfs in the field of Draco dSph \citep{2019A&A...627A.128S}. The flux of the source stays at similar values over the three \xmm\, observations.  In addition, the individual  light curve of each observation shows no flare activity. The X-ray luminosity of the order $10^{29}$ erg\,s$^{-1}$ is consistent with an M2~dwarf with an age of $\sim$0.01 Gyr \citep[see][]{2013MNRAS.431.2063S}.
\subsection{Source No.\,57, symbiotic star in the Wil~1 dSph}
\begin{figure}
\includegraphics[clip, trim={0.0cm 2.8cm 0.0cm 0.0cm},width=0.225\textwidth]{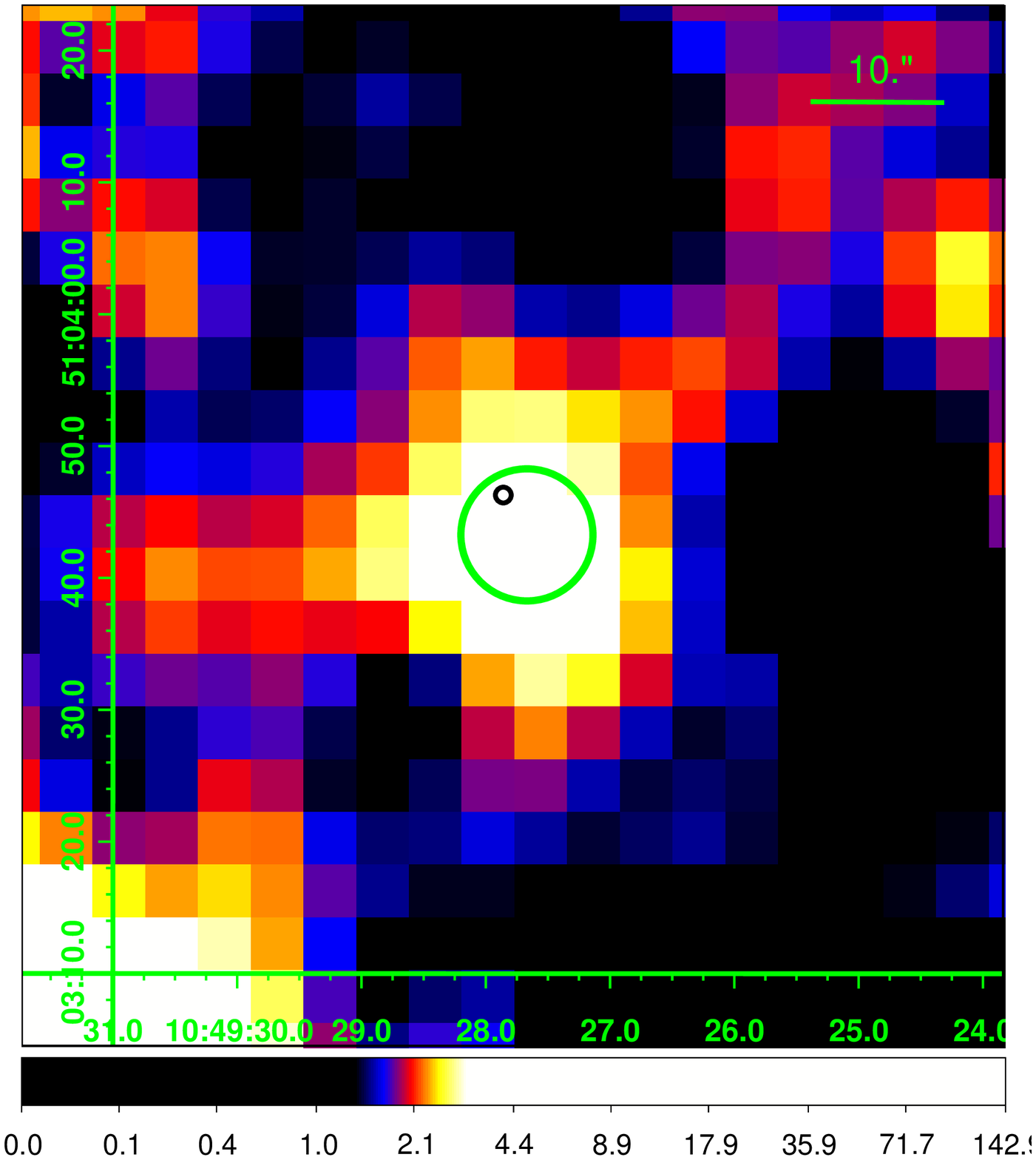}
\includegraphics[clip, trim={0.0cm 2.4cm 0.0cm 0.0cm},width=0.25\textwidth]{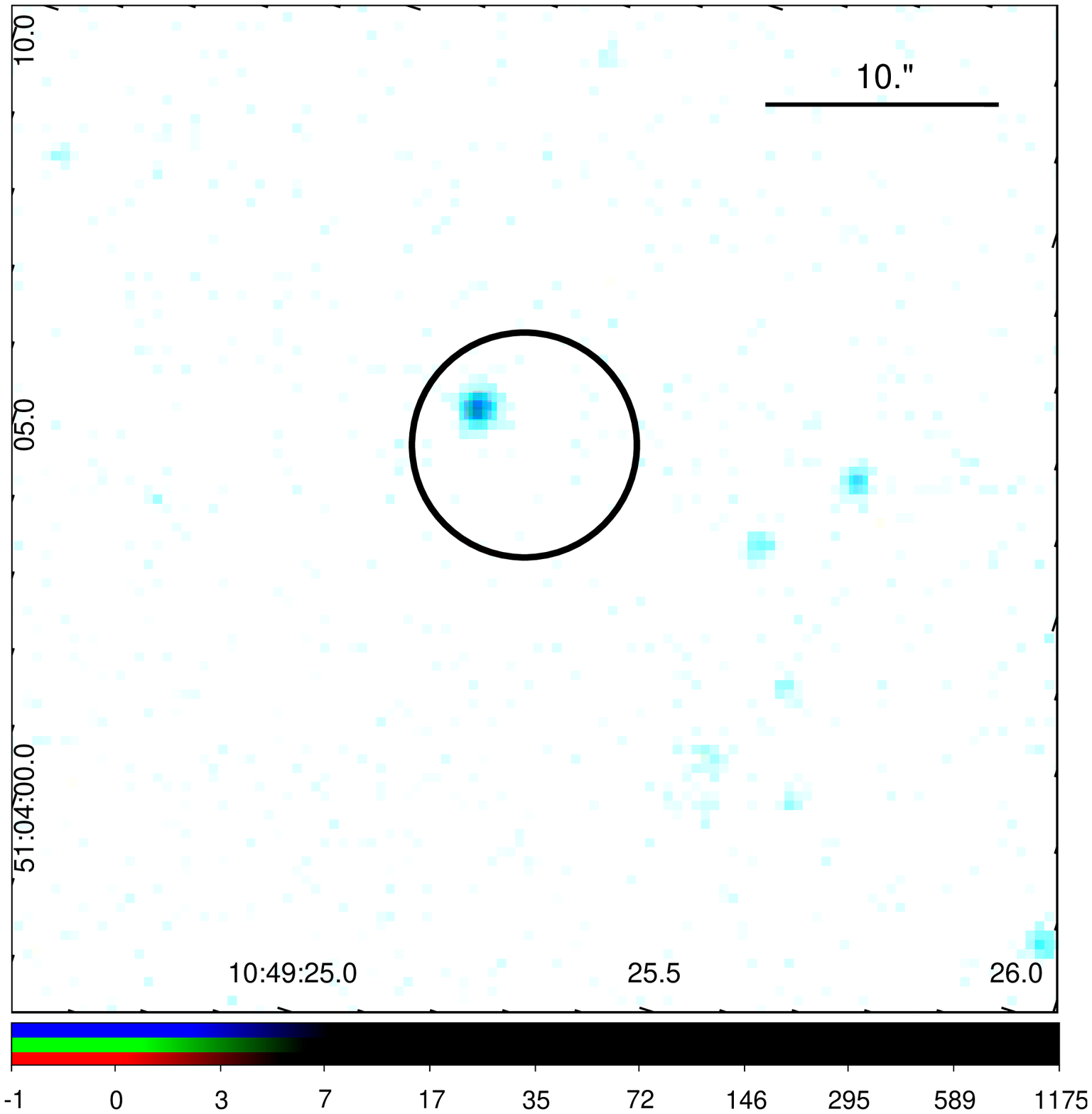}
\caption{{\bf Left}: X-ray mosaic image of Source No.\,57 in the energy range of 0.2-12.0 keV. The green circle shows the 3$\sigma$ X-ray position of the source and the black circle shows the position of the optical counterpart. {\bf Right}: Optical SDSS9 image of the stellar counterpart of source No.57. Black region shows the 3$\sigma$  position of the X-ray source. \label{src57}}
\end{figure}

This source is located in the central region of the Wil~1. Figure~\ref{src57} shows the 3$\sigma$ error circle of the X-ray source position in observation 3, where source No.\,57 had the highest detection likelihood. Within this error circle position the source No.\,57 has an optical counterpart, which is located in the red giant branch of the Wil~1 sources (see Fig.~\ref{opt-counterpart}). In addition, this optical counterpart is classified as a member of the Wil~1 by \citet{2007MNRAS.380..281M} with a radial velocity of -11.7$\pm$2.8 km\,s$^{-1}$ and a metallicity of [Fe/H]=--1.6, which both are in a very good agreement with the systemic velocity (-12.8$\pm$1.0 km\,s$^{-1}$) and the low metallicity of Wil~1 members \citep{2007MNRAS.380..281M, 2011AJ....142..128W}. The source has log$(\frac{F_\text{X}}{F_\text{opt}})\leq$0., similar to stellar objects classified by \citet{2019A&A...627A.128S} in Draco dSph. Its X-ray spectrum is fitted well with an absorbed \texttt{apec} model (see Table~\ref{spectral-Table} and Fig.~\ref{spec.fig} upper right). The temperature of the source of $kT$=$3.60^{+9.59}_{-2.12}$\,keV is similar to the temperature of the $\beta$-type symbiotic \citep{2013A&A...559A...6L, 2019A&A...627A.128S}. The column density of the source is higher than the Galactic absorption in the direction of Wil~1 ($N_{\rm H}$=1.17$\times$10$^{20}$\,cm$^{-2}$). We assumed two absorption model for the source: a Galactic absorption, frozen to the $N_{\rm H}$ in the direction of Wil~1, and the intrinsic absorption of $N_{\rm H}$=0.14$^{+0.17}_{-0.09}\times$10$^{22}$\,cm$^{-2}$ for the source itself, which can be related to colliding winds region of the symbiotic star \citep{2013A&A...559A...6L}. We classify this source as a symbiotic star in Wil~1.
\subsection{Sources with optical counterpart}
Only sources {\bf No.\,50, 78, 84 and 92} are unclassified X-ray sources with an optical counterpart (see Fig.~\ref{opt-counterpart}). Their optical counterpart is calssified as a stellar object in SDSS12 survey \citep{2015ApJS..219...12A}.  Among these four source only source No.\,50 has a WISE infrared counterpart too (see Fig.~\ref{infra-plot}).\\
{\bf Sources No.\,50, 84:} The optical counterpart of these two sources can be a main sequence star in Wil~1 (see Fig.~\ref{opt-counterpart}). The X-ray spectrum of source No.\,50 is fitted with an absorbed \texttt{apec} model with two components (see Fig.~\ref{spec.fig} and Table~\ref{spectral-Table}). The absorption of the source is consistent with the Galactic absorption towards Wil~1 and the abundance of the source is consistent with the low metallicity of Wil~1 members.  The HR of source No.\,84 shows that the main emission of the source is <2.keV (see Table \ref{catalogue-x-ray}) and it has a softer emission in comparison to the known background sources (see Fig.~\ref{hrs-plot} and \ref{log-x-opt}). As Fig.~\ref{log-x-opt} shows both sources have log$(\frac{F_\text{X}}{F_\text{opt}})$ >1. Assuming that source No.\,50 and No.\,84 are located at the distance of Wil~1 ($\sim$40~kpc), their luminosities  are $\sim$7.9$\times10^{33}$\ and 2.0$\times10^{33}$ erg\,s$^{-1}$, respectively. \\
In the following, we discuss about the possibilities of these two sources being candidates for either cataclysmic variables\,(CVs, i.e, accreting white dwarf with a main sequence companion), or low mass X-ray binaries\,(LMXBs). The X-ray emission in CVs depends on the strength of the magnetic field of the white dwarf  and the accretion rate. There are two main classes of magnetic and non-magnetic CVs. In non-magnetic CVs an accretion disk can form around the white dwarf. % The disk, itself, is not hot enough to emit X-rays, while a boundary layer between the disk and surface of the white dwarf is the region, where the X-ray emission is produced \citep[e.g,][]{2017PASP..129f2001M, 2006csxs.book..421K}. If the system is in the quiescent state, the boundary layer is optically thin, which leads to hard X-ray emissions with temperatures of a few keV and X-ray luminosity between 10$^{29}$--10$^{32}$~erg\,s$^{-1}$ \citep[][]{2017PASP..129f2001M, 2006csxs.book..421K}.  
 The  disk -itself- is not hot enough to emit X-rays. However, if the boundary layer between the disk and surface of the white dwarf is optically thin, X-ray emission can be produced. In this case the X-ray emission is hard with a corresponding temperature of a few keV and an X-ray luminosity between 10$^{29}$--10$^{32}$~erg\,s$^{-1}$ \citep[][]{2017PASP..129f2001M, 2006csxs.book..421K}. If the system undergoes the nova eruption, the boundary layer becomes optically thick, therefore the system shows a soft X-ray  emission of 0.1--0.5 keV and higher luminosities \citep{2017PASP..129f2001M}. Magnetic CVs have two sub-classes of polars (with strong magnetic field) and intermediate polars (with weakly magnetic white dwarf). Polars have no accretion disk and materials follows the magnetic lines to reach to the surface of the white dwarf.  Polars are characterised by a dominant soft X-ray emission and a bolometric luminosity of $\sim$10$^{32}$~erg\,s$^{-1}$ \citep{2017PASP..129f2001M}.  There are cases, in which polars show emission up to $\sim$10~keV, however, in these cases the bolometric luminosity decreases around two orders of magnitude \citep[e.g,][]{2004MNRAS.350.1373R}. In case of intermediate polars,  a truncated accretion disk can form. The attachment of the inner part of the accretion disk with the magnetic field of white dwarf forms accretion curtains above the poles of the white dwarfs, where the materials experience strong shocks, and therefore, a hard X-ray emission of 5--50~keV, with a X-ray luminosity up to a few 10$^{33}$~erg\,s$^{-1}$ is produced \citep[][]{2017PASP..129f2001M, 2012MmSAI..83..585B}. Fluorescent Fe~K$\alpha$ lines were observed in most of the intermediate polars \citep[][]{2006csxs.book..421K}. On the basis of the samples of \citet{2017PASP..129f2001M}, CVs have a log$(\frac{F_\text{X}}{F_\text{opt}})$<1.0 and magnetic CVs generally have a log$(\frac{F_\text{X}}{F_\text{opt}})$ larger than that of non-magnetic CVs. According the above characteristics of spectrum and luminosity of different types of the CVs, source  No.\,50 and No.\,84 do not fit to any class of them.\\
On the other hand, The low luminosity of the sources suggests that they are no persistent LMXBs, which normally have X-ray luminosities >$10^{35}$ erg\,s$^{-1}$.  However, they might be transient LMXBs. Observational studies show that transient LMXBs are low luminosity X-ray sources ($\sim10^{31}-10^{33}$~erg\,$s^{-1}$) with soft X-ray emissions\,(<\,5.0~keV) \citep[e.g,][]{2000ApJ...539..191Y,  1998A&ARv...8..279C}.  The models behinds these very faint accretors  are explained in the work of e.g, \citet[][]{2006MNRAS.366L..31K} Simulations predict that transient LMXBs are common X-ray sources in galaxies and can be observed several million years after the star formation \citep{2008ARep...52..299B,  2015A&A...579A..33V}.  Sources No.\,50 and 84 show no pulsation. Also,  their optical counterparts are not confirmed as Wil~1 members in the studies of \citet{2007MNRAS.380..281M} and \citet{2011AJ....142..128W}.  Therefore, these sources remain unclassified being, either transient LMXBs or background objects.\\
{\bf Sources No.\,78, 92:} Source No.\,92 is a variable source (see Fig.~\ref{var}). X-ray to optical flux ratio of both sources are  log$(\frac{F_\text{X}}{F_\text{opt}})$ >1. The position of the optical counterparts of these two sources in the colour-magnitude diagram (see Fig.~\ref{opt-counterpart}) and also the location of them (out of the 5$r_{\rm h}$ of the field of Wil~1; see Fig.~\ref{rgb-image} and Sect.~\ref{intro}) make them rather unlikely members of Wil~1. % However, lacking of the infrared counterpart and also the X-ray spectral information do not allow to confirm them as background objects.
%The spectra of the sources No. 50 , 84 are consistent with the spectrum of known transient LMXBs \citep{}. However, these sources are usually observed with luminosities <$10^{33}$\,erg\,s$^{-1}$ \citep{}. the luminosity of these two sources is comparable to those transients in the disk state (see citep{}). consequently, if these sources are transient LMXBs in the disk state, they are expected to show harder specta  with non themal emmsions \citep{}, while the spectum of source no.50 shows themal emmision and the hardness ratios of sources No.84 is significant <5 keV and shows that the main part of the emission is in the softer bands. Therfore, the possibily that these two sources are LMXBs in the Wil~1 is very low. They seems to be rather background sources. 
\subsection{Sources with WISE infrared counterpart}
Sources {\bf No.\,29, 38, 43, 45, 58, 67, 97} have only an infrared counterpart in  WISE survey, and no optical counterpart is found for theses sources. The position of all these sources in colour-colour WISE diagram shows that they all have infrared colours similar to the background objects (see Fig.~\ref{infra-plot}). The spectral analysis of source No.\,45 and 67 (see Fig.~\ref{spec.fig} and Table.~\ref{spectral-Table}) shows that an absorbed \texttt{power-law} model  is fitted to the spectrum of these two sources. Source No.\,97 is also a hard variable source (see Table~\ref{catalogue-x-ray}, and Fig.~\ref{var}). Source No.\,97 and also 43, which is located outside of the  of the  5$r_{\rm h}$ of the main field of Wil~1 can be candidate for a background source. However, in case of sources No.\,45, and 67 the classification needs more criteria, since they can be the members of Wil~1.\\ % This is consistent with the X-ray spectrum expected from AGNs or background galaxies \citep[e.g,][]{2011A&A...530A..42C,2009MNRAS.394.1741O}.\\
Source No.\,58 has a soft X-ray spectrum. An absorbed \texttt{apec} model with  a temperature of $\sim$1.4 keV (see table. \ref{spectral-Table} and Fig.~\ref{spec.fig}) is fitted to the spectrum. %The soft spectrum and lacking of optical counterpart makes this source a candidate for supernova remnant\,(SNR). The extension of the SNRs should be observable in the distance of Wil~1. Source No.\,58 is too faint in X-rays (rate of $\sim$5$\times$10$^{-3}$ cts\,s$^{-1}$ ) and it is not significantly improved as an extended source. However, if we consider its visual X-ray image radius of $\sim 18\arcsec$ (i.e, $\sim$3~pc assuming the distance of 40~kpc), the size of the source is too smaller than the expected size of the typical SNRs observed in X-rays. Low luminosity SNRs are expected to show a size of $>$10~pc \citep[see e.g,][]{2018ApJ...863..137O, 2014A&A...561A..70A}. Therefore, this source can not be an SNR. 
For the sources No.\,29, 38 the HRs show that their main  emission is observed <2.~keV. They remain unclassified since there is not enough information for their classification.
%The lack of a stellar optical counterpart and the soft X-ray spectrum make this source an SNR candidate. The stellar population of dSphs is old, only Type Ia SNe are expected to accur in this type of galaxies. Therefore, if confirmed, this source would be an SNR of Type Ia. On the other hand, the source is too faint in X-rays (rate of $\sim$5$\times$10$^{-3}$ cts\,s$^{-1}$ ) and its extention is not significantly improved. In \xmm, source extent can only be reliably determined for the bright objects \footnote{\tiny https://xmm-tools.cosmos.esa.int/external/sas/current/doc/emldetect/node3.html}. Therefore, the class of the source is not confirmed .
\subsection{Sources without  counterpart}
The rest of the X-ray sources in the field of Wil~1 have no optical/infrared counterparts. Among them, sources {\bf No.\,22, 55, 61, 62, 76, 81, 93} are classified as hard X-ray sources using the HRs criteria (see Sect.~\ref{hrs}). Also, source No.\, 22, 61, 62, 66, 81, 93 are located within the  5$r_{h}$ of Wil~1. For sources {\bf 61, 66} we have performed a spectral analysis. An absorbed \texttt{power-law} model is fitted to the spectrum of source No.\,61. As for source No.\,66, poor statistics of the spectrum do not allow to determine a significant temperature for the source. Since there is no optical/infrared counterpart for these sources, no advanced classification is possible for them.  
%he remaining sources in the field of Wil~1, which are not mentioned in this section, remains unclassified, due the poor statistics of X-ray data and/or lack of a counterpart in other wavelengths.
\section{Summary}
We have detected and performed a classification of X-ray sources in the field of Wil~1 using  three \xmm\, observations and multi-wavelength studies. Our study shows that most of the X-ray sources in the field of Wil~1 are background galaxies and AGNs. This result is consistent with the results, which have been obtained for other dSphs \citep[e.g,][]{2015MNRAS.451.2735M, 2016A&A...586A..64S, 2005MNRAS.364L..61M}. Only one foreground star (M~dwarf) is detected in the field of Wil~1. \\
Also, we  classified a symbiotic star in Wil 1. This is consistent with the result of our recent deep X-ray study of the Draco dSph \citep{2019A&A...627A.128S}  and confirms the detection of accreting white dwarfs in these old population of dSphs. Moreover, eleven sources are classified as hard sources in the field of Wil~1. \\
 We could not confirm the presence of any LMXB in Wil~1.  So far, no LMXBs have been detected in nearby dSph. However, theoretically, transient LMXBs can exist for a long time in galaxies and be detectable as soft and low luminosity X-ray sources.  Additional X-ray studies of different nearby dSphs with different age, star formation, and distance are necessary to verify the existence of LMXBs in dSphds.

%\begin{enumerate}
%\item The spectrum is consistent with a blackbody spectrum with a temperature \textbf{of $T = (1.8\pm 0.3) \times 10^5$ K}. The X-ray luminosity, the blackbody temperature, and the relatively low long-term variability suggest that the X-ray emission of the system is caused by steady nuclear burning.
%\item 
%\item We estimae a radius of $\sim$110 $R_{\sun}$ for the red-giant companion. In addition, we have shown that the red-giant companion has a mass of $\lesssim$1.5 $M_{\sun}$ assuming Roche-lobe overflow.
%\end{enumerate}

\section*{Acknowledgements}
\scriptsize {This  research  was  funded  by  the DLR research grant BWWI/DLR~500R1907 and DFG~SA~2131/12-1. This study is based onobservations obtained with \xmm, an ESA science mission with instruments  and  contributions  directly  funded  by  ESA  Member  States  and  NASA. This  research  has  made  use  of  the  SIMBAD  and  VIZIER  database,  operatedat  CDS,  Strasbourg,  France,  and  of  the  NASA/IPAC  Extragalactic  Database\,(NED),  which  is  operated  by  the  Jet  Propulsion  Laboratory,  California  Institute  of  Technology,  under  contract  with  the  National  Aeronautics  and  Space Administration. This publication makes use of data products from the Wide field Infrared Survey Explorer, which is a joint project of the University of California, Los Angeles, and the Jet Propulsion Laboratory/California Institute of Technology, funded by the National Aeronautics and Space Administration. This publication has made use of data products from the Two Micron All Sky Survey, which is a joint project of the University of Massachusetts and the Infrared Processing  and  Analysis  Center,  funded  by  the  National  Aeronautics  and  Space Administration  and  the  National  Science  Foundation.  Funding  for  SDSS  and SDSS-III has been provided by the Alfred P. Sloan Foundation, the Participating  Institutions,  the  National  Science  Foundation,  and  the  US  Department  of Energy Office of Science. The SDSS-III web site ishttp://www.sdss3.org/.SDSS-III  is  managed  by  the  Astrophysical  Research  Consortium  for  the  Participating  Institutions  of  the  SDSS-III  Collaboration  including  the  University of  Arizona,  the  Brazilian  Participation  Group,  Brookhaven  National  Laboratory, University of Cambridge, University of Florida, the French Participation Group, the German Participation Group, the Instituto de Astrofisica de Canarias, the Michigan State/Notre Dame/JINA Participation Group, Johns Hopkins University,  Lawrence  Berkeley  National  Laboratory,  Max  Planck  Institute  for Astrophysics, New Mexico State University, New York University, Ohio State University, Pennsylvania State University, University of Portsmouth, Princeton University, the Spanish Participation Group, University of Tokyo, University of Utah, Vanderbilt University, University of Virginia, University of Washington, and Yale University. This research has made use of SAO Image DS9, developed by Smithsonian Astrophysical Observatory.}


\bibliographystyle{mnras}
\bibliography{bibtex} % if your bibtex file is called example.bib

%\begin{thebibliography}{99}
%\end{thebibliography}

%%%%%%%%%%%%%%%%%%%%%%%%%%%%%%%%%%%%%%%%%%%%%%%%%%

%%%%%%%%%%%%%%%%% APPENDICES %%%%%%%%%%%%%%%%%%%%%

\appendix
%Flux (0.2--12.~keV)
%($10^{-13}$ erg\,s$^{-1}$\,cm$^{-2}$)
\onecolumn
\begin{landscape}
\raggedright
\section{Source catalogue}
\footnotesize{
\begin{longtable}{clcccccrrrrcl}
    \caption{X-ray sources in the FOV of Willman~1 dSph. \label{catalogue-x-ray}}\tabularnewline
\hline
Src-NO   &RA              &DEC              &r1$\sigma$                       & \multicolumn{3}{c}{Flux$^{\ast~\ast\ast}$(0.2--12.~keV)}  &\multicolumn{4}{l}{Hardness ratio} &Var& Classification$^{\dagger}$\\
                          &  (J2000)       &  (J2000)        & ($\arcsec$)                     & \multicolumn{3}{c}{($10^{-13}$ erg\,s$^{-1}$\,cm$^{-2}$)}&&&&&&\\
                           &                &                 &                                 & OBS1&          OBS2&          OBS3& HR1&HR2&HR3&HR4&&\\
\hline
\endfirsthead
\multicolumn {13}{r}{Continued from previous page}\\
\hline 
 Src-NO  &RA              &DEC              &r1$\sigma$                       & \multicolumn{3}{c}{Flux$^{\ast~\ast\ast}$(0.2--12.~keV)} &\multicolumn{4}{l}{Hardness ratio} &Var& Classification$^{\dagger}$\\
                     &  (J2000)       &  (J2000)        & ($\arcsec$)                     & \multicolumn{3}{c}{($10^{-13}$ erg\,s$^{-1}$\,cm$^{-2}$)}&&&&&&\\
                     &                &                 &                                 & OBS1&          OBS2&          OBS3& HR1&HR2&HR3&HR4&&\\
\hline
\endhead

\hline
\multicolumn {13}{r}{Continued on the next page}\\

\endfoot

\hline
\endlastfoot
1        &  10 47 47.61 &+51 01 03.3  & 1.73  & 0.61  $\pm$  0.20  &  N              &  0.75  $\pm$  0.15        &              --                           &              --                           &              --                           &               --                           & 1.24  $\pm$  0.47 &\\                                         
2        &  10 47 51.63 &+51 03 42.6  & 2.35  & 5.42  $\pm$  0.93  &  N              &  0.91  $\pm$  0.22      & -0.44  $\pm$  0.17 &              --                           &              --                           &               --                           & 7.48  $\pm$  0.35  &AGN$^{2,5,6}$ \\                                                                    
3        &  10 47 55.99 &+51 03 10.4  & 2.85  &   N            &  0.65  $\pm$  0.20  &  0.40  $\pm$  0.09       &              --                           &              --                           &              --                           &               --                           & 1.05  $\pm$  0.43  &AGN$^{1,4}$  \\                                         
4        &  10 47 57.51 &+51 01 20.9  & 0.40  & 3.46  $\pm$  0.19  &  4.02  $\pm$  0.17  &  5.02  $\pm$  0.17      & -0.11  $\pm$  0.04 & -0.09  $\pm$  0.05 & -0.35  $\pm$  0.06 &               --                           & 4.49  $\pm$  0.08  &AGN$^{2,5,6}$ \\                                                                                                                          
5        &  10 47 58.61 &+51 10 18.3  & 1.60  & 0.58  $\pm$  0.22  &   N             &   N                   &              --                           &              --                           &              --                           &               --                           &              --                       &    \\
\end{longtable}}
\end{landscape}
%\section{Optical and infrared counterparts}
\onecolumn
%\section{Optical  counterparts of X-ray sources of Wil~1}
\footnotesize{
\begin{table*}
\caption{Optical magnitudes of counterparts of X-ray sources of Wil~1 in different energy filters of the SDSS12 survey.  \label{opt-count-table}}
\centering
\begin{tabular}{lccccc}
\hline\hline\
Src-No&$u$ mag&$g$ mag&$r$ mag&$i$ mag&$z$ mag\\
\hline
2   &   21.45 $\pm $  0.16  &   21.09 $\pm $  0.04  &   20.76 $\pm $  0.05  &   20.24 $\pm $  0.05  &   20.18 $\pm $  0.17  \\ 
3   &   21.58 $\pm $  0.11  &   20.75 $\pm $  0.02  &   20.51 $\pm $  0.03  &   20.44 $\pm $  0.04  &   20.25 $\pm $  0.11  \\ 
4   &   20.58 $\pm $  0.06  &   20.24 $\pm $  0.02  &   19.67 $\pm $  0.02  &   19.22 $\pm $  0.02  &   18.77 $\pm $  0.04  \\ 
6   &   22.17 $\pm $  0.21 &   21.64 $\pm $  0.05  &   21.45 $\pm $  0.05  &   21.34 $\pm $  0.08  &   20.44 $\pm $  0.13  \\ 
7   &   23.71 $\pm $  0.64  &   22.80 $\pm $  0.14  &   22.38 $\pm $  0.15  &   22.32 $\pm $  0.23  &   21.59 $\pm $  0.37  \\ 
8   &   22.11 $\pm $  0.20  &   21.37 $\pm $  0.04  &   20.82 $\pm $  0.03  &   20.62 $\pm $  0.04  &   20.61 $\pm $  0.15  \\ 
9   &   22.26 $\pm $  0.24  &   21.68 $\pm $  0.06  &   21.28 $\pm $  0.07  &   20.50 $\pm $  0.05  &   20.29 $\pm $  0.15  \\
\hline                                                                                                                         
\end {tabular}                                                                                                                   
\end {table*}} 
\onecolumn
%\section{Infrared counterparts of X-ray sources of Wil~1}
\footnotesize{
  \begin{table*}
\caption{Infrared magnitudes of counterparts of X-ray sources of Wil~1 in different energy filters 2MASS and WISE surveys.
\label{inf-count}
}
\centering
\begin{tabular}{lccccccc}
\hline\hline\
Src-No&$J$ mag&$H$ mag&$K$ mag&$W1$ mag&$W2$ mag&$W3$ mag&$W4$ mag\\
\hline
 2   &   &  &  &   16.77 $\pm $ 0.08  &   15.98 $\pm $  0.15  &   <12.69$^{\ast}$     &    <9.06$^{\ast}$     \\
 4   &   &  &  &   15.34 $\pm $ 0.03  &   14.59 $\pm $  0.05  &   12.29 $\pm $  0.33   &   <8.53$^{\ast}$       \\
 6   &   &  &  &   16.33 $\pm $ 0.06  &   15.94 $\pm $  0.13  &   <12.40$^{\ast}$     &    <8.34$^{\ast}$     \\
 8   &   &  &  &   16.40 $\pm $ 0.06  &   15.36 $\pm $  0.08  &   <12.29$^{\ast}$    &    <8.79$^{\ast}$    \\
 9   &   &  &  &   16.23 $\pm $ 0.06  &   15.68 $\pm $  0.12  &   <12.18$^{\ast}$    &    <8.83$^{\ast}$     \\
14   &   &  &  &   16.65 $\pm $ 0.07  &   15.89 $\pm $  0.12  &   <12.83$^{\ast}$     &    <9.32$^{\ast}$     \\
17   &   &  &  &   16.54 $\pm $ 0.07  &   15.82 $\pm $  0.12  &   <12.63$^{\ast}$     &    <9.13$^{\ast}$   \\
\end{tabular}  
%\vspace{2.cm} 
\end{table*}}
\pagebreak

\section{Image of optical SDSS9 counterparts}
%The optical image of the counterpart of the X-ray sources from the SDSS9 survey. For each source, the imges from right to left are the SDSS survey bands of $u$, $g$, $r$, $i$, and $z$, respectively. Images shows 3$\sigma$ circle error of X-ray sources (black), optical counterparts (blue) and infrared counterpart (red). \\
\label{SDSS-image}
\begin{figure*}
\raggedright{The optical image of the counterpart of the X-ray sources from the SDSS9 survey. The images are the $r$ band of SDSS survey. Images shows 3$\sigma$ circle error of X-ray sources (black), optical counterparts (blue) and infrared counterpart (red).} \\
  \centering 
%\vspace{-0.5cm}
  \subfloat[Src No.1]{\includegraphics[clip, trim={0.0cm 2.cm 0.cm 0.0cm},width=0.20\textwidth]{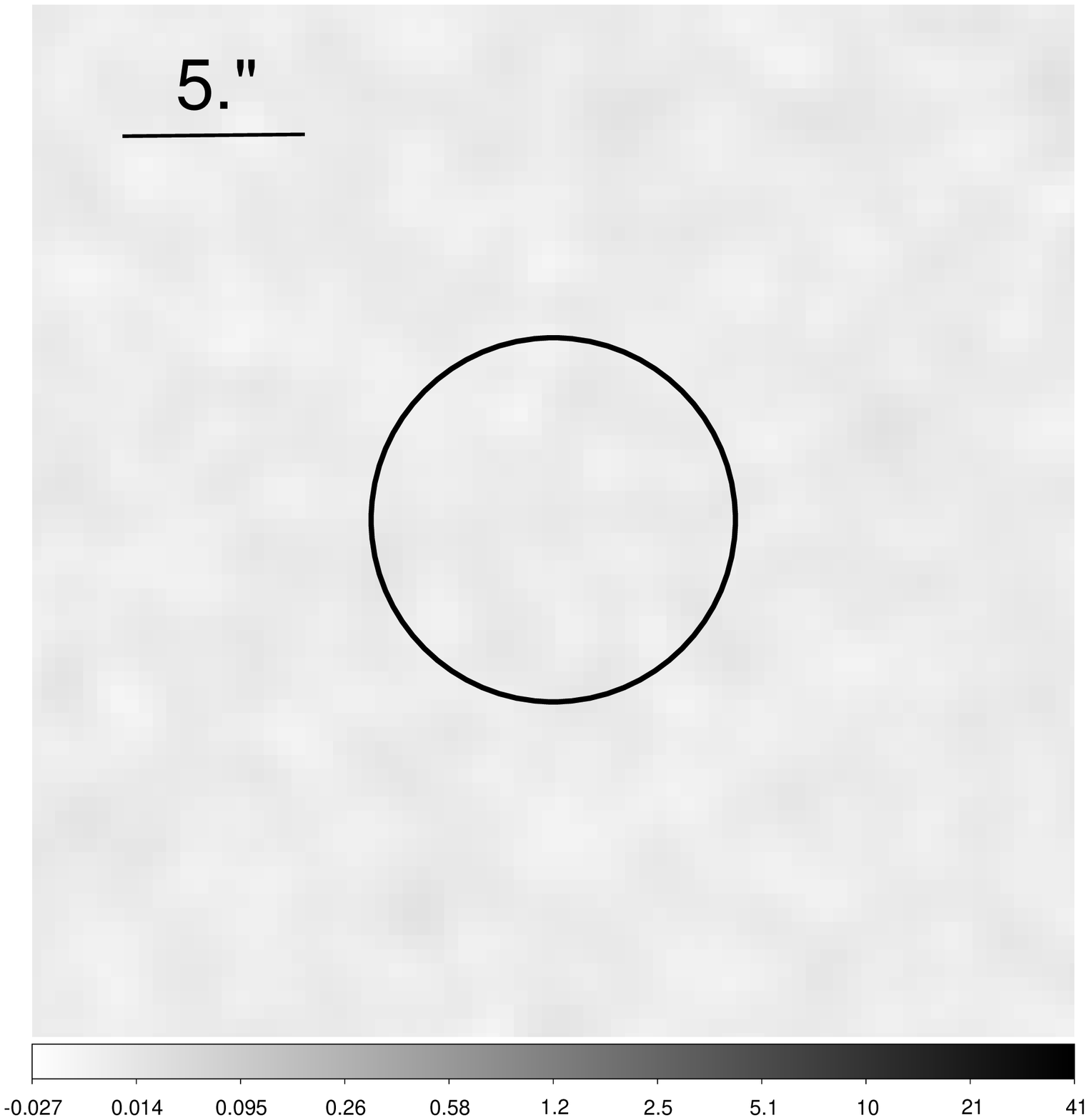}}
%\caption{\scriptsize{Src No.1}}
  \subfloat[Src No.2]{\includegraphics[clip, trim={0.0cm 2.cm 0.cm 0.0cm},width=0.20\textwidth]{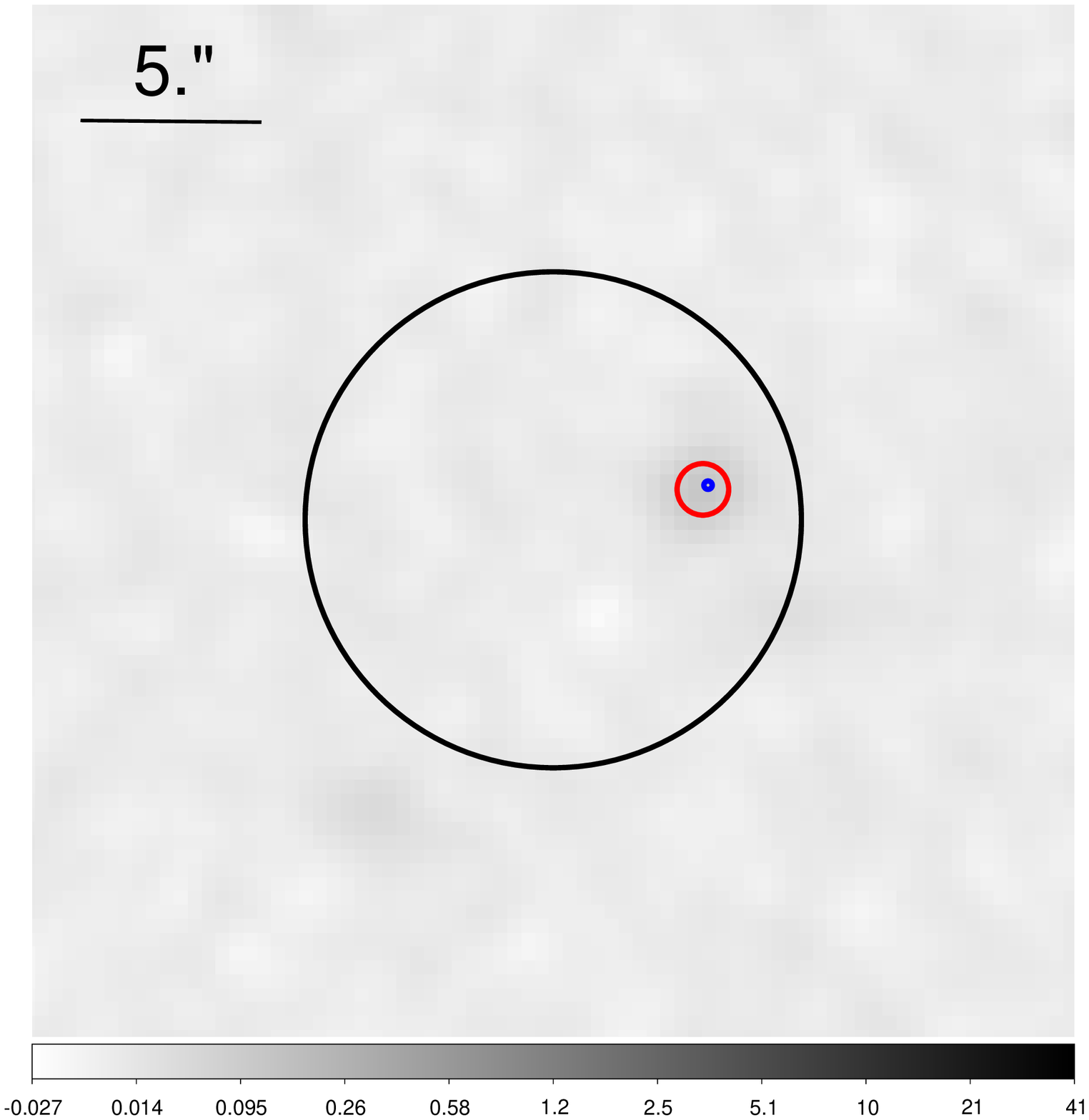}}
%\caption{\scriptsize{Src No.2}}
  \subfloat[Src No.3]{\includegraphics[clip, trim={0.0cm 2.cm 0.cm 0.0cm},width=0.20\textwidth]{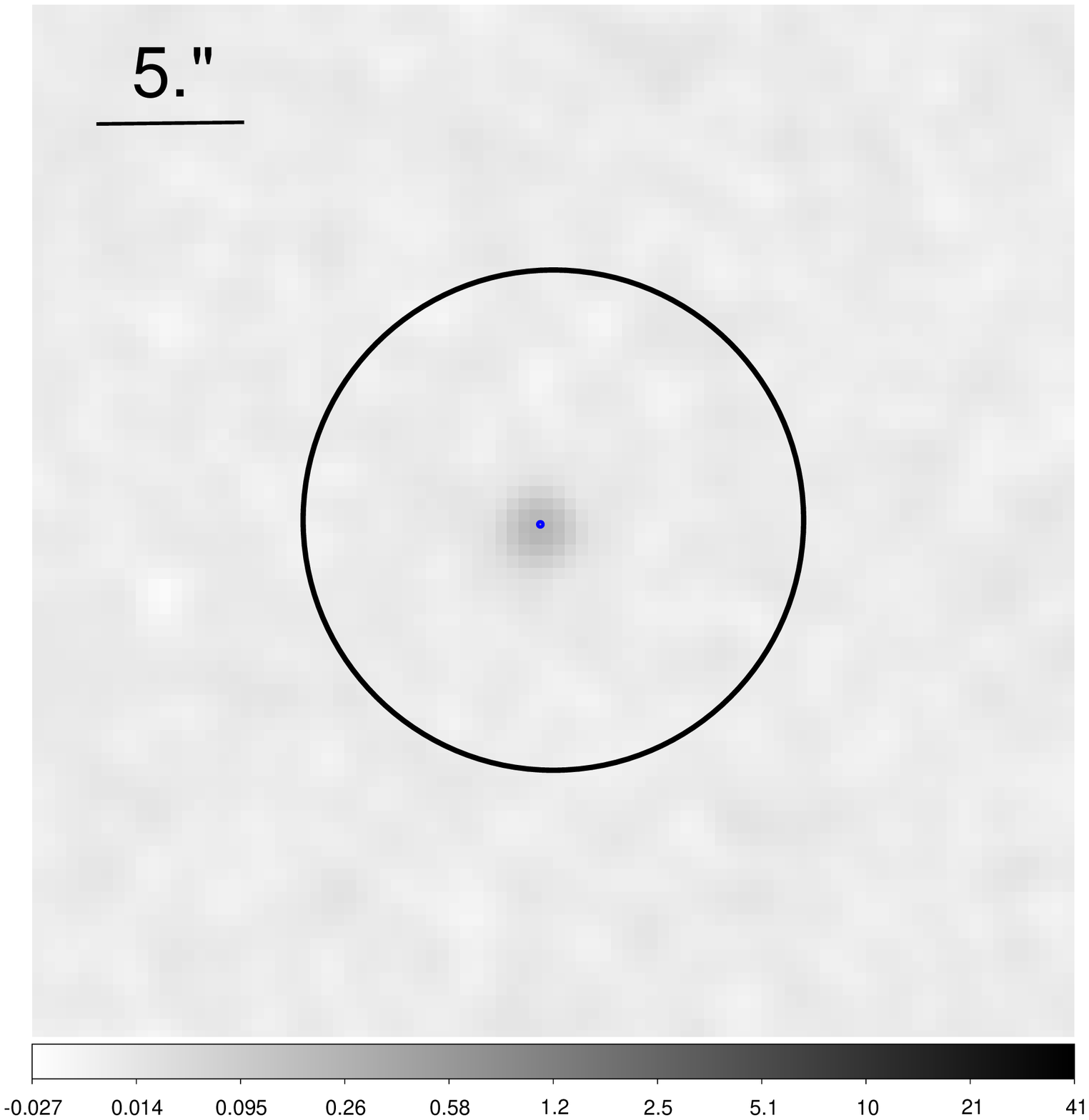}}
%\caption{\scriptsize{Src No.3}}
  \subfloat[Src No.4]{\includegraphics[clip, trim={0.0cm 2.cm 0.cm 0.0cm},width=0.20\textwidth]{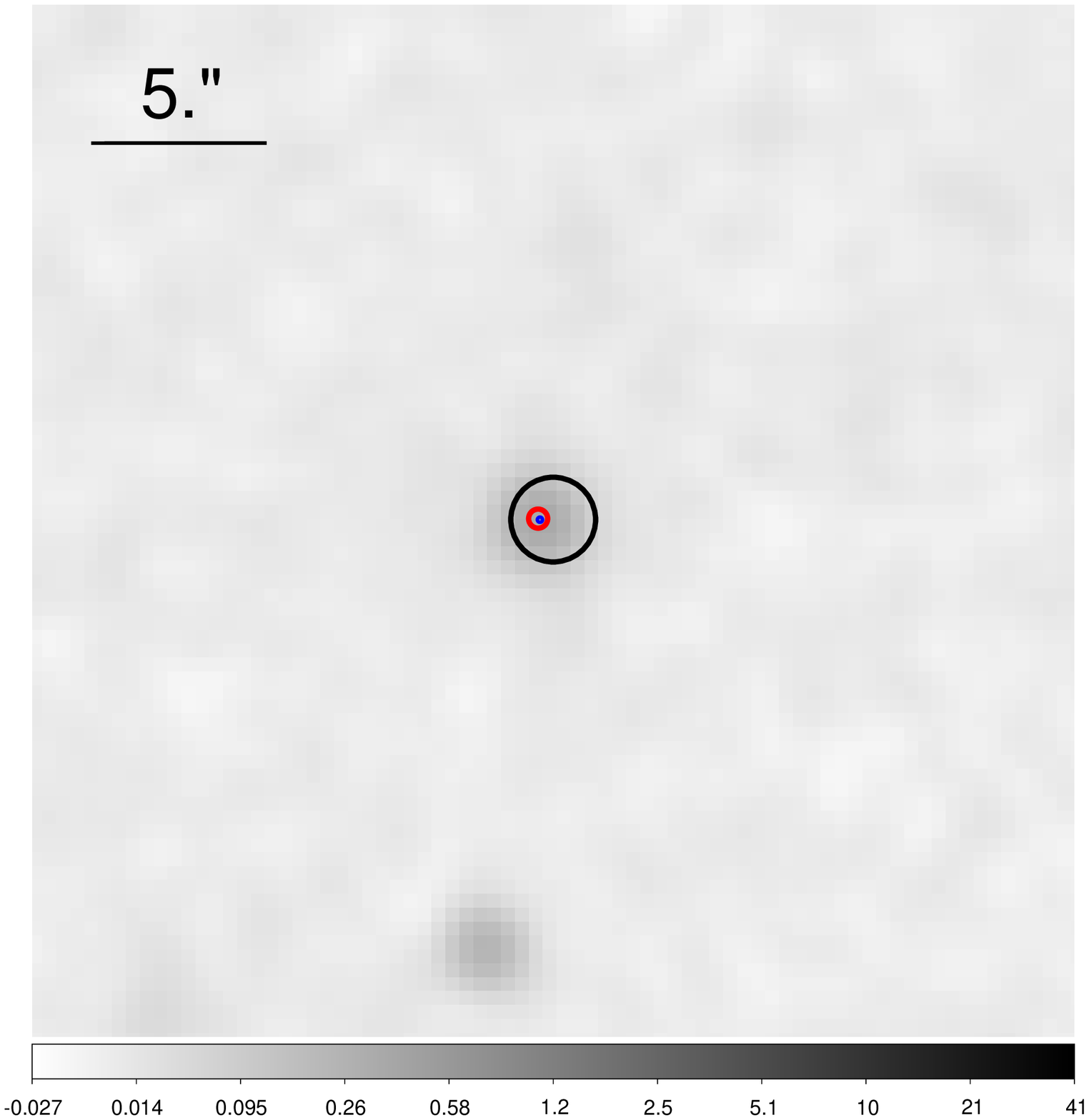}}\\
%\caption{\scriptsize{Src No.4}}
  \subfloat[Src No.5]{\includegraphics[clip, trim={0.0cm 2.cm 0.cm 0.0cm},width=0.20\textwidth]{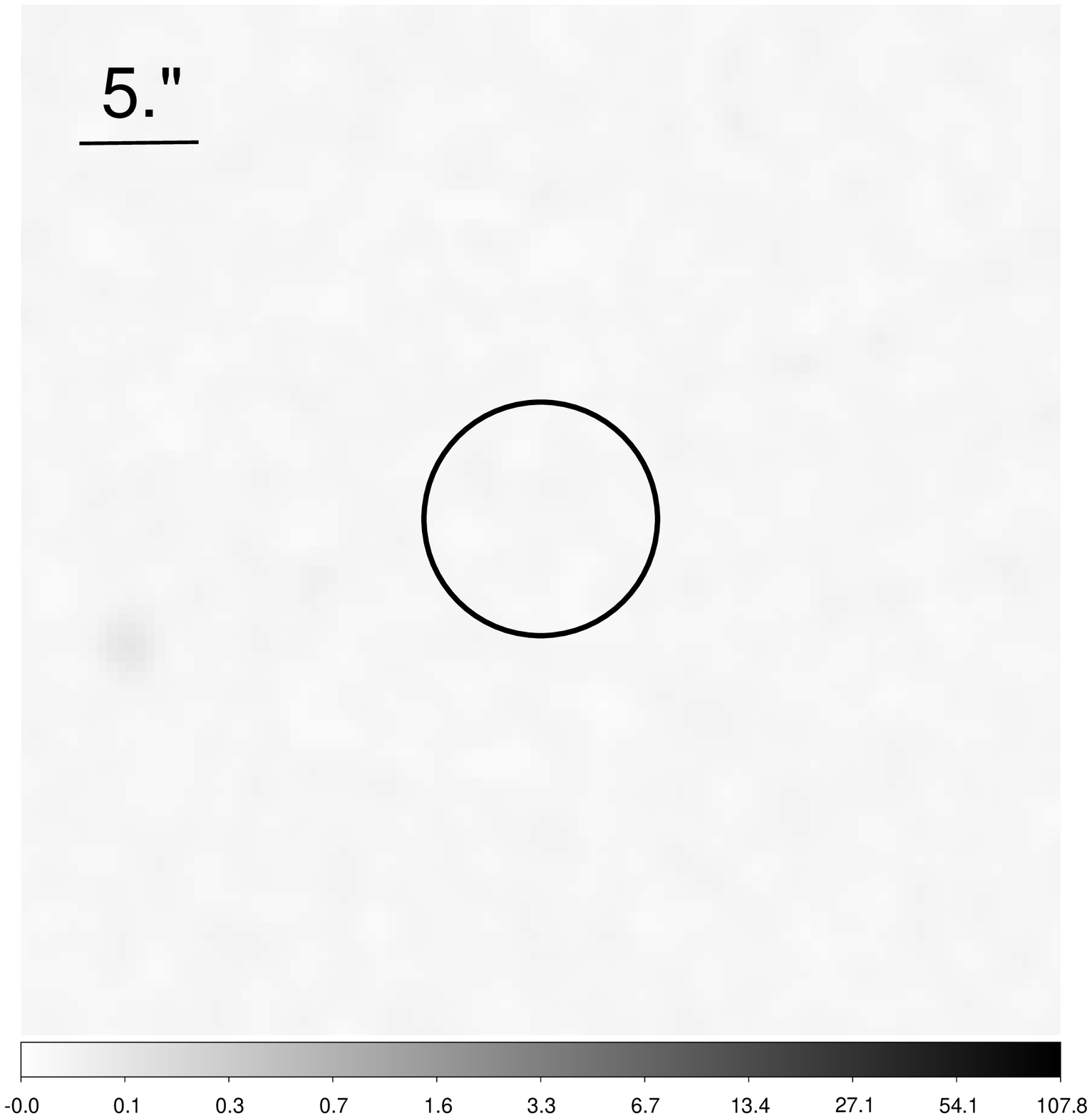}}
%\caption{\scriptsize{Src No.5}}
  \subfloat[Src No.6]{\includegraphics[clip, trim={0.0cm 2.cm 0.cm 0.0cm},width=0.20\textwidth]{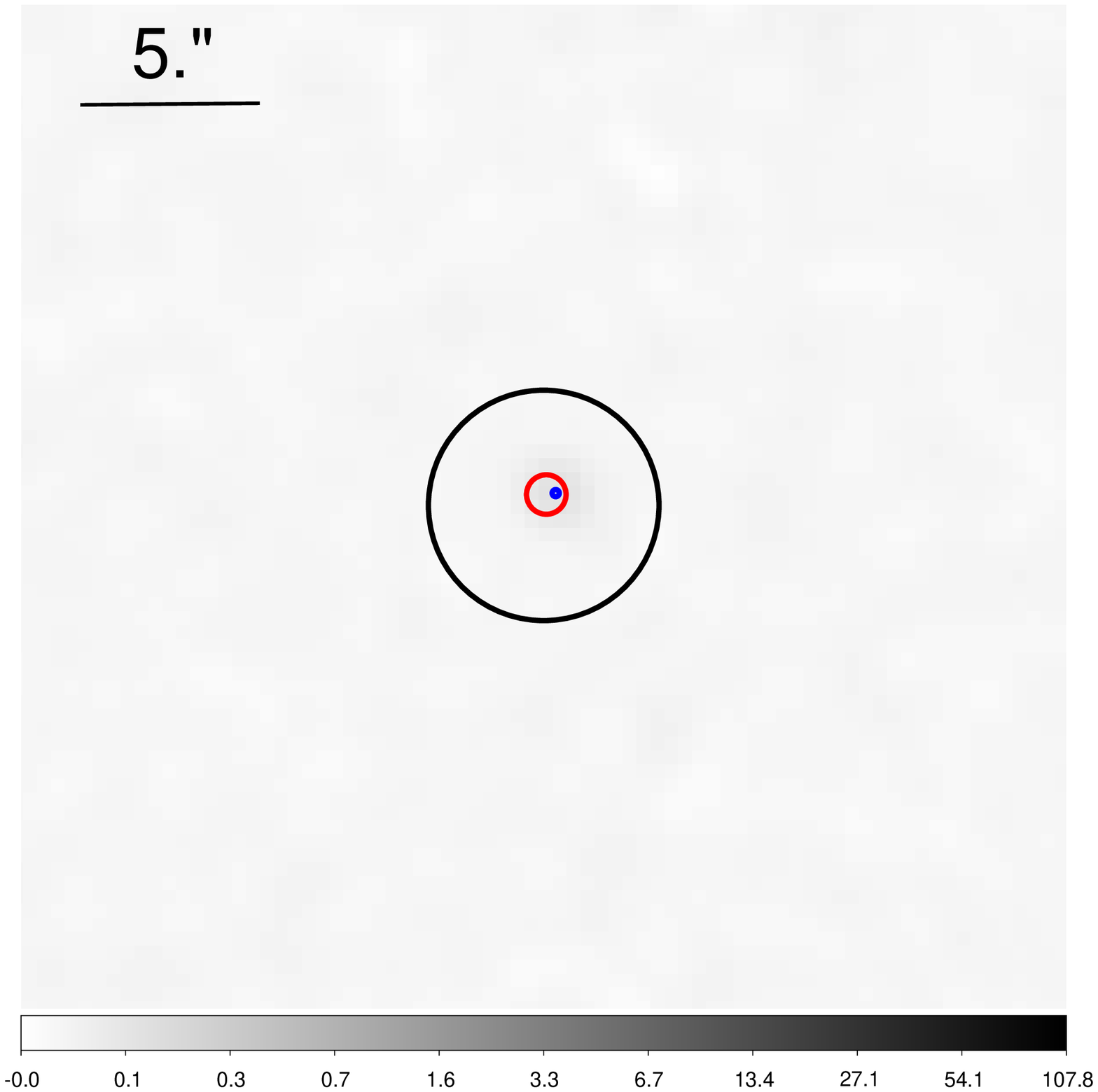}}
%\caption{\scriptsize{Src No.6}}
  \subfloat[Src No.7]{\includegraphics[clip, trim={0.0cm 2.cm 0.cm 0.0cm},width=0.20\textwidth]{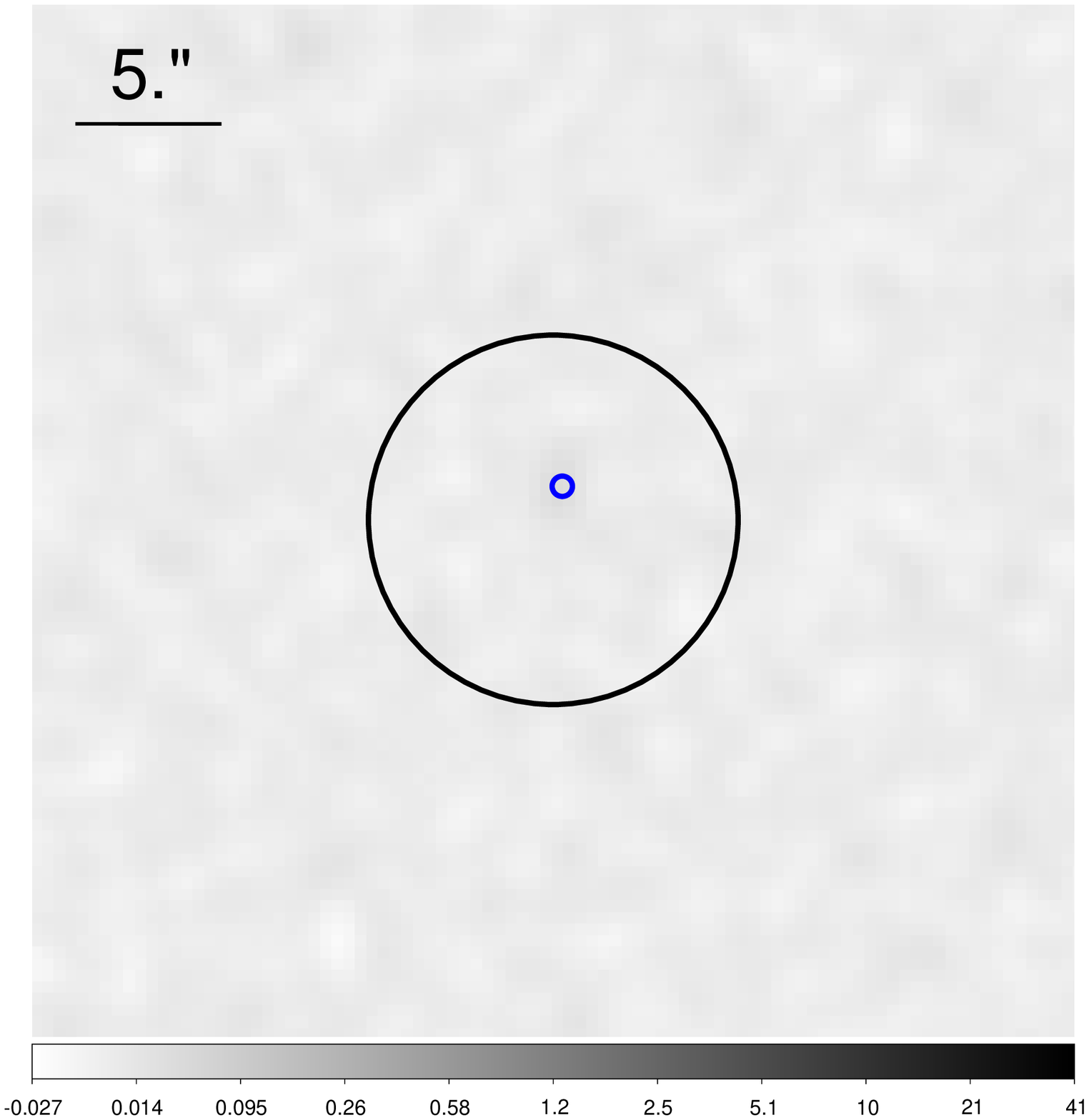}}
%\caption{\scriptsize{Src No.7}}
  \subfloat[Src No.8]{\includegraphics[clip, trim={0.0cm 2.cm 0.cm 0.0cm},width=0.20\textwidth]{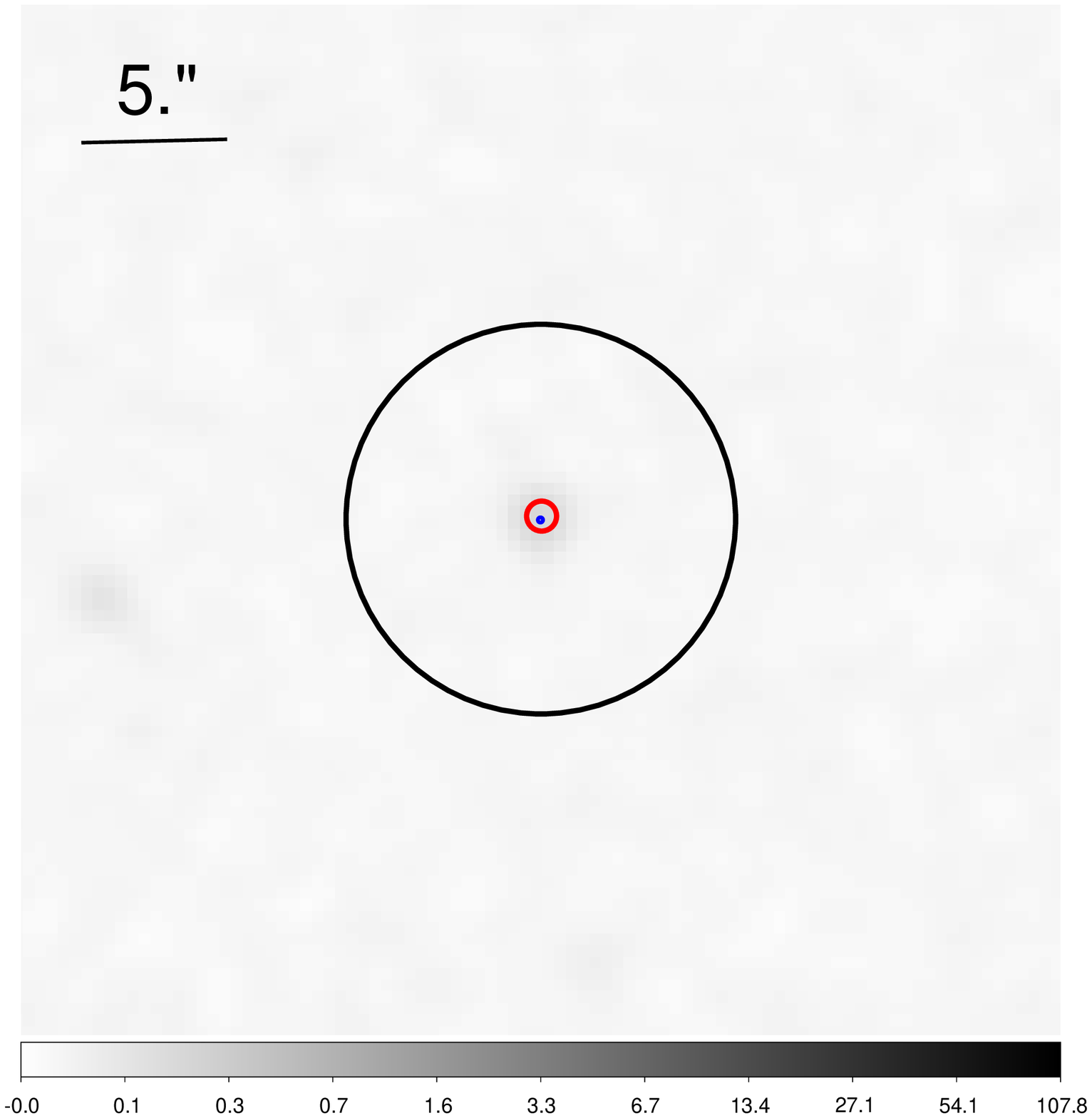}}\\
%\caption{\scriptsize{Src No.8}}
  \subfloat[Src No.9]{\includegraphics[clip, trim={0.0cm 2.cm 0.cm 0.0cm},width=0.20\textwidth]{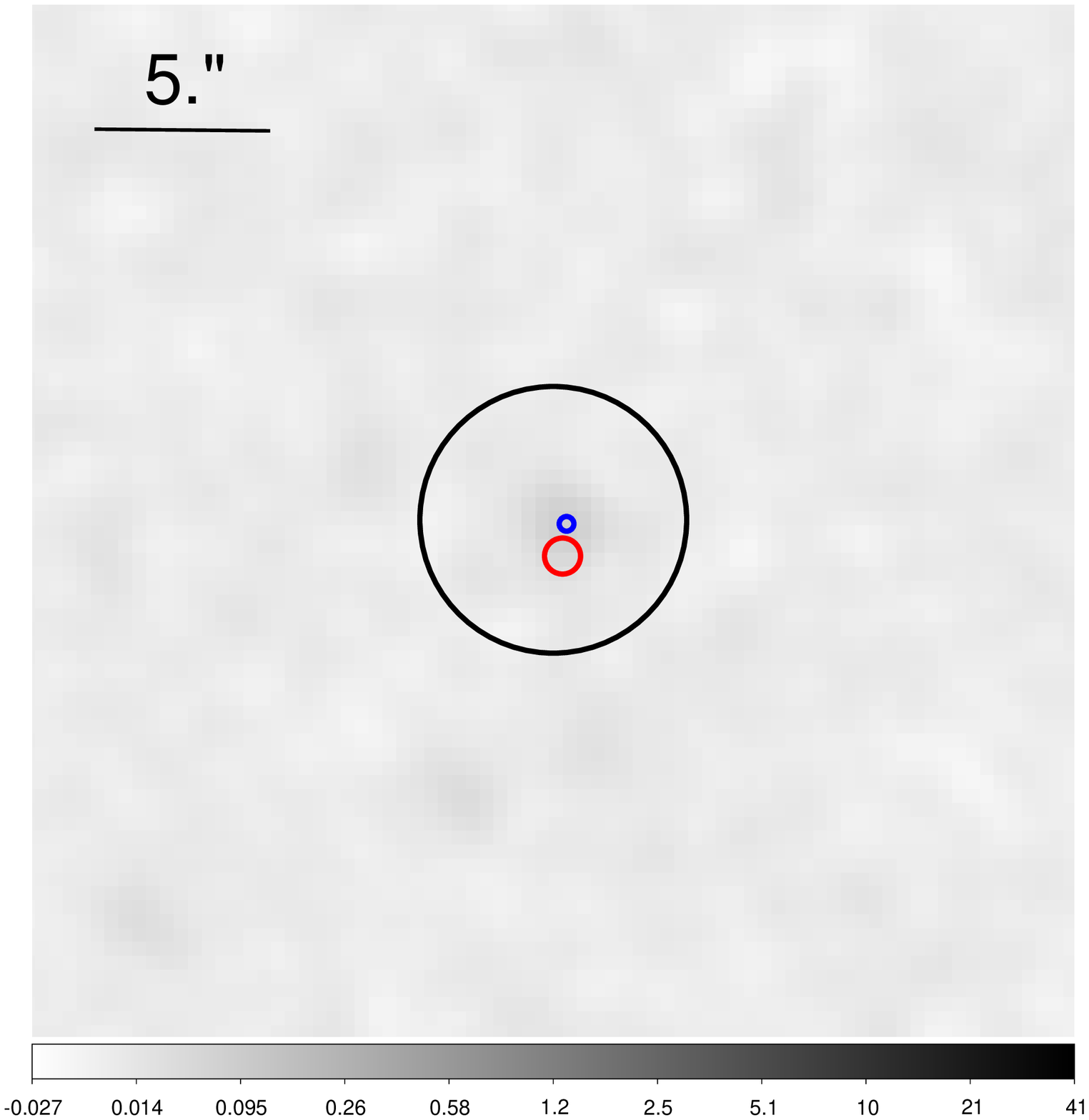}}
%\caption{\scriptsize{Src No.9}}
  \subfloat[Src No.10]{\includegraphics[clip, trim={0.0cm 2.cm 0.cm 0.0cm},width=0.20\textwidth]{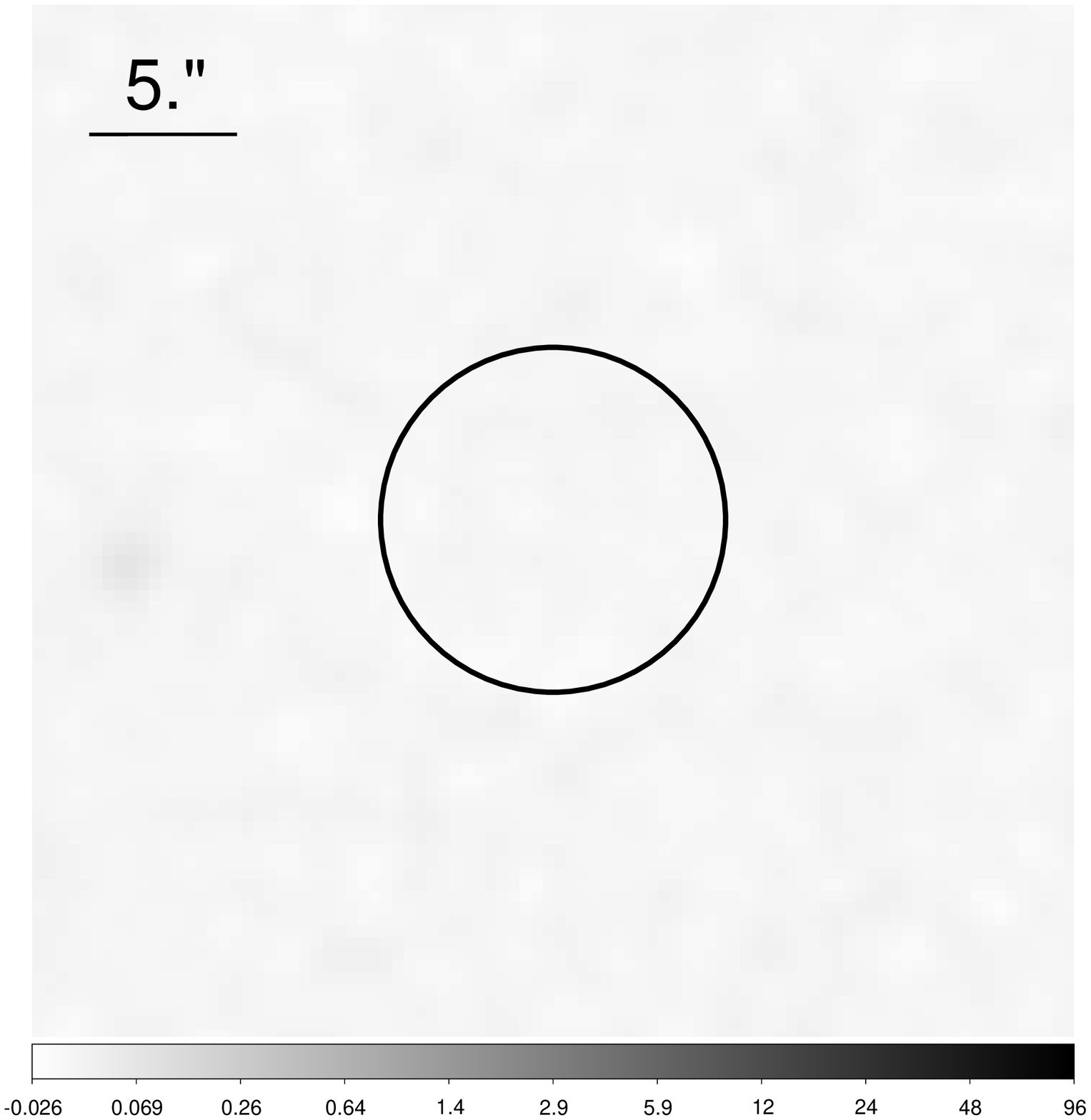}}
%\caption{\scriptsize{Src No.10}}
  \subfloat[Src No.11]{\includegraphics[clip, trim={0.0cm 2.cm 0.cm 0.0cm},width=0.20\textwidth]{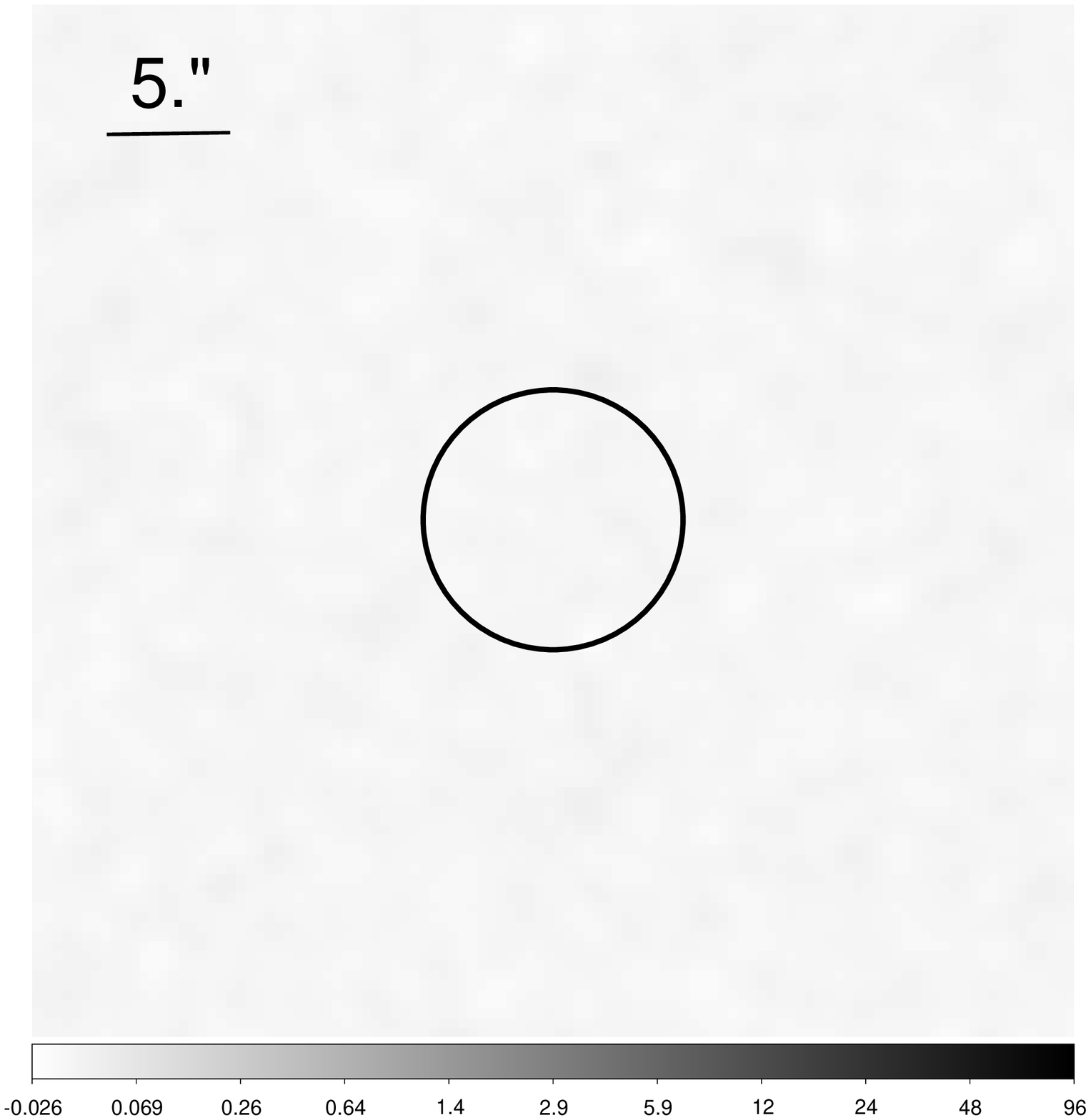}}
%\caption{\scriptsize{Src No.11}}
  \subfloat[Src No.12]{\includegraphics[clip, trim={0.0cm 2.cm 0.cm 0.0cm},width=0.20\textwidth]{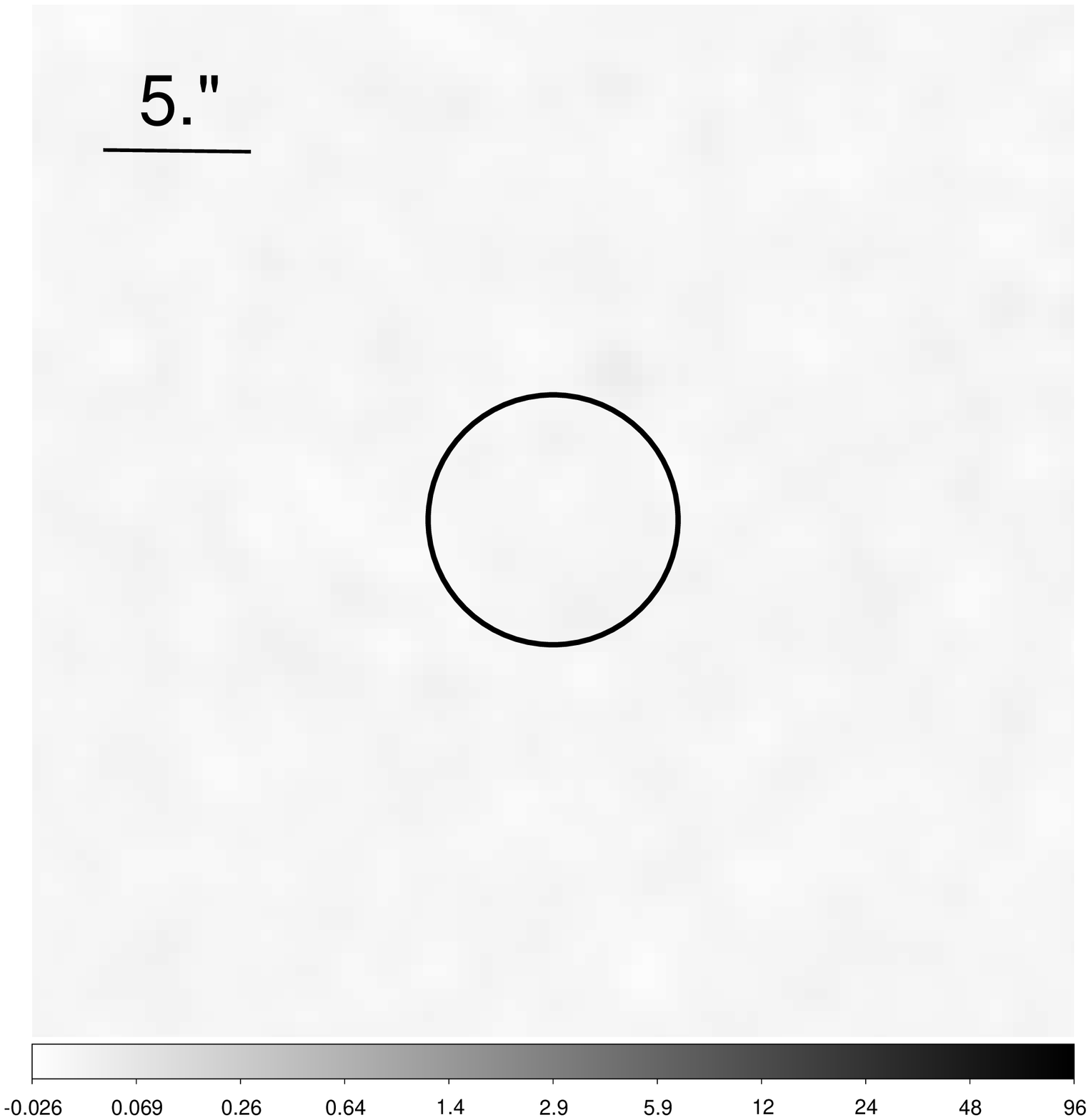}}\\
%\caption{\scriptsize{Src No.12}}
  \subfloat[Src No.13]{\includegraphics[clip, trim={0.0cm 2.cm 0.cm 0.0cm},width=0.20\textwidth]{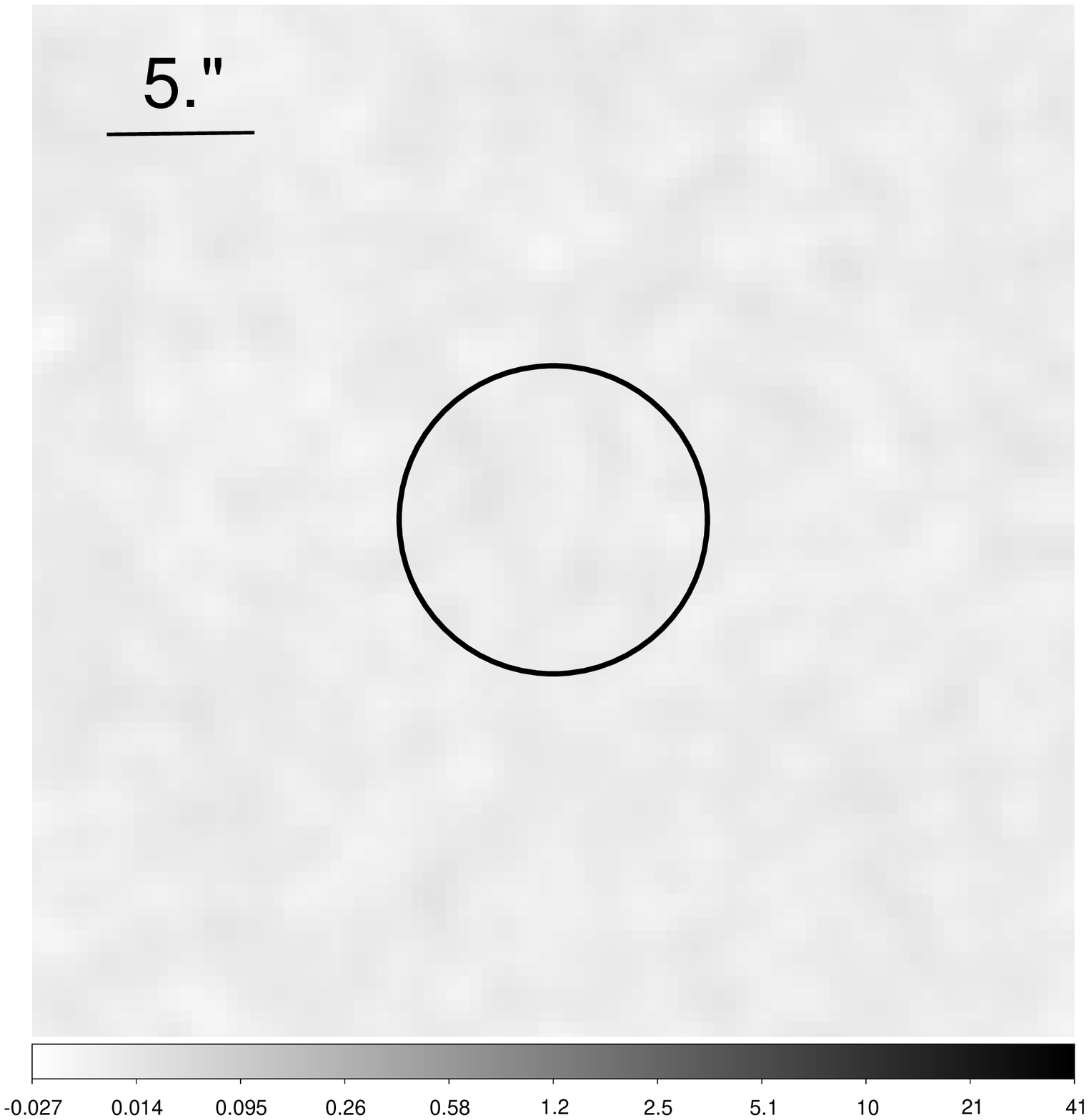}}
%\caption{\scriptsize{Src No.13}}
 \subfloat[Src No.14]{\includegraphics[clip, trim={0.0cm 2.cm 0.cm 0.0cm},width=0.20\textwidth]{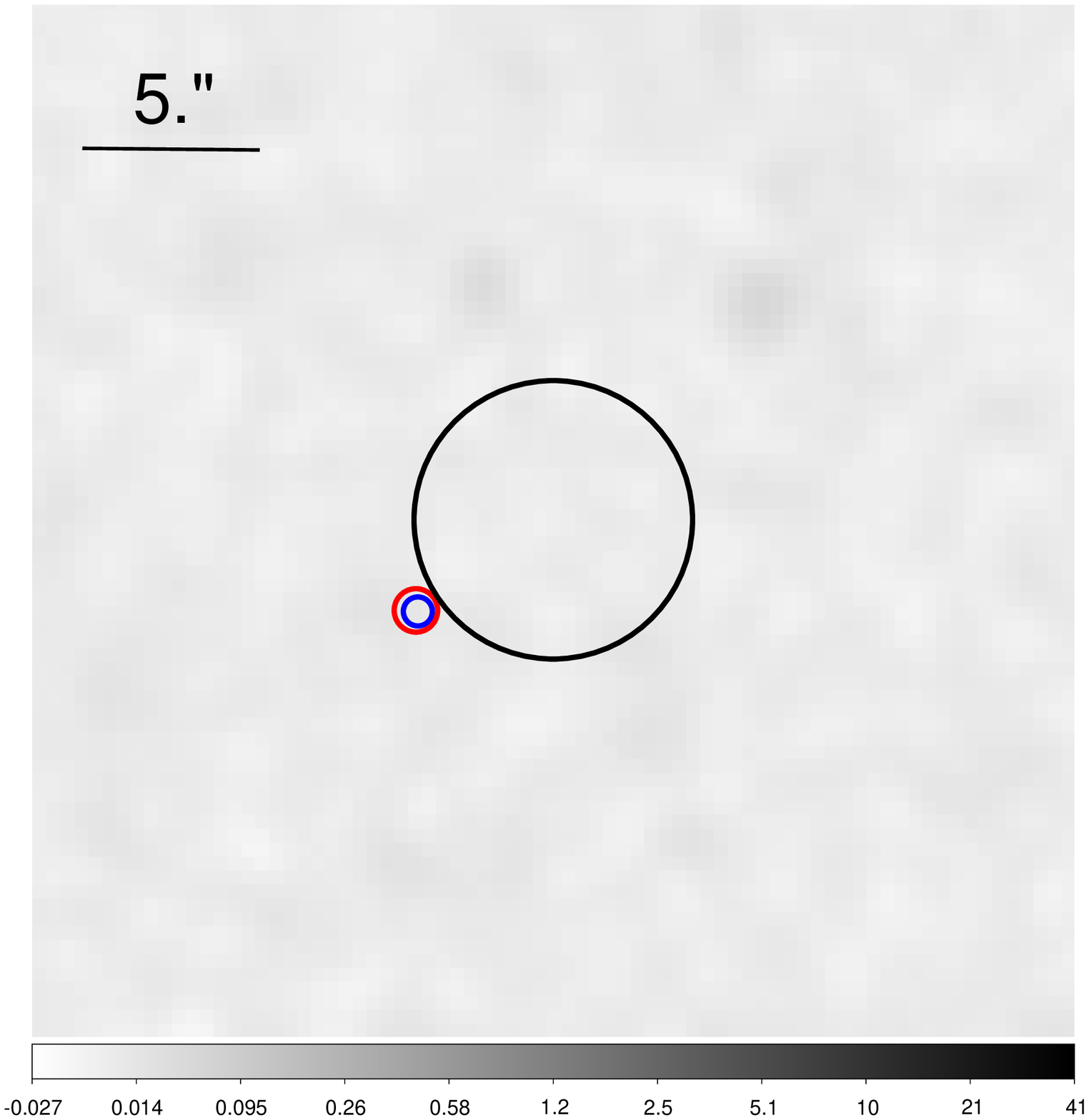}}
%\caption{\scriptsize{Src No.14}}
  \subfloat[Src No.15]{\includegraphics[clip, trim={0.0cm 2.cm 0.cm 0.0cm},width=0.20\textwidth]{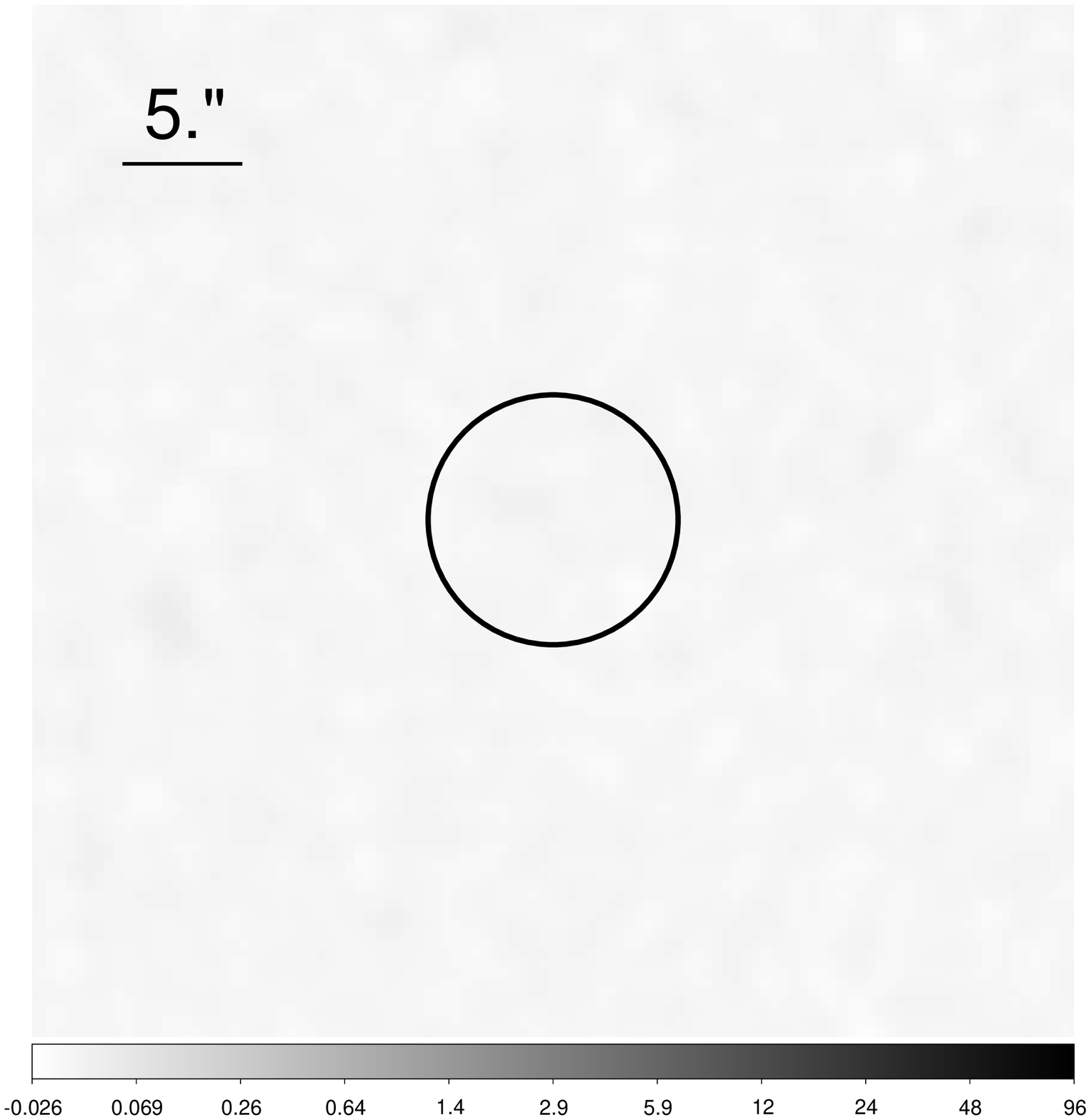}}
%\caption{\scriptsize{Src No.15}}
  \subfloat[Src No.16]{\includegraphics[clip, trim={0.0cm 2.cm 0.cm 0.0cm},width=0.20\textwidth]{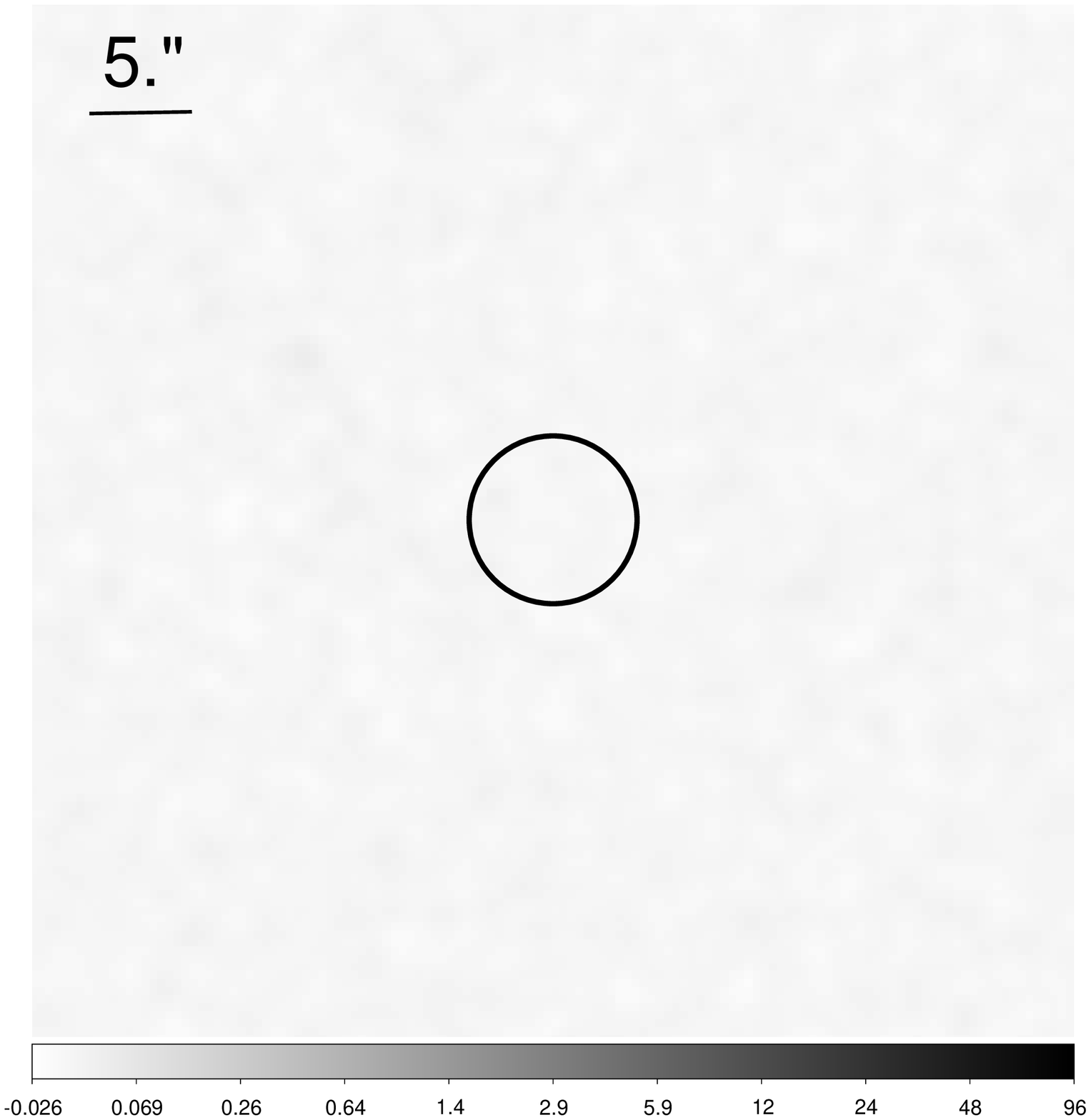}}\\
%\caption{\scriptsize{Src No.16}}
  \subfloat[Src No.17]{\includegraphics[clip, trim={0.0cm 2.cm 0.cm 0.0cm},width=0.20\textwidth]{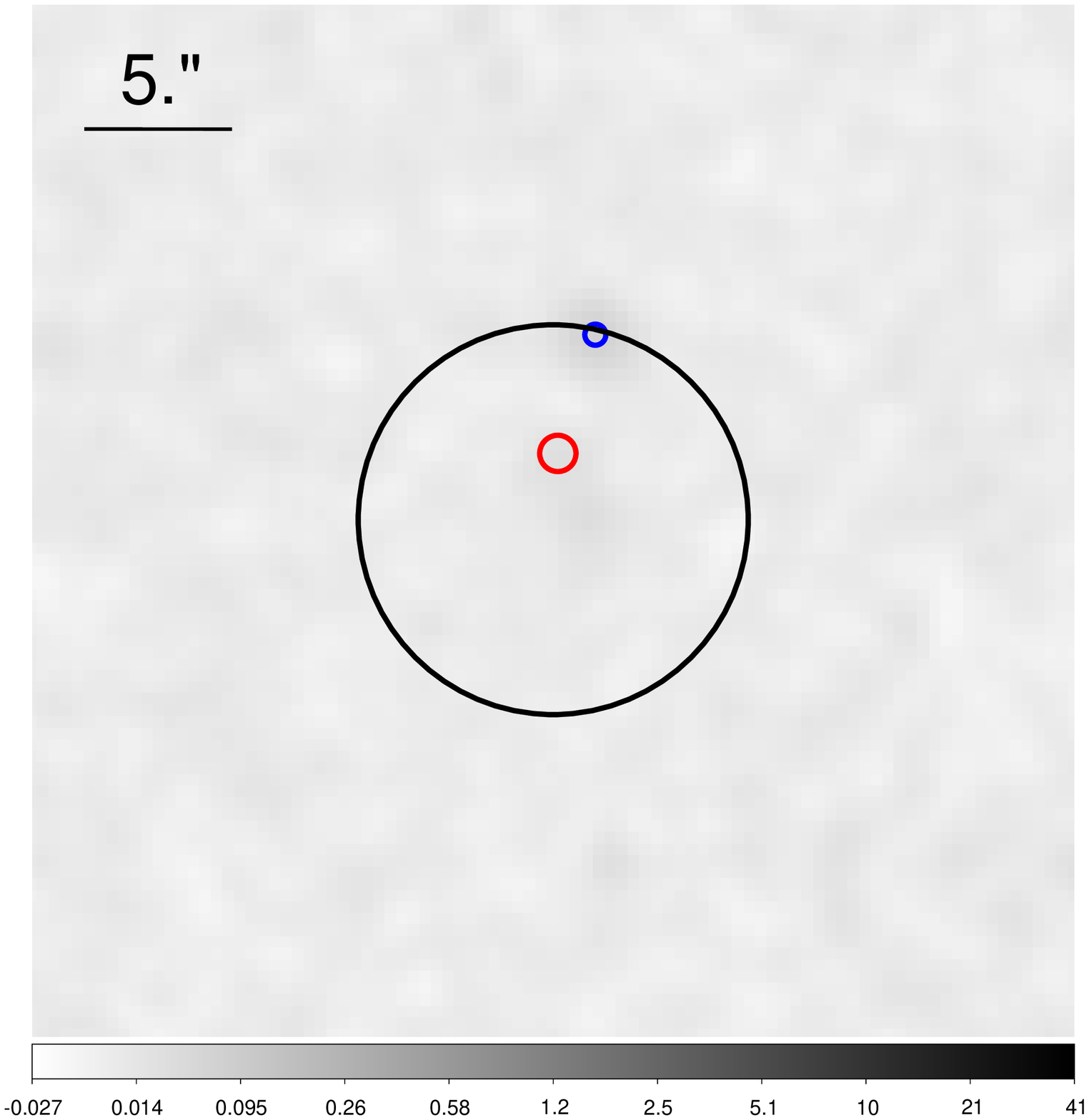}}
%\caption{\scriptsize{Src No.17}}
\subfloat[Src No.18]{\includegraphics[clip, trim={0.0cm 2.cm 0.cm 0.0cm},width=0.20\textwidth]{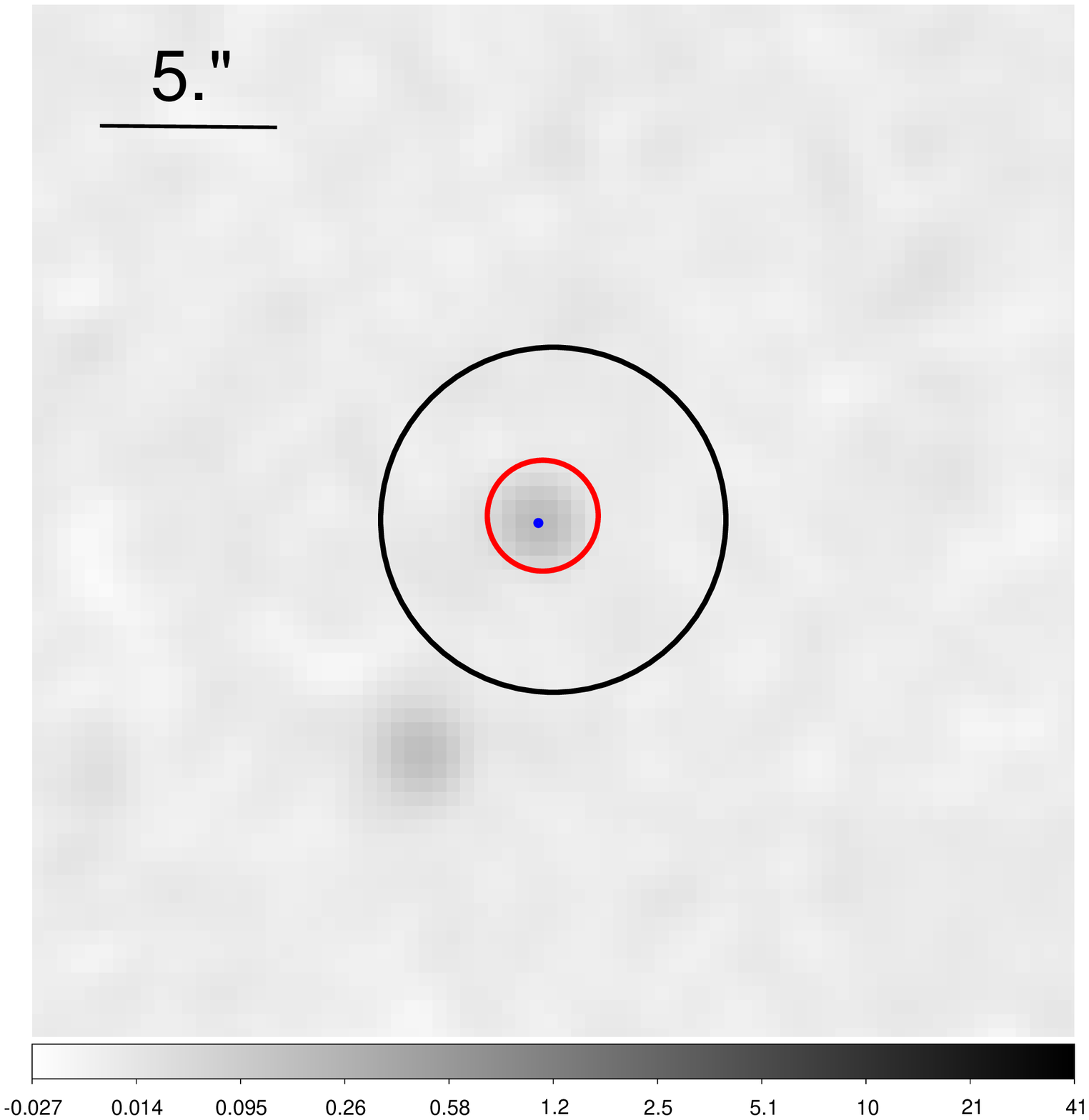}}
 %\caption{\scriptsize{Src No.18}}
 \subfloat[Src No.19]{\includegraphics[clip, trim={0.0cm 2.cm 0.cm 0.0cm},width=0.20\textwidth]{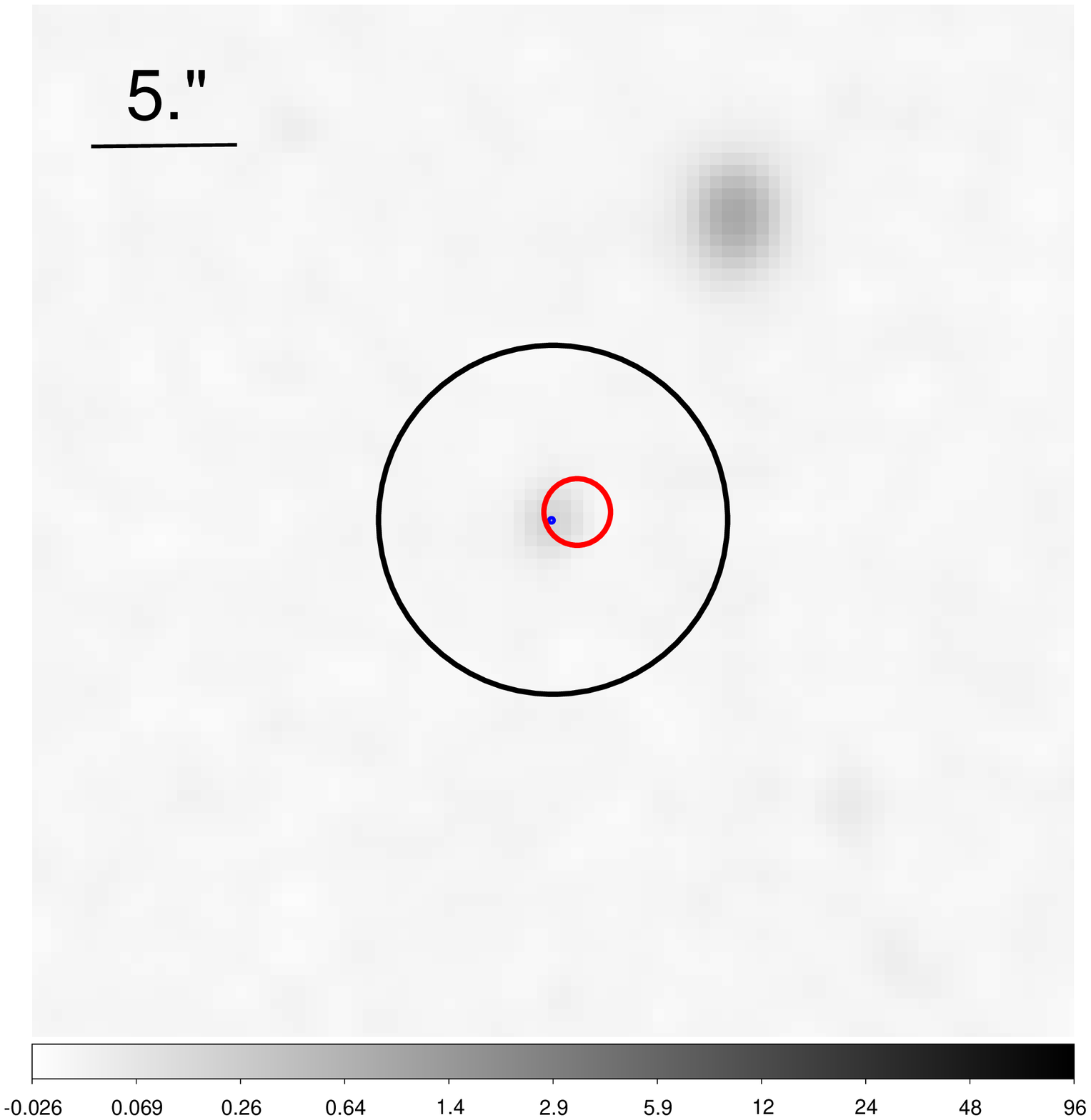}}
 %\caption{\scriptsize{Src No.19}}
 \subfloat[Src No.20]{\includegraphics[clip, trim={0.0cm 2.cm 0.cm 0.0cm},width=0.20\textwidth]{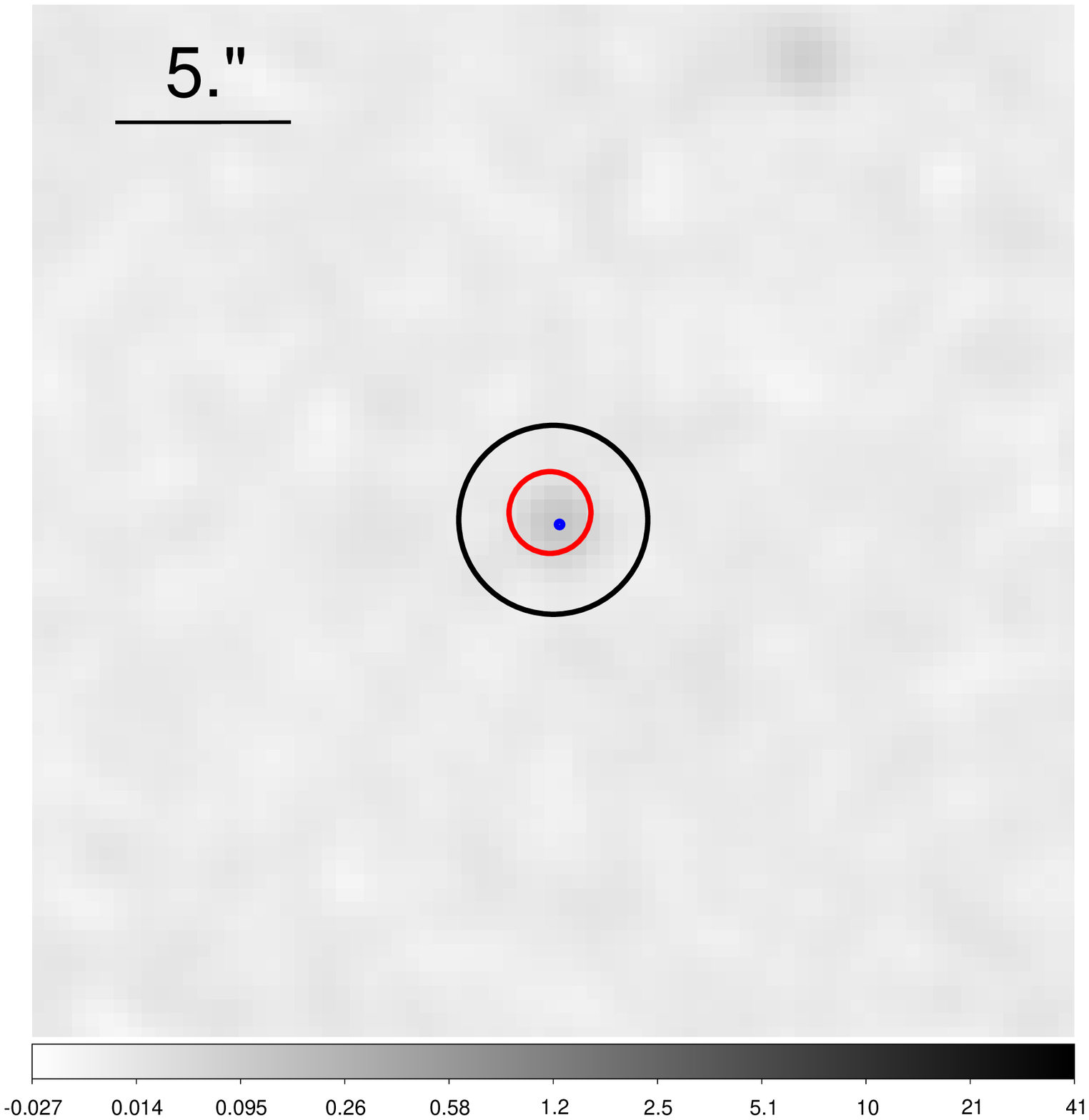}}\\
 %\caption{\scriptsize{Src No.20}}
  \end{figure*}
\pagebreak
\clearpage
%\hspace{0.3cm}Appendix B continued: Image of optical SDSS9 counterparts
\begin{figure*}
\vspace{-0.3cm}
\subfloat[Src No.21]{\includegraphics[clip, trim={0.0cm 2.cm 0.cm 0.0cm},width=0.19\textwidth]{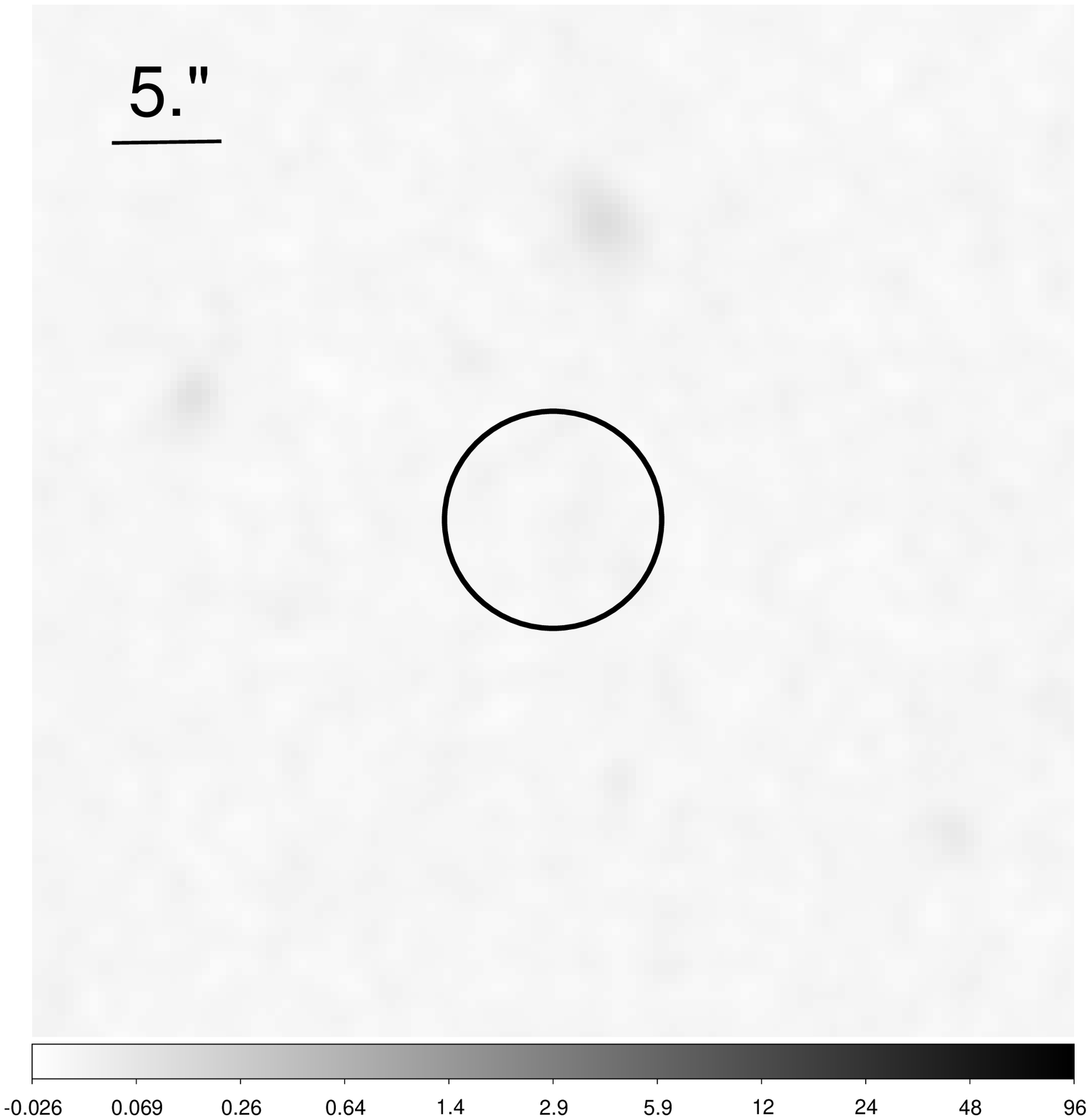}}
  %\caption{\scriptsize{Src No.21}}
 \subfloat[Src No.22]{\includegraphics[clip, trim={0.0cm 2.cm 0.cm 0.0cm},width=0.19\textwidth]{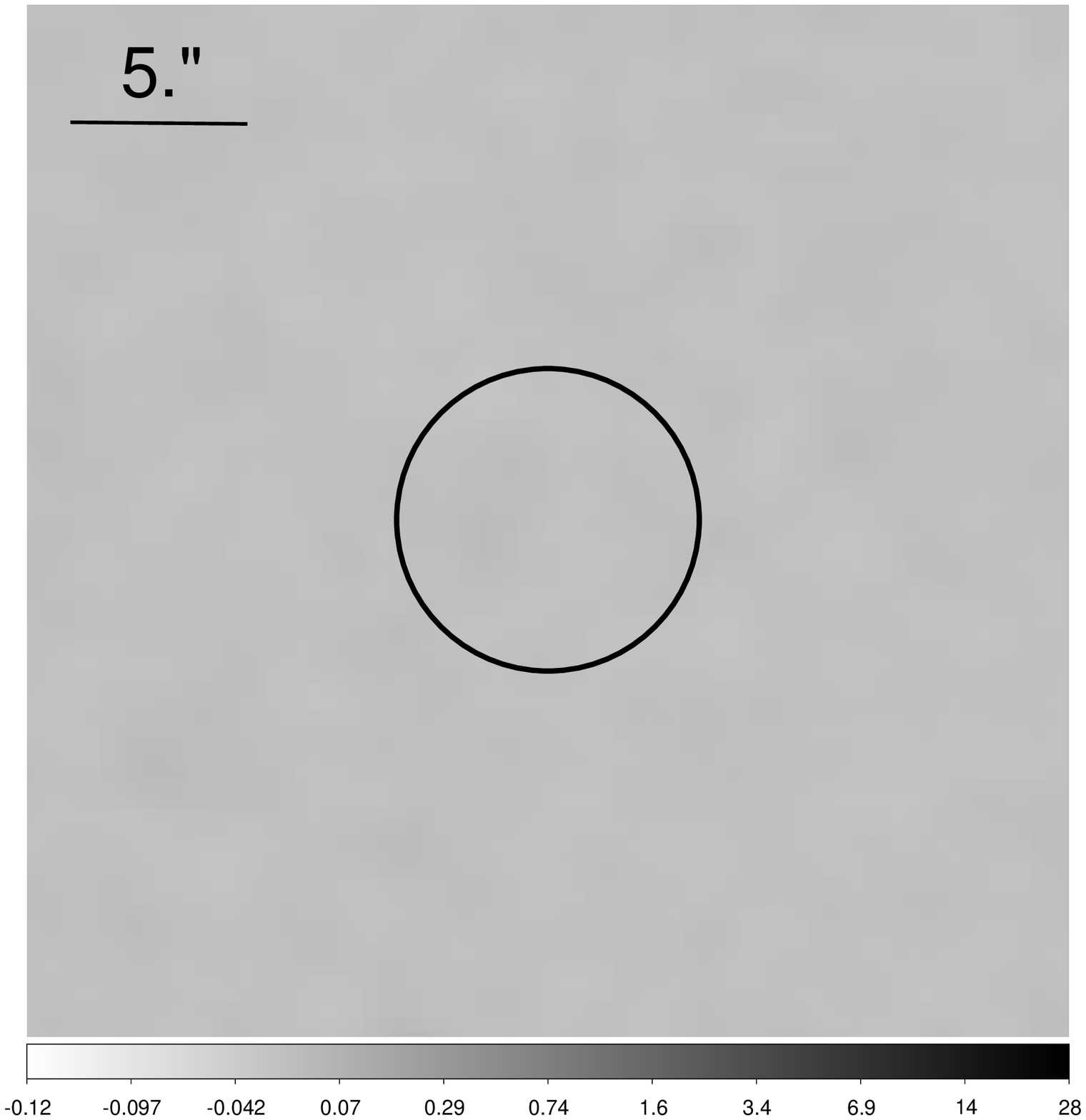}}
  %\caption{\scriptsize{Src No.22}}
   \subfloat[Src No.23]{\includegraphics[clip, trim={0.0cm 2.cm 0.cm 0.0cm},width=0.19\textwidth]{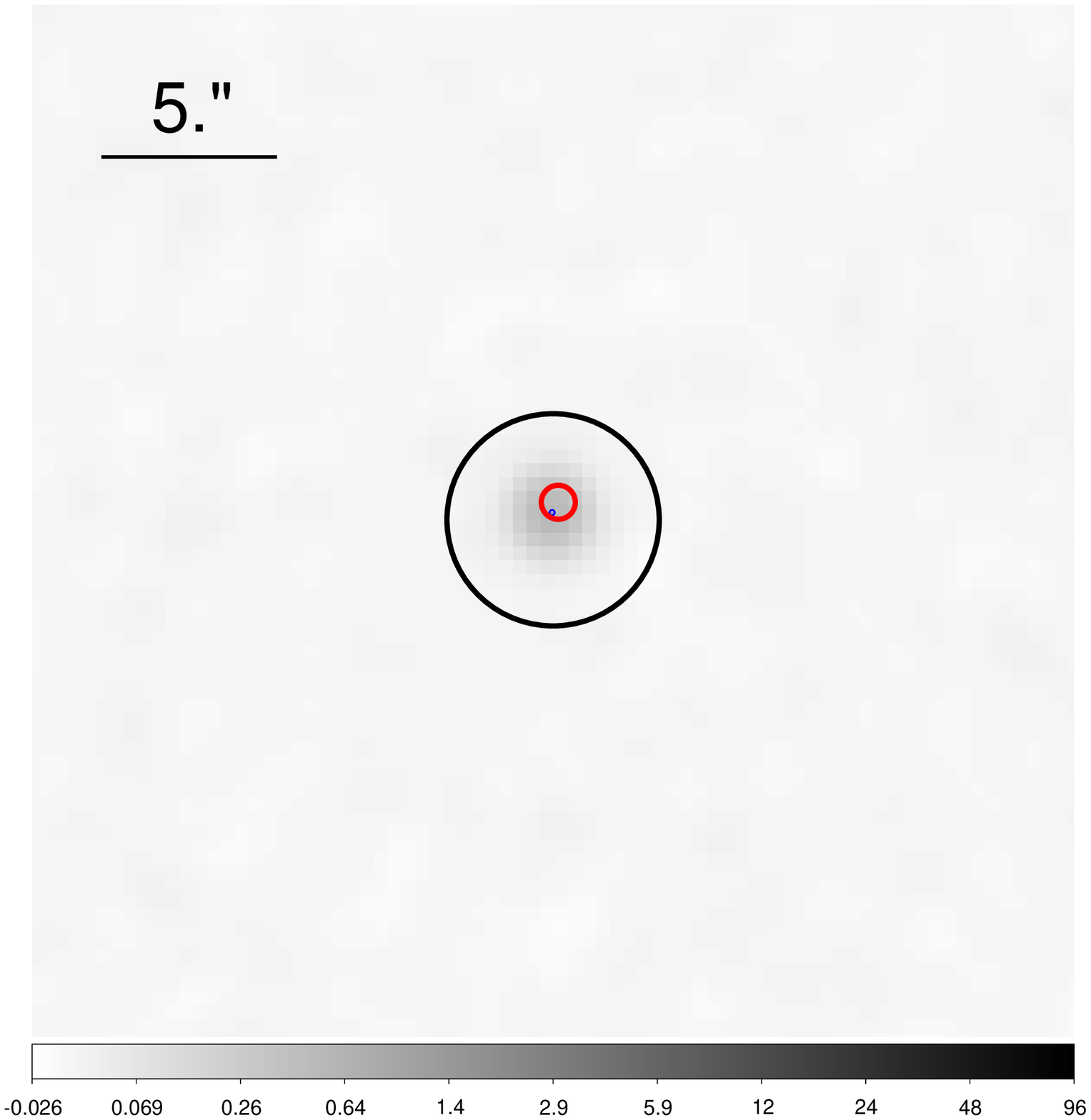}}
 %\caption{\scriptsize{Src No.23}}
   \subfloat[Src No.24]{\includegraphics[clip, trim={0.0cm 2.cm 0.cm 0.0cm},width=0.19\textwidth]{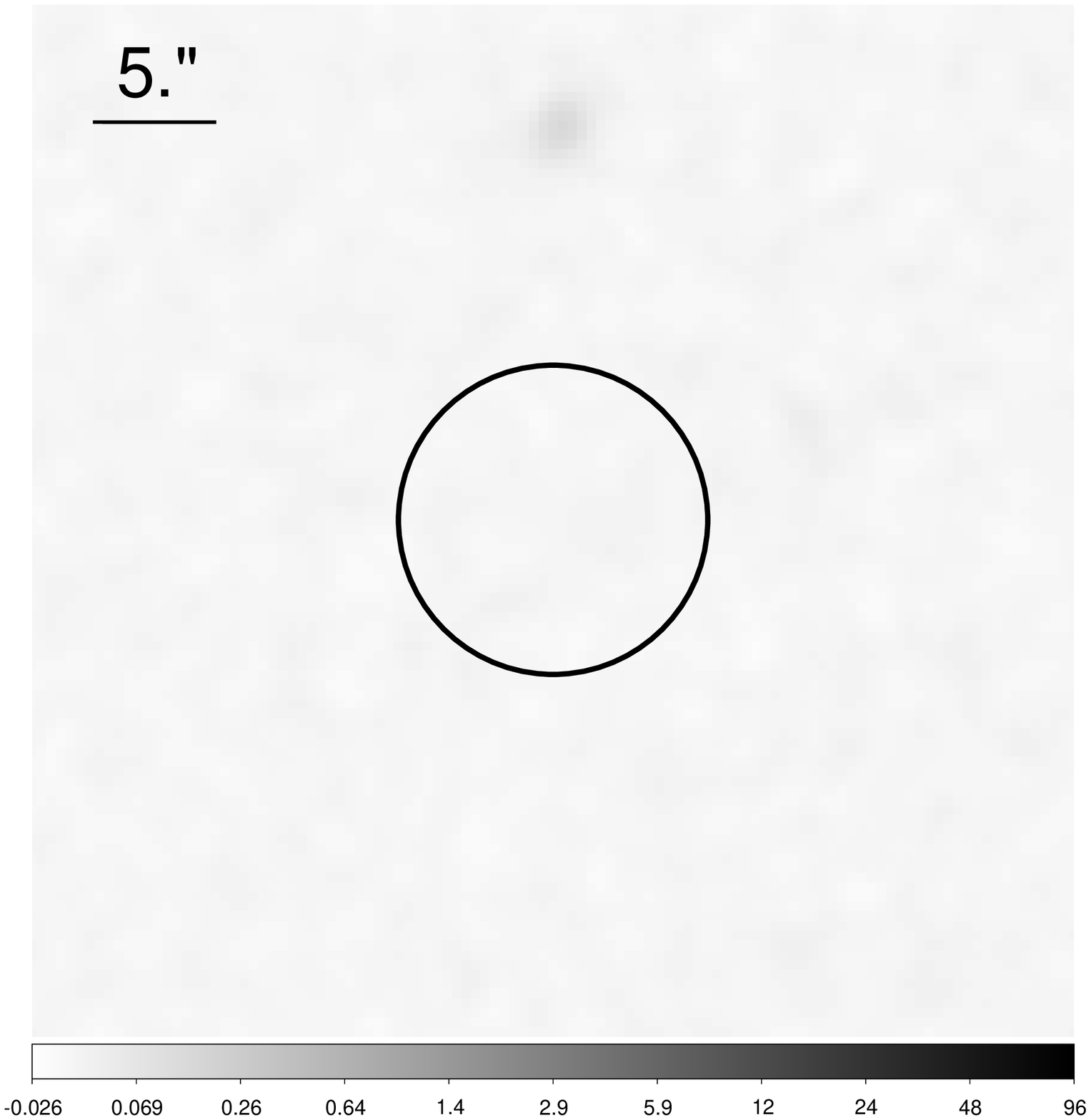}}\\
%\caption{\scriptsize{Src No.24}}
  \subfloat[Src No.25]{\includegraphics[clip, trim={0.0cm 2.cm 0.cm 0.0cm},width=0.19\textwidth]{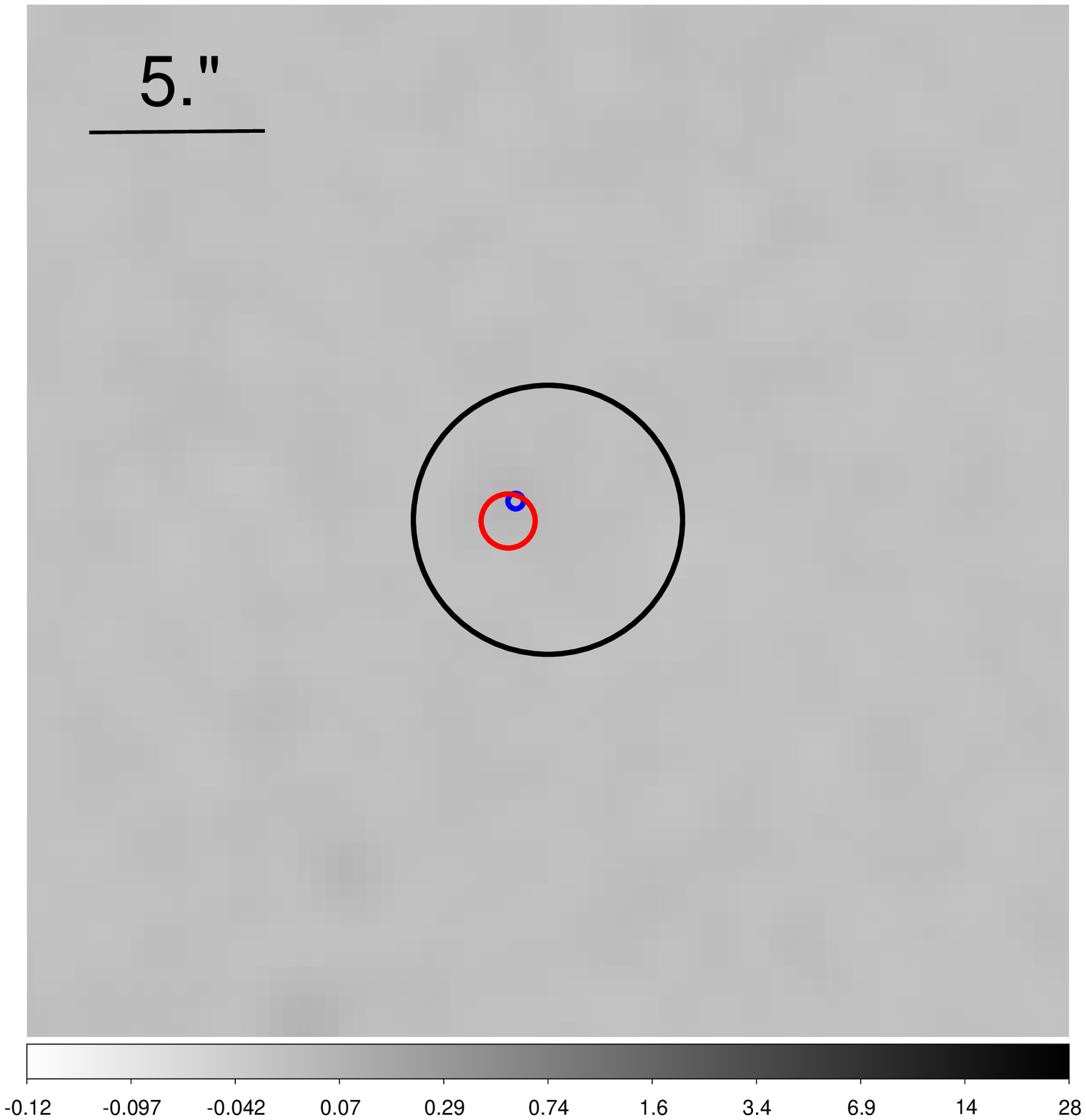}}
  %\caption{\scriptsize{Src No.25}}
 \subfloat[Src No.26]{\includegraphics[clip, trim={0.0cm 2.cm 0.cm 0.0cm},width=0.19\textwidth]{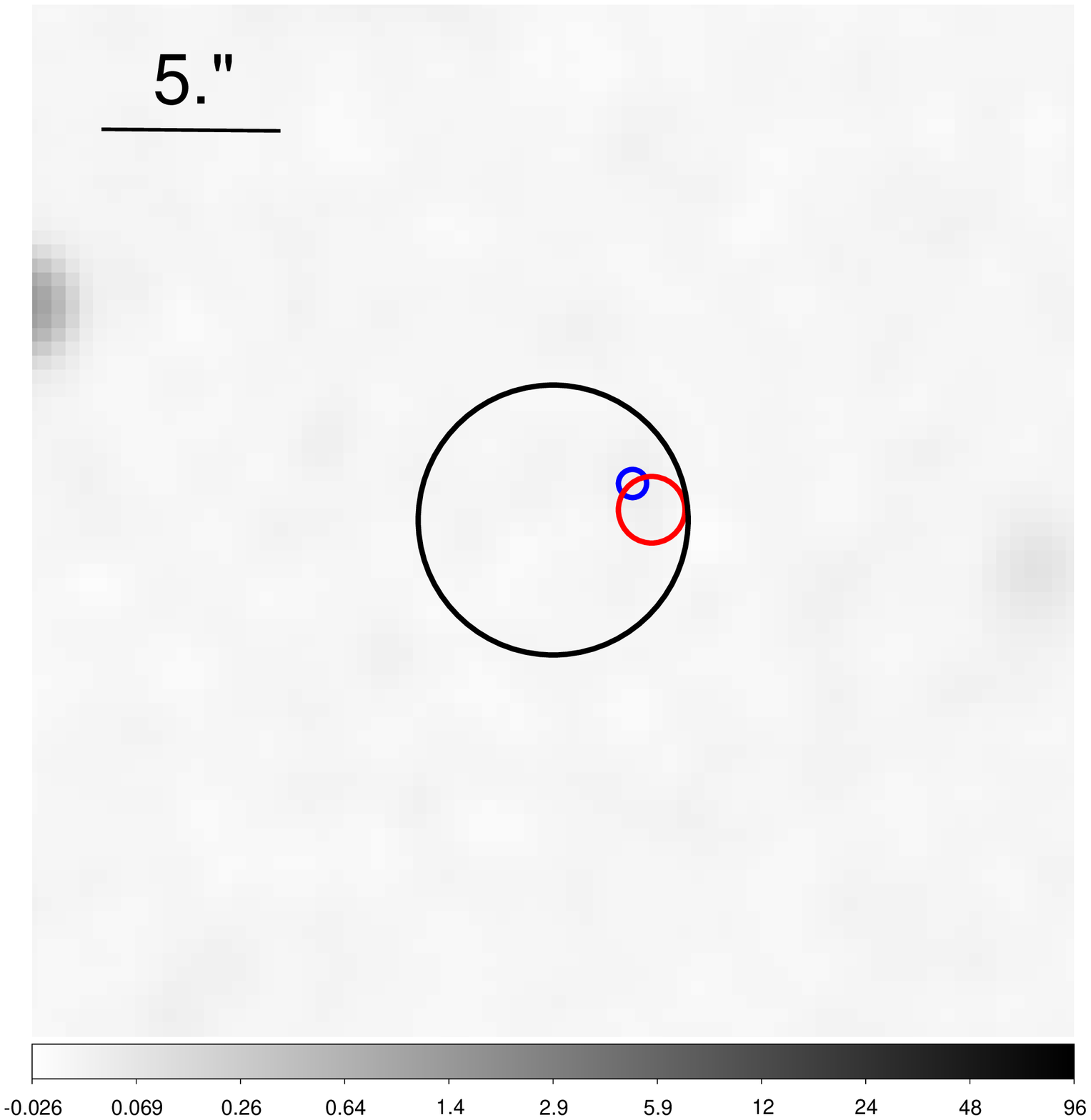}}
  %\caption{\scriptsize{Src No.26}}
 \subfloat[Src No.27]{\includegraphics[clip, trim={0.0cm 2.cm 0.cm 0.0cm},width=0.19\textwidth]{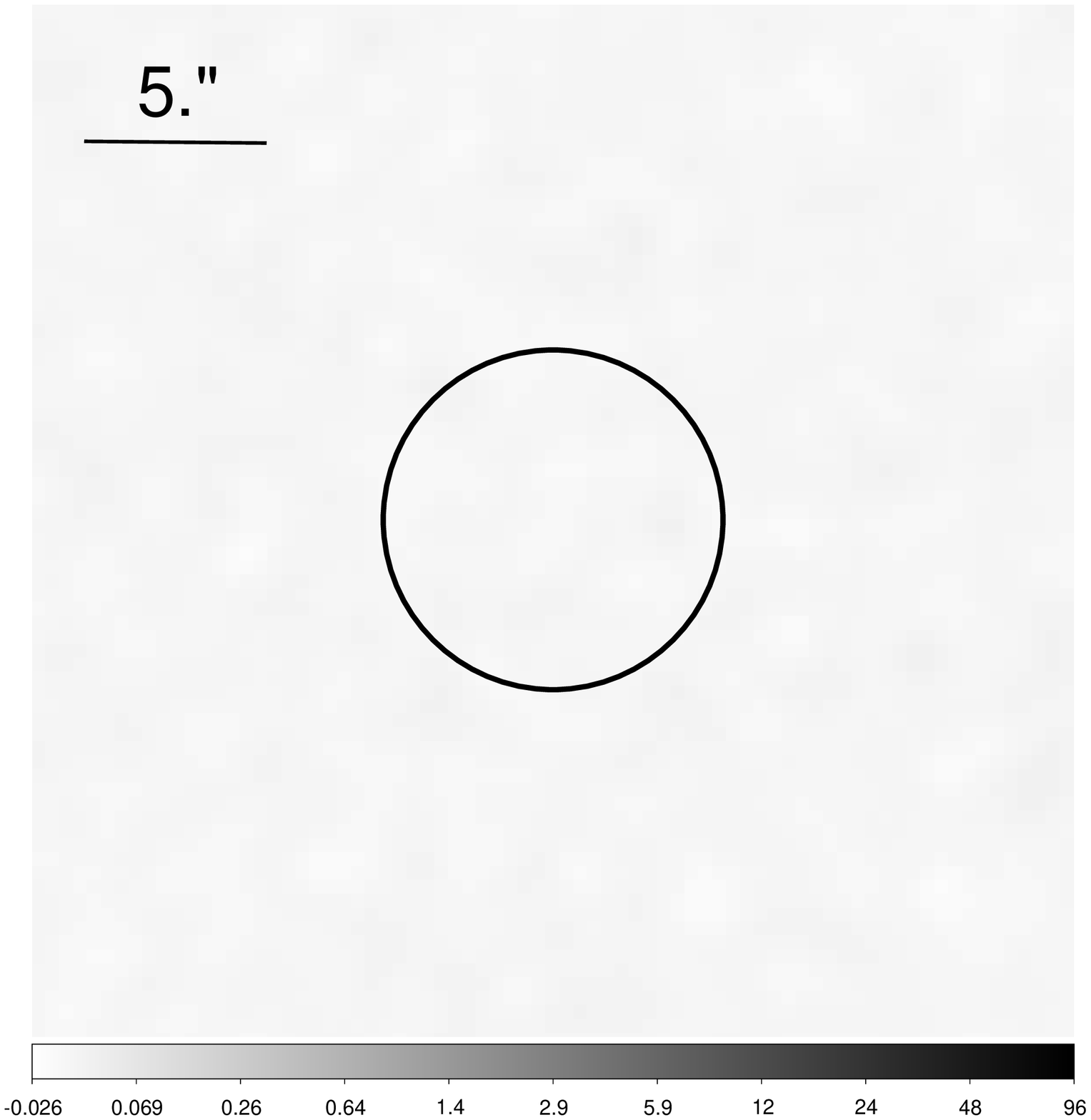}}
  %\caption{\scriptsize{Src No.27}} 
  \subfloat[Src No.28]{\includegraphics[clip, trim={0.0cm 2.cm 0.cm 0.0cm},width=0.19\textwidth]{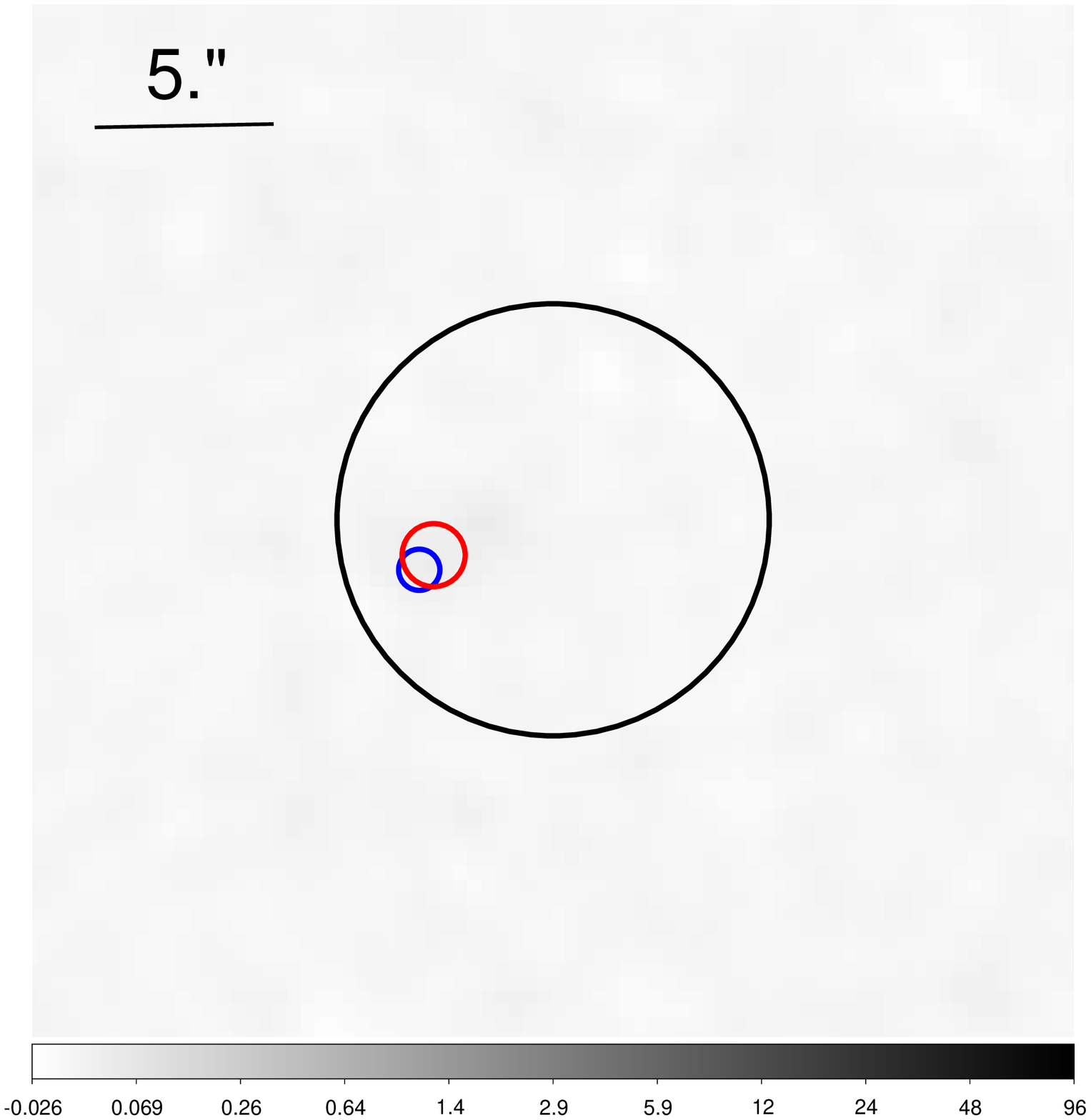}}\\
 %\caption{\scriptsize{Src No.28}}
  \subfloat[Src No.29]{\includegraphics[clip, trim={0.0cm 2.cm 0.cm 0.0cm},width=0.19\textwidth]{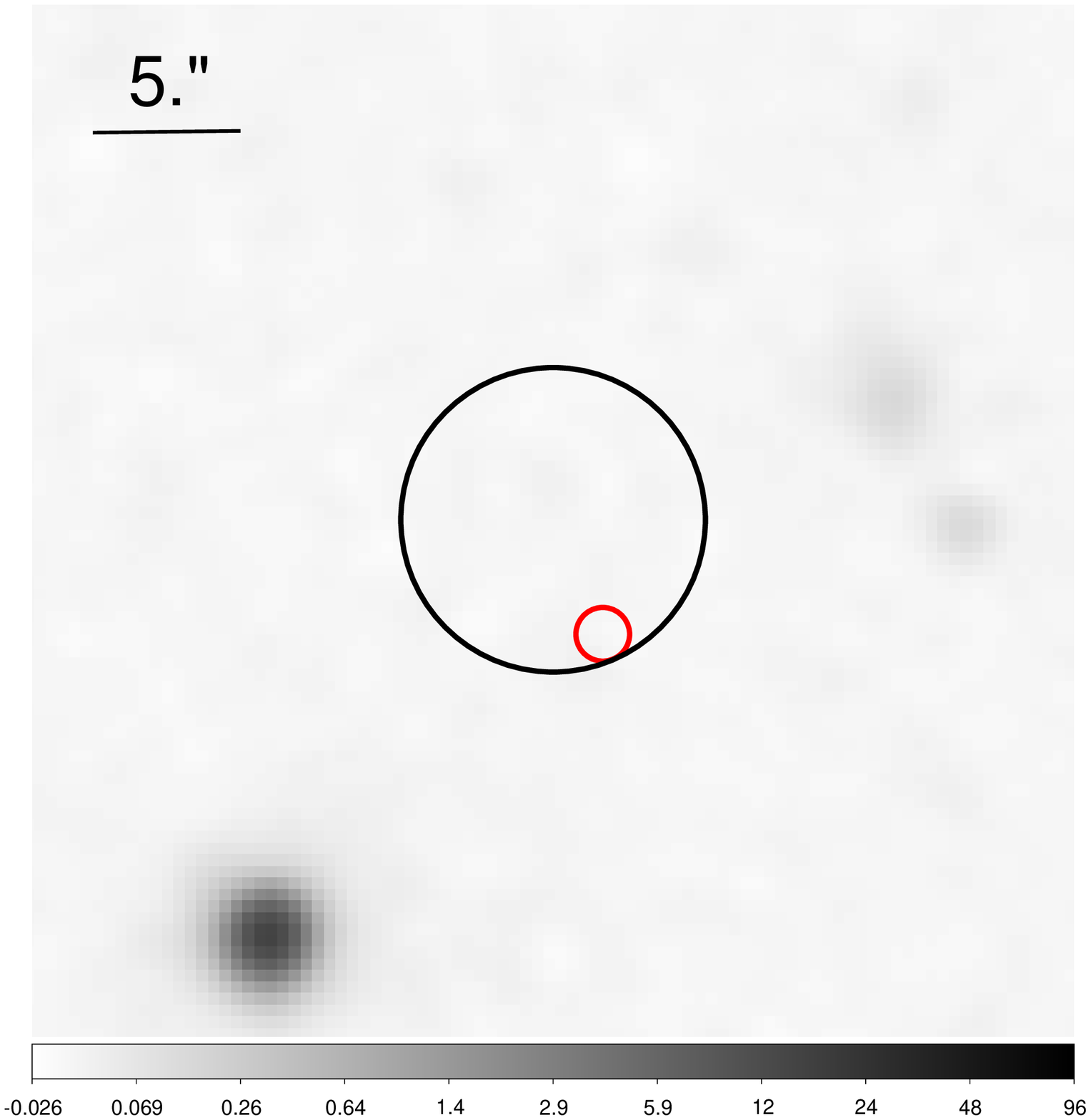}}
%\caption{\scriptsize{Src No.29}}
  \subfloat[Src No.30]{\includegraphics[clip, trim={0.0cm 2.cm 0.cm 0.0cm},width=0.19\textwidth]{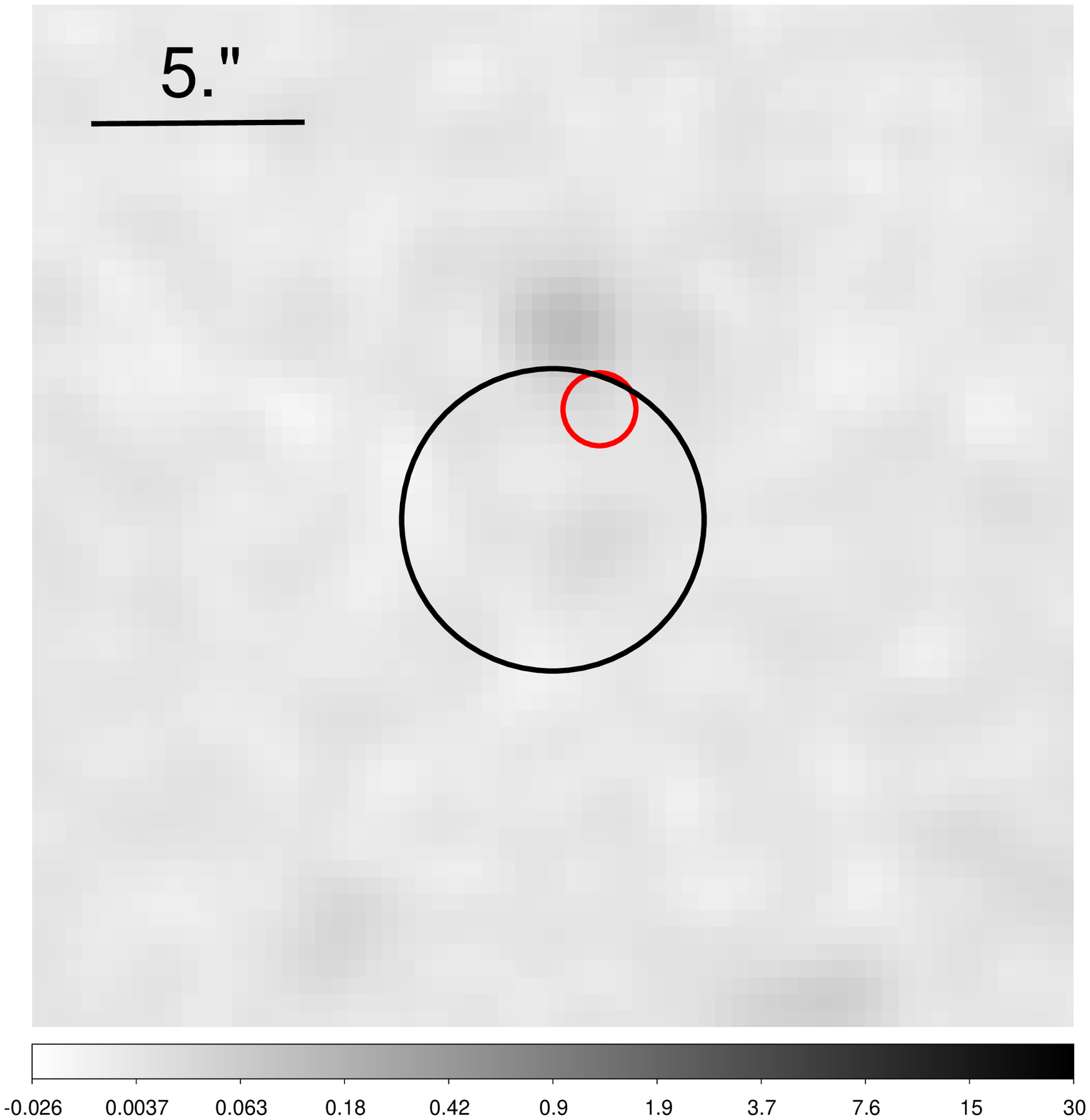}}
%\caption{\scriptsize{Src No.30}}
  \subfloat[Src No.31]{\includegraphics[clip, trim={0.0cm 2.cm 0.cm 0.0cm},width=0.19\textwidth]{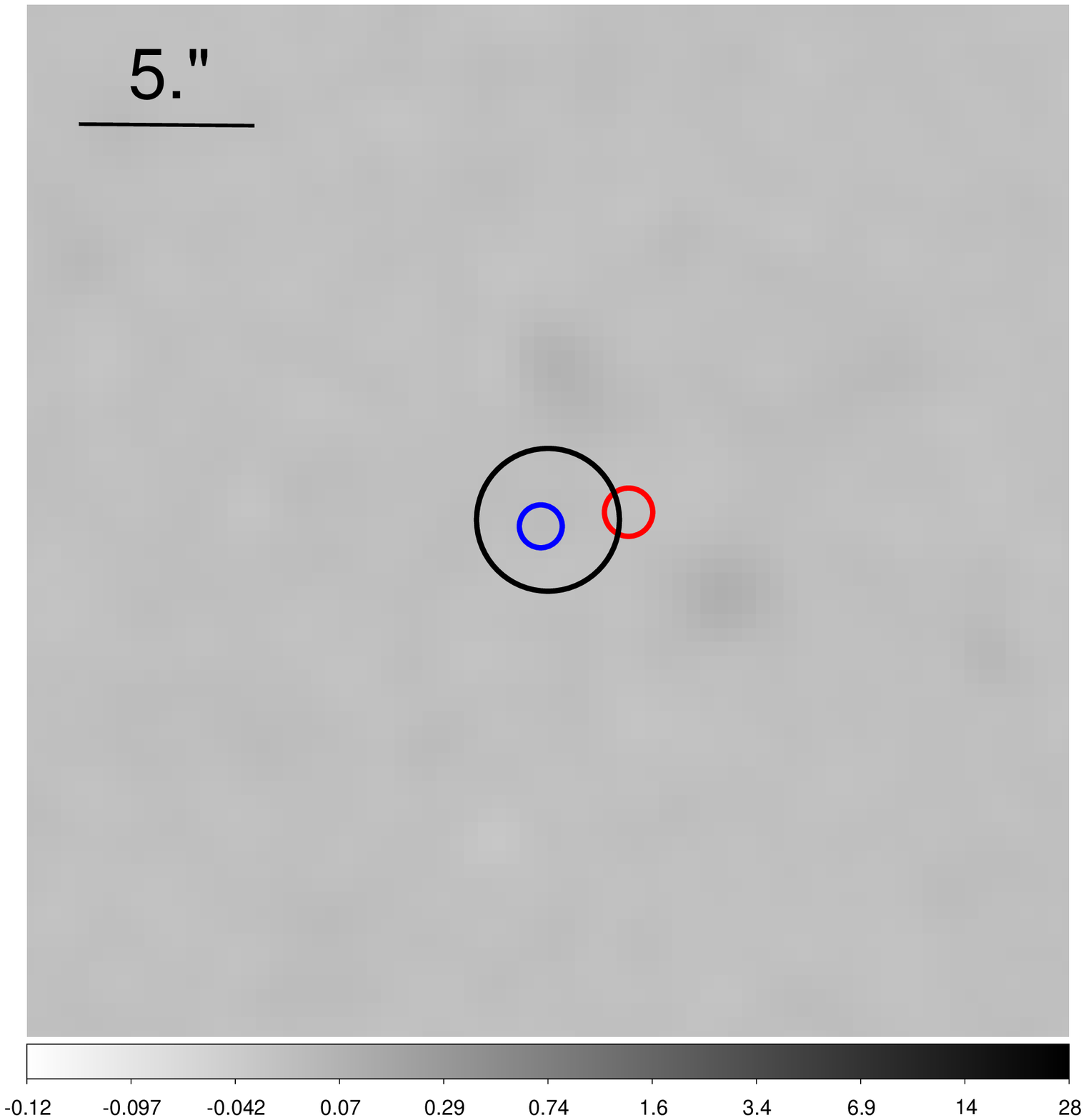}}
%\caption{\scriptsize{Src No.31}}
  \subfloat[Src No.32]{\includegraphics[clip, trim={0.0cm 2.cm 0.cm 0.0cm},width=0.19\textwidth]{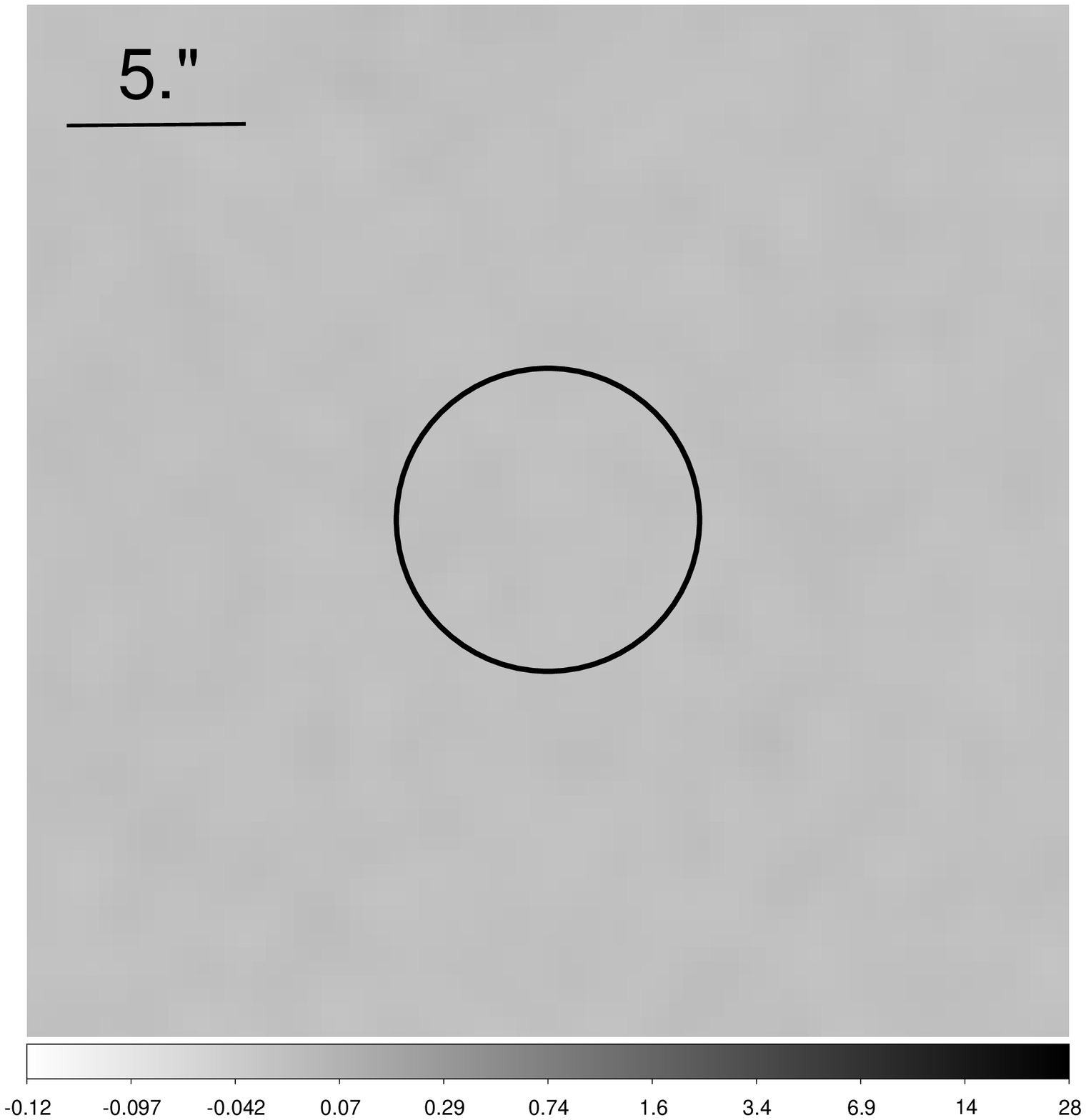}}\\
%\caption{\scriptsize{Src No.32}}
  \subfloat[Src No.33]{\includegraphics[clip, trim={0.0cm 2.cm 0.cm 0.0cm},width=0.19\textwidth]{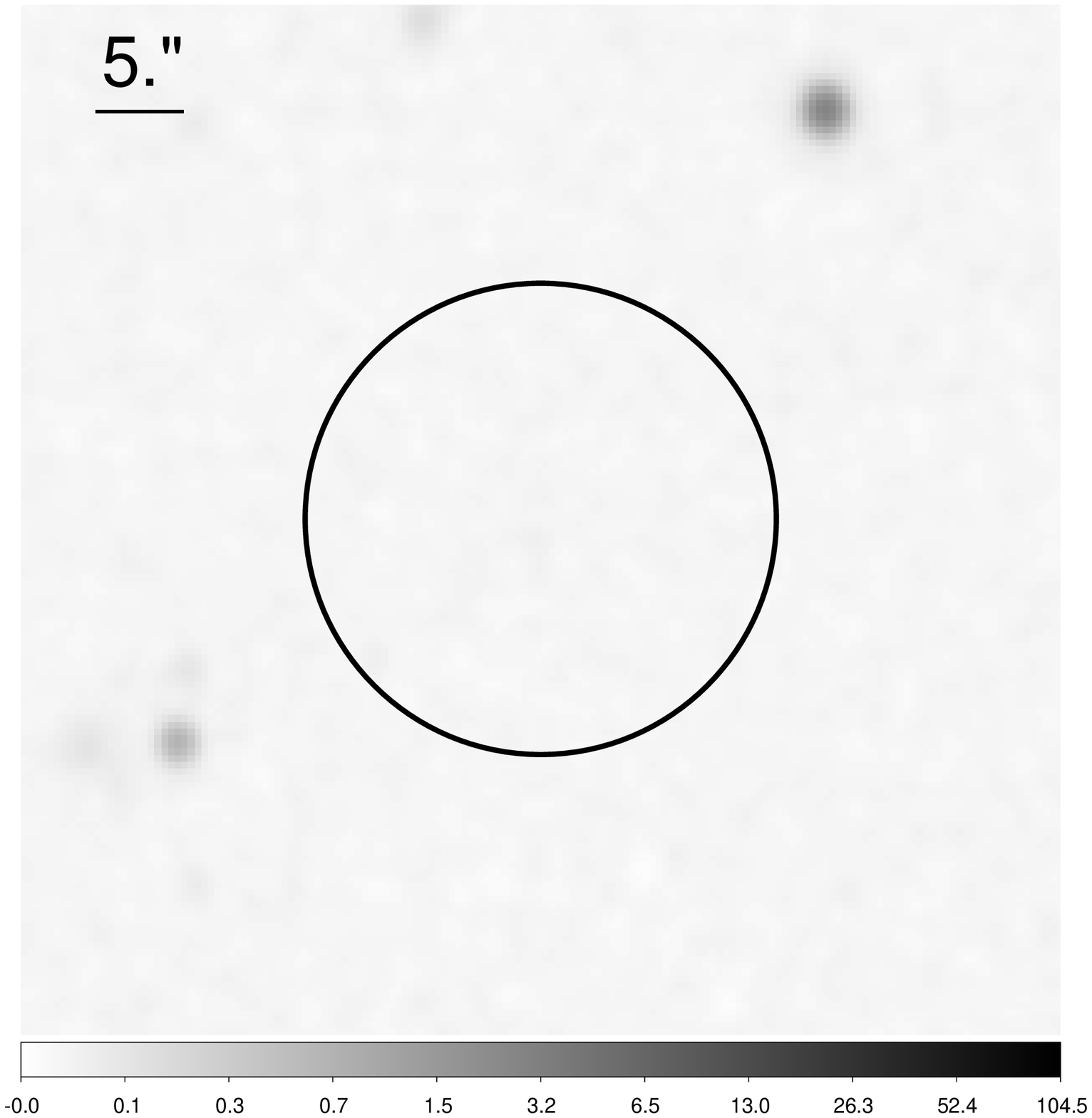}}
%\caption{\scriptsize{Src No.33}}
  \subfloat[Src No.34]{\includegraphics[clip, trim={0.0cm 2.cm 0.cm 0.0cm},width=0.19\textwidth]{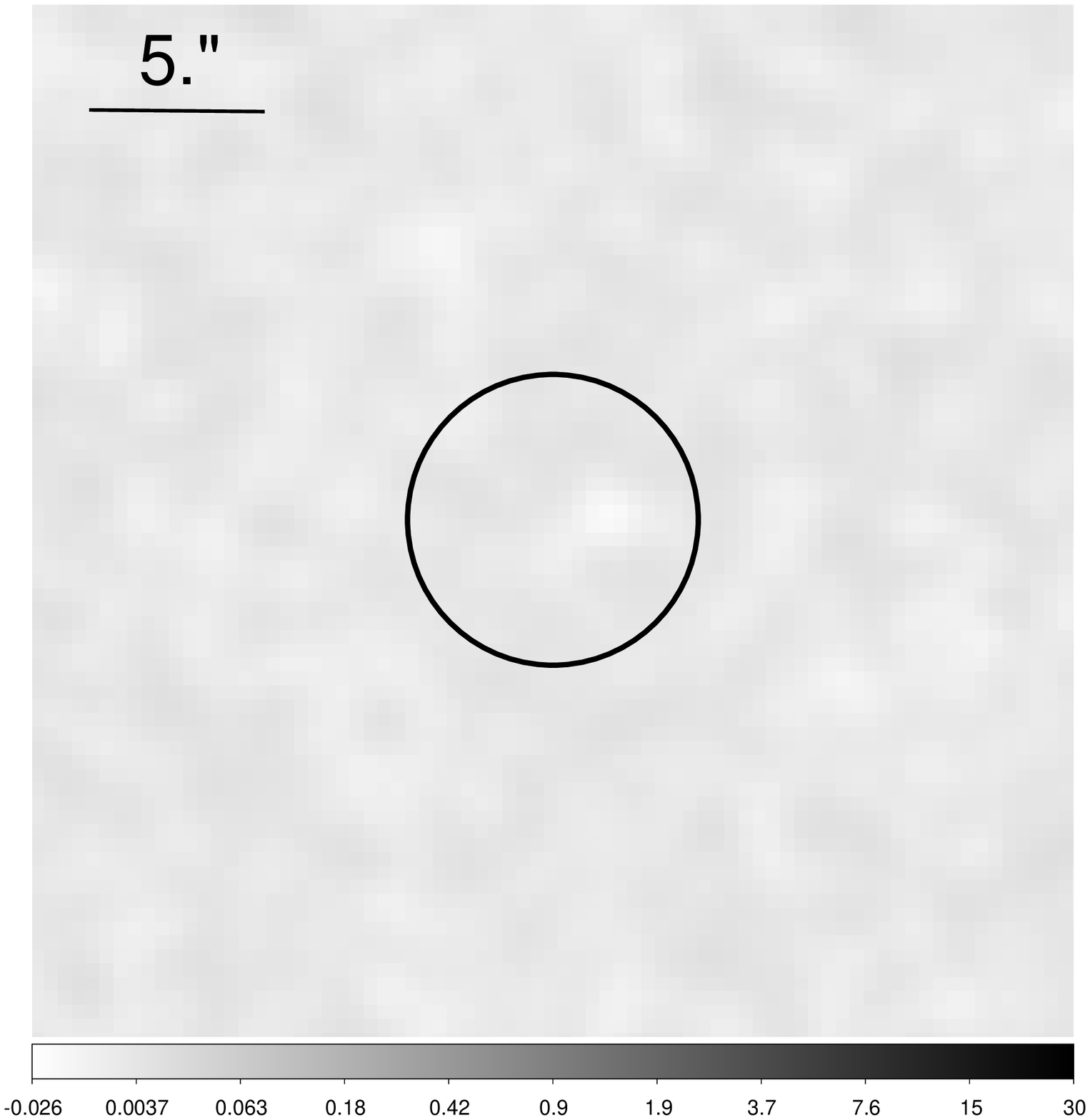}}
%\caption{\scriptsize{Src No.34}}
  \subfloat[Src No.35]{\includegraphics[clip, trim={0.0cm 2.cm 0.cm 0.0cm},width=0.19\textwidth]{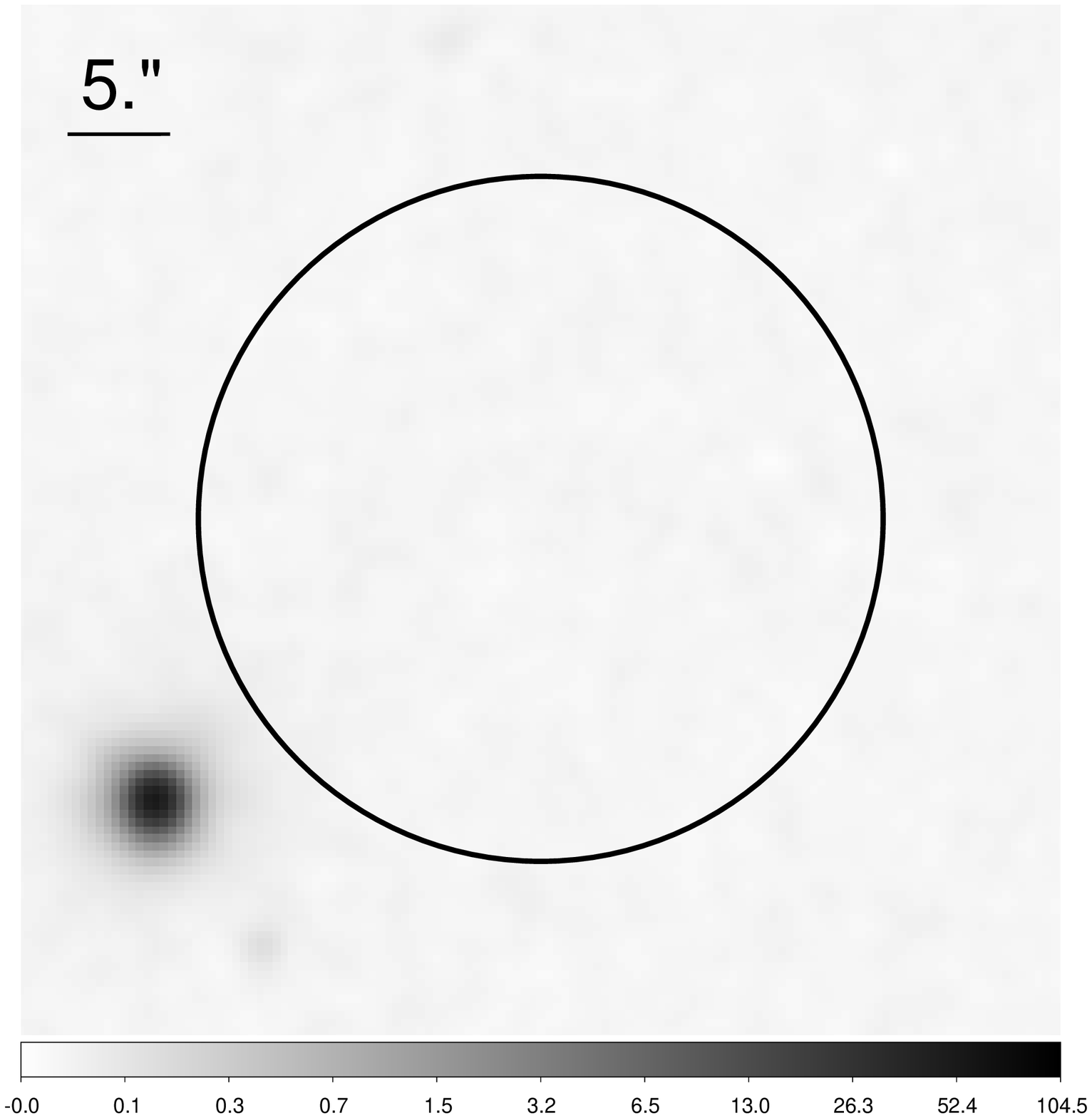}}
%\caption{\scriptsize{Src No.35}}
  \subfloat[Src No.36]{\includegraphics[clip, trim={0.0cm 2.cm 0.cm 0.0cm},width=0.19\textwidth]{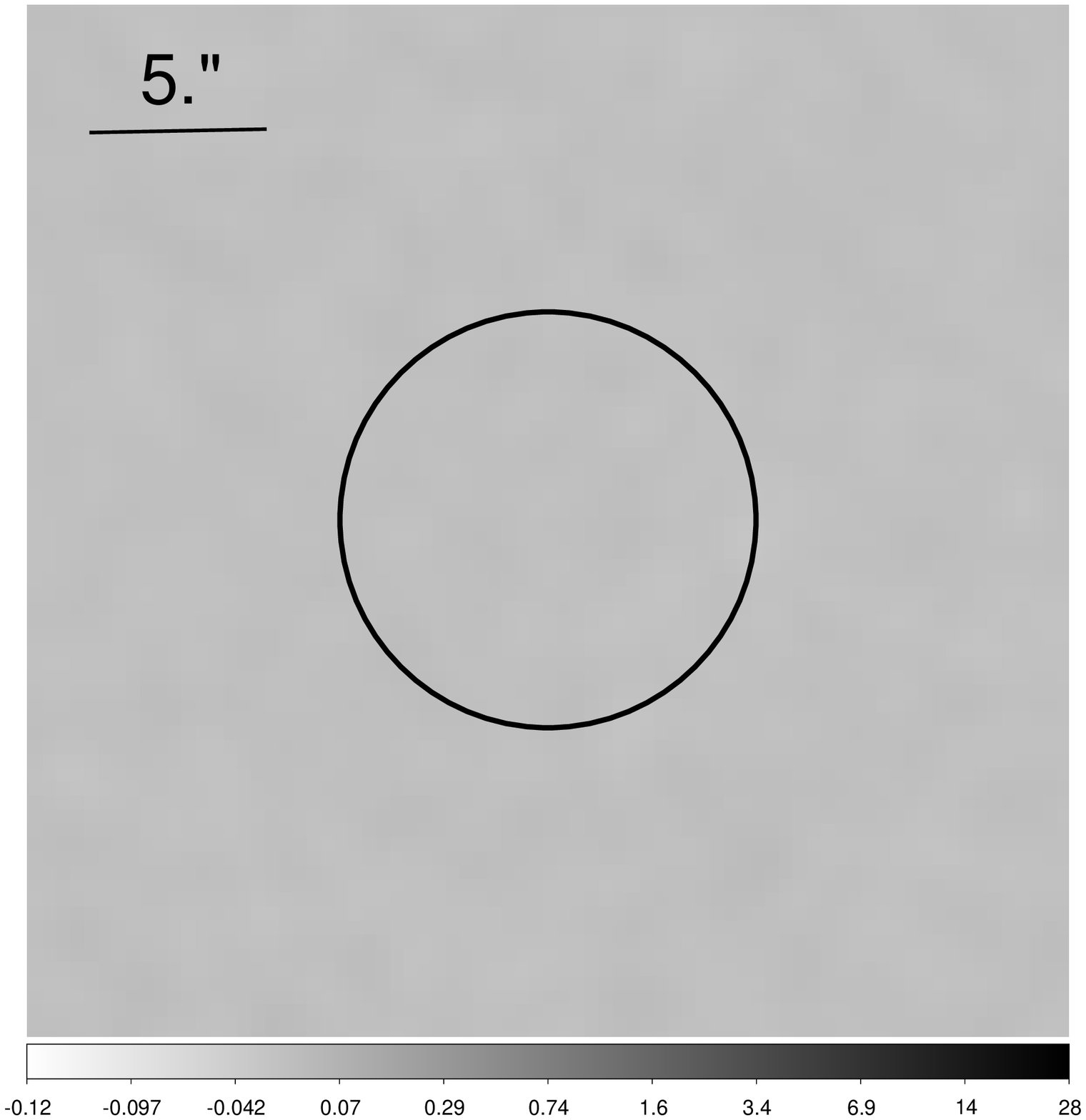}}\\
%\caption{\scriptsize{Src No.36}}
  \subfloat[Src No.37]{\includegraphics[clip, trim={0.0cm 2.cm 0.cm 0.0cm},width=0.19\textwidth]{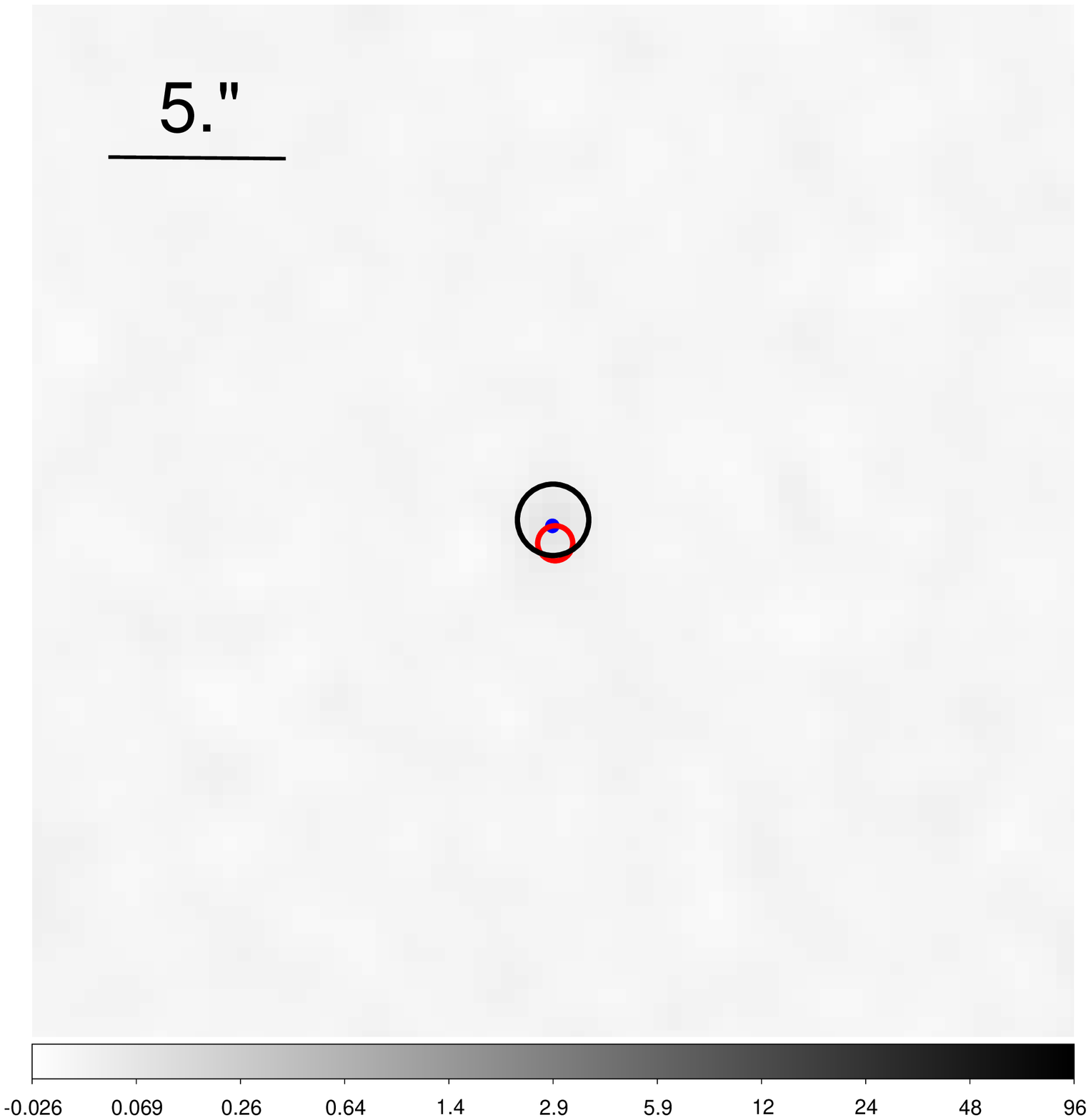}}
%\caption{\scriptsize{Src No.37}}
 \subfloat[Src No.38]{\includegraphics[clip, trim={0.0cm 2.cm 0.cm 0.0cm},width=0.19\textwidth]{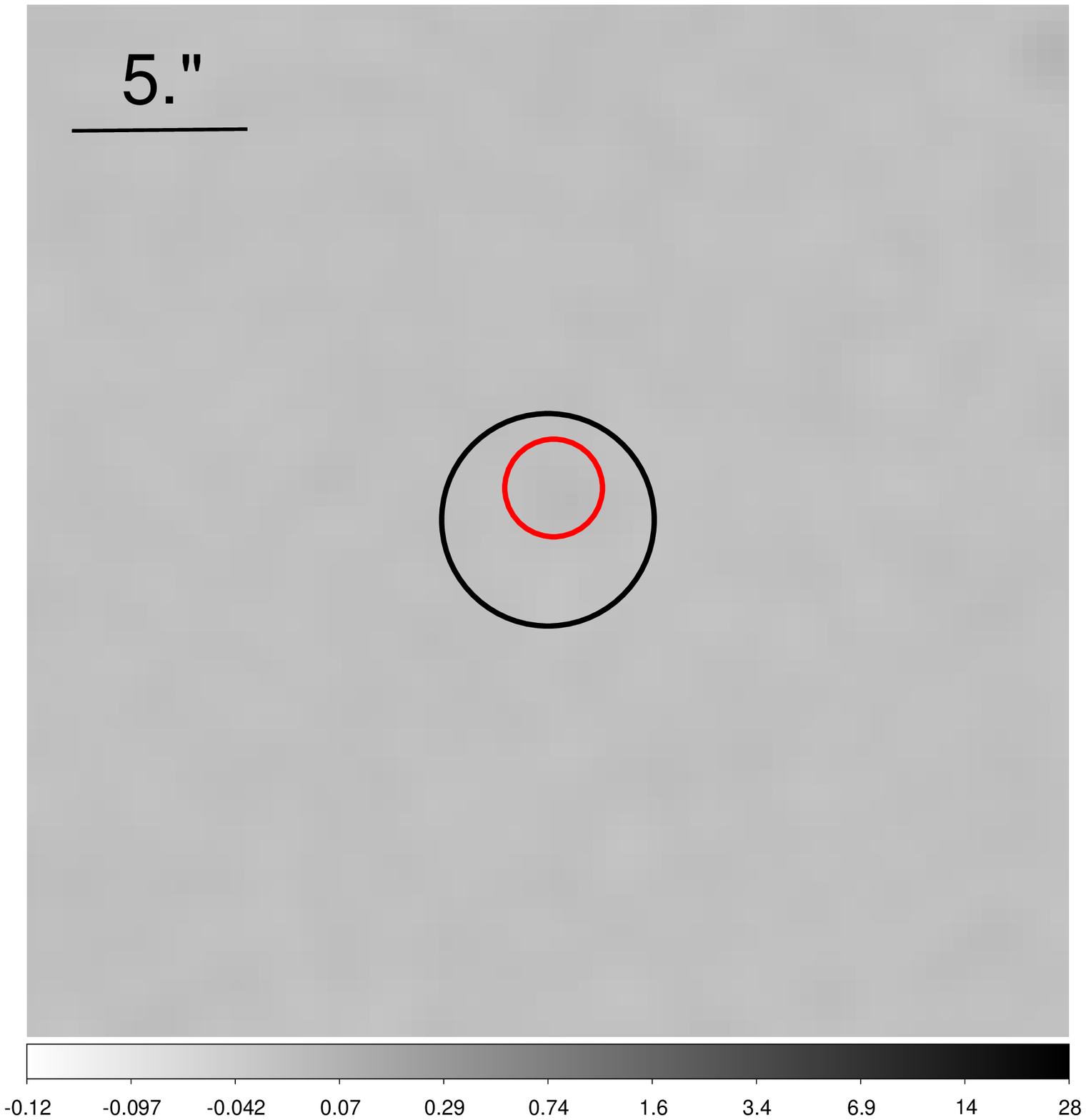}}
%\caption{\scriptsize{Src No.38}}
  \subfloat[Src No.39]{\includegraphics[clip, trim={0.0cm 2.cm 0.cm 0.0cm},width=0.19\textwidth]{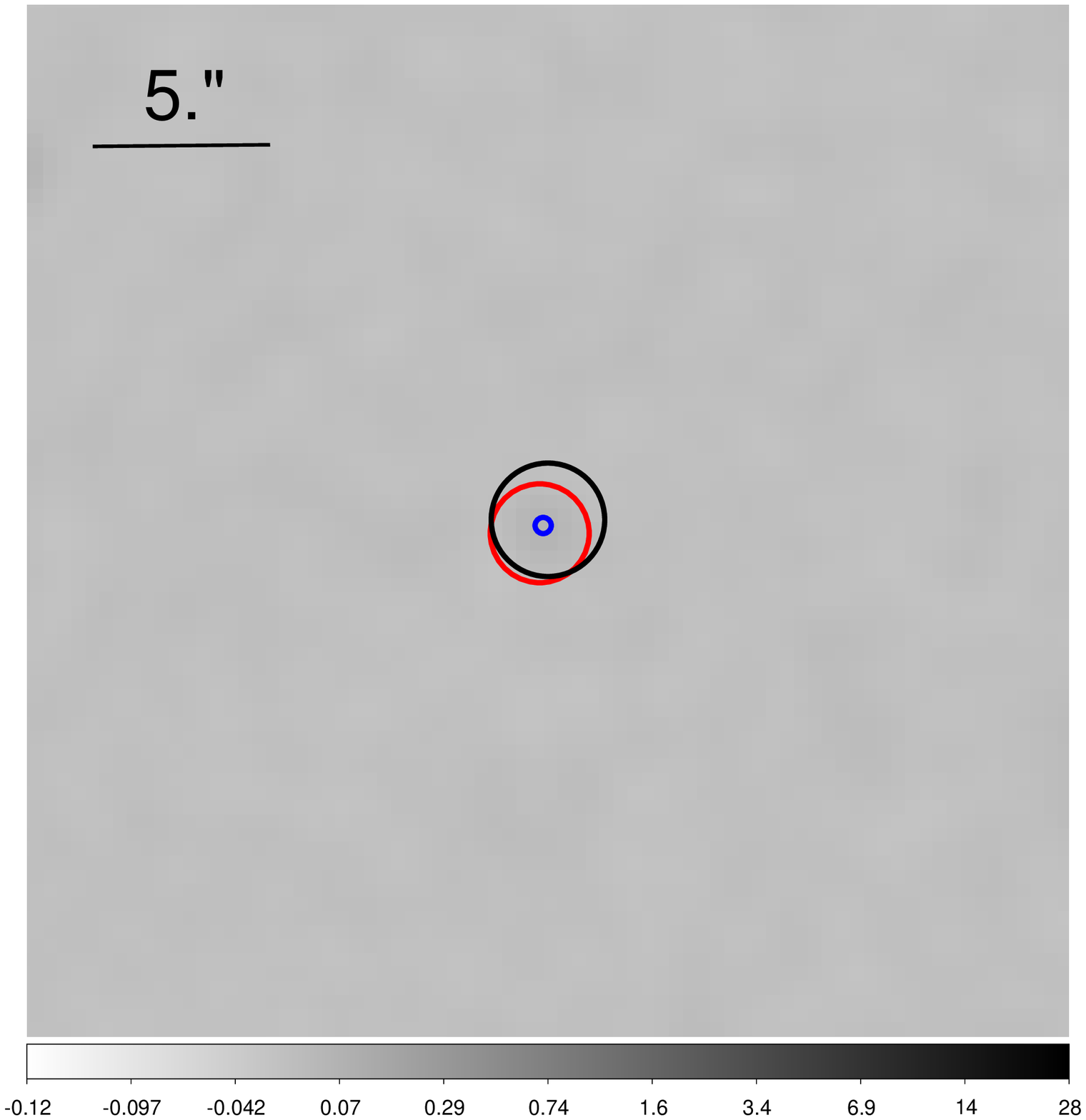}}
%\caption{\scriptsize{Src No.39}}
  \subfloat[Src No.40]{\includegraphics[clip, trim={0.0cm 2.cm 0.cm 0.0cm},width=0.19\textwidth]{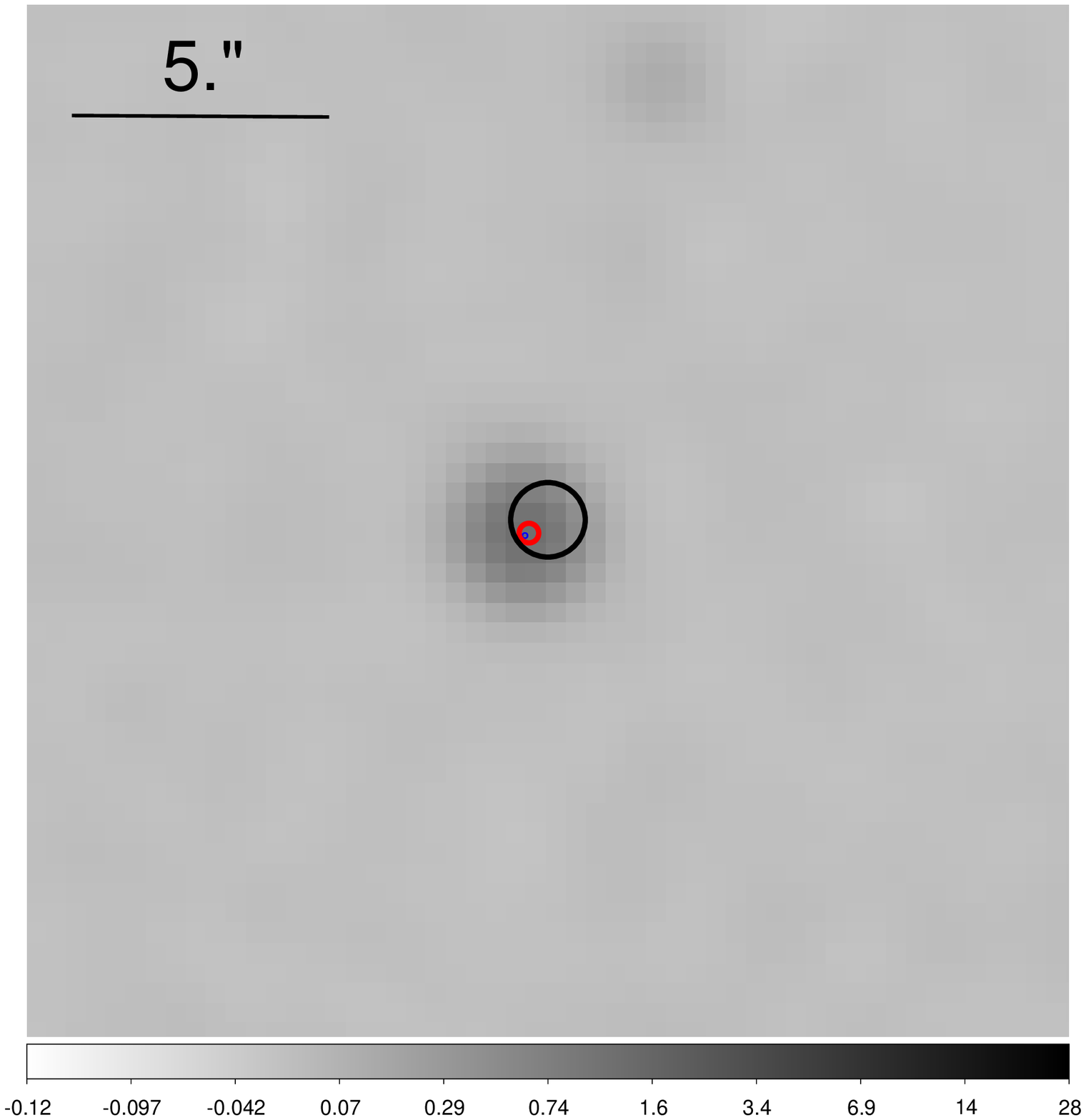}}\\
%\caption{\scriptsize{Src No.40}}
\subfloat[Src No.41]{\includegraphics[clip, trim={0.0cm 2.cm 0.cm 0.0cm},width=0.19\textwidth]{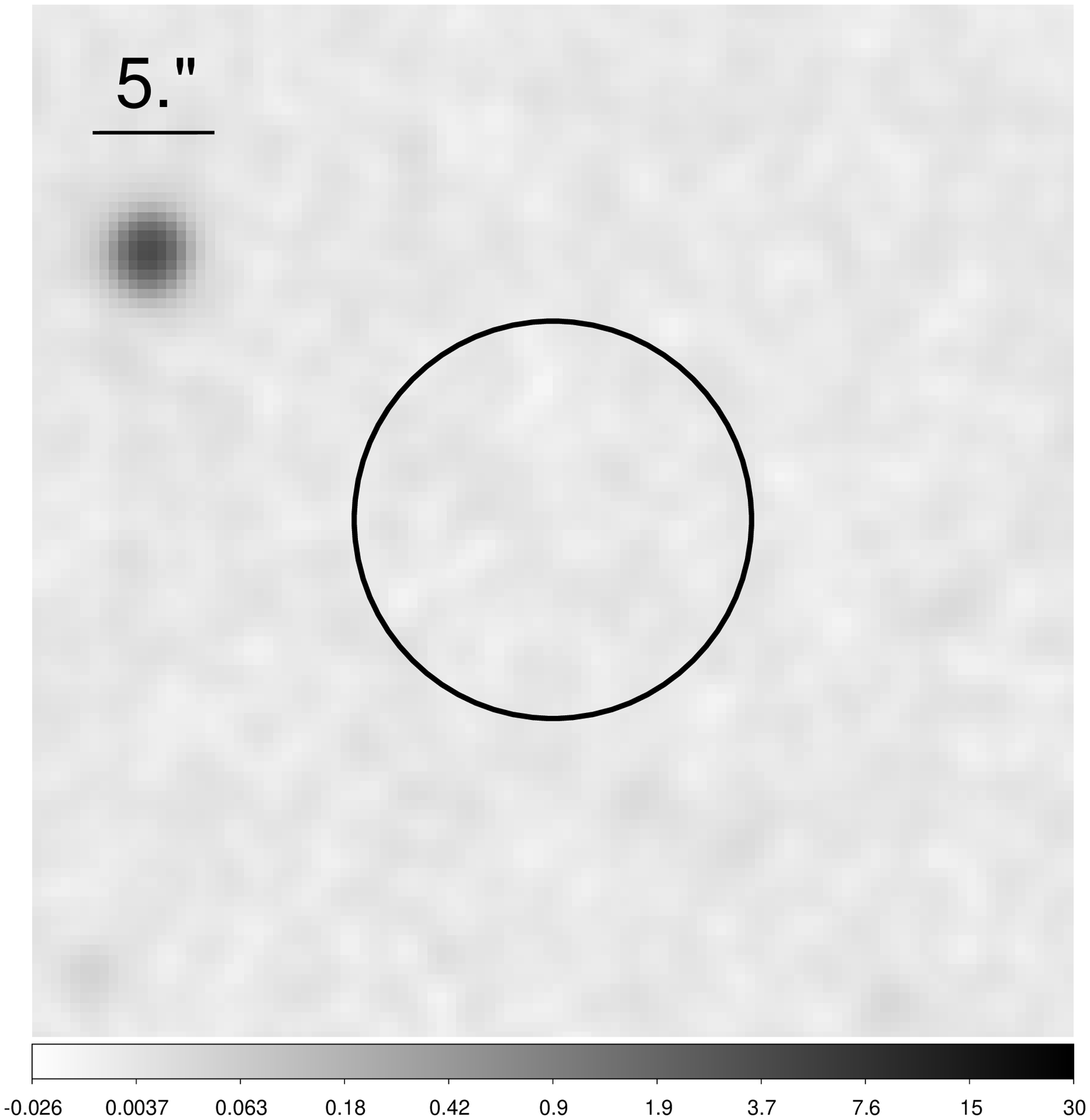}}
%\caption{\scriptsize{Src No.41}}
  \subfloat[Src No.42]{\includegraphics[clip, trim={0.0cm 2.cm 0.cm 0.0cm},width=0.19\textwidth]{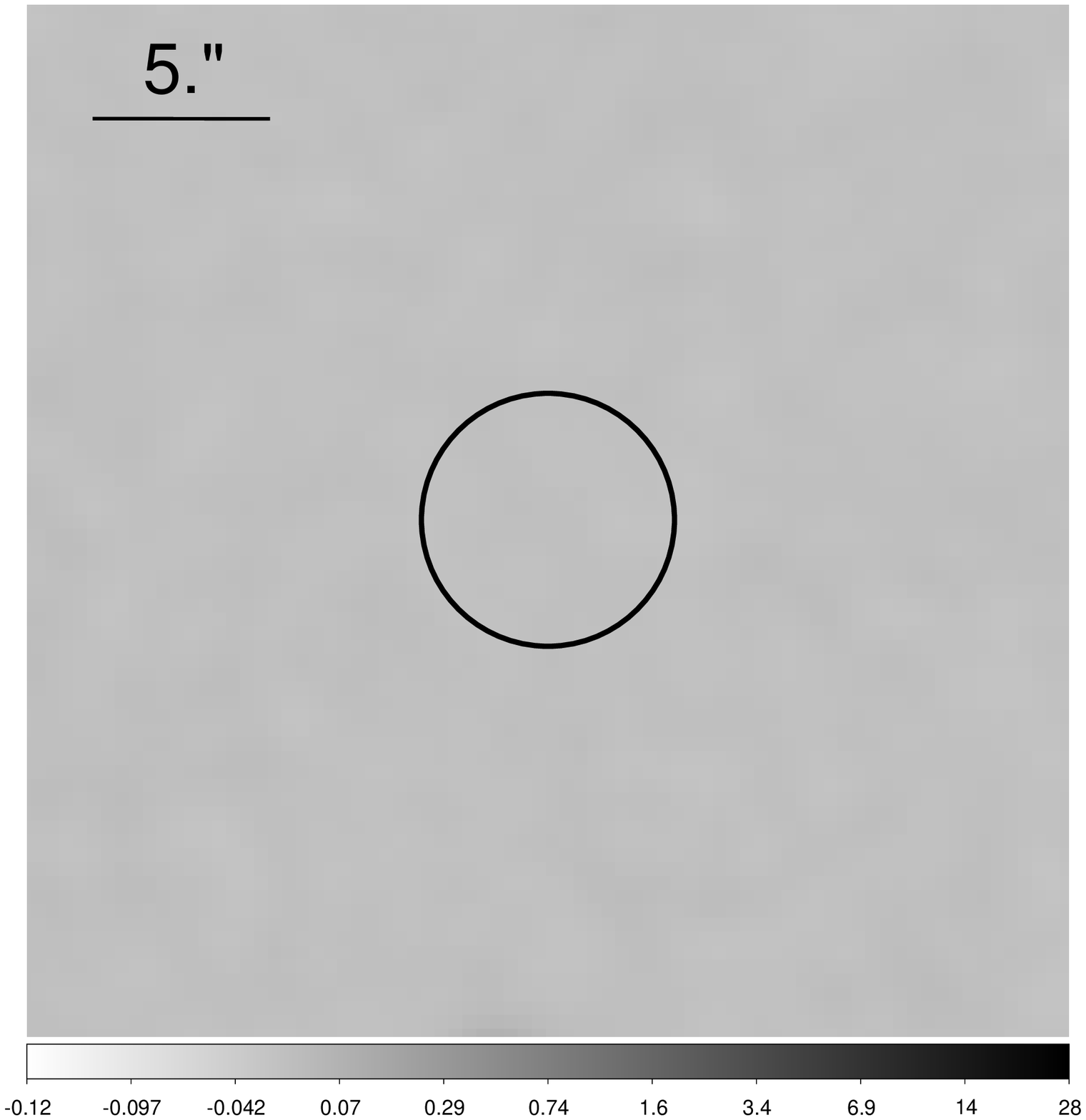}}
%\caption{\scriptsize{Src No.42}}
  \subfloat[Src No.43]{\includegraphics[clip, trim={0.0cm 2.cm 0.cm 0.0cm},width=0.19\textwidth]{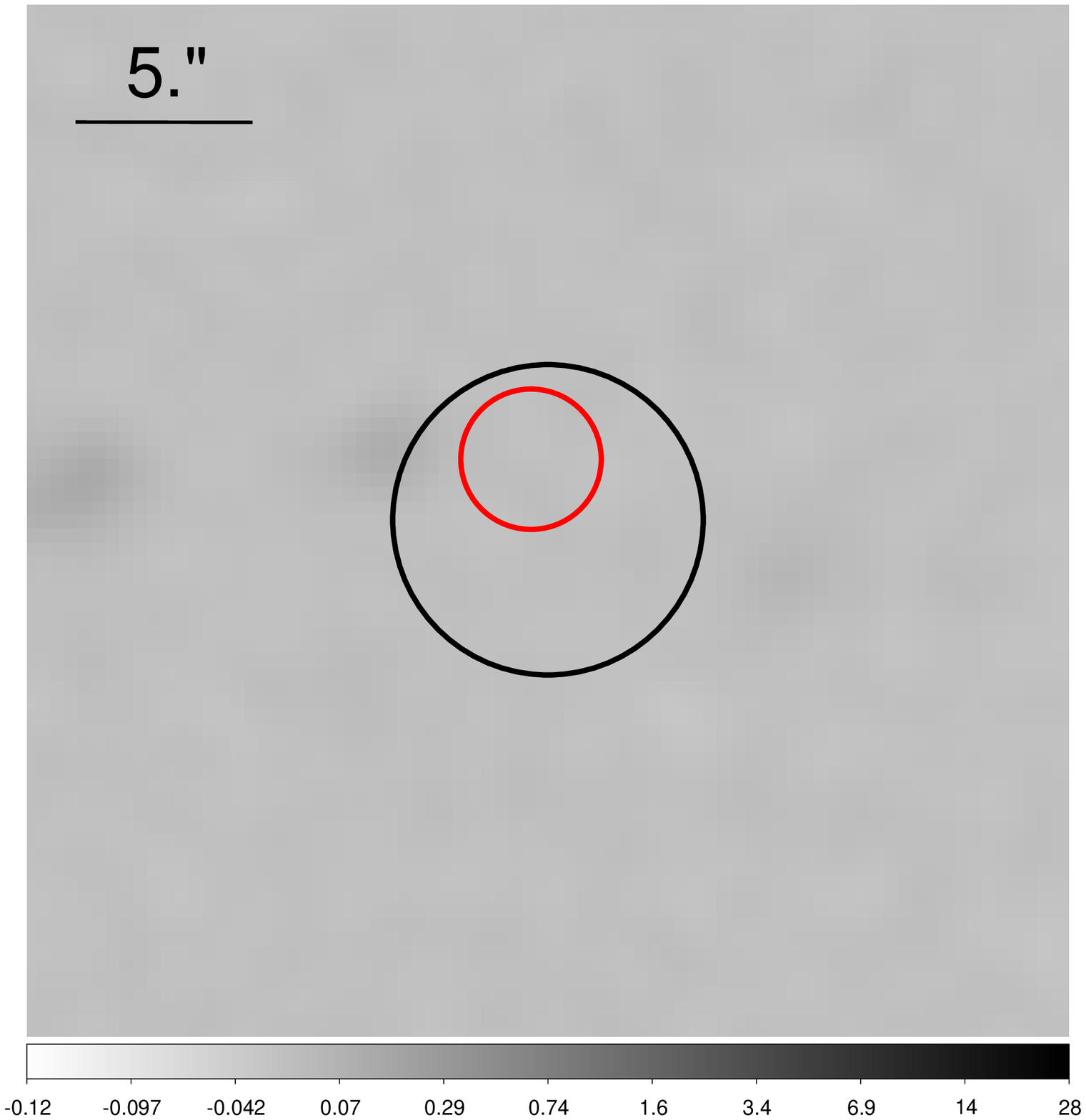}}
%\caption{\scriptsize{Src No.43}}
  \subfloat[Src No.44]{\includegraphics[clip, trim={0.0cm 2.cm 0.cm 0.0cm},width=0.19\textwidth]{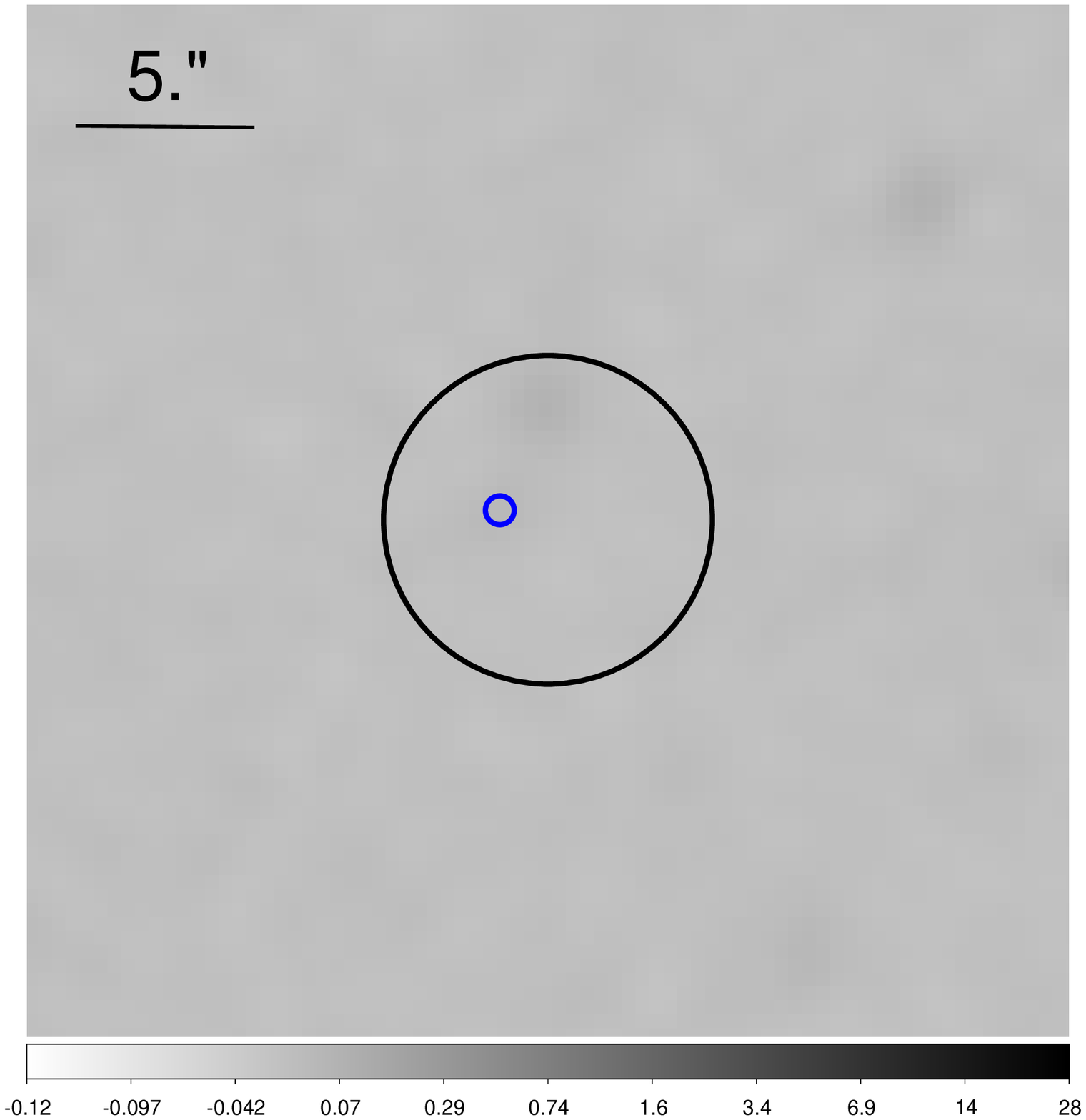}}\\
%\caption{\scriptsize{Src No.44}}
\end{figure*}
\pagebreak
\clearpage
%\hspace{0.3cm}Appendix B continued: Image of optical SDSS9 counterparts
\begin{figure*}
\vspace{-0.5cm}
  \subfloat[Src No.45]{\includegraphics[clip, trim={0.0cm 2.cm 0.cm 0.0cm},width=0.19\textwidth]{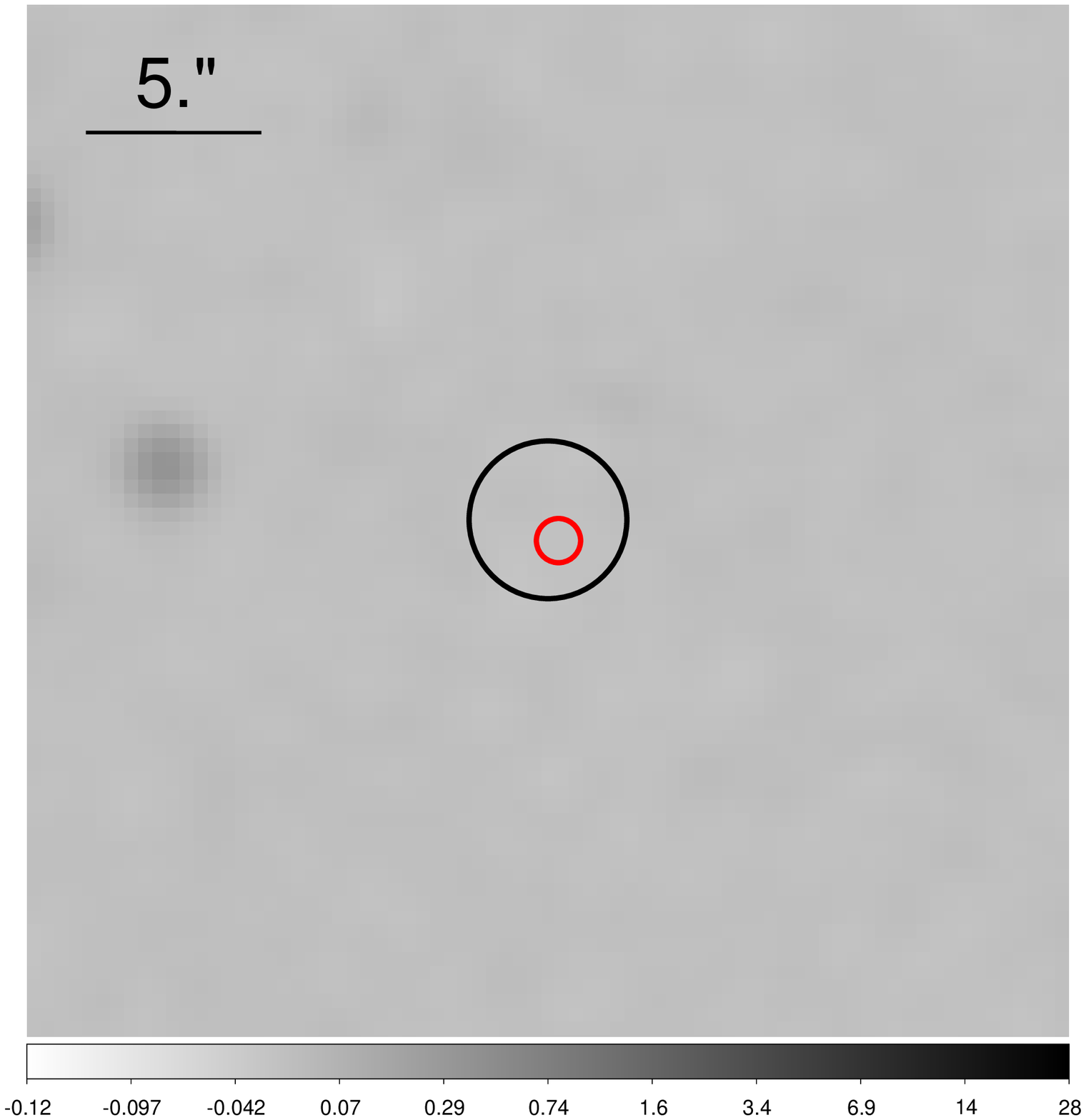}}
%\caption{\scriptsize{Src No.45}}
  \subfloat[Src No.46]{\includegraphics[clip, trim={0.0cm 2.cm 0.cm 0.0cm},width=0.19\textwidth]{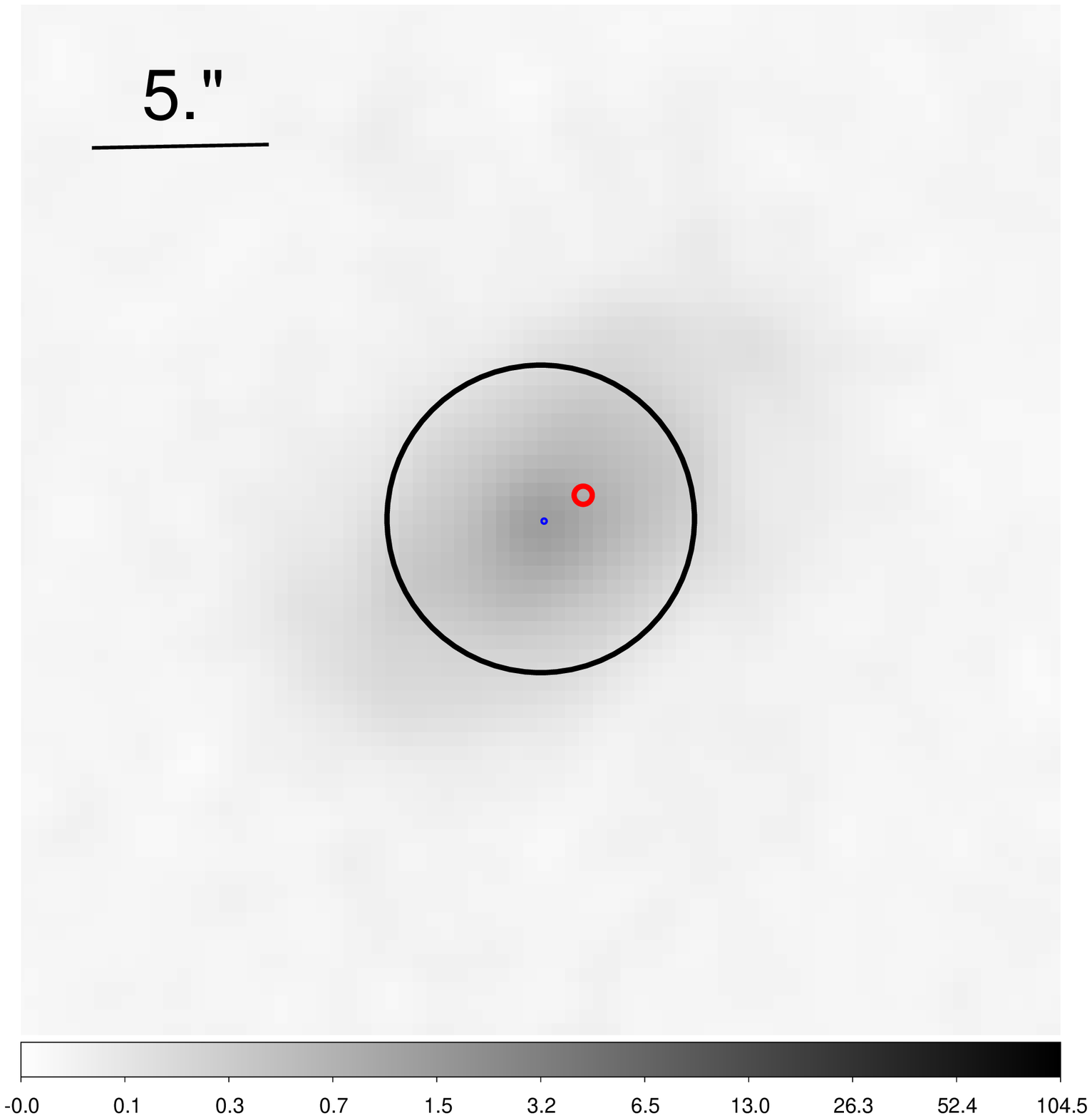}}
%\caption{\scriptsize{Src No.46}}
  \subfloat[Src No.47]{\includegraphics[clip, trim={0.0cm 2.cm 0.cm 0.0cm},width=0.19\textwidth]{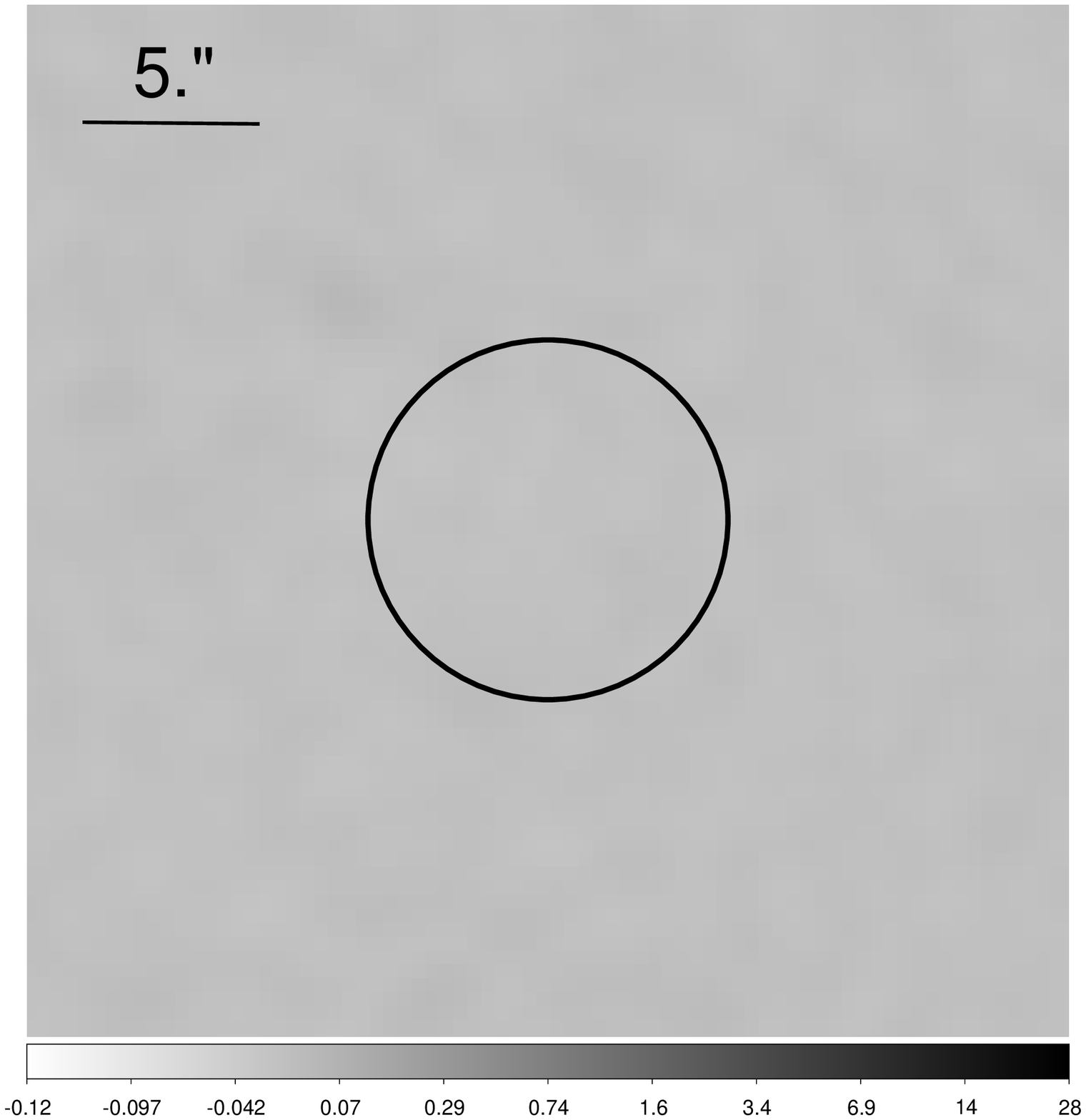}}
%\caption{\scriptsize{Src No.47}}
  \subfloat[Src No.48]{\includegraphics[clip, trim={0.0cm 2.cm 0.cm 0.0cm},width=0.19\textwidth]{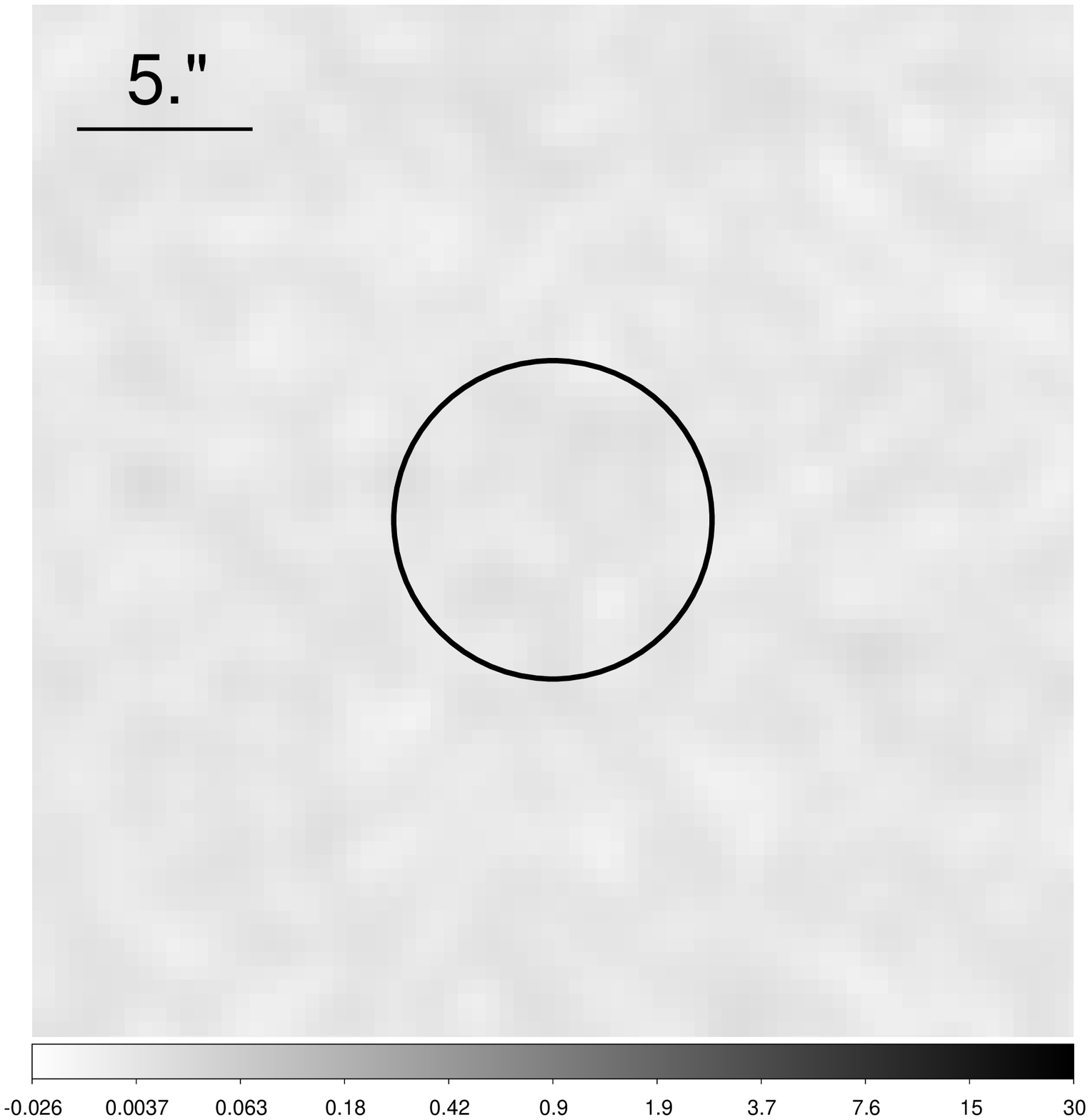}}\\
%\caption{\scriptsize{Src No.48}}
  \subfloat[Src No.49]{\includegraphics[clip, trim={0.0cm 2.cm 0.cm 0.0cm},width=0.19\textwidth]{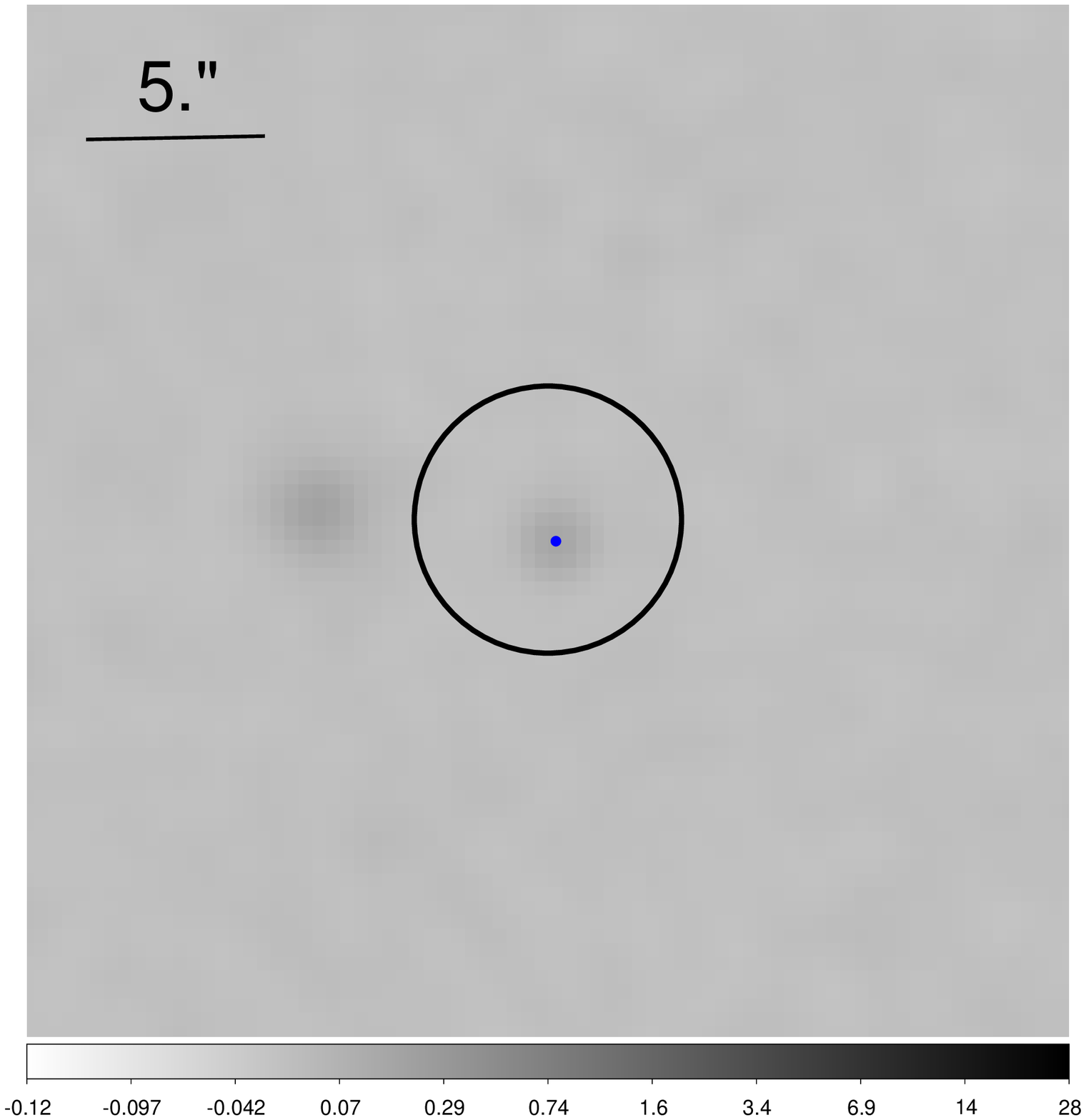}}
%\caption{\scriptsize{Src No.49}}
  \subfloat[Src No.50]{\includegraphics[clip, trim={0.0cm 2.cm 0.cm 0.0cm},width=0.19\textwidth]{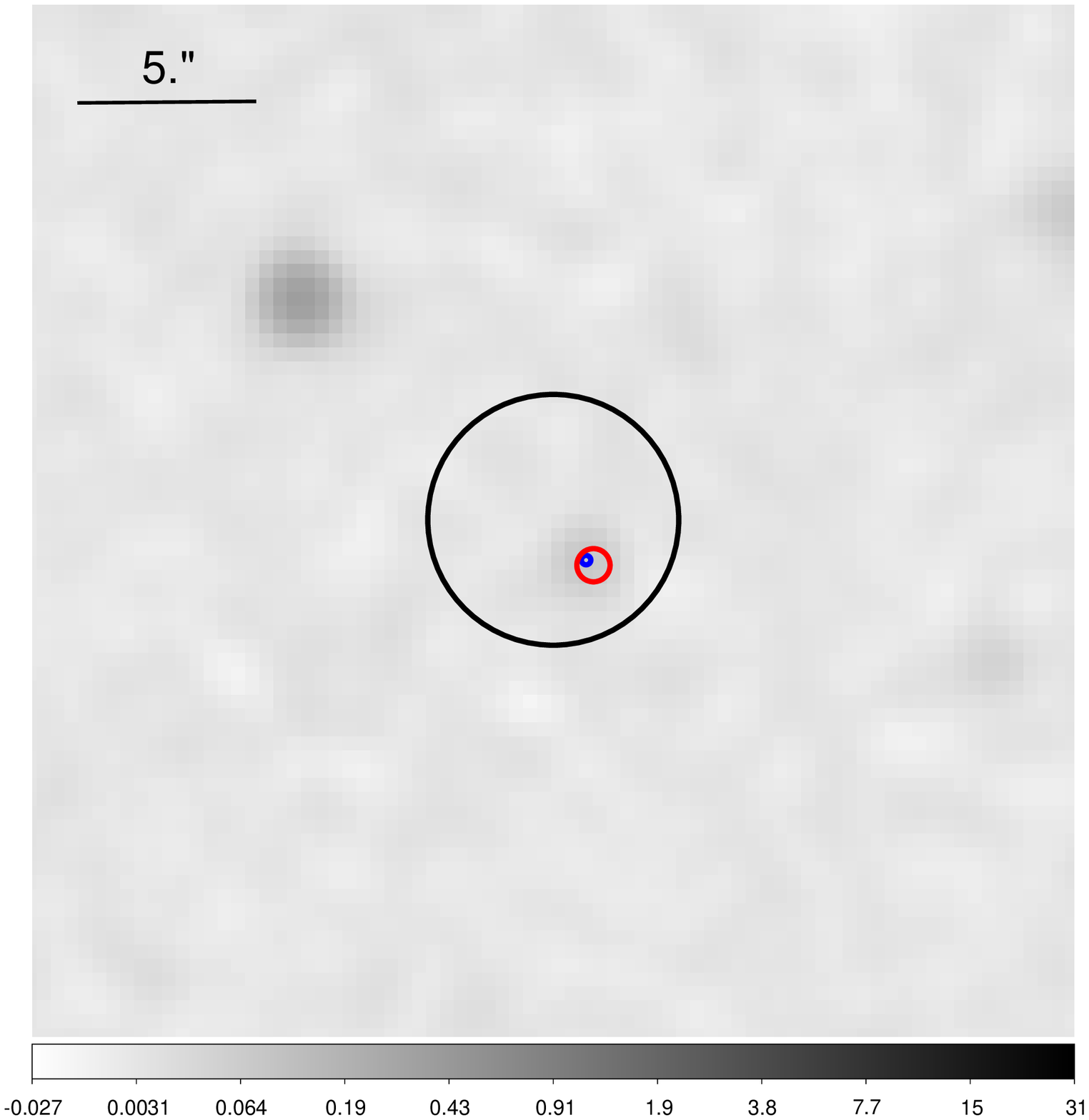}}
%\caption{\scriptsize{Src No.50}}
  \subfloat[Src No.51]{\includegraphics[clip, trim={0.0cm 2.cm 0.cm 0.0cm},width=0.19\textwidth]{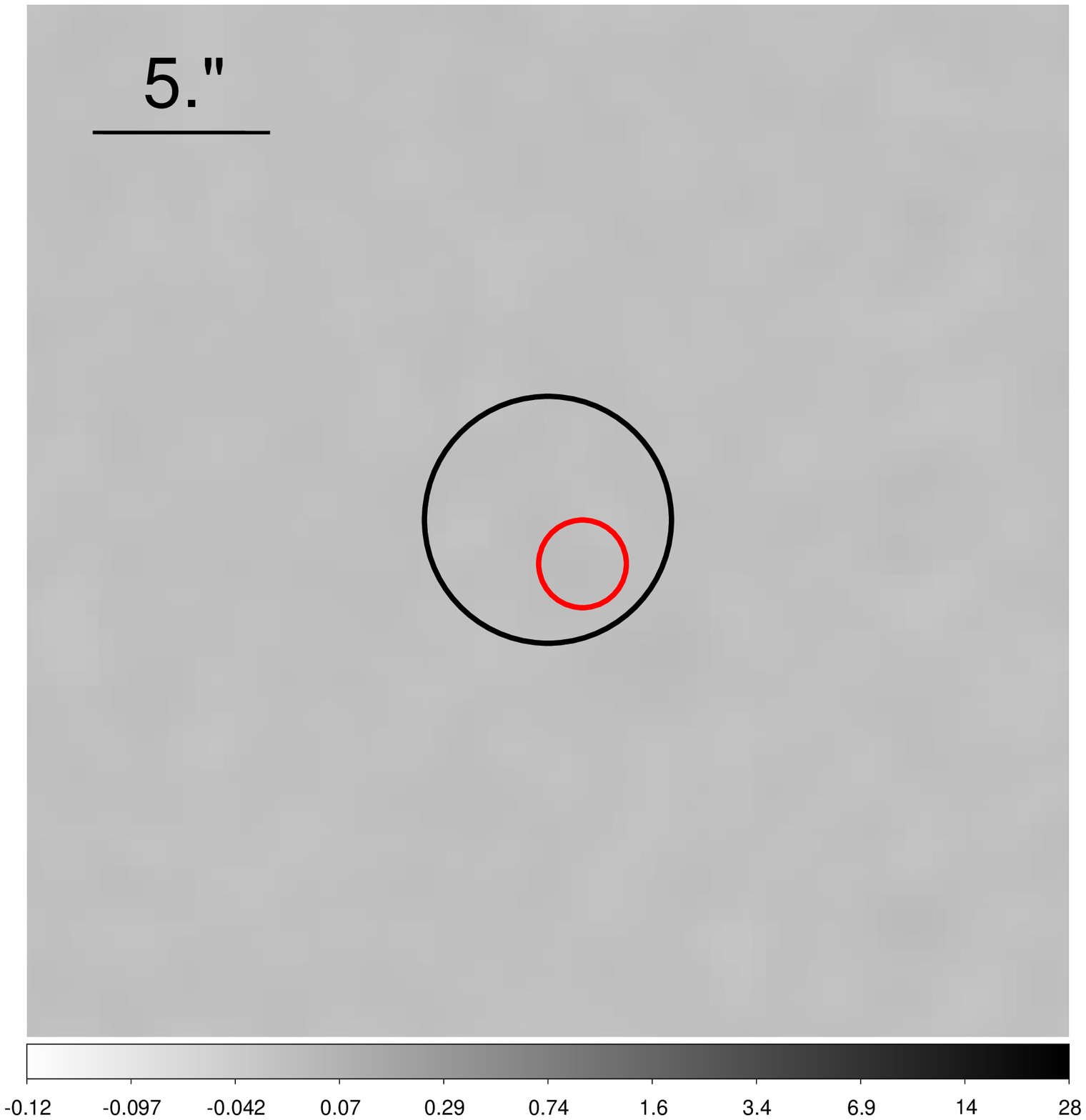}}
%\caption{\scriptsize{Src No.51}}
  \subfloat[Src No.52]{\includegraphics[clip, trim={0.0cm 2.cm 0.cm 0.0cm},width=0.19\textwidth]{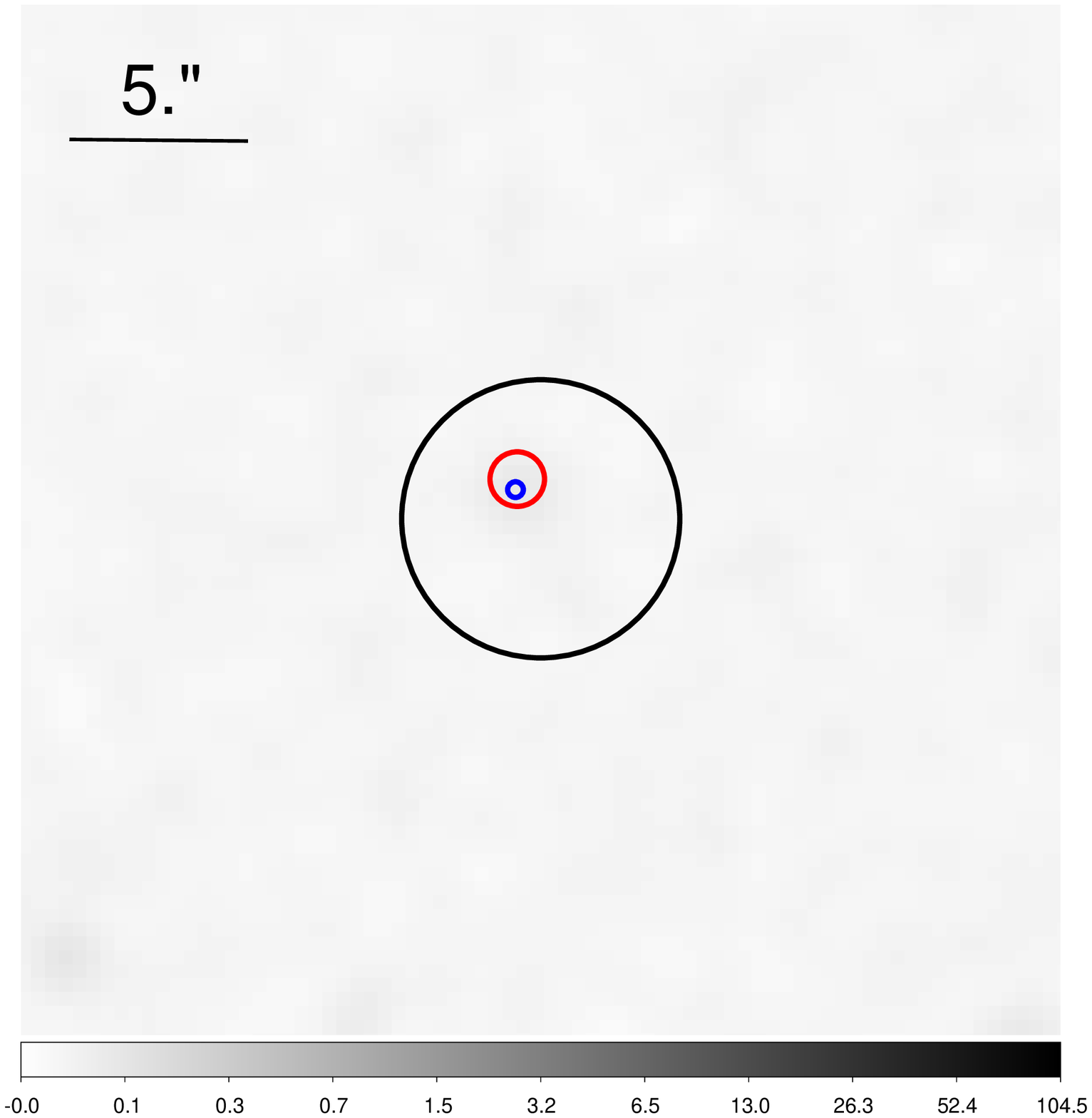}}\\
%\caption{\scriptsize{Src No.52}}
  \subfloat[Src No.53]{\includegraphics[clip, trim={0.0cm 2.cm 0.cm 0.0cm},width=0.19\textwidth]{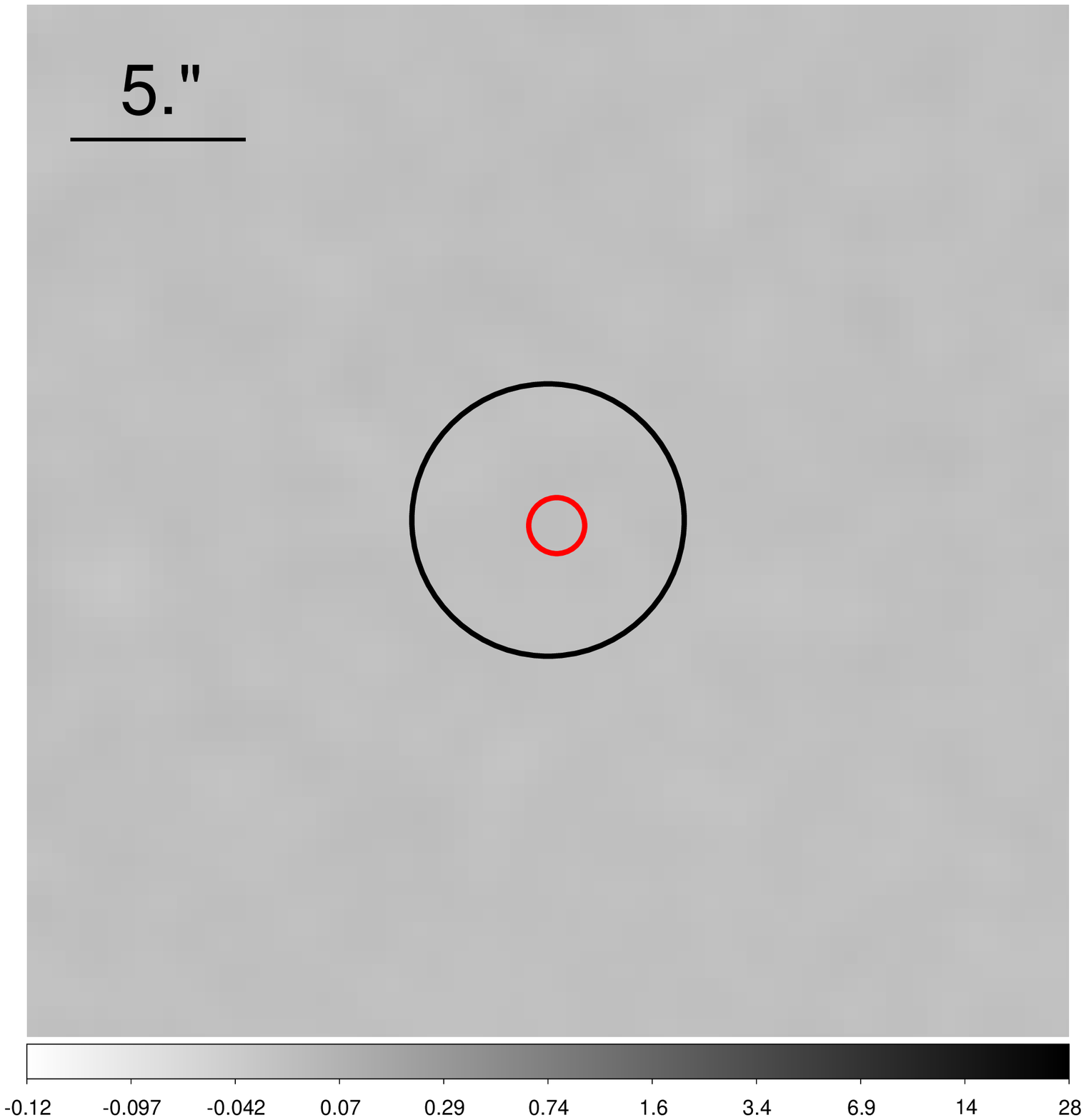}}
%\caption{\scriptsize{Src No.53}}
  \subfloat[Src No.54]{\includegraphics[clip, trim={0.0cm 2.cm 0.cm 0.0cm},width=0.19\textwidth]{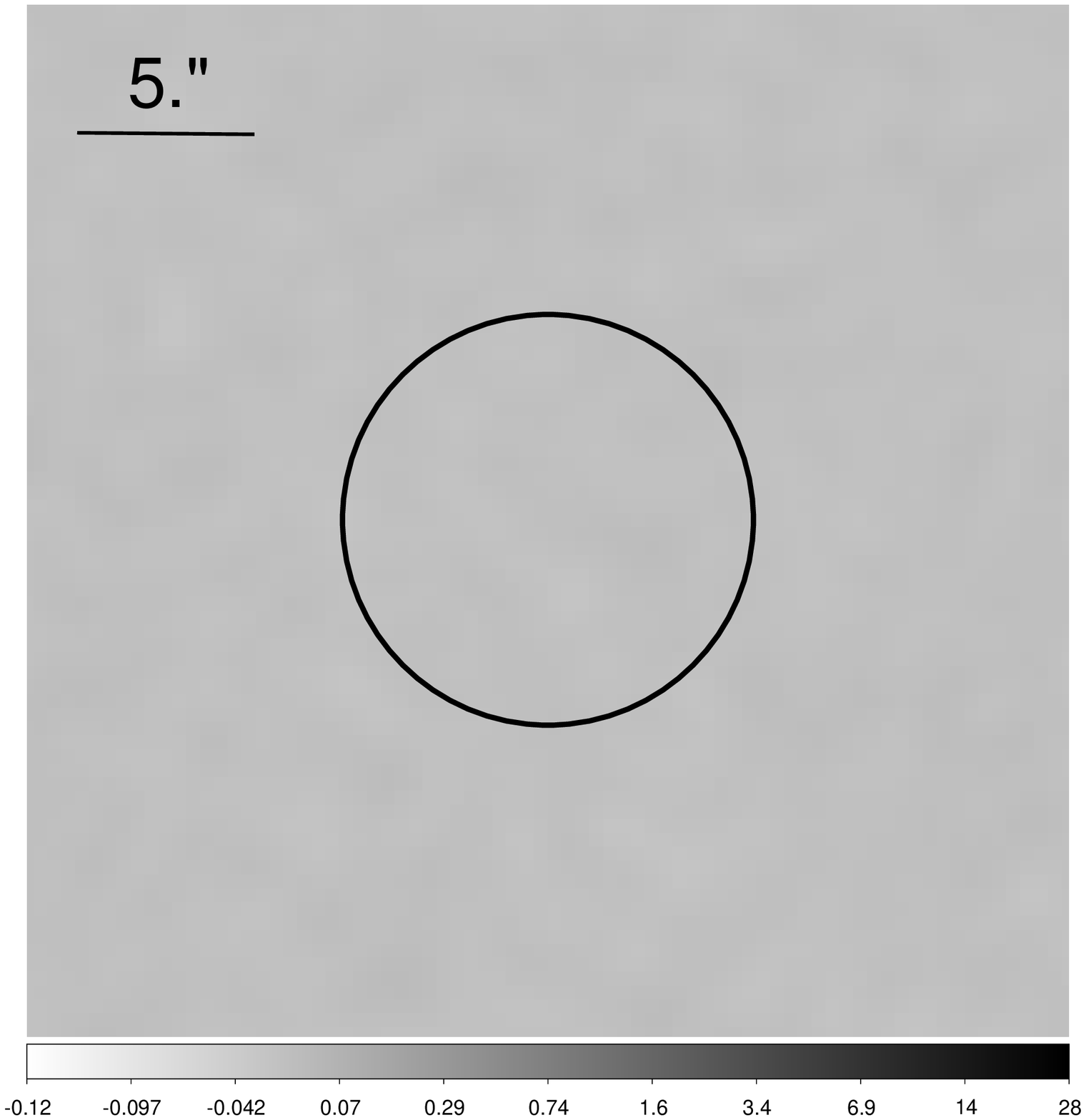}}
%\caption{\scriptsize{Src No.54}}
  \subfloat[Src No.55]{\includegraphics[clip, trim={0.0cm 2.cm 0.cm 0.0cm},width=0.19\textwidth]{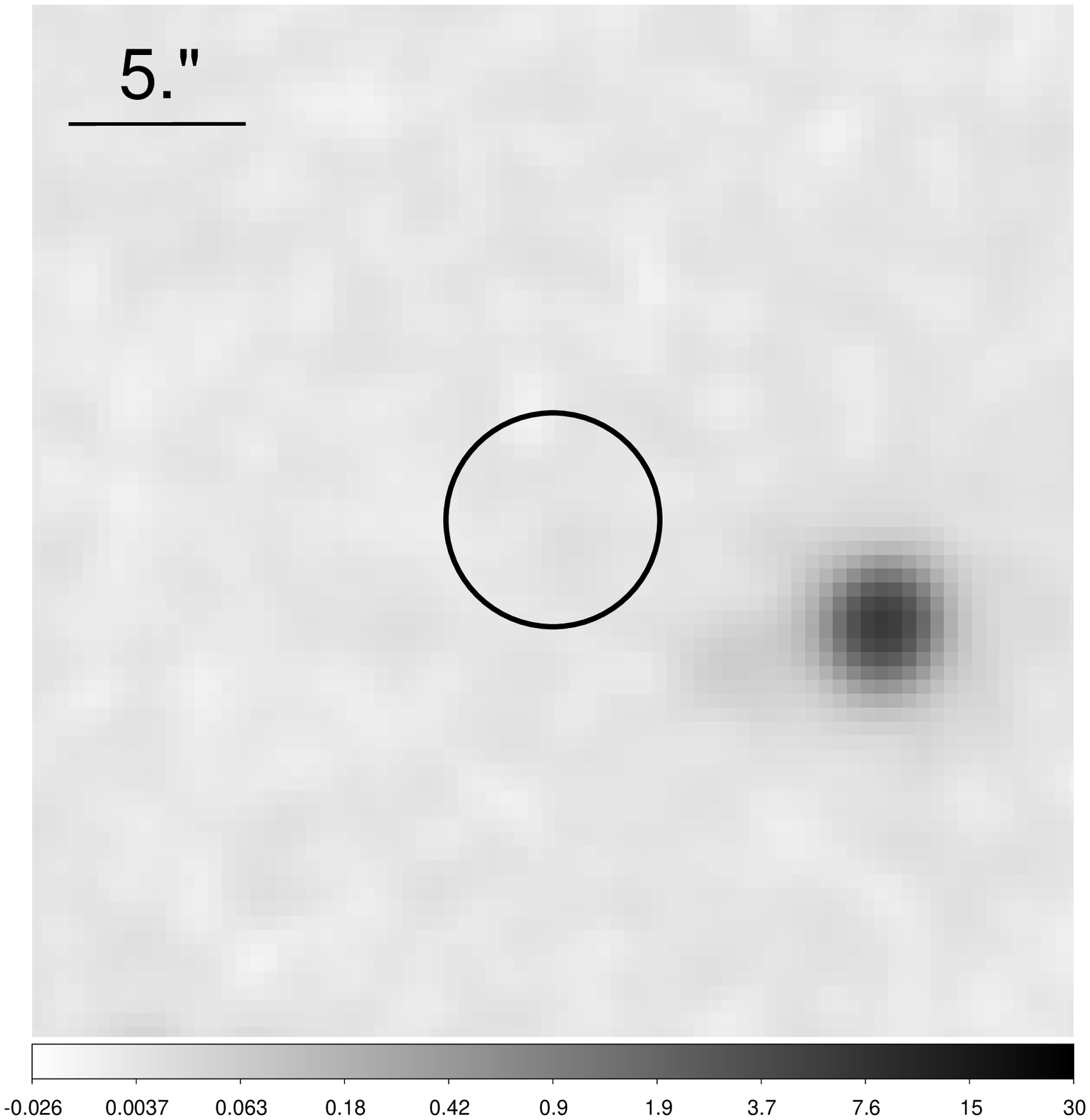}}
%\caption{\scriptsize{Src No.55}}
  \subfloat[Src No.56]{\includegraphics[clip, trim={0.0cm 2.cm 0.cm 0.0cm},width=0.19\textwidth]{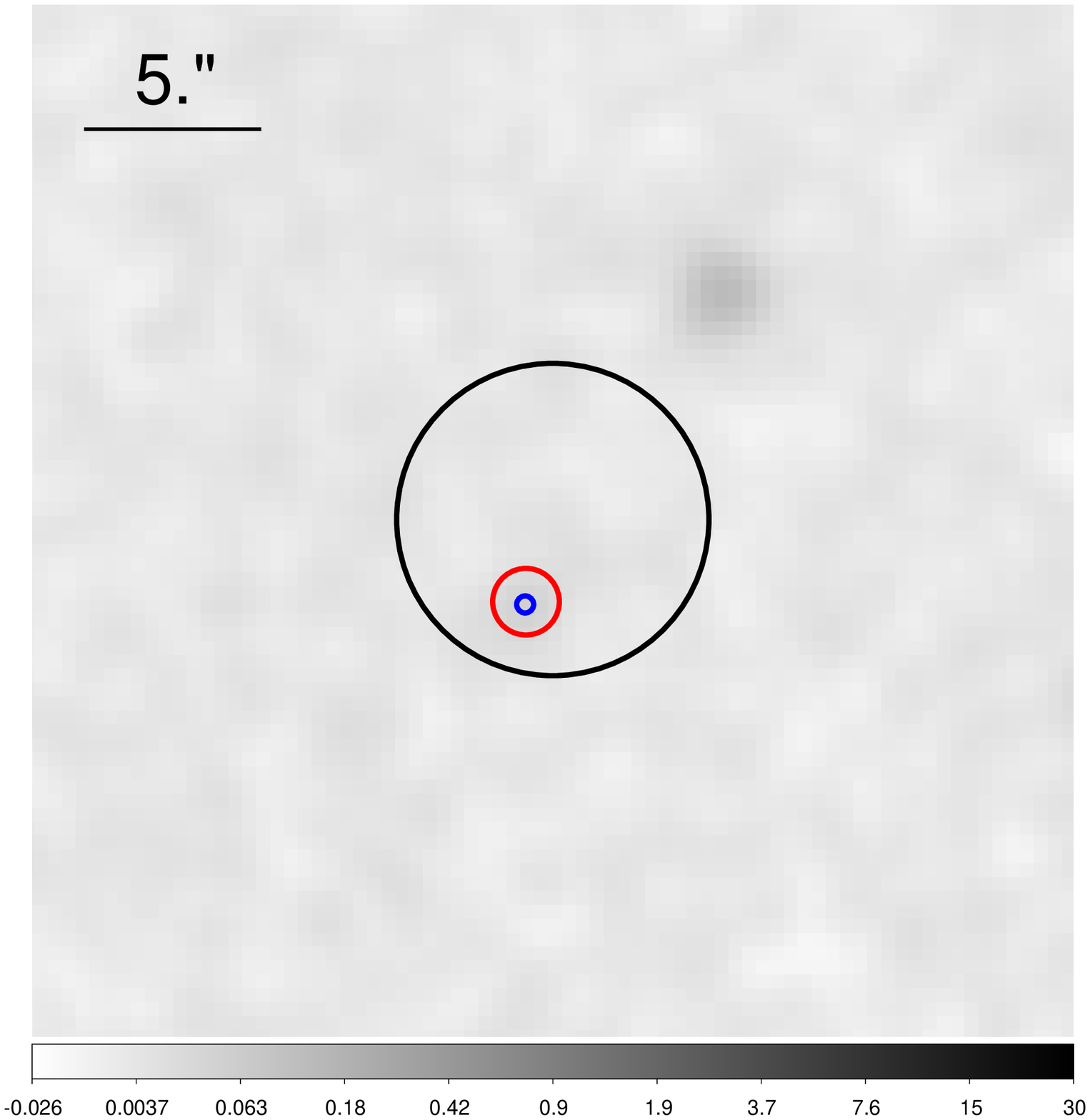}}\\
%\caption{\scriptsize{Src No.56}}
  \subfloat[Src No.57]{\includegraphics[clip, trim={0.0cm 2.cm 0.cm 0.0cm},width=0.19\textwidth]{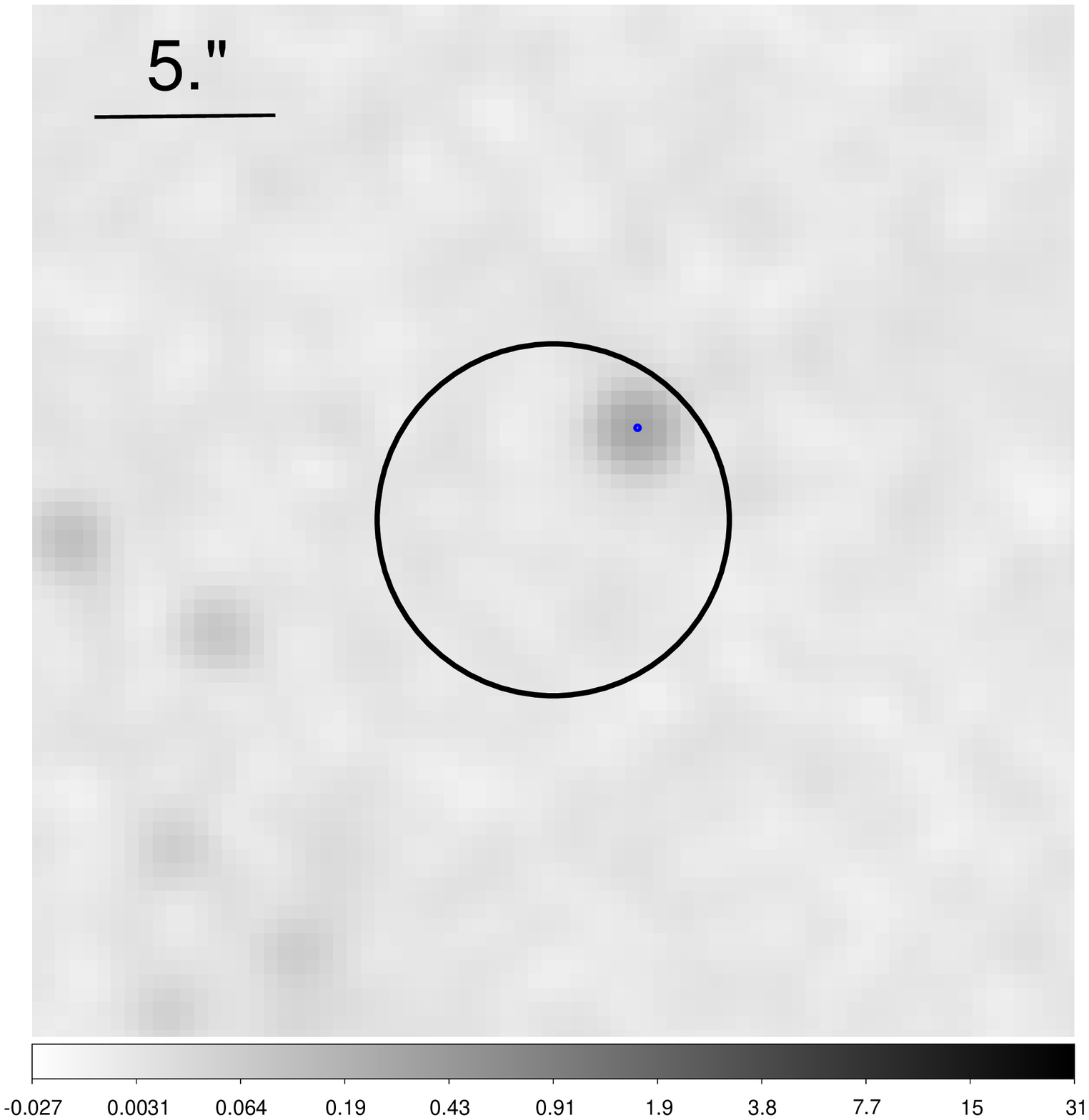}}
%\caption{\scriptsize{Src No.57}}
  \subfloat[Src No.58]{\includegraphics[clip, trim={0.0cm 2.cm 0.cm 0.0cm},width=0.19\textwidth]{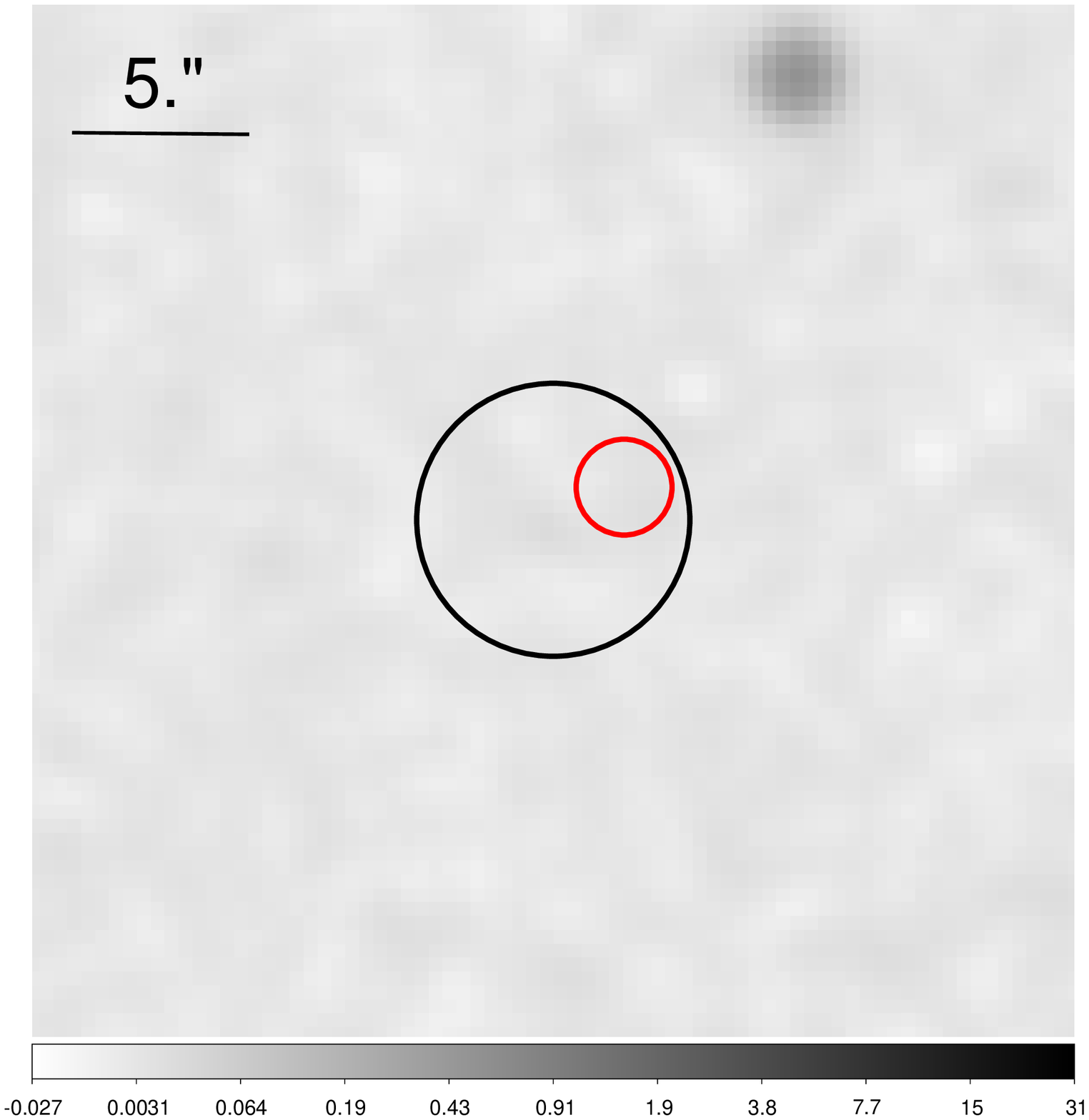}}
%\caption{\scriptsize{Src No.58}}
  \subfloat[Src No.59]{\includegraphics[clip, trim={0.0cm 2.cm 0.cm 0.0cm},width=0.19\textwidth]{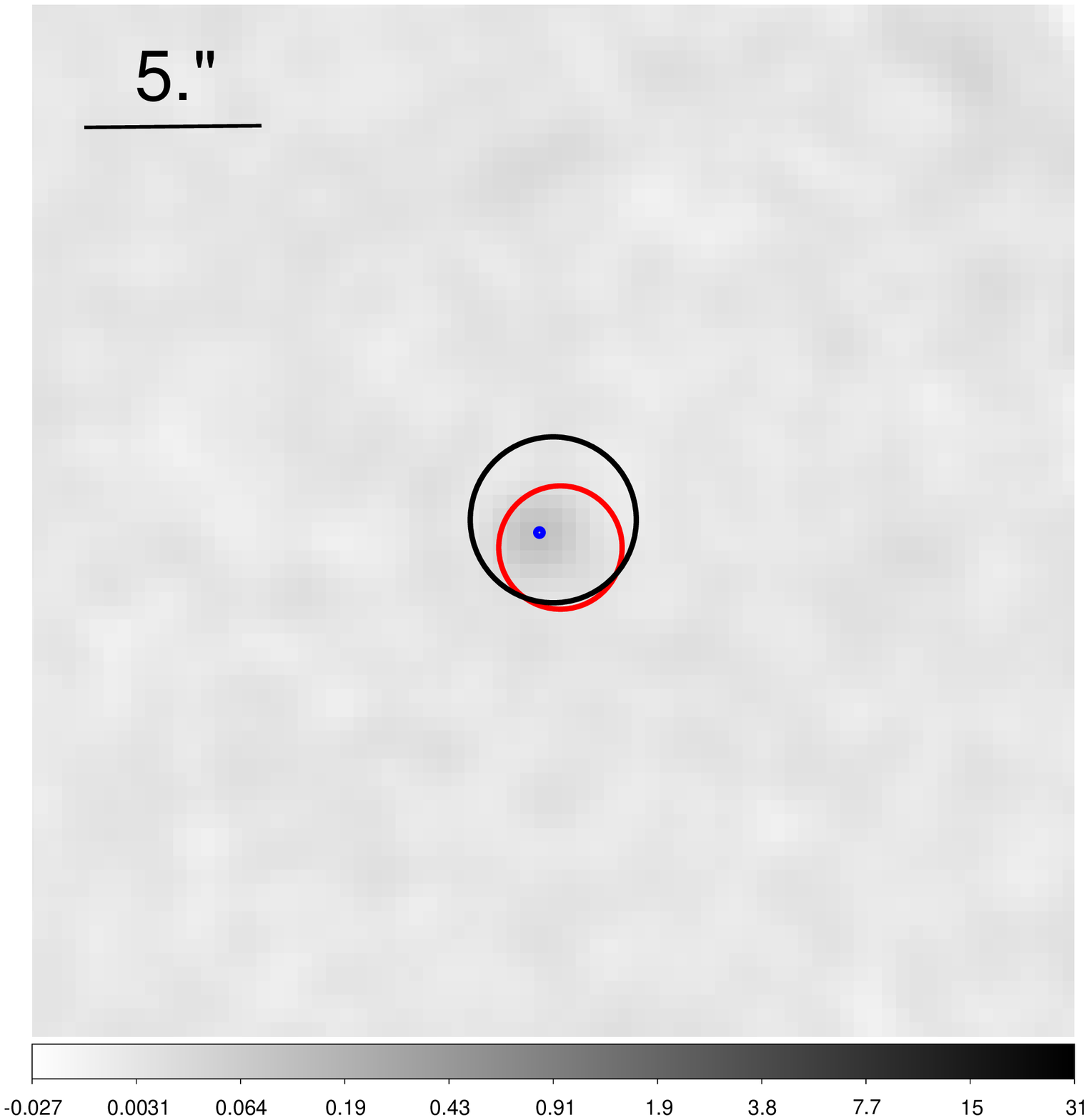}}
%\caption{\scriptsize{Src No.59}}
  \subfloat[Src No.60]{\includegraphics[clip, trim={0.0cm 2.cm 0.cm 0.0cm},width=0.19\textwidth]{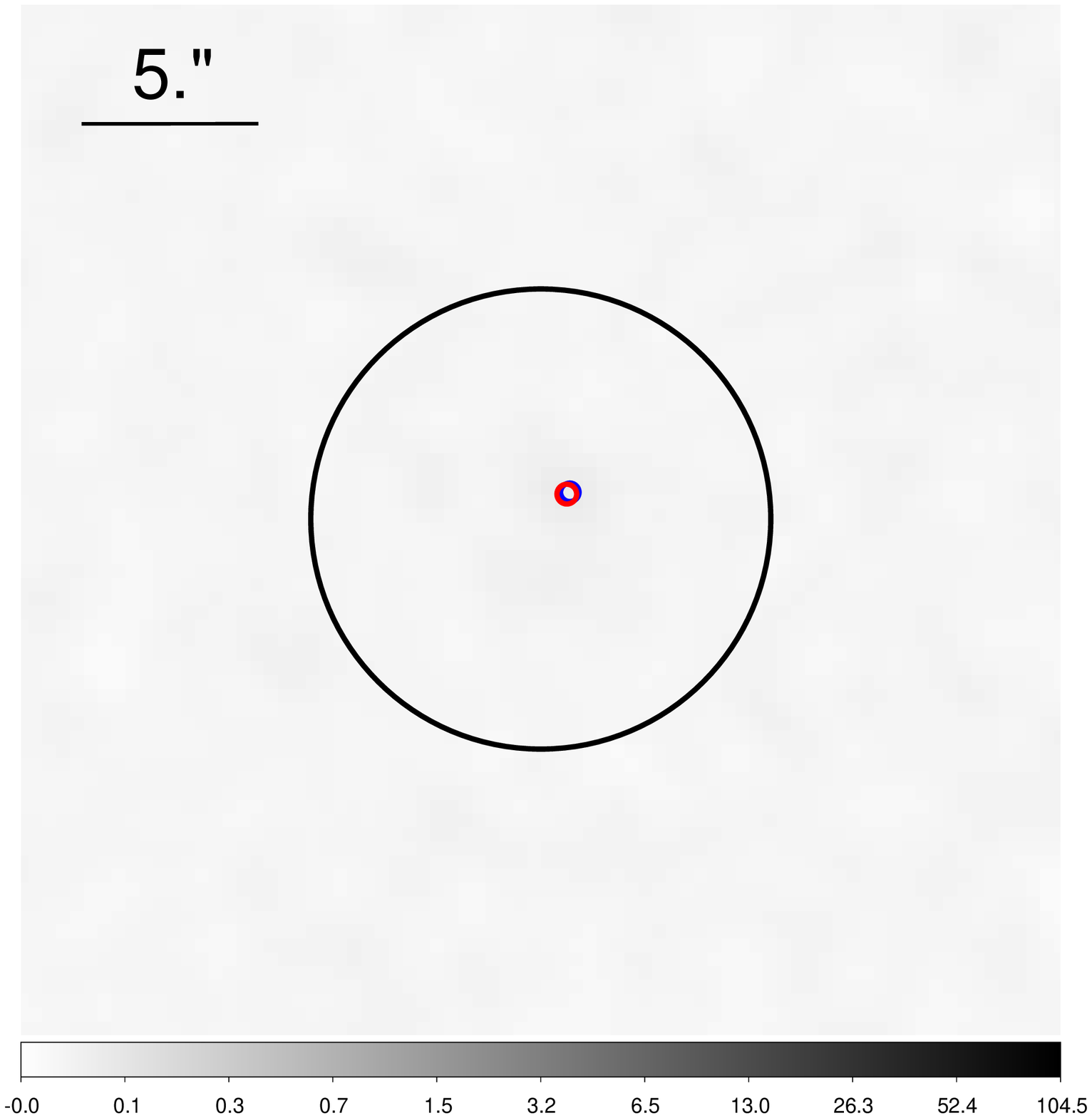}}\\
%\caption{\scriptsize{Src No.60}}
 \subfloat[Src No.61]{\includegraphics[clip, trim={0.0cm 2.cm 0.cm 0.0cm},width=0.19\textwidth]{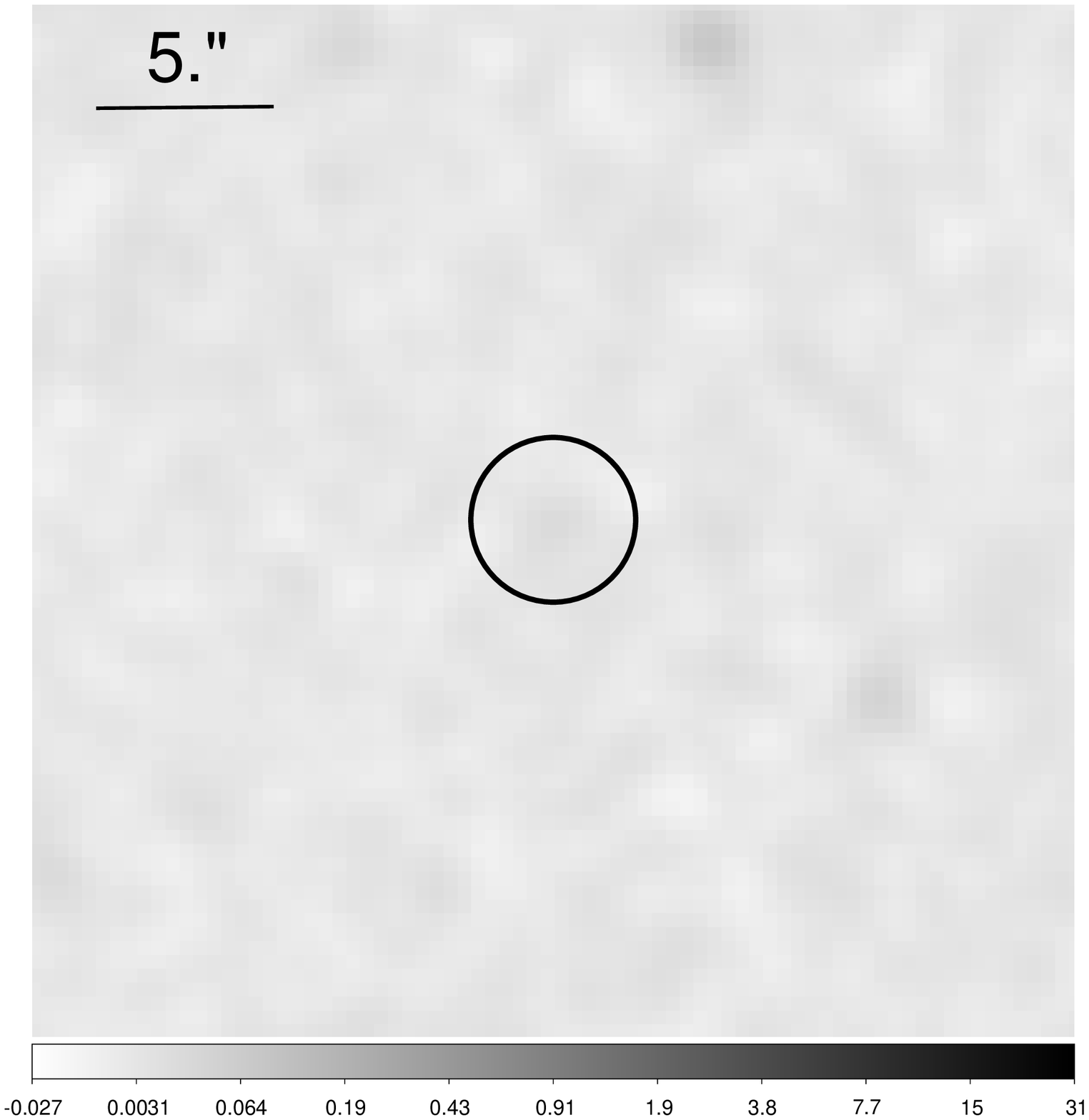}}
%\caption{\scriptsize{Src No.61}}
  \subfloat[Src No.62]{\includegraphics[clip, trim={0.0cm 2.cm 0.cm 0.0cm},width=0.19\textwidth]{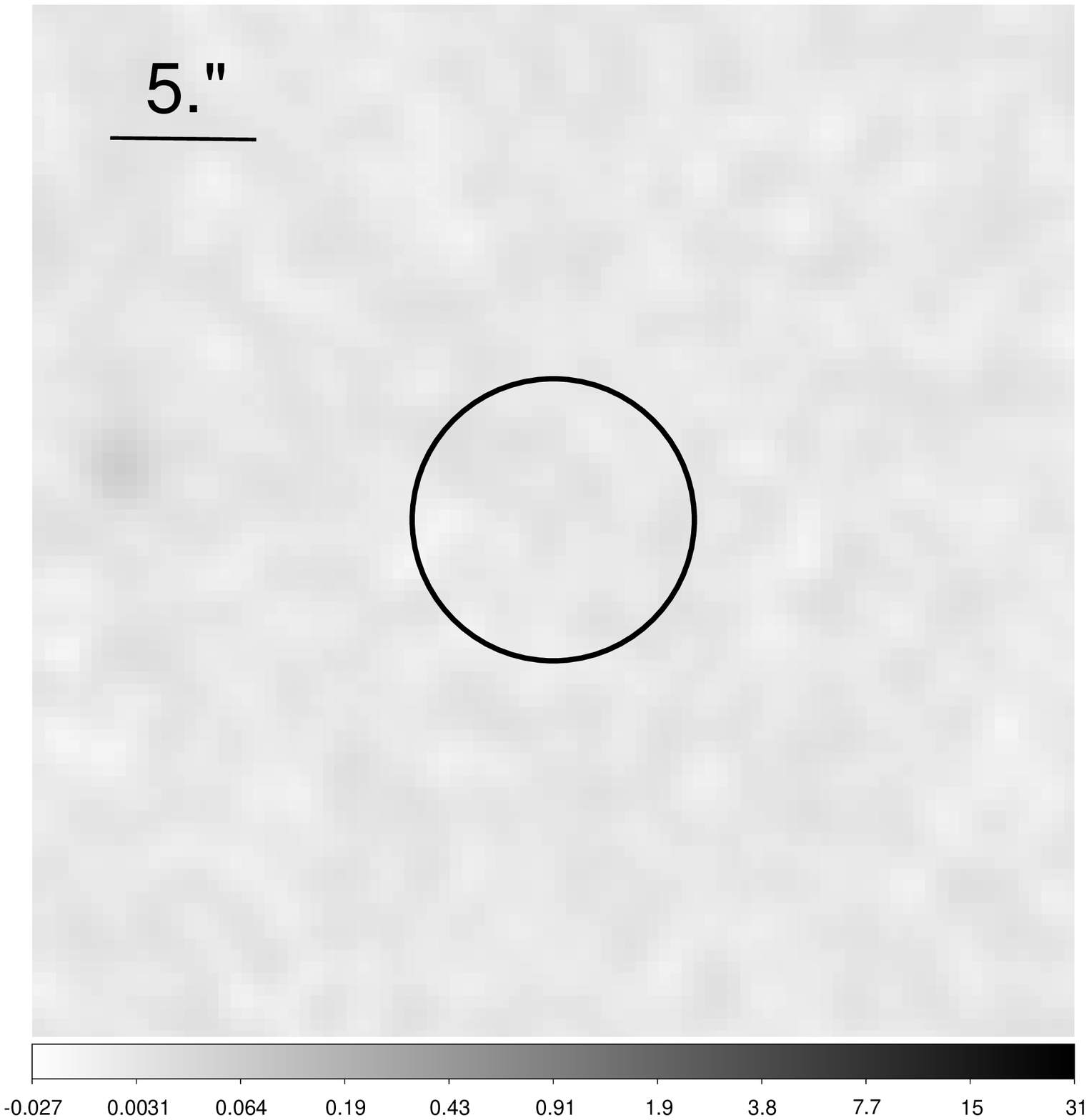}}
%\caption{\scriptsize{Src No.62}}
  \subfloat[Src No.63]{\includegraphics[clip, trim={0.0cm 2.cm 0.cm 0.0cm},width=0.19\textwidth]{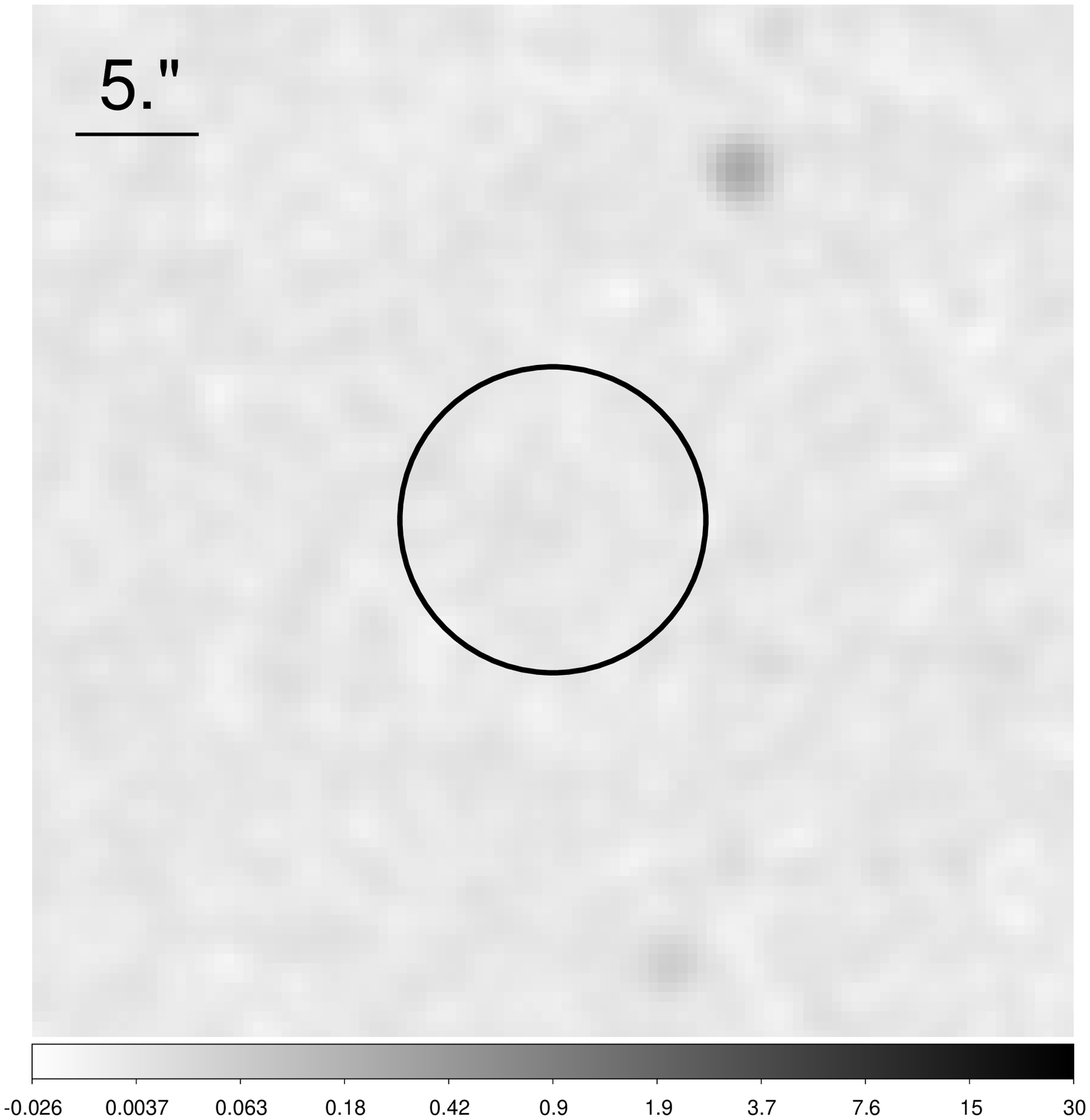}}
%\caption{\scriptsize{Src No.63}}
  \subfloat[Src No.64]{\includegraphics[clip, trim={0.0cm 2.cm 0.cm 0.0cm},width=0.19\textwidth]{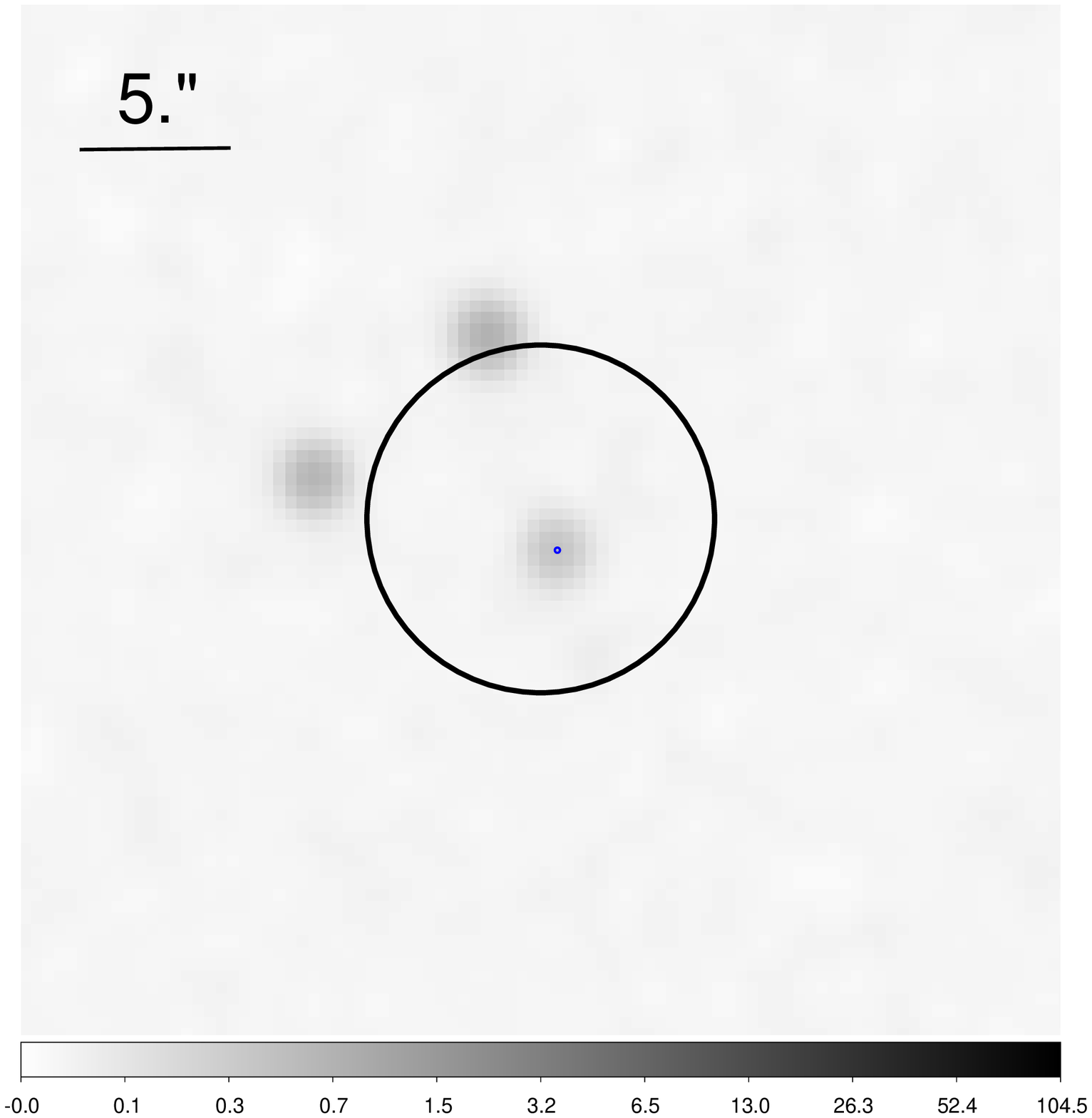}}\\
%\caption{\scriptsize{Src No.64}}
\subfloat[Src No.65]{\includegraphics[clip, trim={0.0cm 2.cm 0.cm 0.0cm},width=0.19\textwidth]{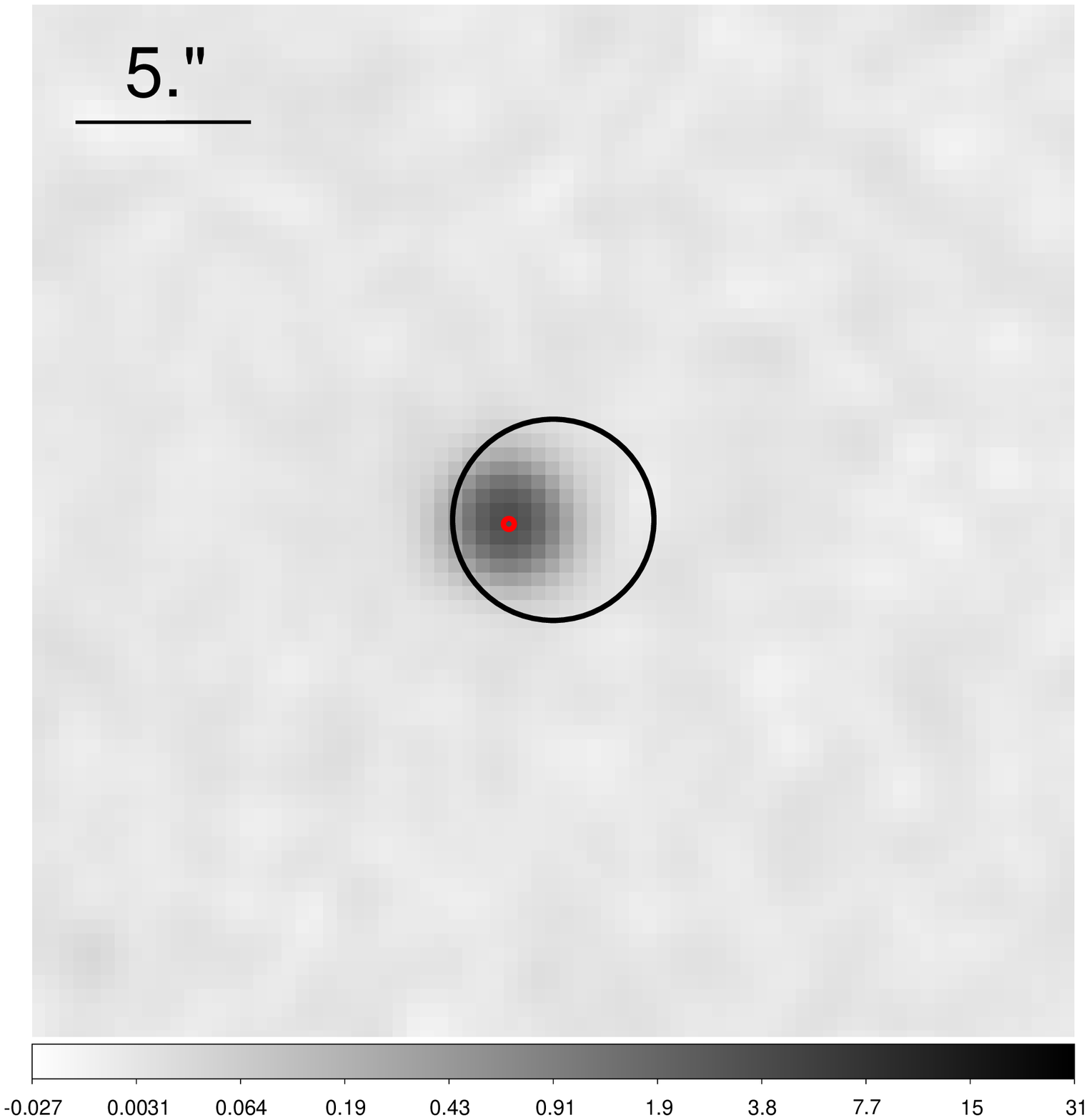}}
%\caption{\scriptsize{Src No.65}}
  \subfloat[Src No.66]{\includegraphics[clip, trim={0.0cm 2.cm 0.cm 0.0cm},width=0.19\textwidth]{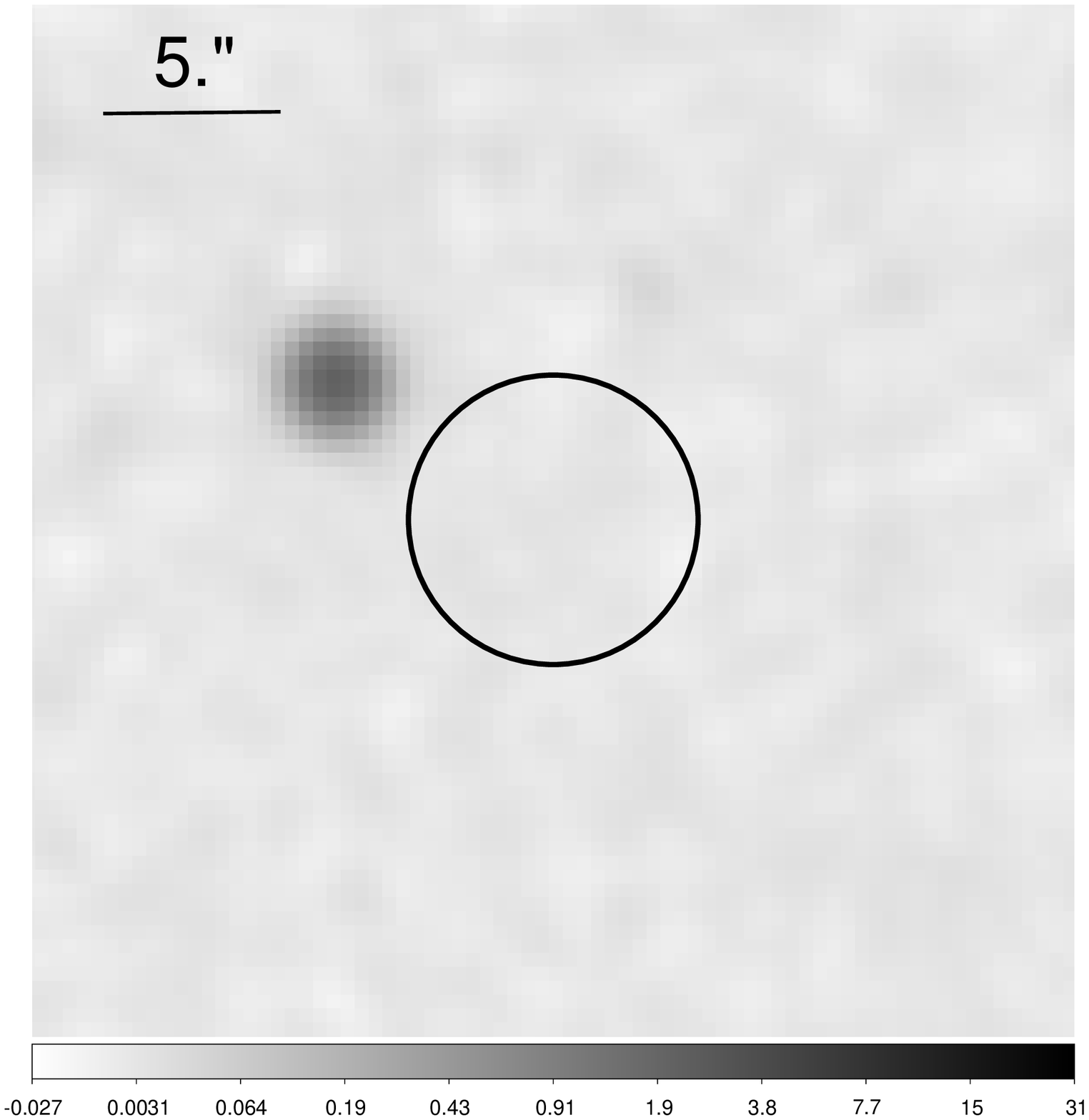}}
%\caption{\scriptsize{Src No.66}}
  \subfloat[Src No.67]{\includegraphics[clip, trim={0.0cm 2.cm 0.cm 0.0cm},width=0.19\textwidth]{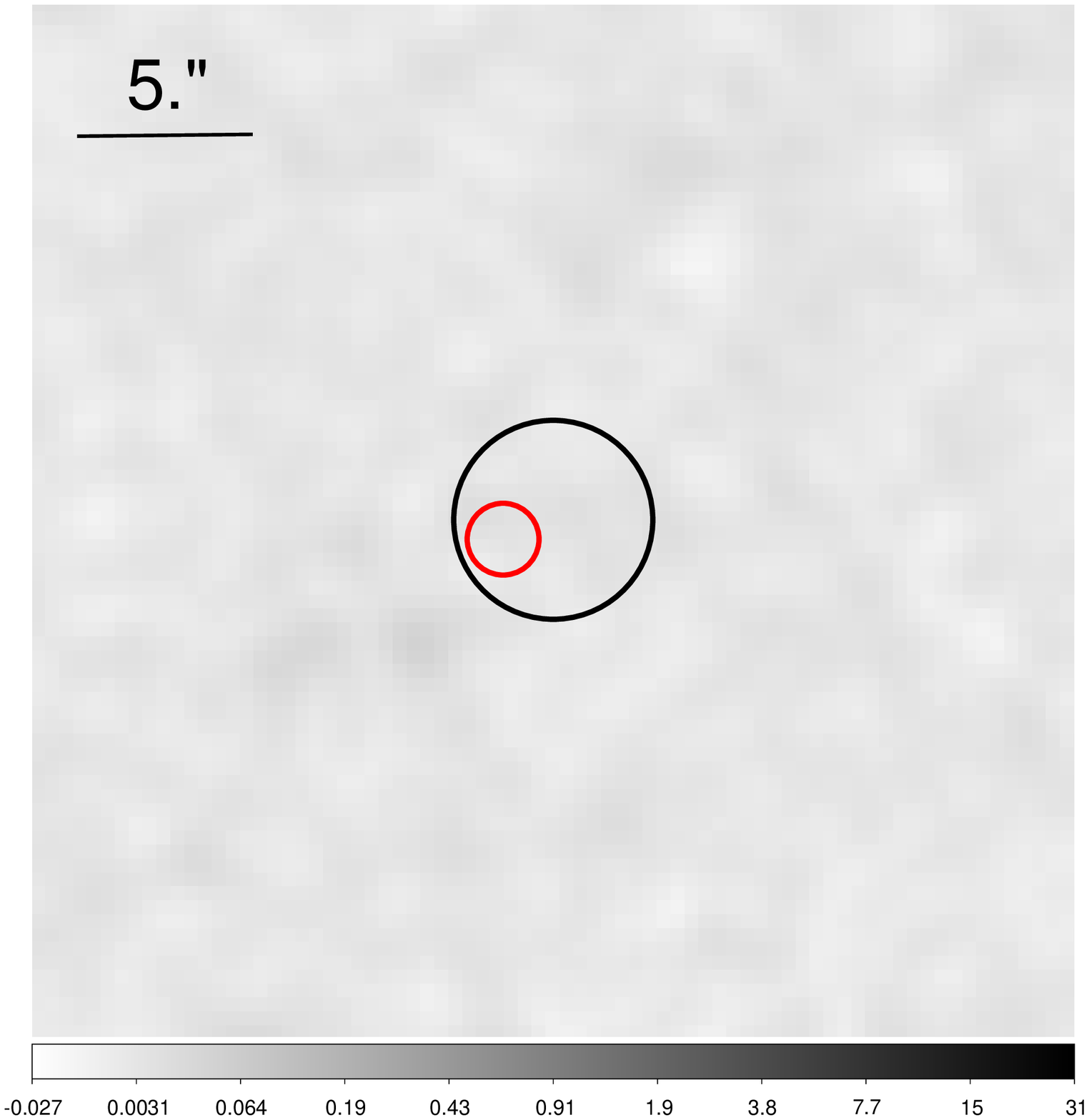}}
%\caption{\scriptsize{Src No.67}}
  \subfloat[Src No.68]{\includegraphics[clip, trim={0.0cm 2.cm 0.cm 0.0cm},width=0.19\textwidth]{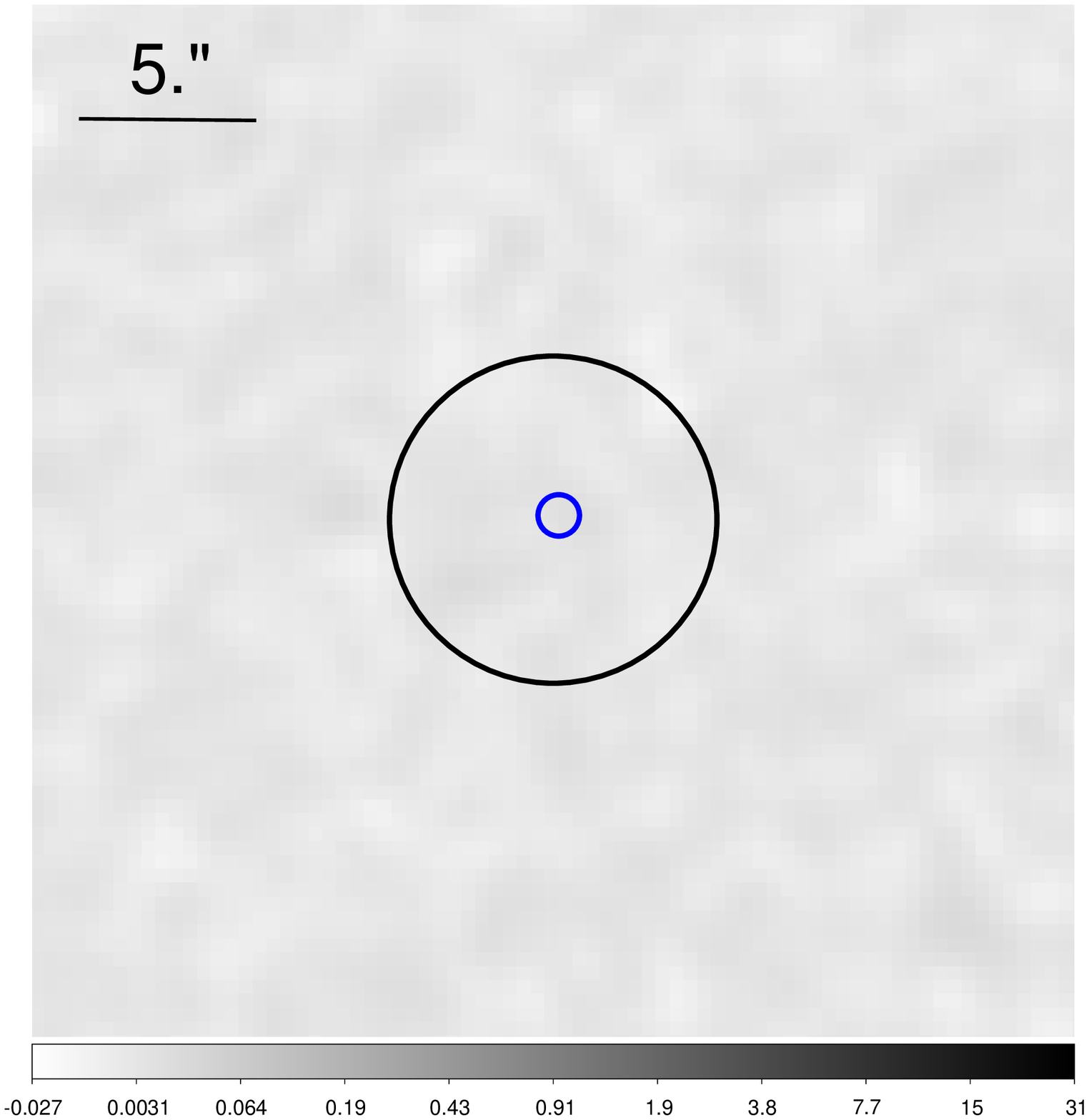}}\\
%\caption{\scriptsize{Src No.68}}

\end{figure*}
\pagebreak
\clearpage
%\hspace{0.3cm}Appendix B continued: Image of optical SDSS9 counterparts
\begin{figure*}
\vspace{-0.5cm}
  \subfloat[Src No.69]{\includegraphics[clip, trim={0.0cm 2.cm 0.cm 0.0cm},width=0.19\textwidth]{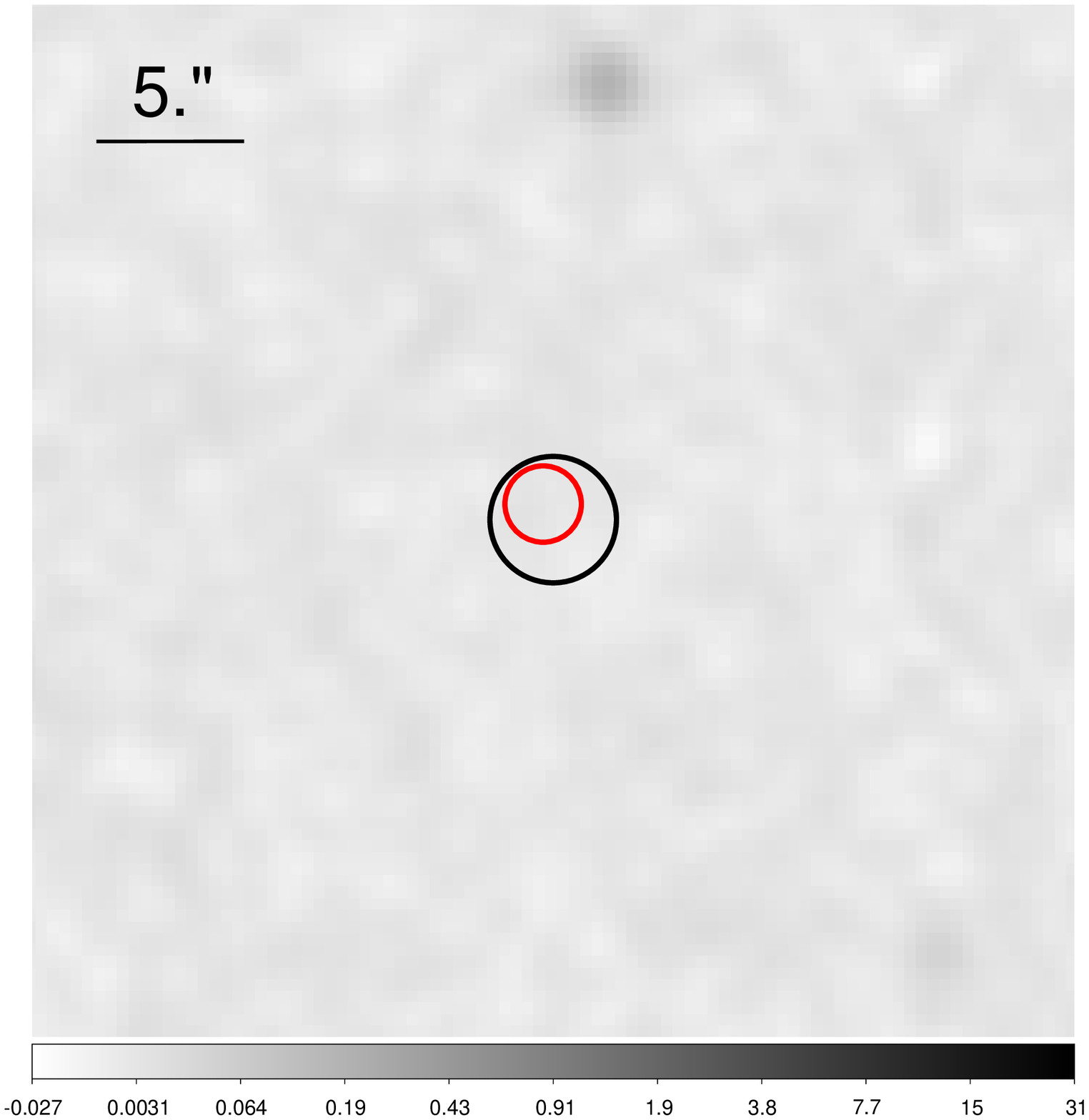}}
%\caption{\scriptsize{Src No.69}}
  \subfloat[Src No.70]{\includegraphics[clip, trim={0.0cm 2.cm 0.cm 0.0cm},width=0.19\textwidth]{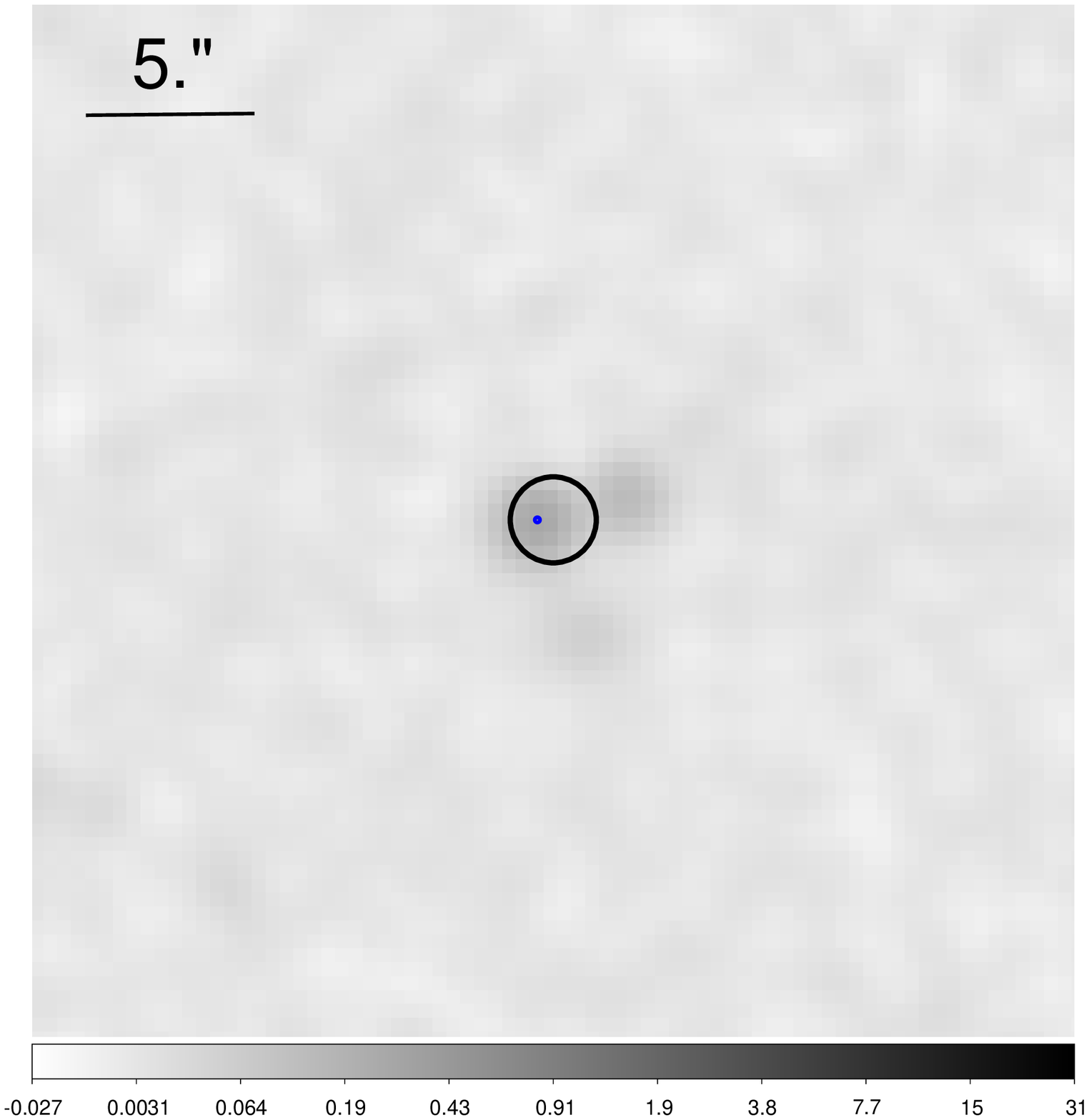}}
%\caption{\scriptsize{Src No.70}}
  \subfloat[Src No.71]{\includegraphics[clip, trim={0.0cm 2.cm 0.cm 0.0cm},width=0.19\textwidth]{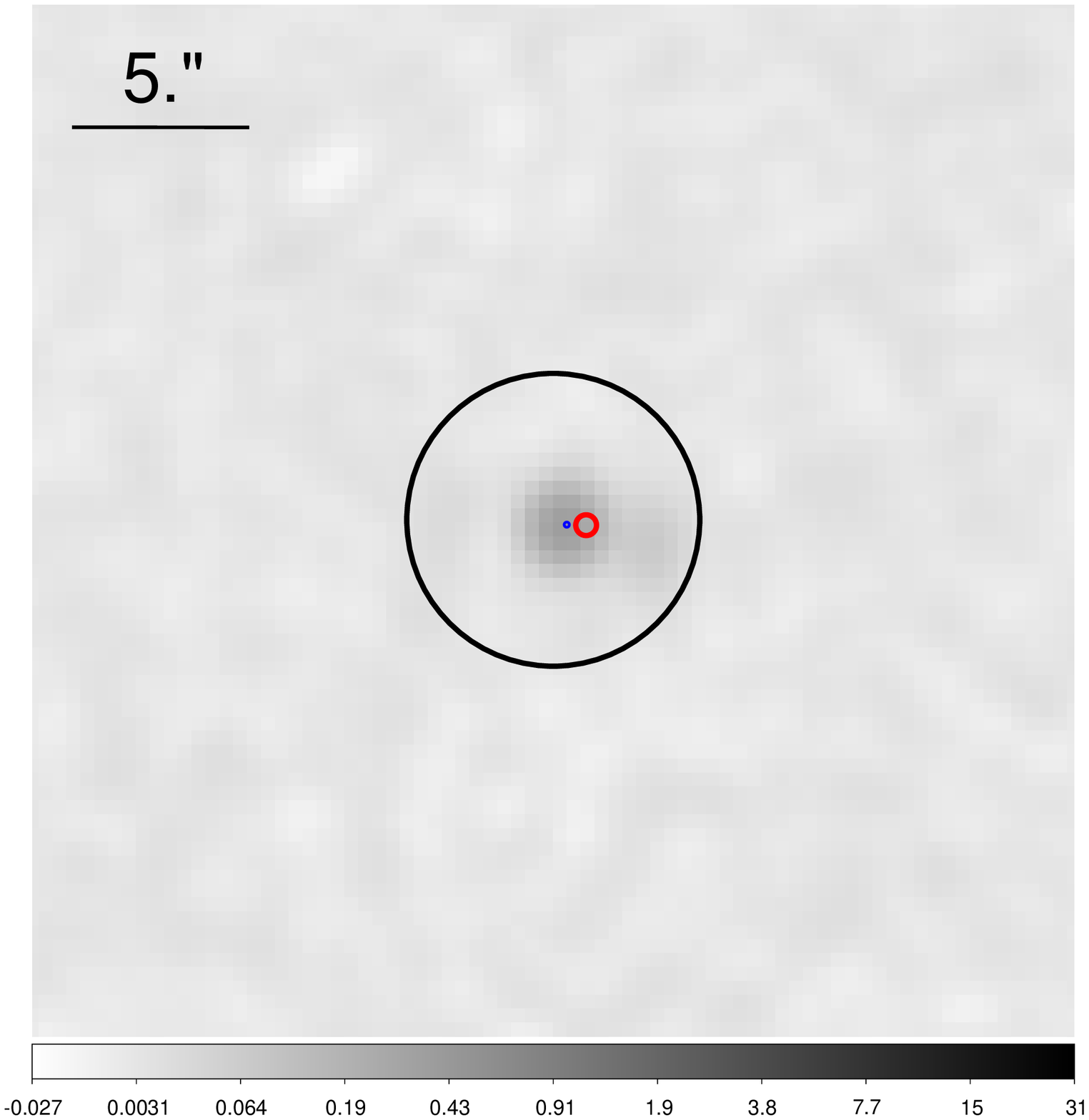}}
%\caption{\scriptsize{Src No.71}}
  \subfloat[Src No.72]{\includegraphics[clip, trim={0.0cm 2.cm 0.cm 0.0cm},width=0.19\textwidth]{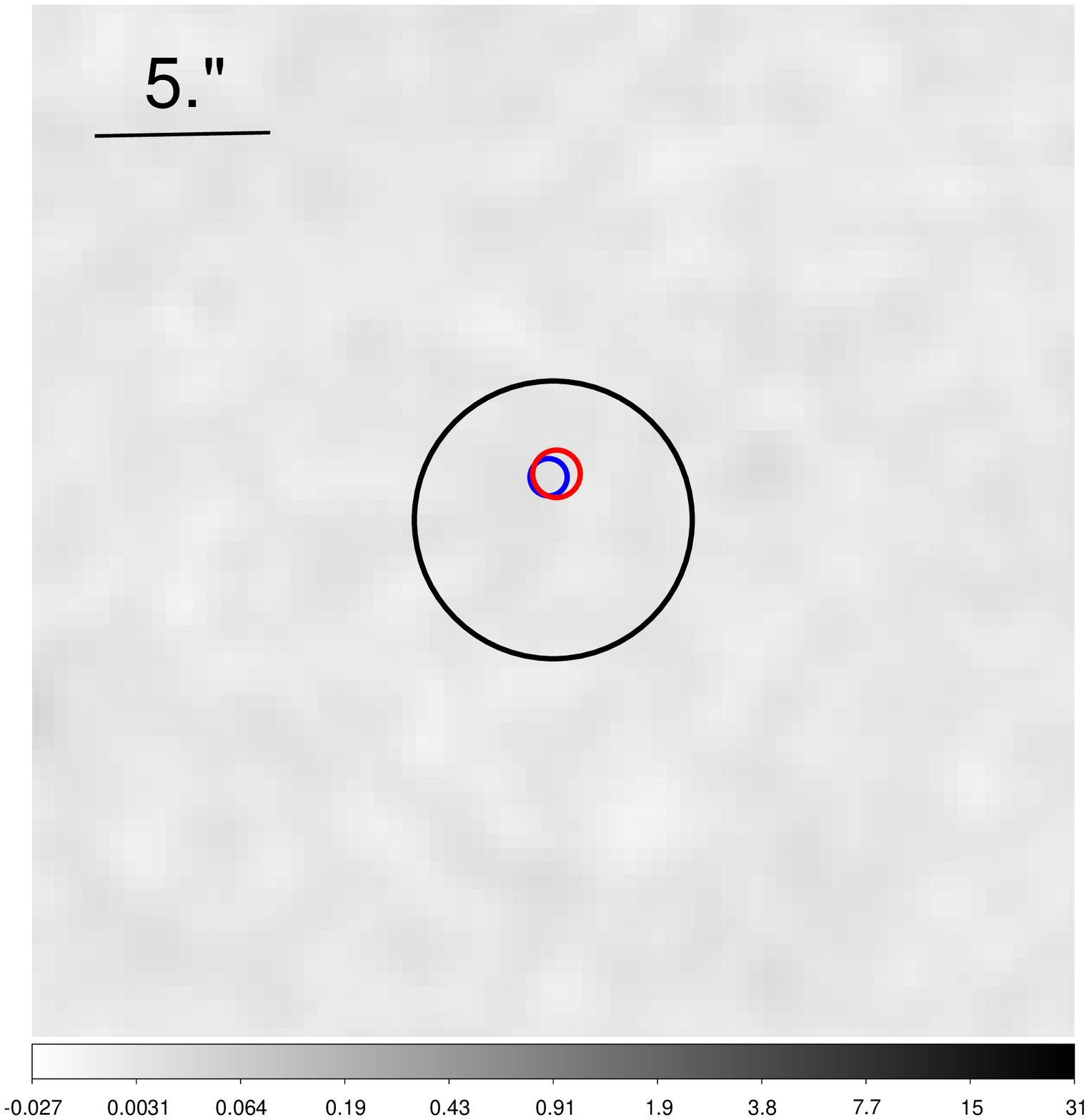}}\\
%\caption{\scriptsize{Src No.72}}
  \subfloat[Src No.73]{\includegraphics[clip, trim={0.0cm 2.cm 0.cm 0.0cm},width=0.19\textwidth]{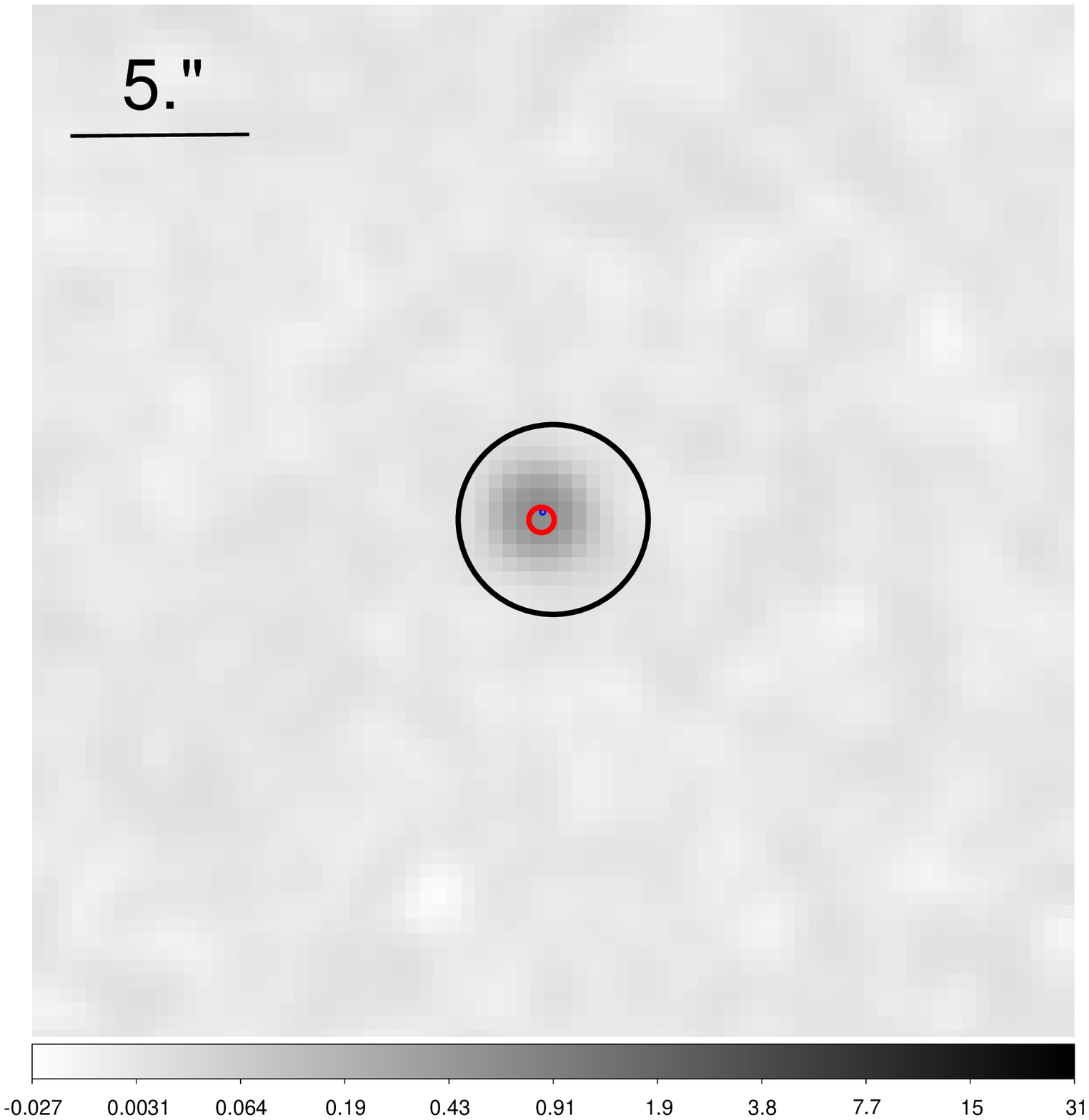}}
%\caption{\scriptsize{Src No.73}}
  \subfloat[Src No.74]{\includegraphics[clip, trim={0.0cm 2.cm 0.cm 0.0cm},width=0.19\textwidth]{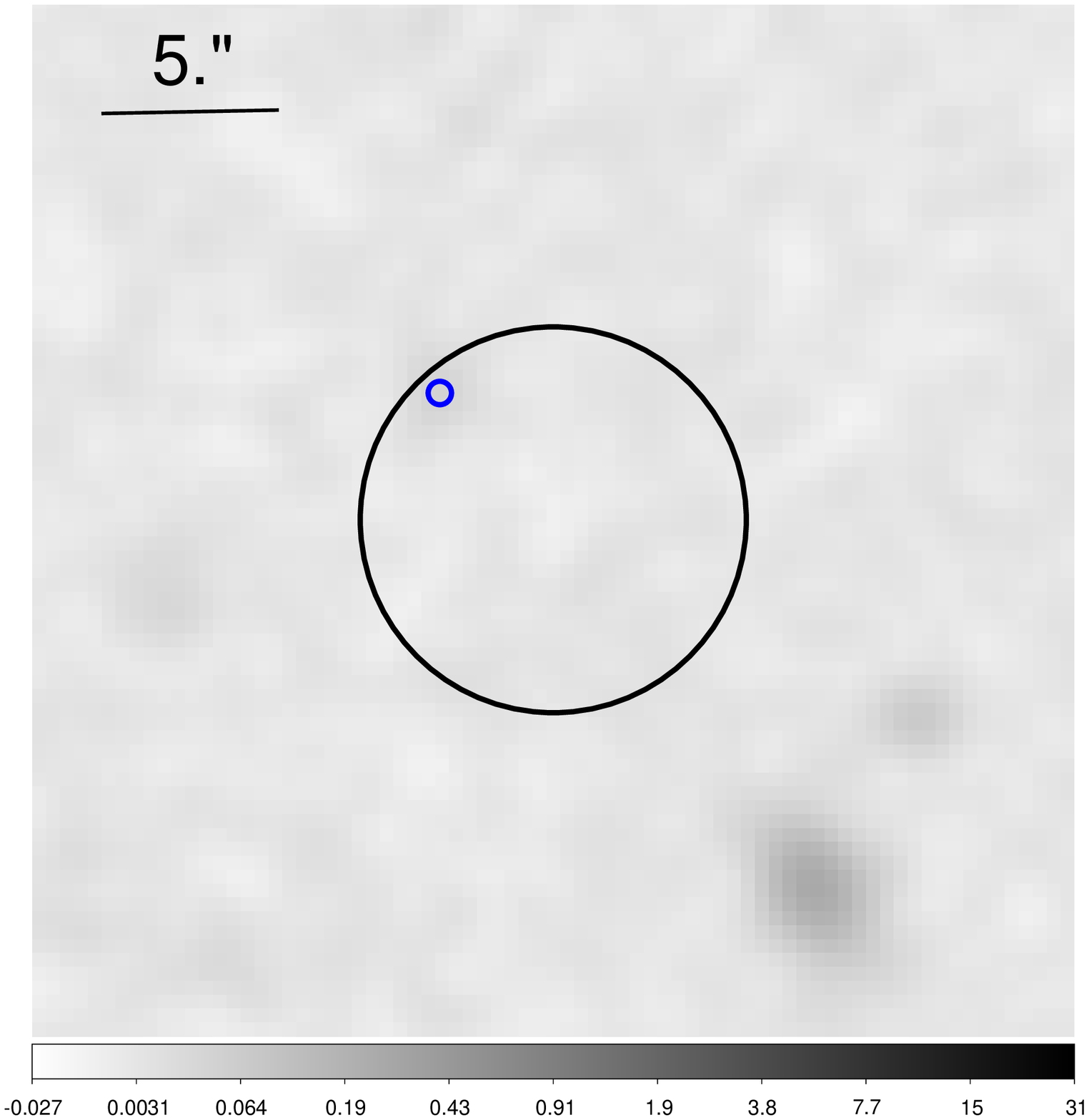}}
%\caption{\scriptsize{Src No.74}}
  \subfloat[Src No.75]{\includegraphics[clip, trim={0.0cm 2.cm 0.cm 0.0cm},width=0.19\textwidth]{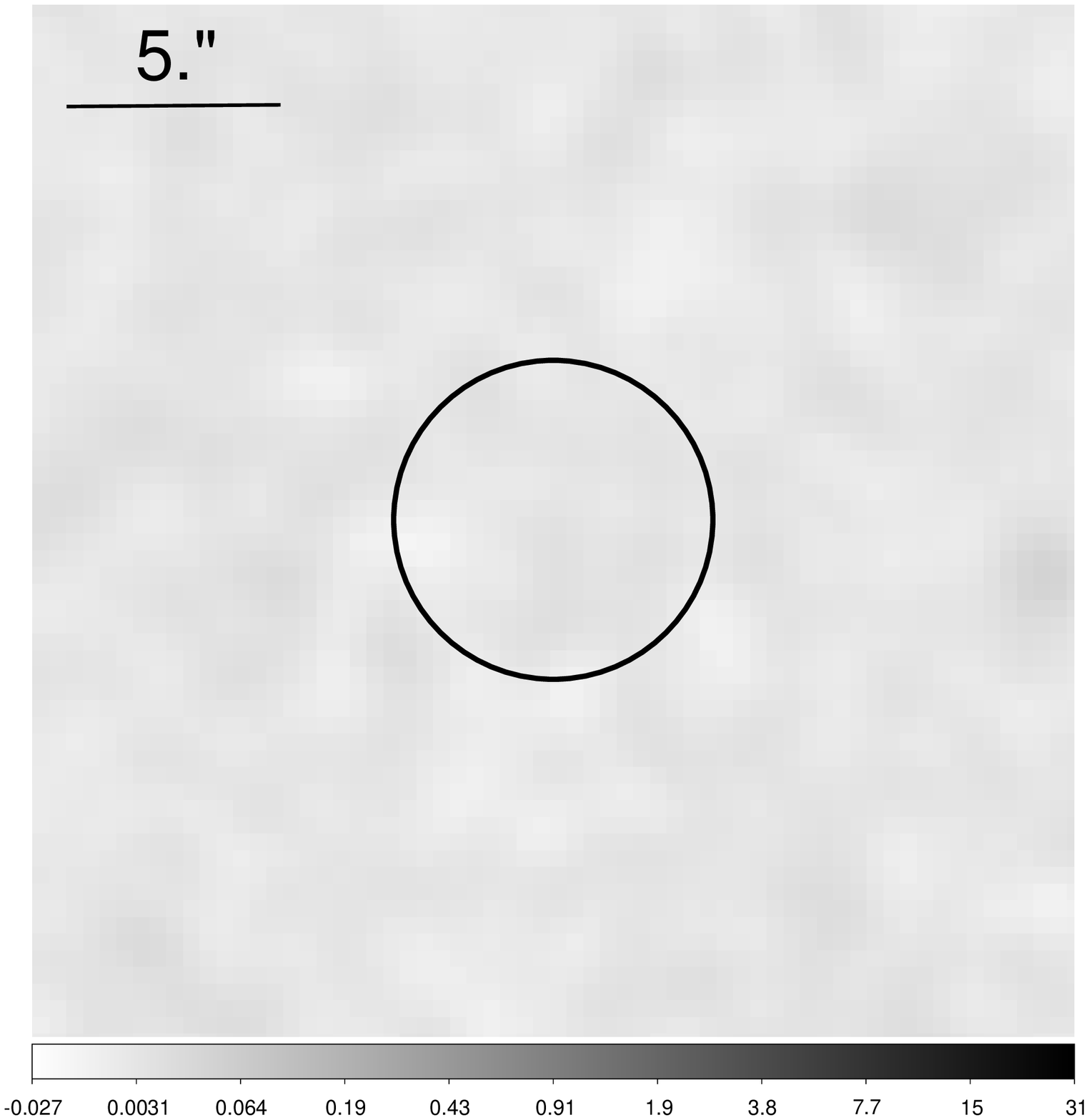}}
%\caption{\scriptsize{Src No.75}}
  \subfloat[Src No.76]{\includegraphics[clip, trim={0.0cm 2.cm 0.cm 0.0cm},width=0.19\textwidth]{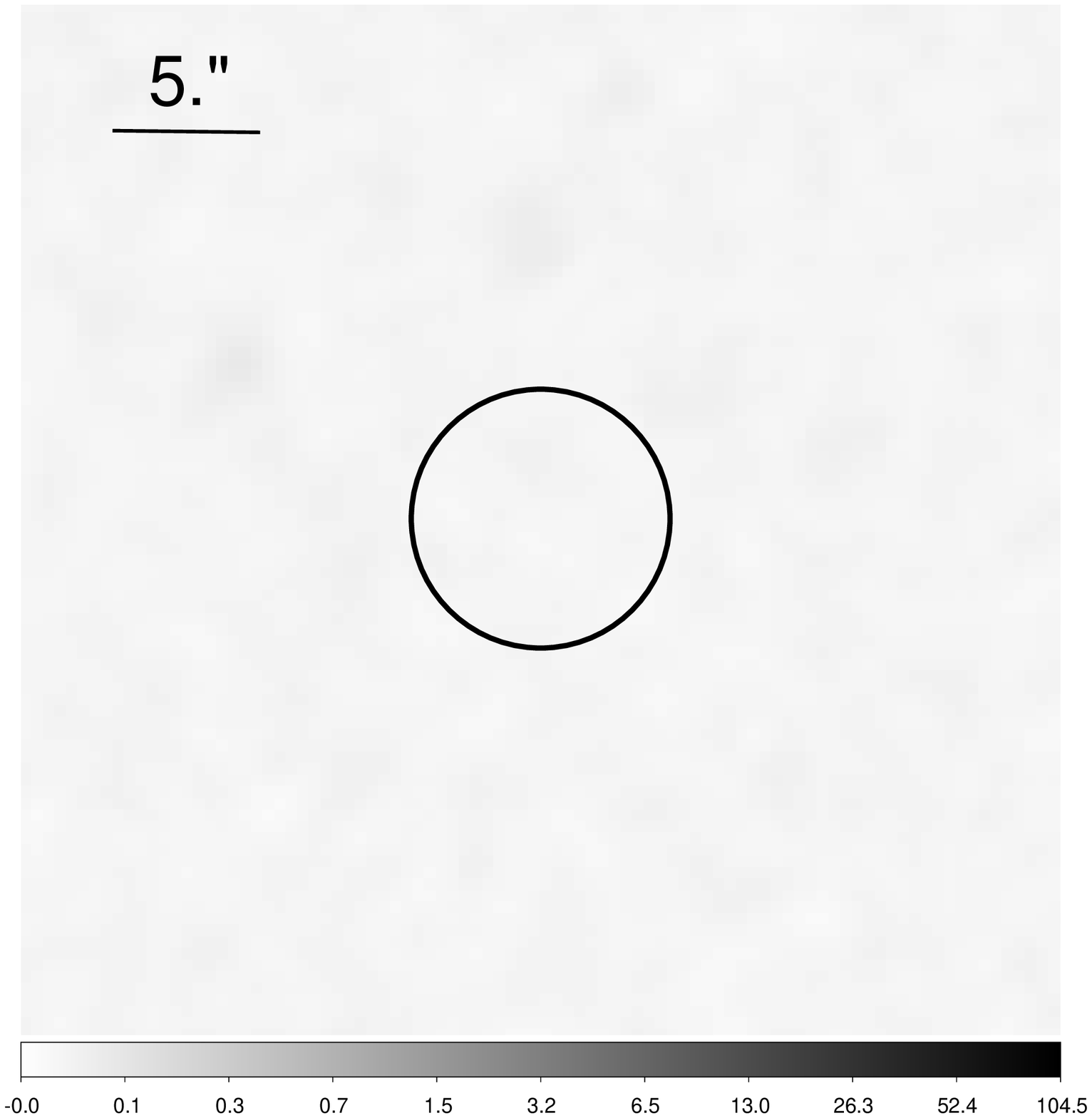}}\\
%\caption{\scriptsize{Src No.76}}
  \subfloat[Src No.77]{\includegraphics[clip, trim={0.0cm 2.cm 0.cm 0.0cm},width=0.19\textwidth]{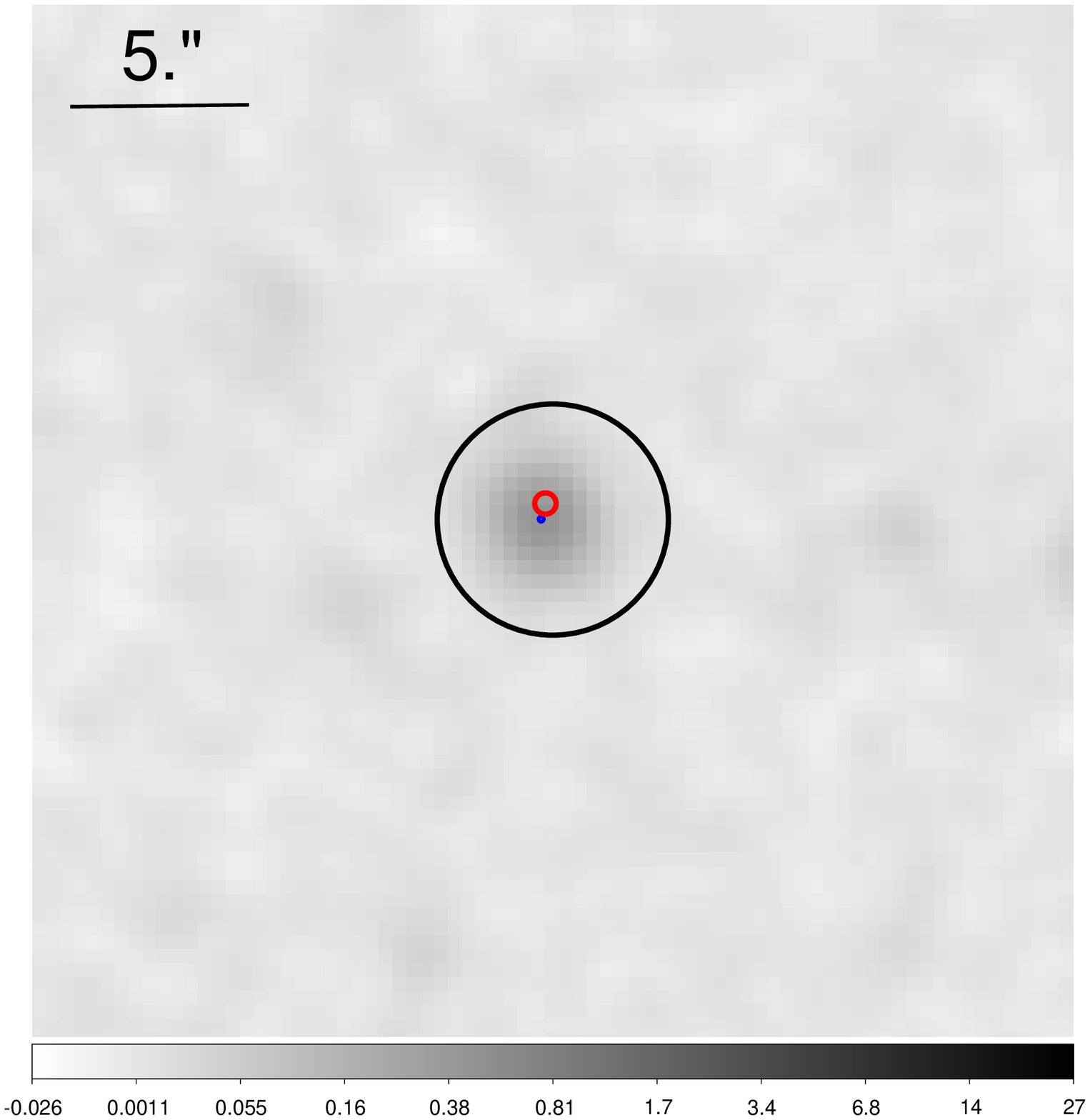}}
%\caption{\scriptsize{Src No.77}}
  \subfloat[Src No.78]{\includegraphics[clip, trim={0.0cm 2.cm 0.cm 0.0cm},width=0.19\textwidth]{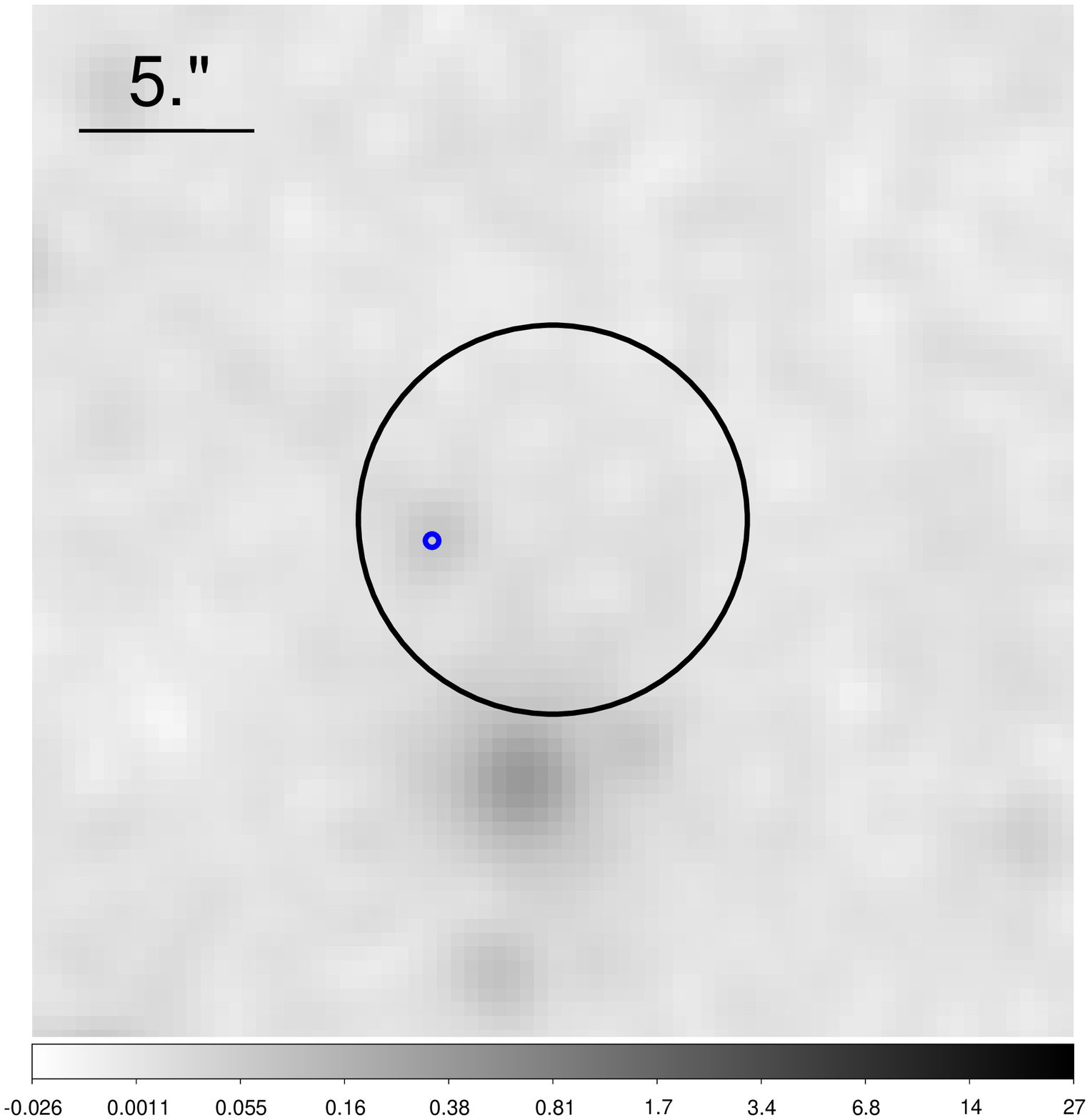}}
%\caption{\scriptsize{Src No.78}}
  \subfloat[Src No.79]{\includegraphics[clip, trim={0.0cm 2.cm 0.cm 0.0cm},width=0.19\textwidth]{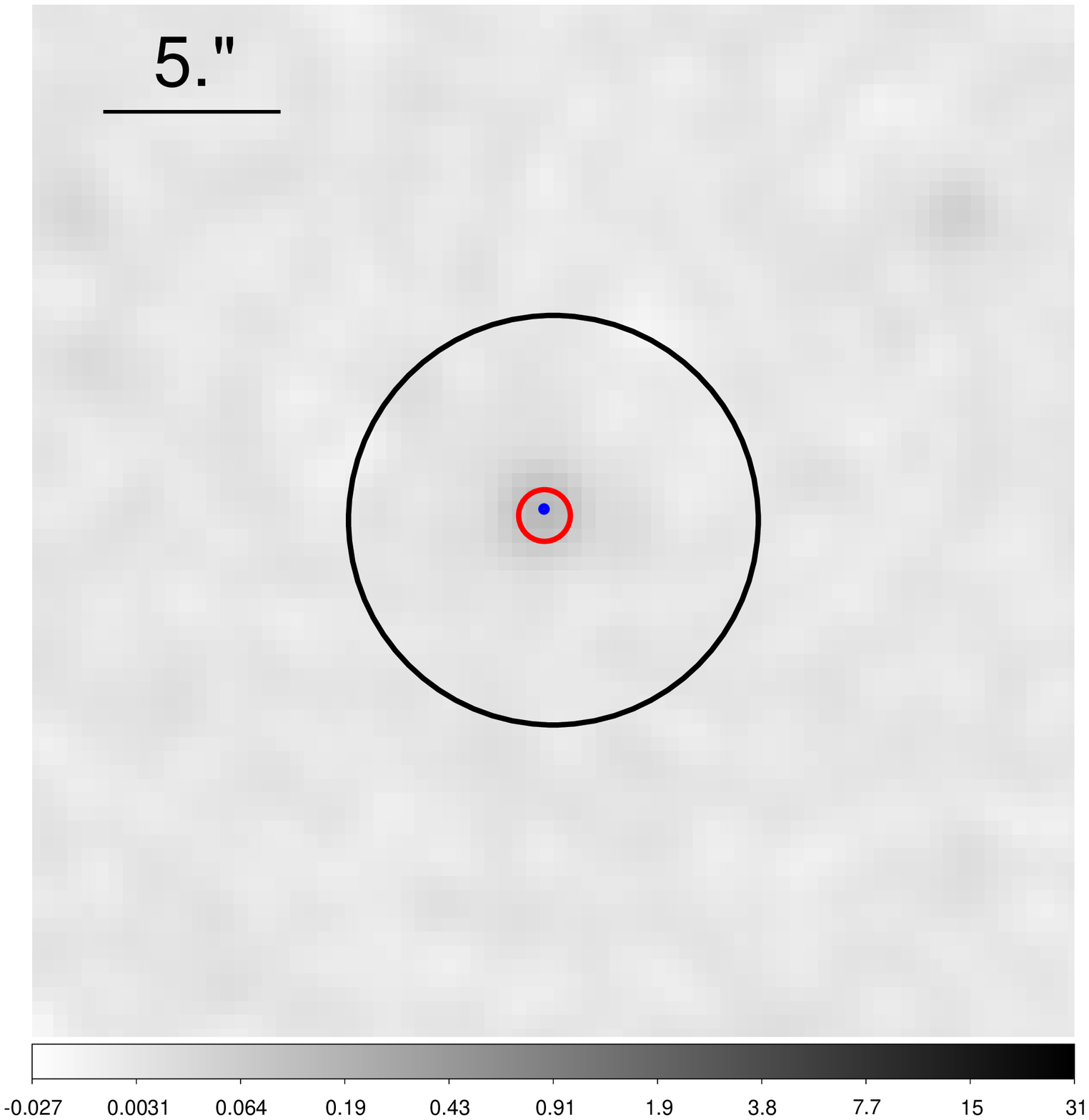}}
%\caption{\scriptsize{Src No.79}}
  \subfloat[Src No.80]{\includegraphics[clip, trim={0.0cm 2.cm 0.cm 0.0cm},width=0.19\textwidth]{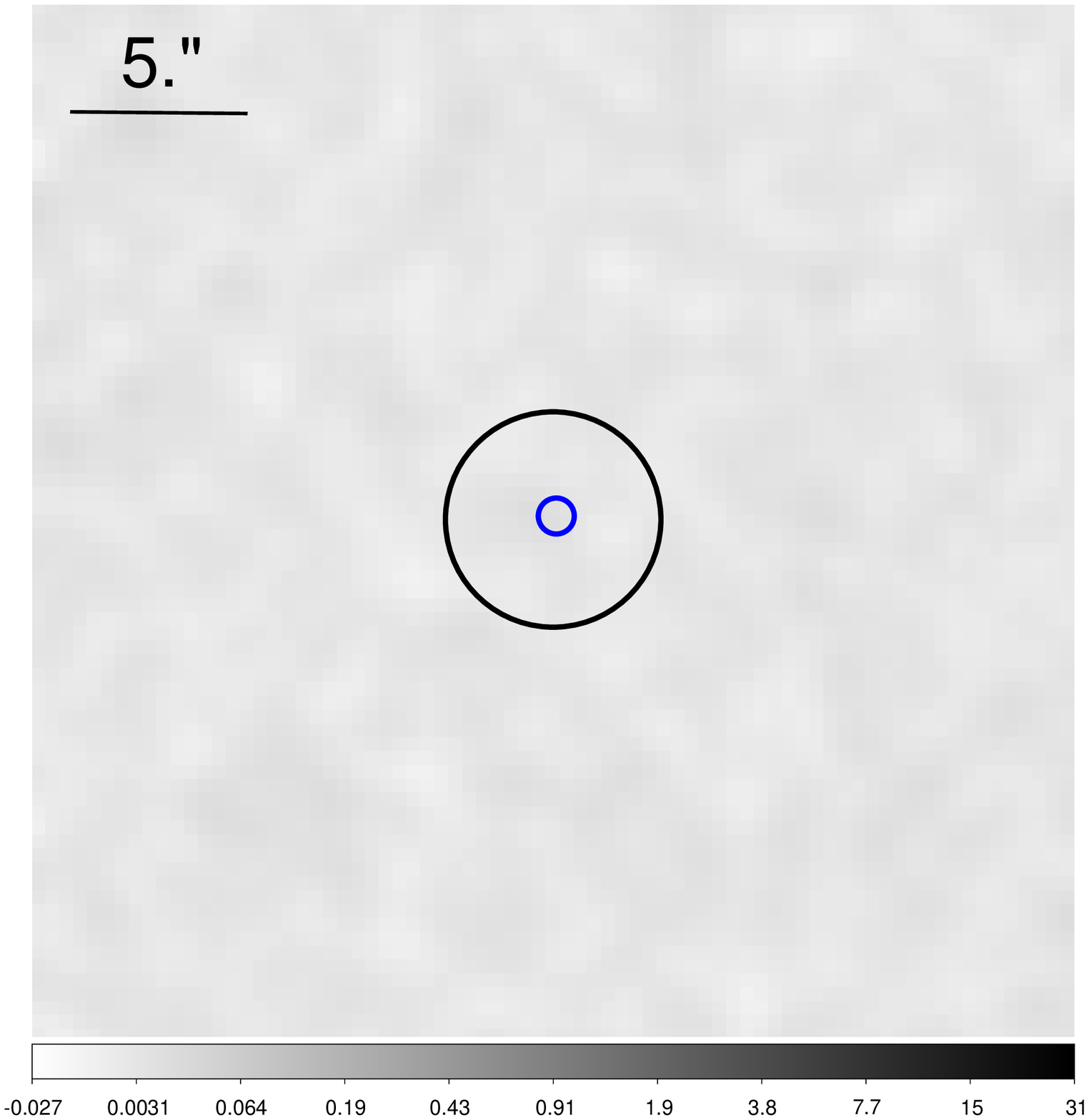}}\\
%\caption{\scriptsize{Src No.80}}
 \subfloat[Src No.81]{\includegraphics[clip, trim={0.0cm 2.cm 0.cm 0.0cm},width=0.19\textwidth]{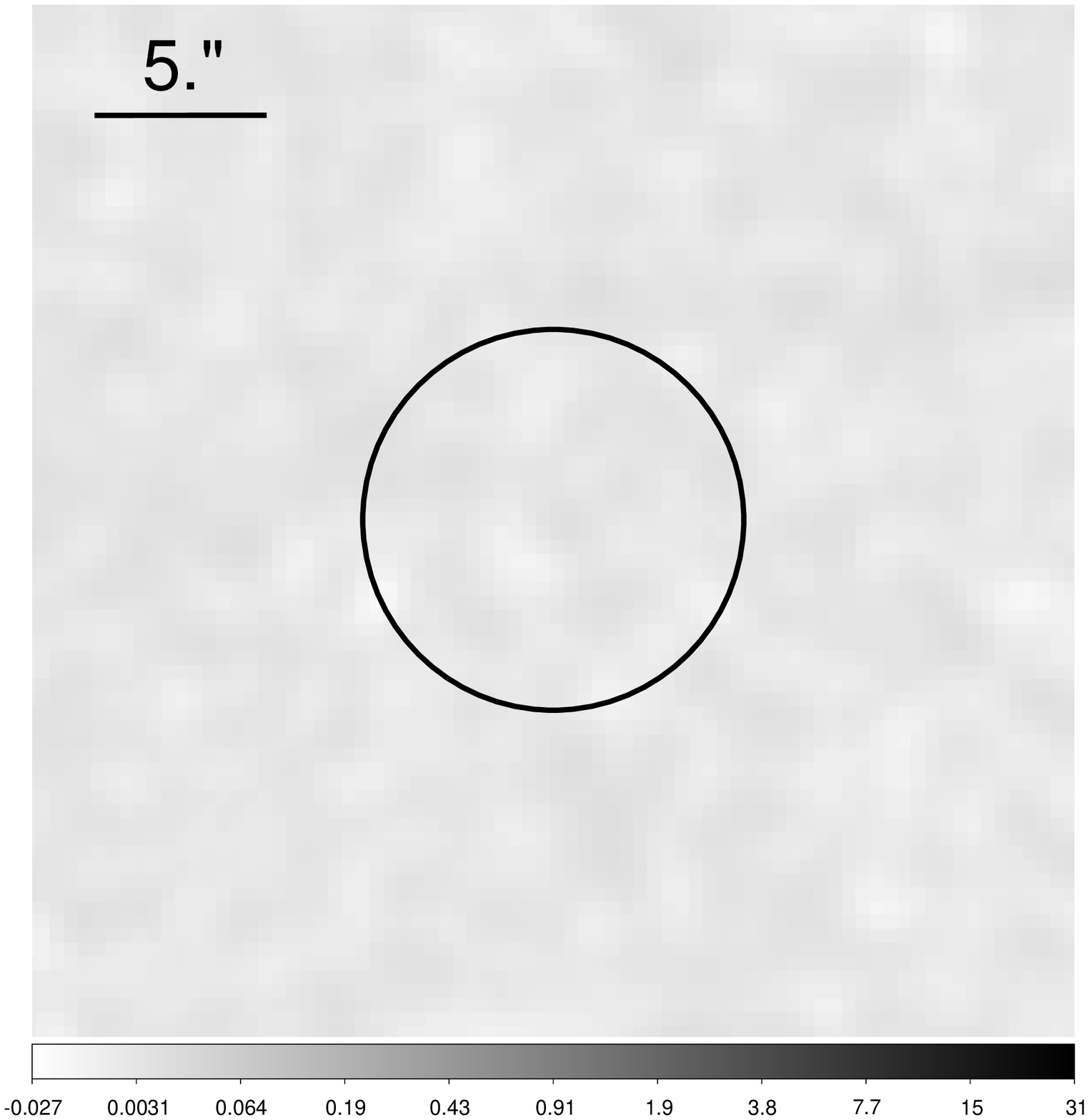}}
%\caption{\scriptsize{Src No.81}}
 \subfloat[Src No.82]{\includegraphics[clip, trim={0.0cm 2.cm 0.cm 0.0cm},width=0.19\textwidth]{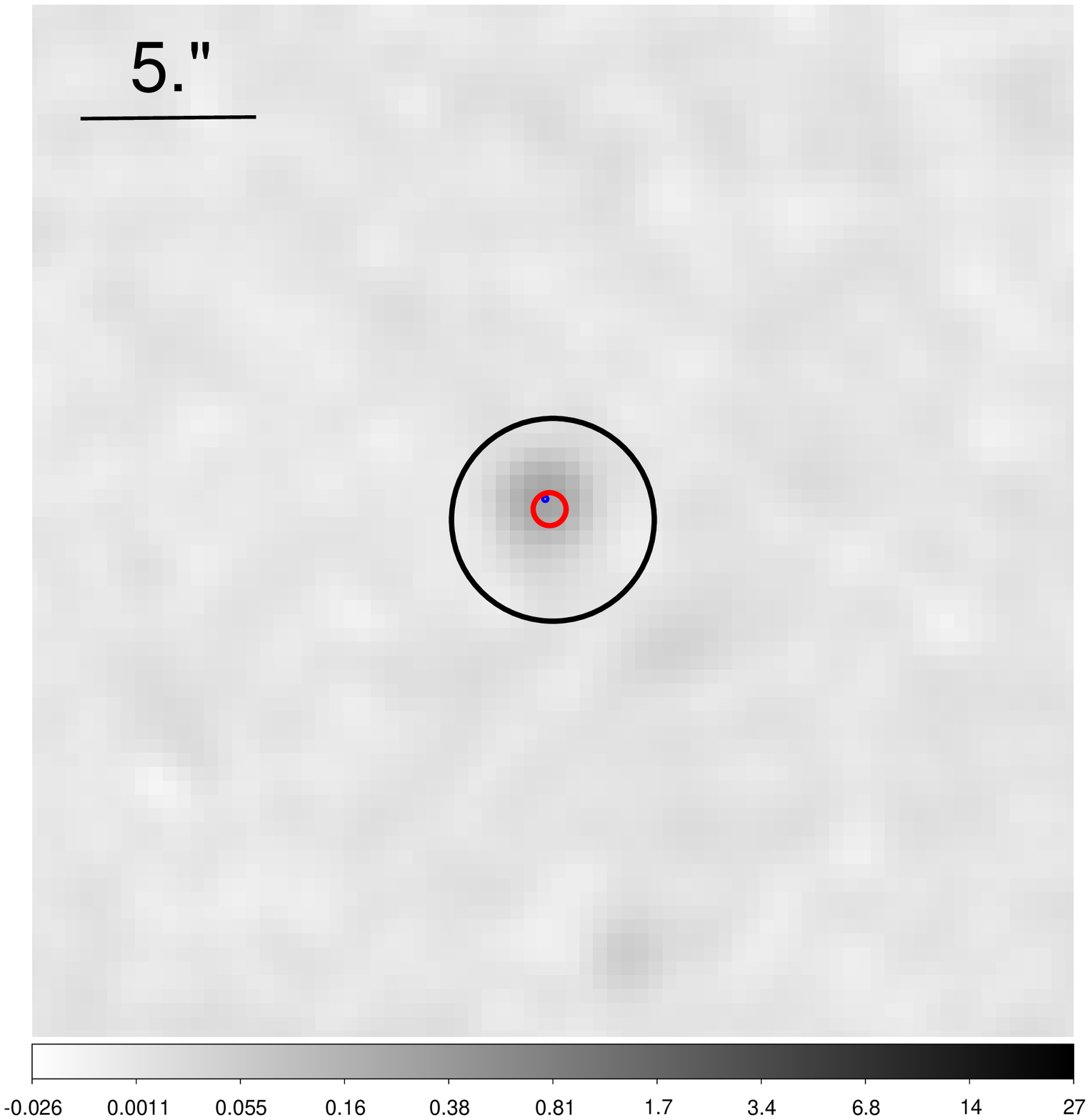}}
%\caption{\scriptsize{Src No.82}}
  \subfloat[Src No.83]{\includegraphics[clip, trim={0.0cm 2.cm 0.cm 0.0cm},width=0.19\textwidth]{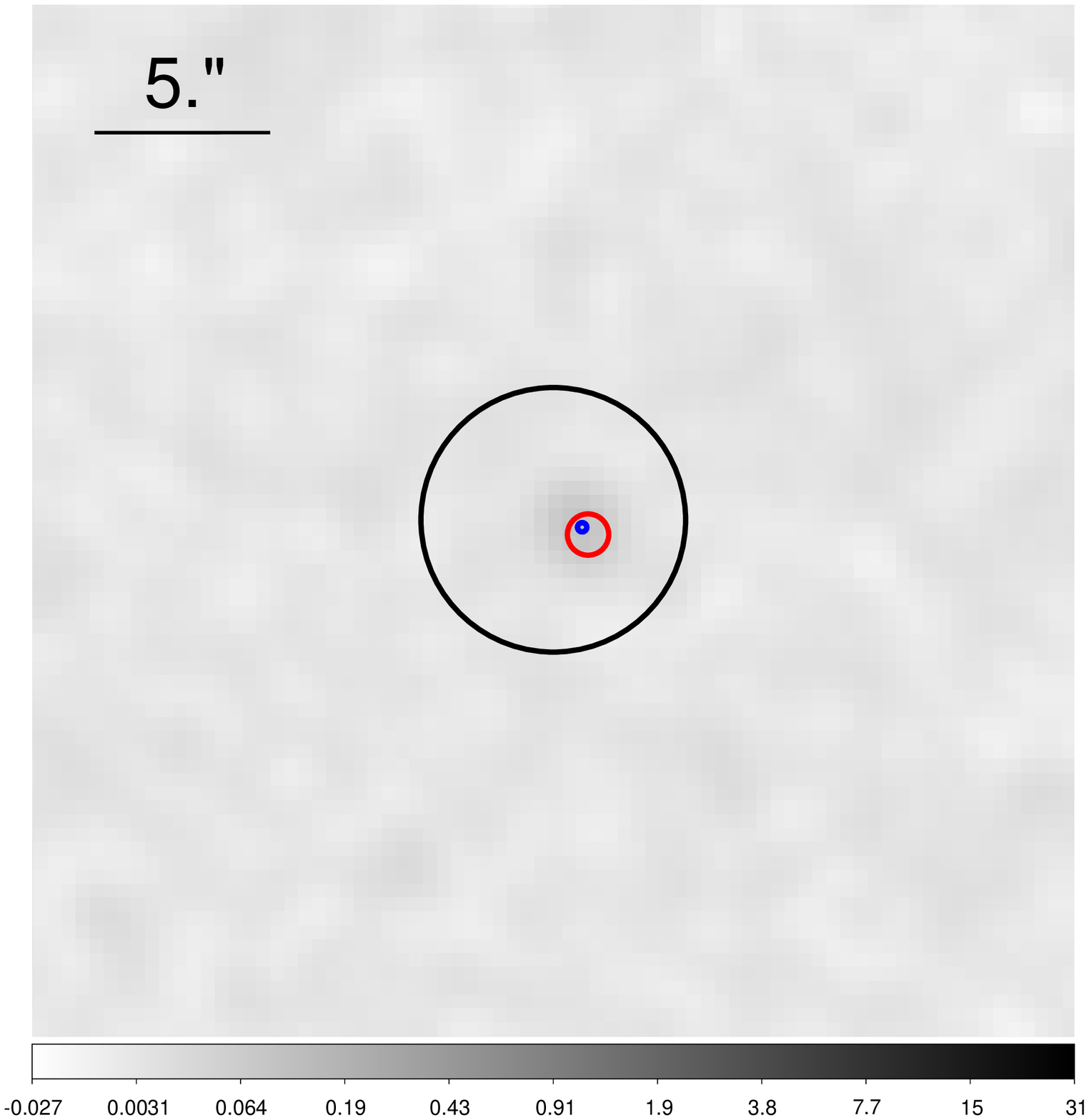}}
%\caption{\scriptsize{Src No.83}}
  \subfloat[Src No.84]{\includegraphics[clip, trim={0.0cm 2.cm 0.cm 0.0cm},width=0.19\textwidth]{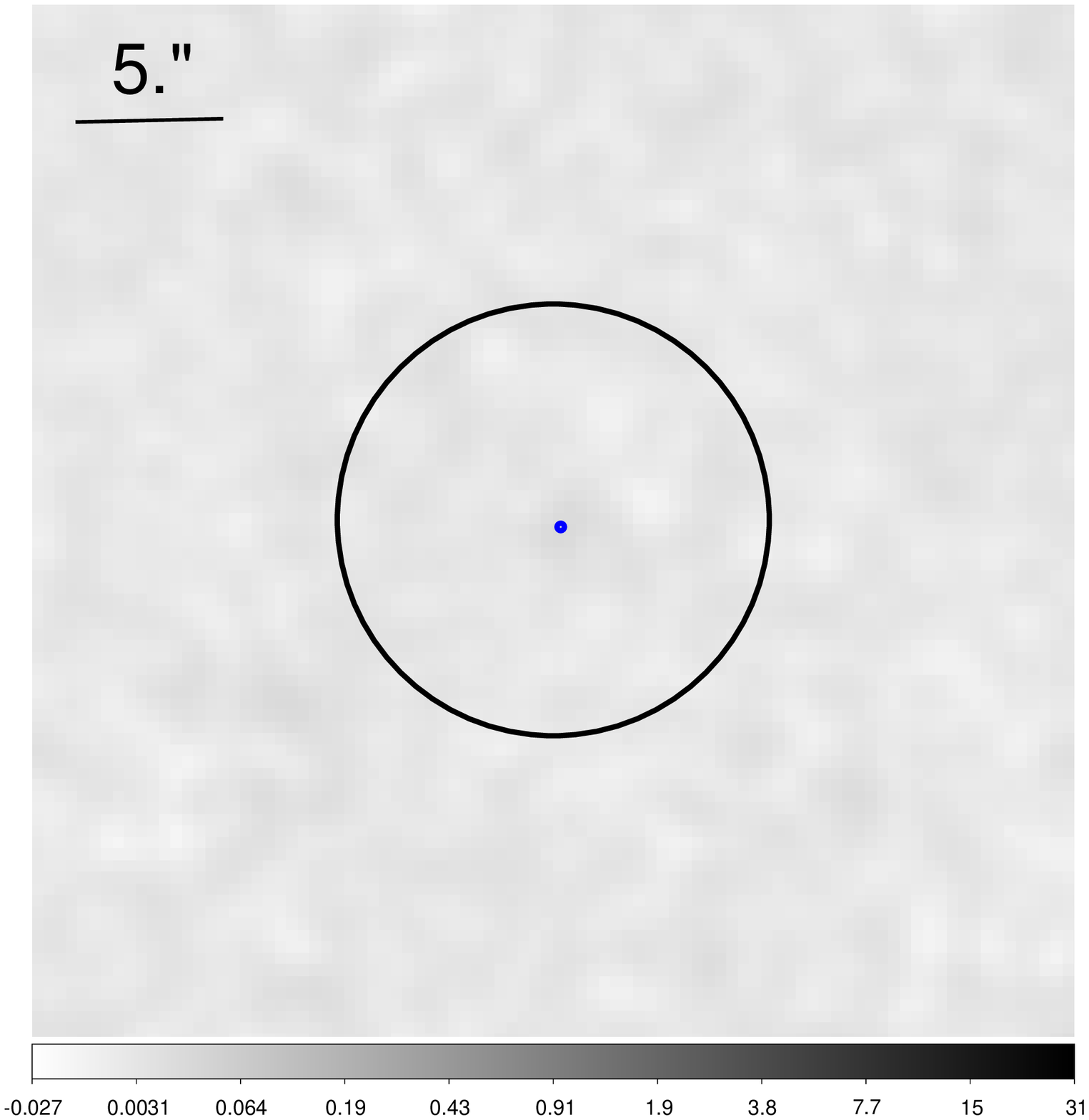}}\\
%\caption{\scriptsize{Src No.84}}
  \subfloat[Src No.85]{\includegraphics[clip, trim={0.0cm 2.cm 0.cm 0.0cm},width=0.19\textwidth]{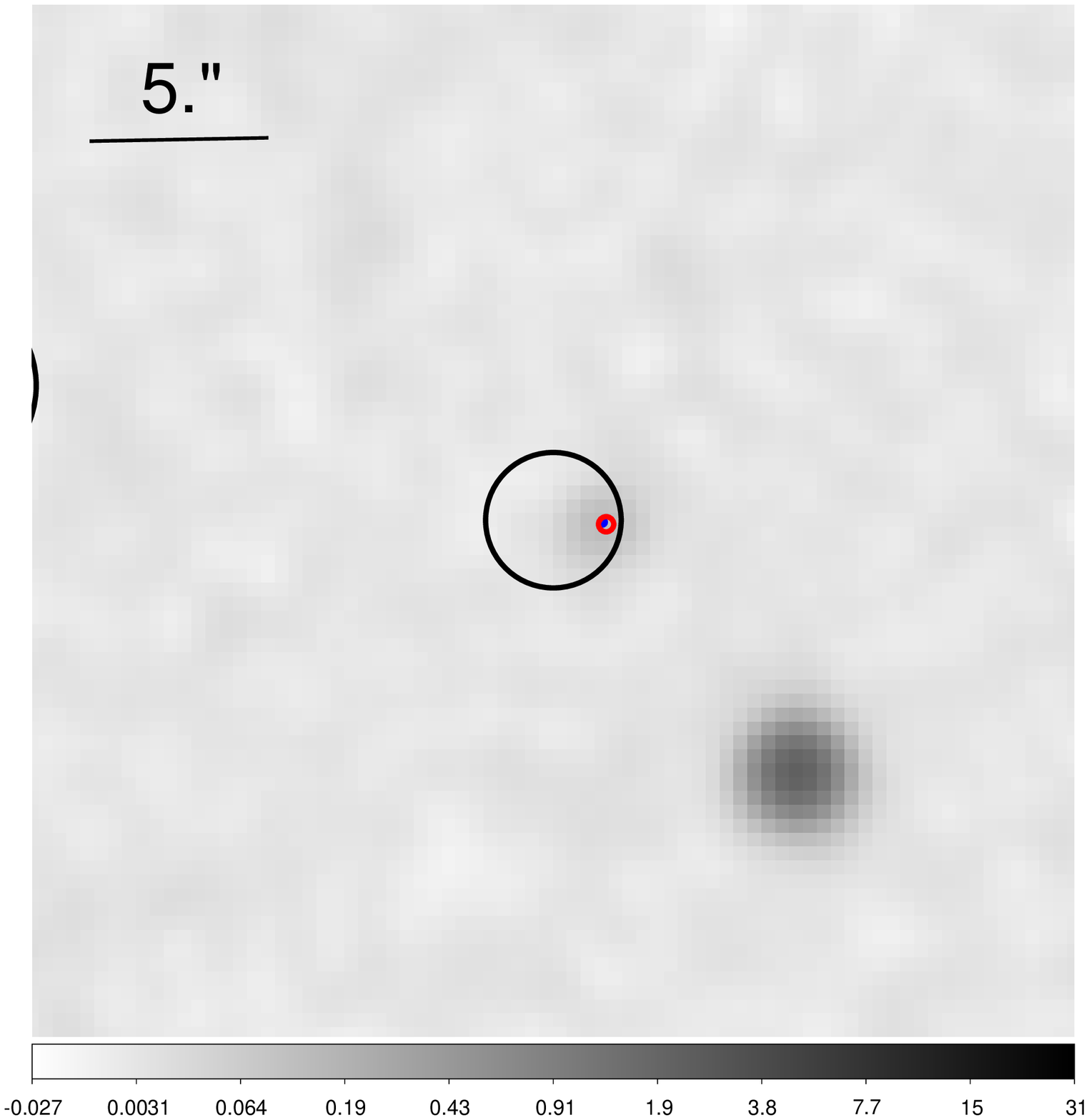}}
%\caption{\scriptsize{Src No.85}}
  \subfloat[Src No.86]{\includegraphics[clip, trim={0.0cm 2.cm 0.cm 0.0cm},width=0.19\textwidth]{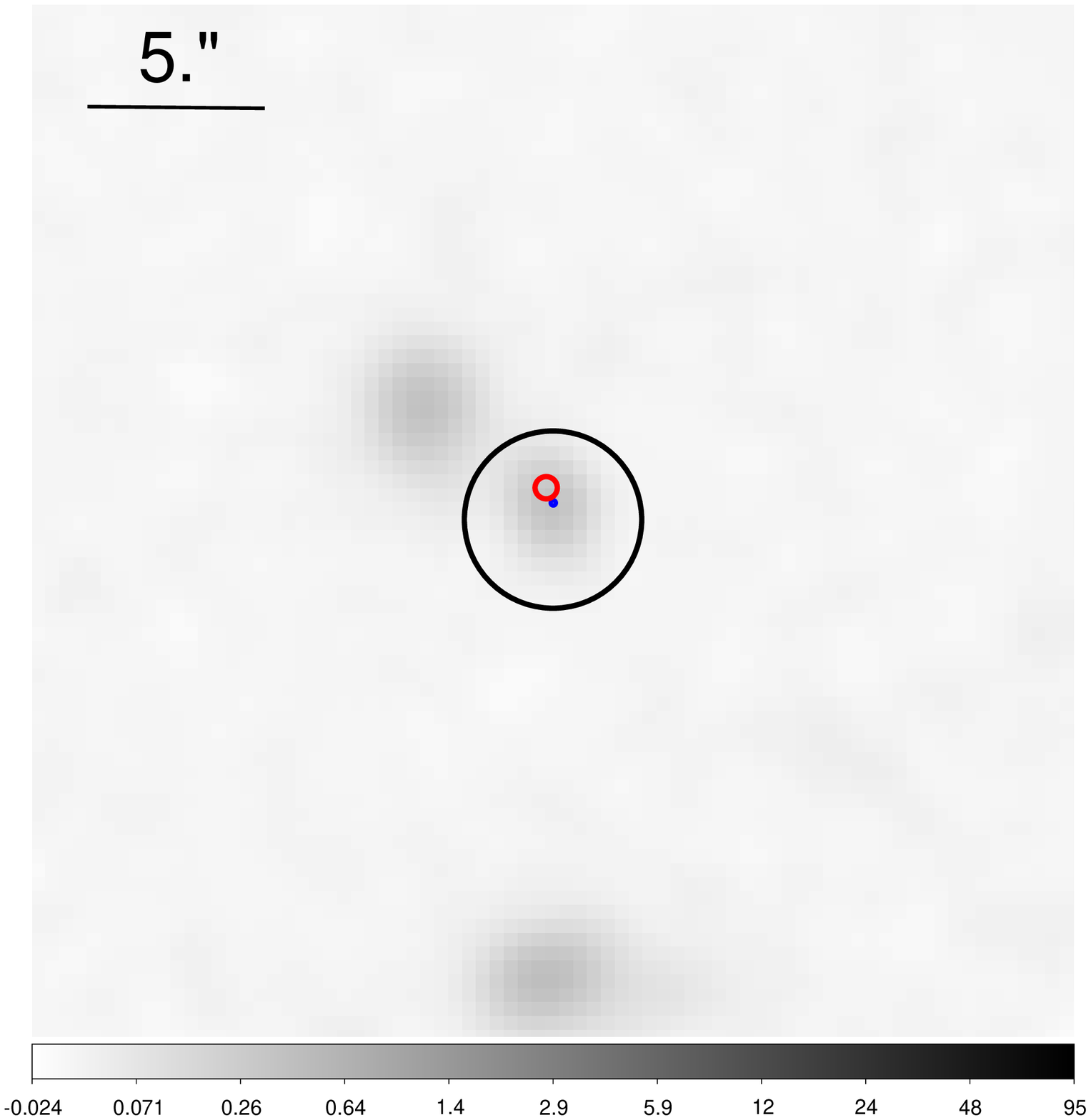}}
%\caption{\scriptsize{Src No.86}}
  \subfloat[Src No.87]{\includegraphics[clip, trim={0.0cm 2.cm 0.cm 0.0cm},width=0.19\textwidth]{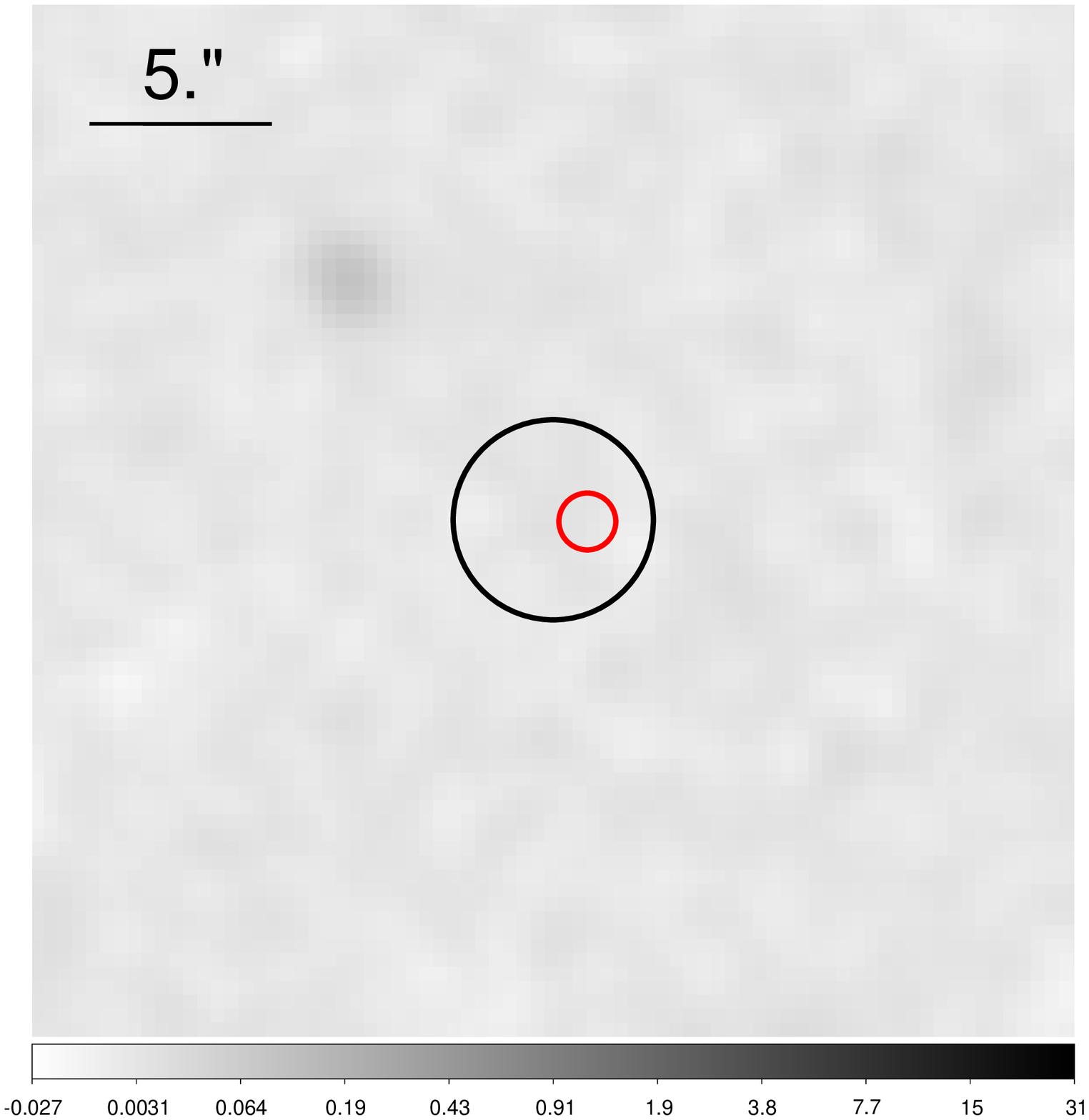}}
%\caption{\scriptsize{Src No.87}}
  \subfloat[Src No.88]{\includegraphics[clip, trim={0.0cm 2.cm 0.cm 0.0cm},width=0.19\textwidth]{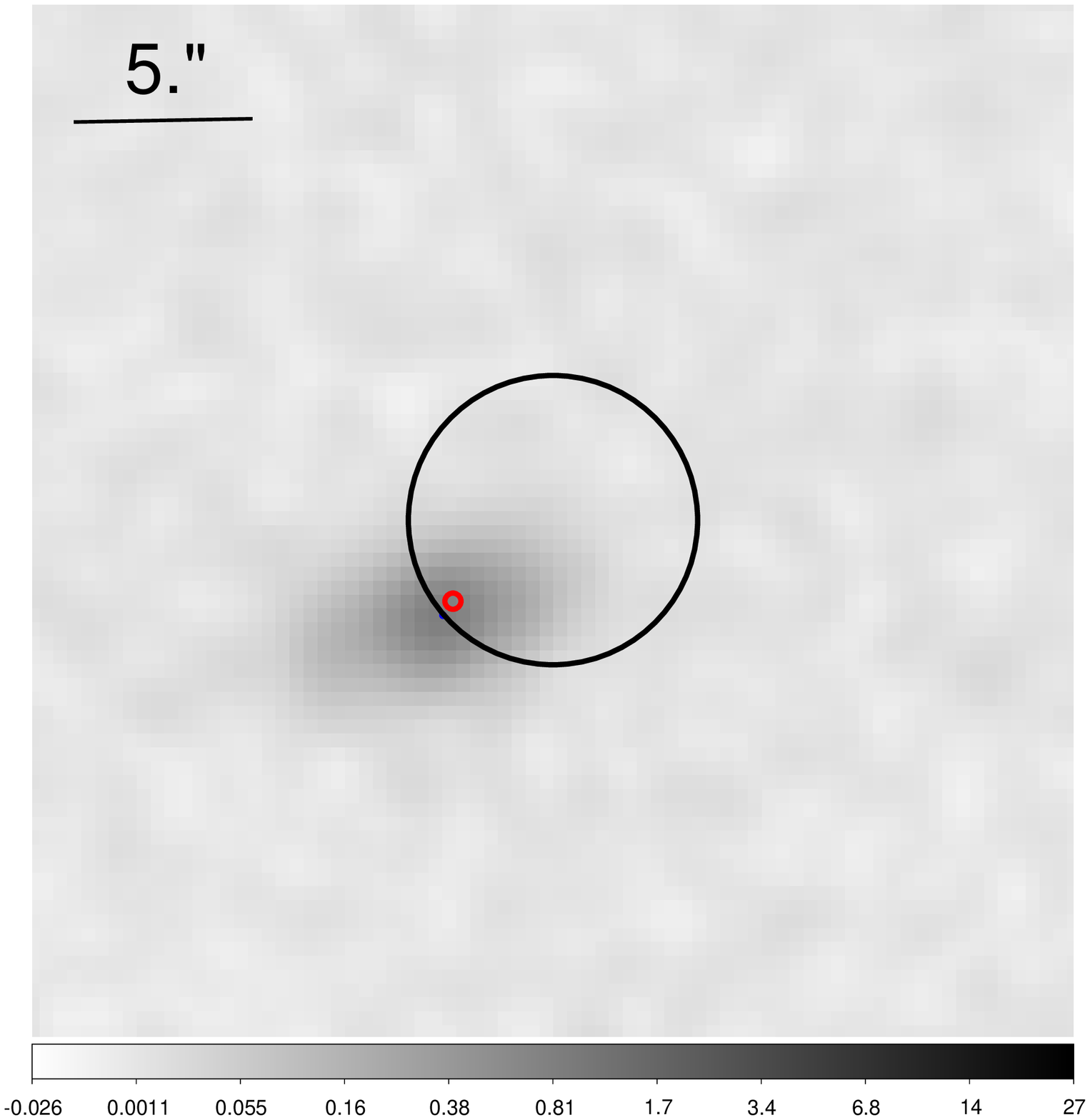}}\\
%\caption{\scriptsize{Src No.88}}
\subfloat[Src No.89]{\includegraphics[clip, trim={0.0cm 2.cm 0.cm 0.0cm},width=0.19\textwidth]{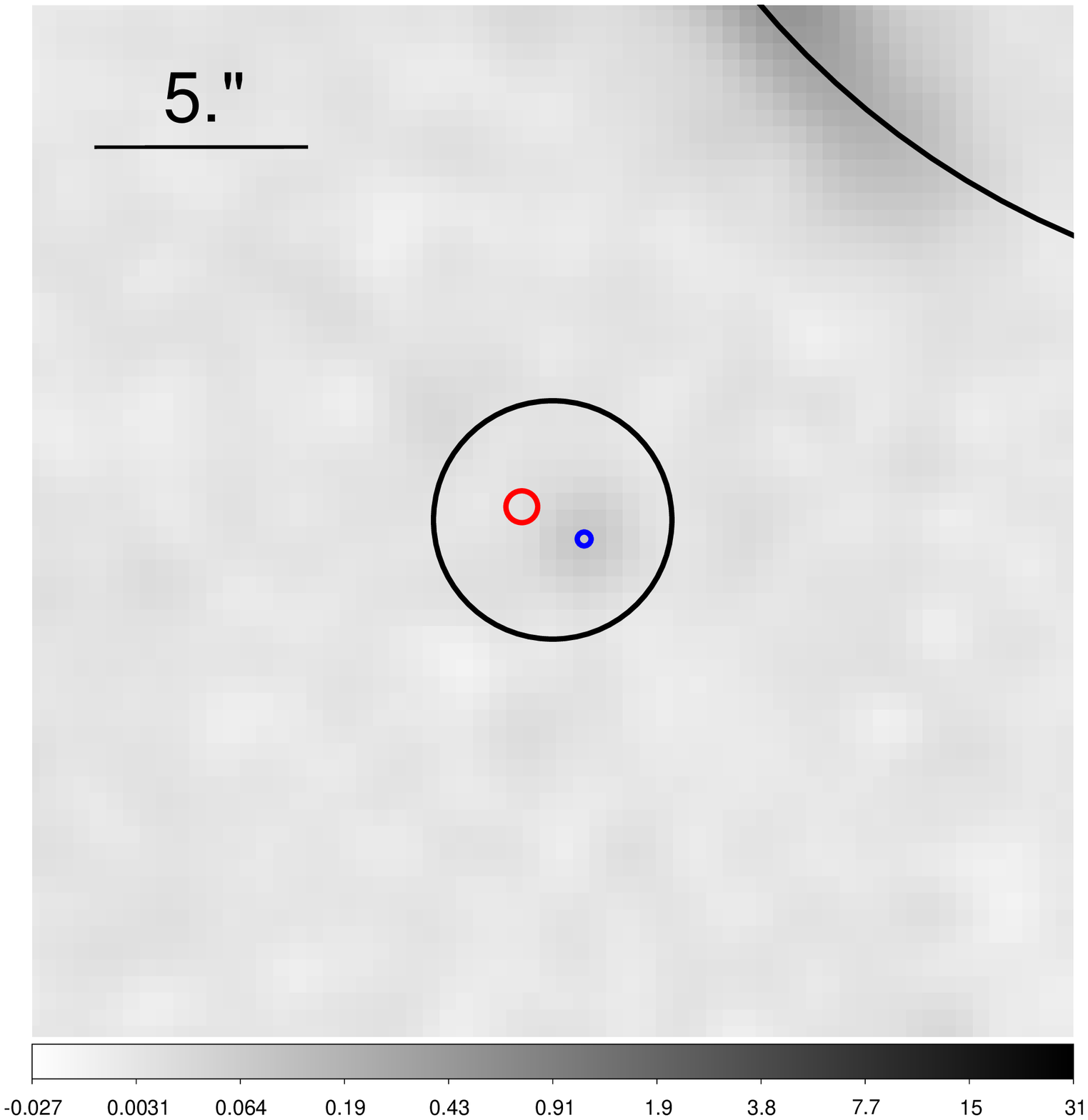}}
%\caption{\scriptsize{Src No.89}}
  \subfloat[Src No.90]{\includegraphics[clip, trim={0.0cm 2.cm 0.cm 0.0cm},width=0.19\textwidth]{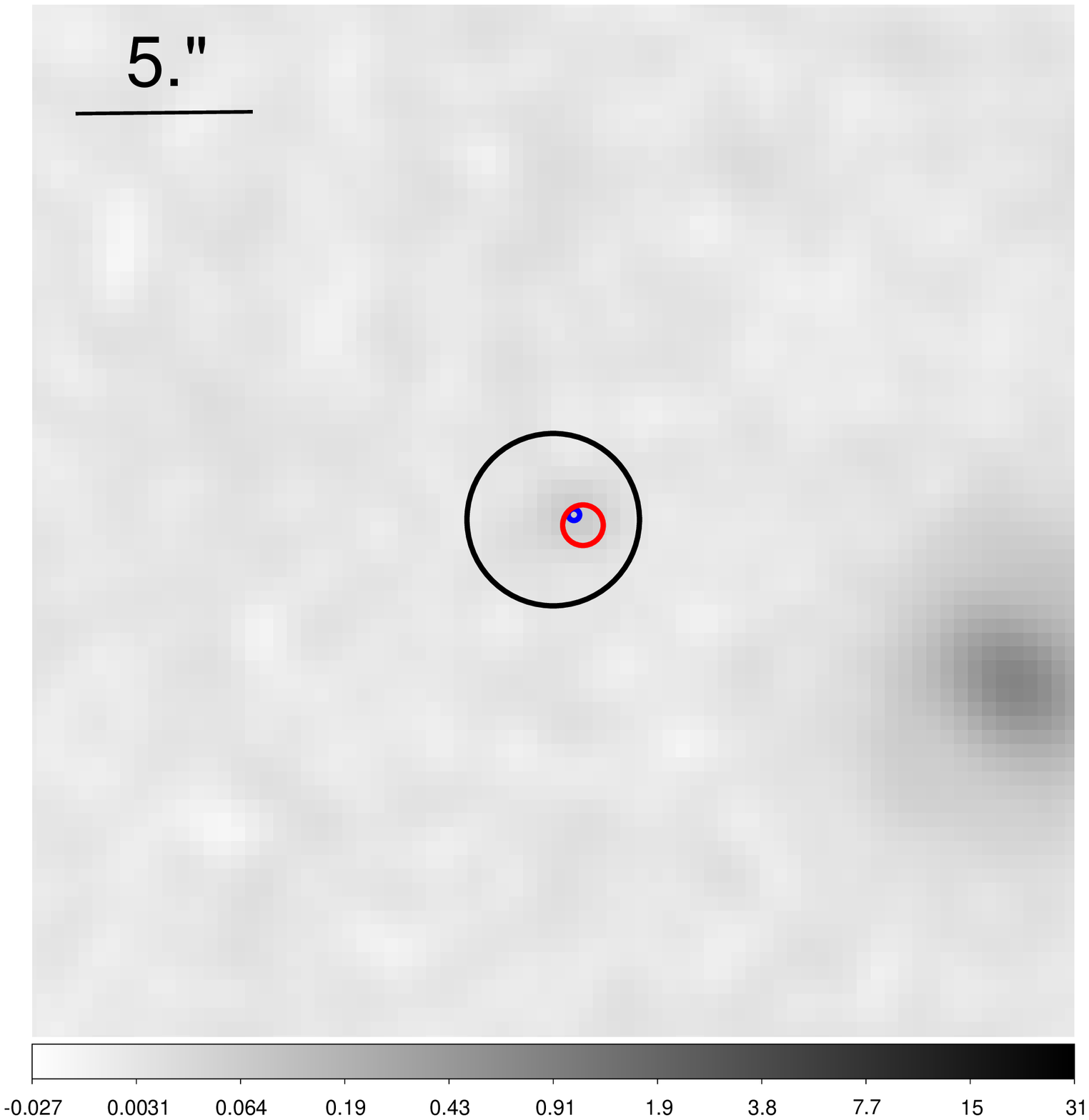}}
%\caption{\scriptsize{Src No.90}}
  \subfloat[Src No.91]{\includegraphics[clip, trim={0.0cm 2.cm 0.cm 0.0cm},width=0.19\textwidth]{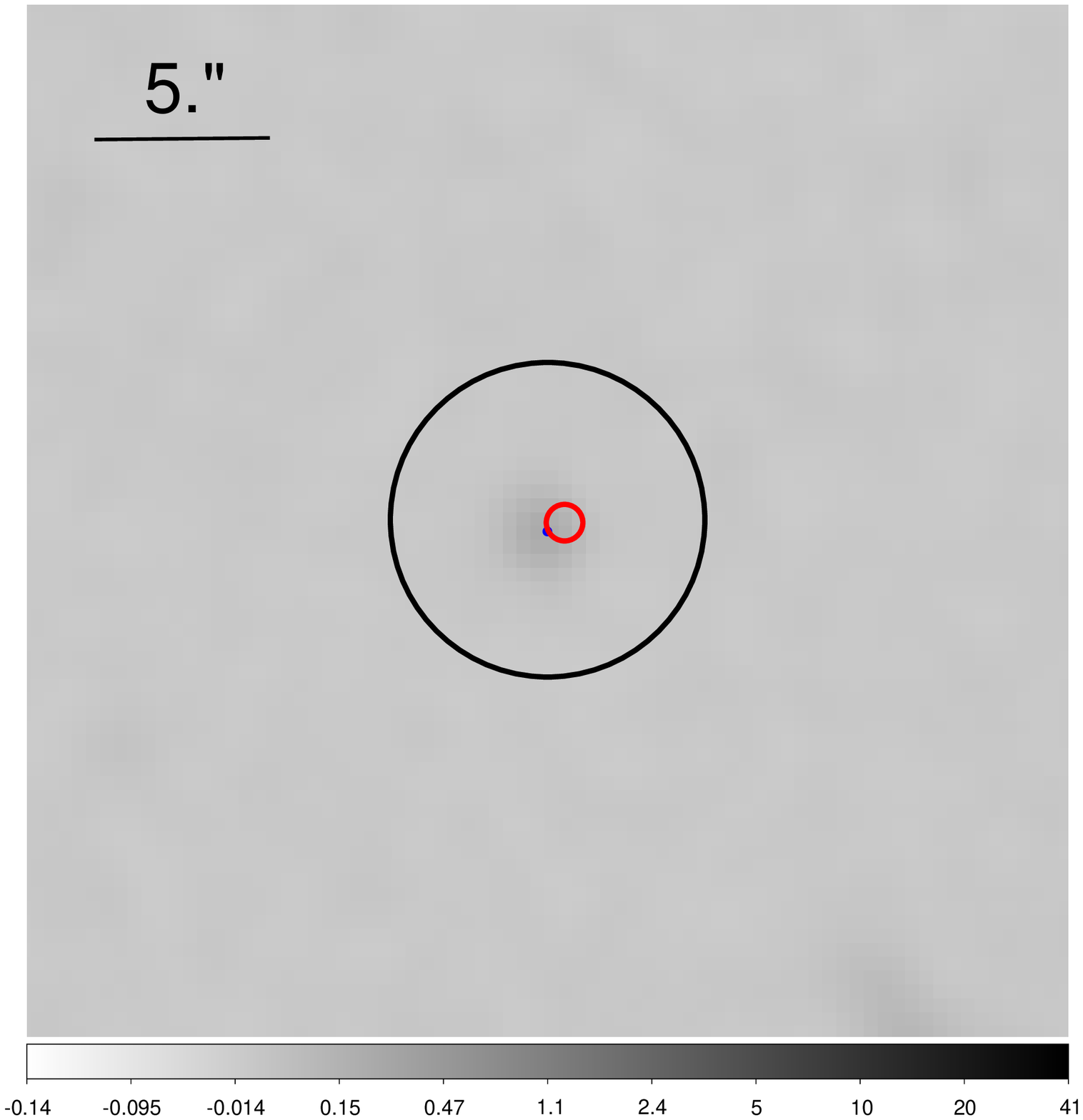}}
%\caption{\scriptsize{Src No.91}}
  \subfloat[Src No.92]{\includegraphics[clip, trim={0.0cm 2.cm 0.cm 0.0cm},width=0.19\textwidth]{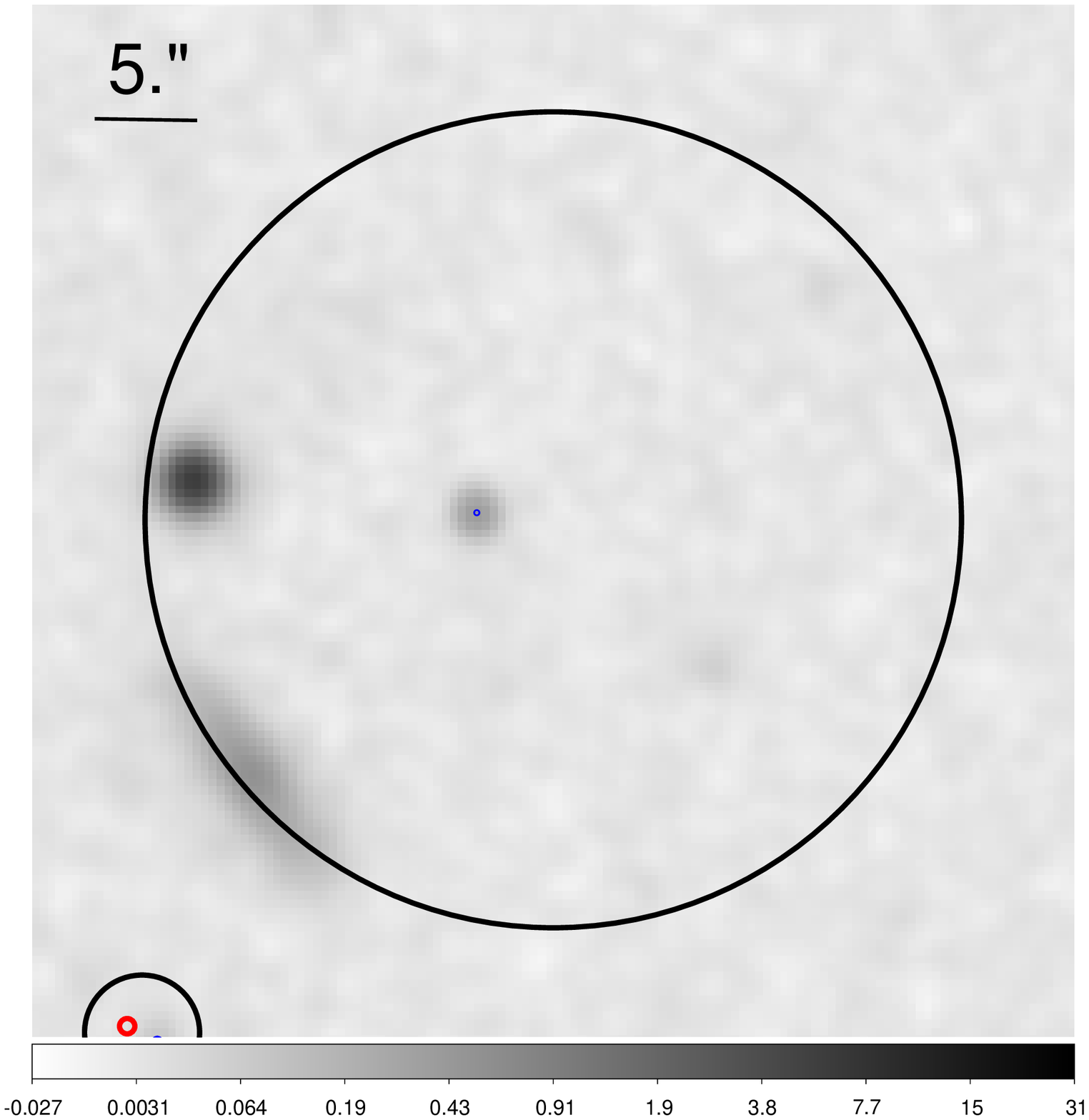}}\\
%\caption{\scriptsize{Src No.92}}
\end{figure*}
\pagebreak
\clearpage
%\hspace{0.3cm}Appendix B continued: Image of optical SDSS9 counterparts
\begin{figure*}
\vspace{-0.5cm}

  \subfloat[Src No.93]{\includegraphics[clip, trim={0.0cm 2.cm 0.cm 0.0cm},width=0.19\textwidth]{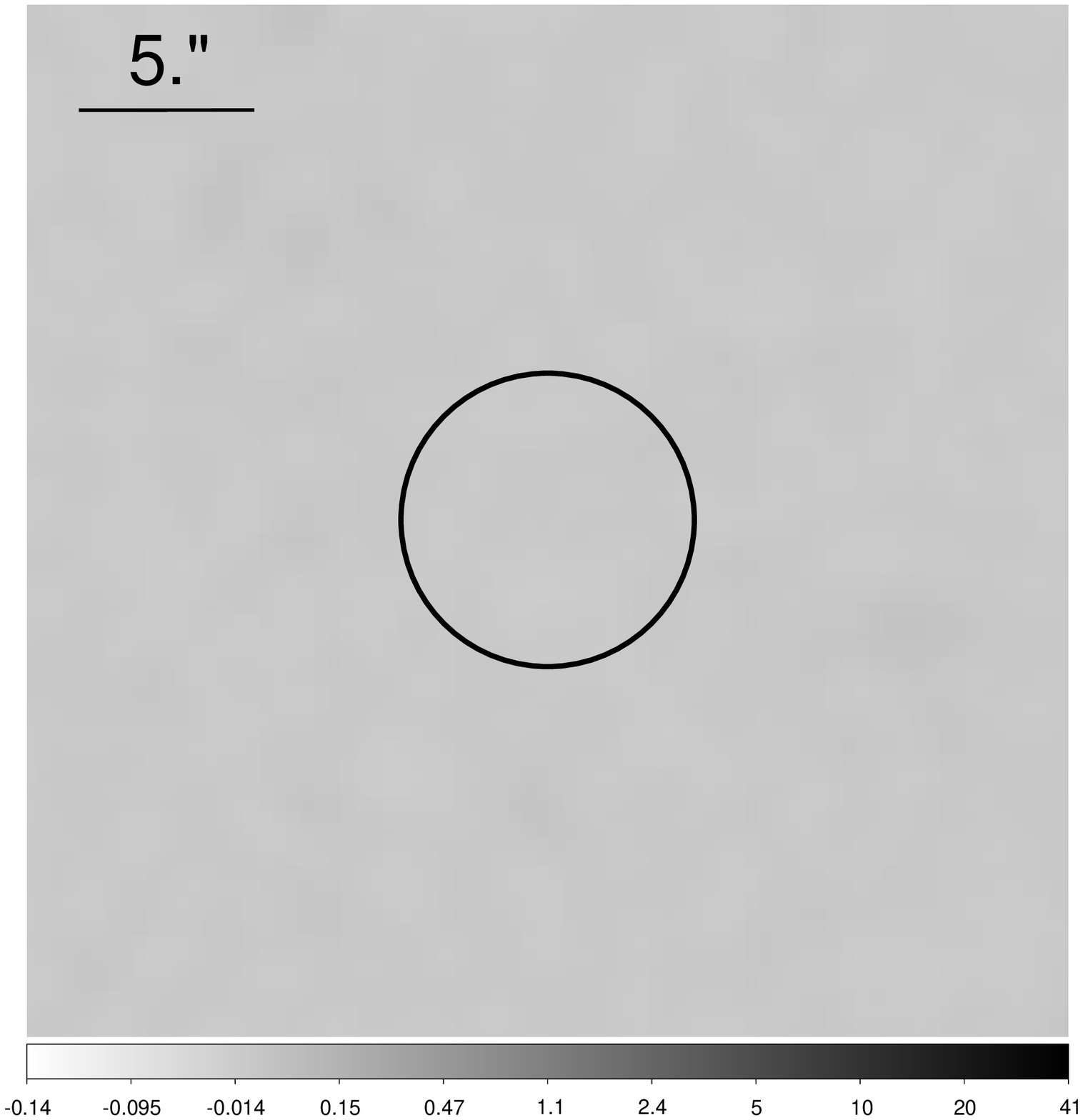}}
%\caption{\scriptsize{Src No.93}}
  \subfloat[Src No.94]{\includegraphics[clip, trim={0.0cm 2.cm 0.cm 0.0cm},width=0.19\textwidth]{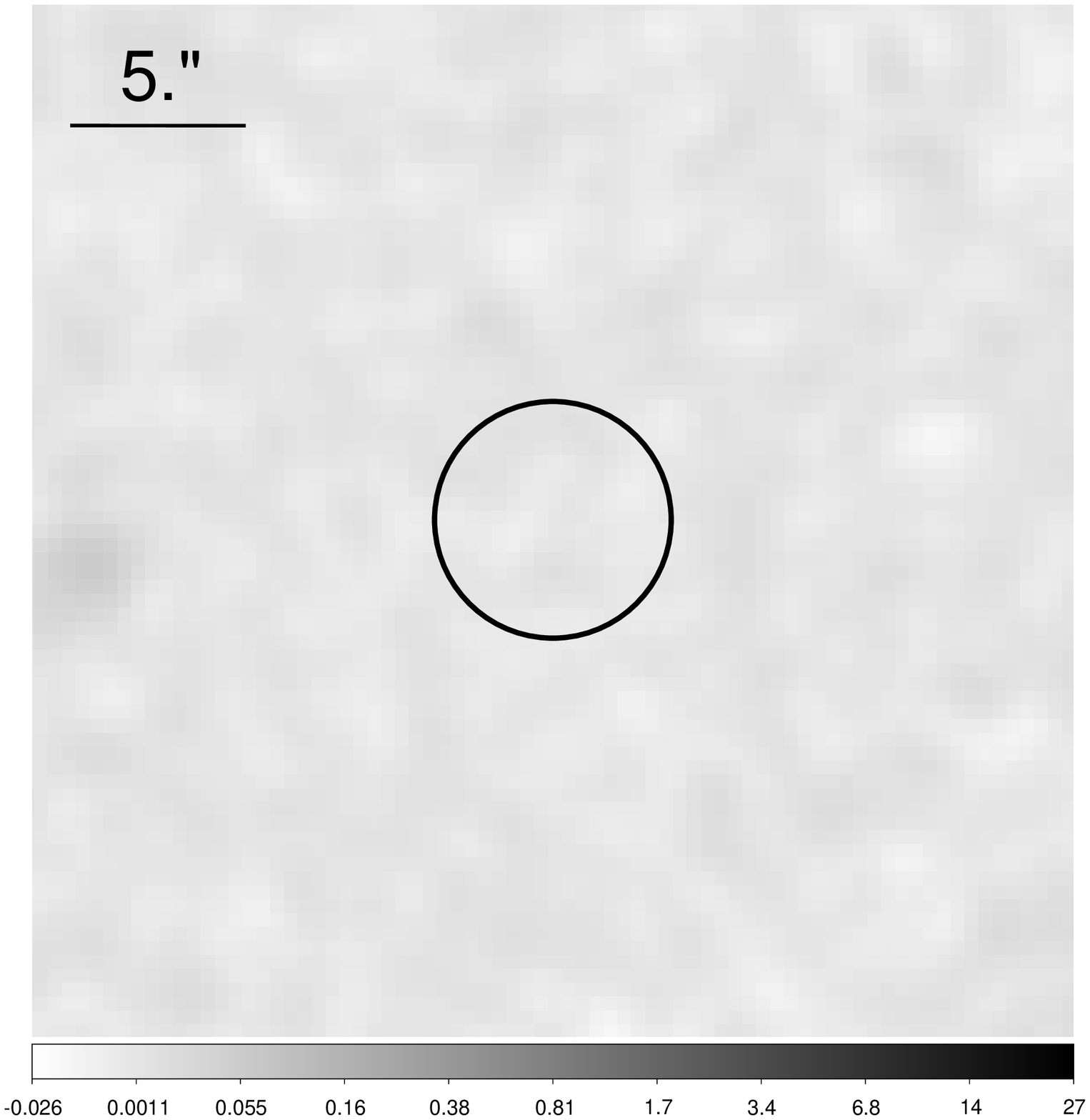}}
%\caption{\scriptsize{Src No.94}}
  \subfloat[Src No.95]{\includegraphics[clip, trim={0.0cm 2.cm 0.cm 0.0cm},width=0.19\textwidth]{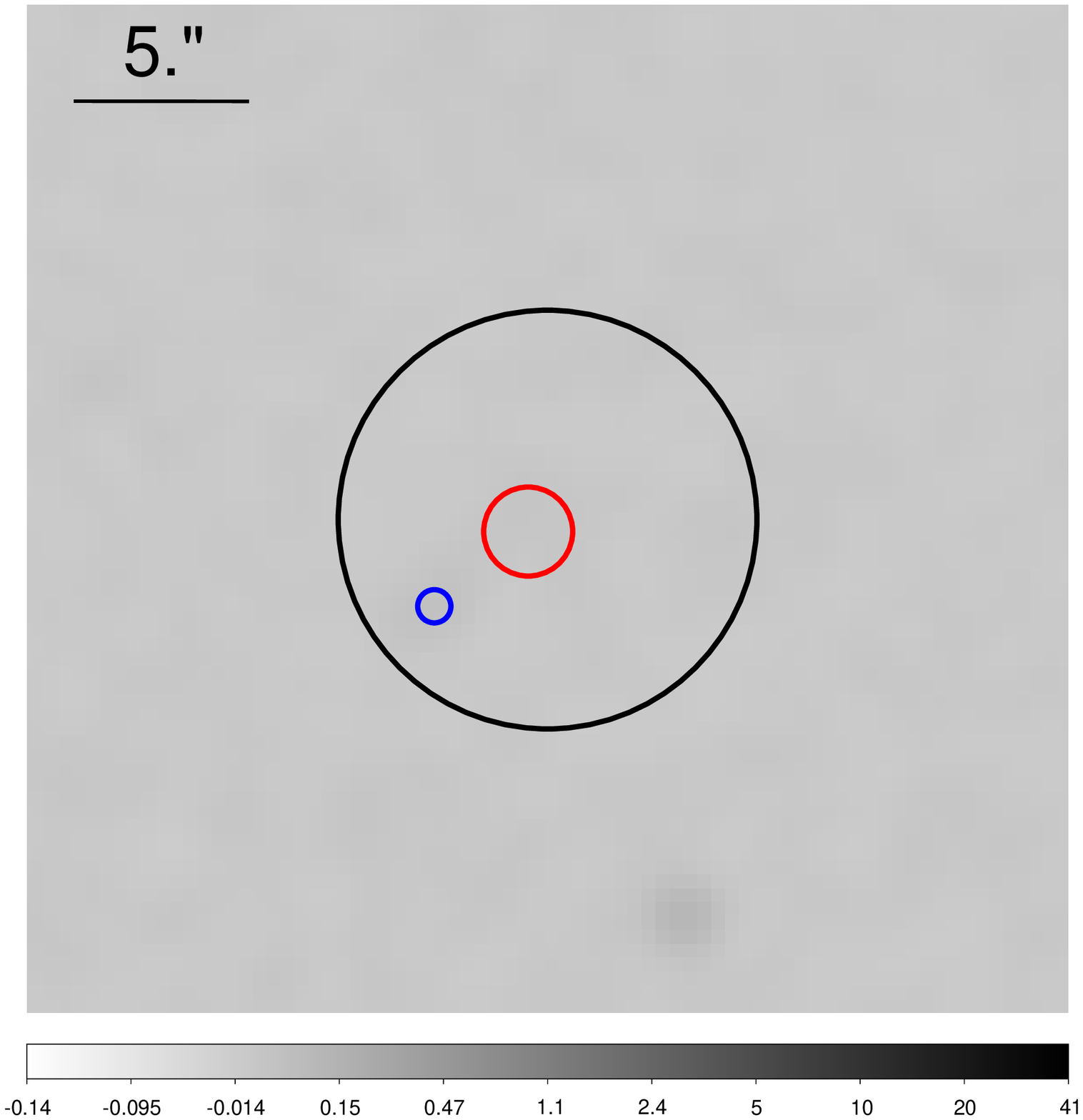}}
%\caption{\scriptsize{Src No.95}}
  \subfloat[Src No.96]{\includegraphics[clip, trim={0.0cm 2.cm 0.cm 0.0cm},width=0.19\textwidth]{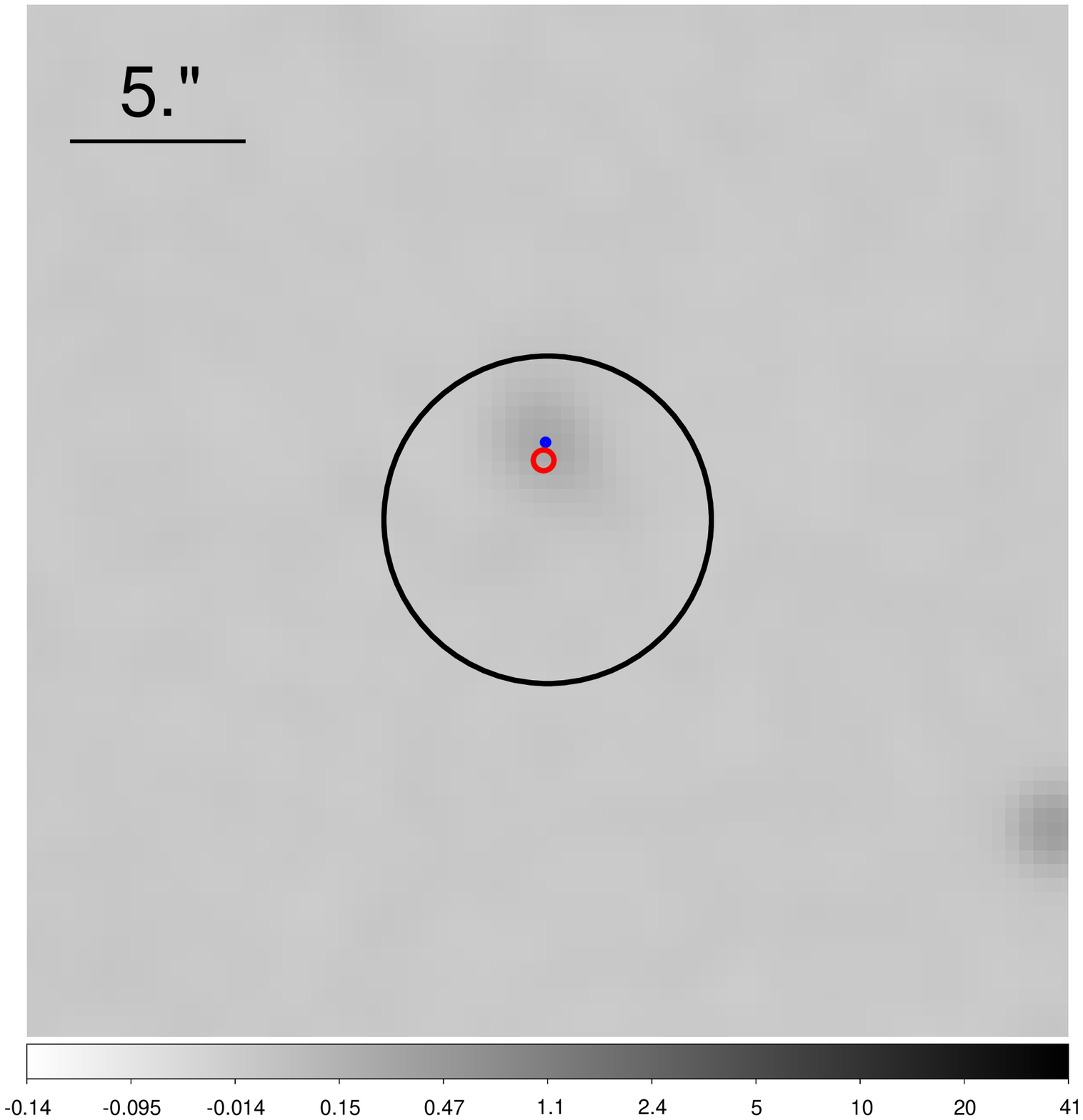}}\\
%\caption{\scriptsize{Src No.96}}
  \subfloat[Src No.97]{\includegraphics[clip, trim={0.0cm 2.cm 0.cm 0.0cm},width=0.19\textwidth]{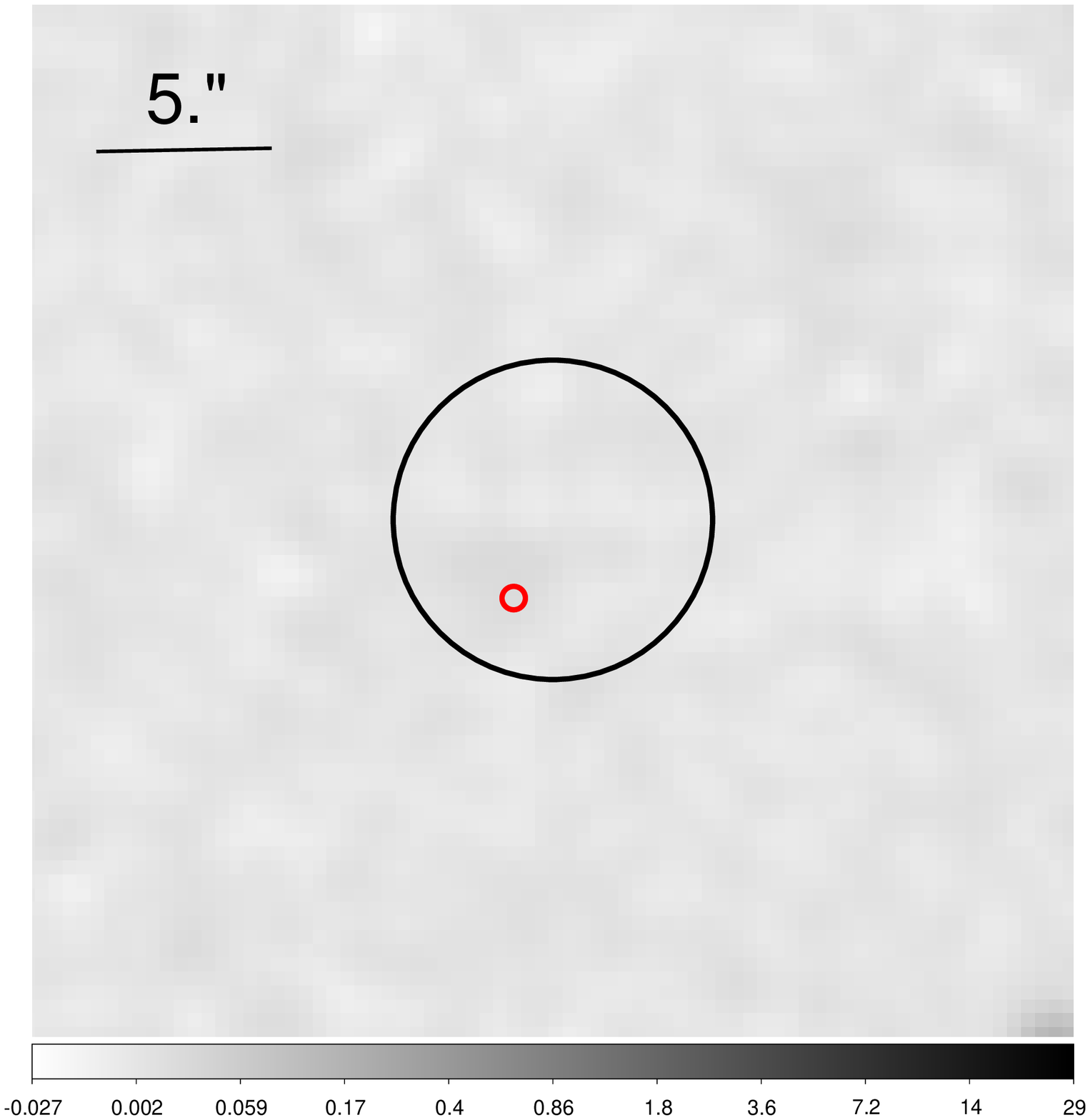}}
%\caption{\scriptsize{Src No.97}}
\end{figure*}


%%%%%%%%%%%%%%%%%%%%%%%%%%%%%%%%%%%%%%%%%%%%%%%%%%

% Don't change these lines
\bsp	% typesetting comment
\label{lastpage}
\end{document}